\documentclass[usegraphicx,useAMS,usenatbib, a4paper]{mn2e}
\usepackage{times}
\usepackage{amssymb}
\usepackage{amsmath}
\usepackage{graphicx}
\usepackage{euscript}
\usepackage{rotating}
\usepackage{epsfig}
\usepackage{url}
\def\ergps{{\rm\thinspace erg~s^{-1}}}

\def\kpc{{\rm\thinspace kpc}}
\def\km{\hbox{$\rm\thinspace km$}}

\def\kev{{\rm\thinspace keV}}

\def\Ms{{\rm\thinspace Ms}}
\def\pcm2{{\rm\thinspace cm^{-2}}}

\def\gx339{GX~339-4}
\def\pca{{\it PCA}}
\def\hexte{{\it HEXTE}}
\def\rxte{{\it RXTE}}

\newcommand{\eg}{e.g.\thinspace}

\begin{document}

\title{A Global Study of the Behaviour of Black Hole X-ray Binary Discs}
\author[Dunn, Fender, K\"ording, Belloni \& Merloni]% \& Cabanac]
{\parbox[]{6.in} {R.~J.~H.~Dunn$^{1}$\thanks{E-mail:
      robert.dunn@ph.tum.de}\thanks{Alexander von Humboldt Fellow}, R.~P.~Fender$^2$,
    E.~G.~K\"ording$^{3}$, T.~Belloni$^{4}$ and A. Merloni$^{1,5}$\\
    \footnotesize
    $^1$Excellence Cluster Universe, Technische Universit\"at M\"unchen, Garching, 85748, Germany\\
    $^2$School of Physics and Astronomy, Southampton, University of Southampton, SO17
    1BJ, UK,\\
    $^3$AIM - Unit\'e Mixte de Recherche CEA - CNRS - Universit\'e
    Paris VII - UMR 7158, CEA-Saclay, Service d'Astrophysique, F-91191
    Gif-sur-Yvette Cedex, France\\
    $^4$INAF-Osservatorio Astronomico di Brera, Via E. Bianchi 46, I-23807 Merate (LC), Italy\\ 
    $^5$Max Planck Institut f\"ur Extraterrestrische Physik,
Giessenbachstra\ss e, 85471 Garching, Germany\\}}

\maketitle

\begin{abstract}
We investigate the behaviour of the accretion discs in
the outbursts of the low-mass black-hole X-ray binaries (BHXRB), an
overview of which we have presented previously.  Almost all of the
systems in which there are 
sufficient observations in the most disc dominated states show a 
variation of the disc luminosity with temperature close to
$L\,\tilde\propto\,T^4$.  This in turn implies that in these states,
the disc radius, $R_{\rm in}$, and 
the colour correction factor, $f_{\rm col}$, are almost constant.
Deviations away from the $T^4$ law are observed at the
beginning and end of the most disc dominated states, during the
intermediate states.  Although these could be explained by an inward
motion of the accretion disc, they are more likely to be the result of
an increase in the value of $f_{\rm col}$ as the
disc fraction decreases.  By comparing the expected and observed disc
luminosities, we place approximate limits on the allowed distances and masses of the
BHXRB system.  In a number of cases, the measured distances
and masses of the BHXRB system indicate that it is possible that the
black hole may be spinning.
\end{abstract}


\section{Introduction}\label{sec:intro}

The outbursts of black hole X-ray binaries (BHXRBs) are dramatic and intriguing
events.  They have the potential for allowing the study of the
physical and emission processes close to the event horizon.  The
accretion process and associated intermittent jet-production results
in emission across the electromagnetic spectrum.  In the study
presented here, we focus on the X-rays, as these arise from the inner
parts of the accretion disc and flow.  At the other end of the
spectrum, the radio emission is thought to arise from a synchrotron
emitting jet.  Therefore the radio emission is a good tracer of
whether a jet is active or not, and the X-rays are good at determining
the state of the accretion flow.

In the now commonly-accepted picture of the
changes that occur in the BHXRB system, the BHXRB spends most of its
time in a quiescent state.  There the total luminosity of the BHXRB is
very low, in all bands.  As the outburst starts, the X-rays are characterised by a
hard emission spectrum with a powerlaw slope of $\Gamma \sim
1.5$ - the ``hard state''.  As the X-ray luminosity rises, the radio
luminosity rises in 
step \citep{Corbel00, Corbel03}.  The radio spectrum also indicates
the presence of a steady jet emitting synchrotron radiation.  As the
outburst progresses, the disc spectrum becomes
increasingly dominant, eventually softening
the entire X-ray spectrum as the BHXRB enters the ``soft-state''.
This transition is very fast compared to the speed at which the
luminosity rose.  Over the course of weeks or even months, the disc luminosity and
temperature decay \citep{Gierlinski04}, as the disc dominance
decreases \citep{Dunn10}.  Eventually the source returns to the ``hard-state'' and
the luminosity continues to fade.  For further details on this picture
of the progression of BHXRB outbursts see \citet{Fender04, Done03, Homan05,
  Remillard06, Done07, Belloni10}.  

The accretion onto the compact object is what drives
their luminosity in X-rays, which arise from the disc, the corona and
even the jet \citep{Russell10}.  Studying emission from the disc allows the accretion
process, and also the behaviour of the material within the disc as it
approaches close to the compact object to be investigated.  The standard theory of
accretion discs from \citet{Shakura73} shows that the accreting
material will form a geometrically thin, but optically thick disc, the
inner extent of which depends on the spin of the black hole.

Previous studies on the behaviours of the BHXRB accretion discs have
selected those observations where the disc was dominant, in order to
ensure that the disc parameters were well determined
\citep{Gierlinski04, Done07}.  This allowed a detailed study of the
disc emission from well characterised BHXRBs to be carried out.  Using
the full archival coverage of the \rxte\footnote{Rossi X-ray Timing Explorer} satellite, which has been
observing BHXRBs for 13 years, we 
present a study of {\it all} observations in which a disc was detected using
the analysis of \citet{Dunn10}.  This allows us to investigate the
properties of the disc in the transition periods, between the fully
disc-dominated and powerlaw-dominated states, as well as in the disc
dominated states.  

In Sections
\ref{sec:DR} and \ref{sec:obj}
we recap the data reduction proceedure presented in \citet{Dunn08,
  Dunn10} and the final BHXRBs which were selected for this study.
The behaviour of the disc's temperature and luminosity are discussed
in Section \ref{sec:L-T}.  The deviations from the expected behaviour
of the disc luminosity and temperature are presented in Section
\ref{sec:discinner}, where the inner radius of the disc is
investigated, and Section \ref{sec:spurs}, where larger departures are
linked to the colour temperature correction.  In Section
\ref{sec:Dists} we investigate the limits which can be placed on the
distances, masses and spins of the BHXRBs from the observations.  The
degeneracy of the broken powerlaw model with the disc model, and the
evolution of the powerlaw in the hard state are presented in Sections
\ref{sec:bpl} and \ref{sec:PLevol}.  

\section{Data Reduction}\label{sec:DR}

In the analysis of the disc properties of the sample of
BHXRBs presented in this work we use the analysis of \rxte\ data
detailed in \citet{Dunn10}.  We recap the main points, but refer to
\citet{Dunn08, Dunn10} for more details.

We use all the available data publicly available in the
\rxte\ archive\footnote{The cut-off date used was the 4 August 2009,
  as in \citet{Dunn10}.}.
This gave a baseline of around 13 years to study the 
evolution of the disc properties during the numerous
outbursts observed within that time.  All data were subjected to the
same data reduction procedure, in order to minimise differences
arising from different data reduction routines.

Both the Proportional Counter Array (\pca) and High Energy X-ray
Timing Experiment (\hexte) data were required when fitting the
spectra, as the \hexte\ data allows the powerlaw to be constrained at
high energy when the \pca\ data are dominated by the disc.  We
followed the procedure outlined in the
\rxte\ Cookbook\footnote{\url{http://rxte.gsfc.nasa.gov/docs/xte/recipes/cook_book.html}} 
using the tools from
HEASOFT\footnote{\url{http://heasarc.gsfc.nasa.gov/lheasoft/}} version
6.6.2. 

To reduce variations the between observations further, we only use the data
from Proportional Counter Unit (PCU) 2 on the \pca\ as this has been
on throughout the 
\rxte\ mission.  Our analysis concentrates on the bright periods when
the BHXRBs are in outburst, and so we use the bright model background
for all data.  Lower count rates are more likely in the inter-outburst
periods, and so this choice of a single background is unlikely to bias
our results.

In order to proceed with the spectral fitting, we require a \pca\ observation with at least 1000
background subtracted counts, and a \hexte\ observation with either
Cluster A or B (or both) with at least 2000 background subtracted
counts.  The other \hexte\ cluster has to have at least a positive
number of counts\footnote{The background subtraction procedure for
  \hexte\ can result in negative numbers of foreground counts for low
  fluxes.}  This count restriction is in place to try to ensure that the
spectra which are fitted are of good quality and fit within a
reasonable time with well-constrained parameters.

The spectra were fitted in {\scshape xspec} (v12.5.0an).  In order to
study the disc parameters in detail, we needed to analyse the spectra
to the lowest energies possible.  The relation between the channel
numbers of the \pca\ instrument and the energies they correspond to
has drifted over the 13 years of the mission.  However, all channels
below number 7 are not well calibrated for spectral analysis.  We therefore choose to ignore
\pca\ channels $\leq 6$, which corresponds to around $3\kev$, but the
exact energy has drifted over time (see the \rxte\ documentation).  We also
ignore \pca\ data $>25\kev$, and \hexte\ data $<25\kev$ and $>250\kev$.

To be able to characterise the state of the BHXRBs as they go through an outburst
we fit three
types of base model - unbroken powerlaw ({\scshape power}, PL), broken
powerlaw ({\scshape bknpower}, BPL) and powerlaw + disc ({\scshape
  power + diskbb}, DPL).  These allow the study of
the non-thermal component using the {\scshape power/bknpower} parameters, and the disc
using the {\scshape diskbb} parameters.  To study the presence and
change in the iron line we add an optional $6.4\kev$ gaussian feature
to all these spectra, giving in total six models which were fitted.  The
low energy sensitivity of \rxte\ is insufficient to allow the $N_{\rm
  H}$ to be determined from the spectra, and so we fix this value to
the accepted value for each BHXRB (see Table \ref{tab:obj}).

From the six fitted models, we select the best fitting one on $\chi^2$
terms.  However if this is not the simplest model, we then determine
whether the increase in complexity of the model is significant using an
$F$-test with $\mathcal{P}<0.001$ as the significance level.  For the
complete routine see \citet{Dunn10}, but a quick outline is described
below.  When the best fitting model is
complex but contains no 
gaussian component, we test this best fitting model against the simple
powerlaw result.  If the best fitting model is complex and contains a
gaussian component we first test whether the underlying complex
continuum model is an improvement over the simple powerlaw, and if it
is we test whether a line is required in this complex model.  When the
complex continuum is not an improvement over the simple powerlaw a
number of further steps are performed, as detailed in \citet{Dunn10}.

Once the best fitting model has been selected, we further cut the
observation number by removing any observation whose $3-10\kev$ flux is less
than $1\times 10^{-11} \ergps$, where the flux was not well
determined or where the powerlaw was not well constrained (even
if the disc was).  The flux cut was performed to focus on
the periods in which the BHXRBs are in outburst, and so streamline the
data reduction process.  We also removed those fits whose $\chi^2>5.0$ as
these are spectra which are not well fit by any of the models
available within our automated procedure.  The distribution of the
$\chi^2$ of the best fitting models is shown in \citet{Dunn10}
Fig. 2.  The majority of fits are clustered around $\chi^2\sim 1$, but there
is a large tail to higher values.  As the spectral fitting in this work
has been automated, such large tail is expected.

\subsection{Model fitting issues}\label{sec:model}

The relatively high lower energy bound for the \rxte\ response limits
our ability to detect discs when they are not dominant.  The maximum
power emitted by the {\scshape diskbb} model occurs around $2.4
kT_{\rm Disc}$
which is usually around the lower limit of the \pca\ bandpass (for
discs at $\sim 1 \kev$).  Therefore we
rarely detect the peak of the disc emission, and more usually observe
the Wien tail.  Using a
simple powerlaw to model the non-thermal continuum, even when
including \hexte\ data, does not allow for small breaks or curvature
within this component.  If a disc component was
included in these observations, it was found to try and fit these small curvatures in the
powerlaw rather than any true underlying disc component, resulting in
unphysical disc parameters.  We therefore
limited the minimum temperature for the disc during the fitting to
$k_{\rm B}T=0.1\kev$.  Furthermore we then penalise the $\chi^2$ of
any model which has a $k_{\rm B}T <0.4\kev$ when selecting the best
fitting model.  We note that in doing this we are limiting our
sensitivity to low temperature discs, in the intermediate and hard
states for example, and are probably excluding a few accurate disc
fits.  We investigate further degeneracies between the disc and broken
powerlaw models in Section \ref{sec:bpl}.

More complex models, for example Comptonization, would in principle
give more information on the state of the system in these disc
dominated states, as it links the non-thermal emission to the disc
temperature.  However, in order to freely fit all the parameters of
the Comptonization models a high signal-to-noise observation is
required.  Not all of our observations have sufficient counts to be
able to do this; in fact very few would allow all parameters to be
determined from the observations.  Although fixing some parameters
would allow these models to fit successfully, this goes against the
methodology of this work, by {\em a priori} constraining parameters differently for
different states.

\section{Selected Objects}\label{sec:obj}

After all the data reduction, dead-time and selection the $15\Ms$ of raw
\rxte\ data was trimmed to $\sim 10\Ms$ in 3919 observations, with
well fitted spectra and high enough fluxes and counts.  The sample of
objects was not designed to be complete in any way.  We selected
objects which were well known BHXRBs in the literature as well as
those which were known to have outbursts which had been well monitored
by \rxte.  The set of BHXRBs analysed in this sample, along with their
physical parameters (where known) and the final number of observations
used in this study are shown in Tables \ref{tab:obj} and
\ref{tab:obstimes}.  

There are two notable BHXRBs which were
purposely not included in this study (\eg Cyg X1 and GRS 1915-105).
These two sources were not included for a number of reasons.  One was
a purely practical one resulting from the shear amount of data
available for these sources.  The reduction of all the observations in
the scheme outlined above would have dominated any of the global
studies presented both here and in \citet{Dunn10}, and selecting
certain parts would have gone against the philosophy of the study,
by not including all of the available data.  Secondly, the behaviour
of these sources is not easily explained by the outburst model
presented in \citet{Fender04}.  In the following sections, we use this
outburst scheme and the states it describes to explain the behaviour
of the disc and powerlaw components.  As the behaviour of these two
well studied BHXRBs do not easily fall fit into this scheme, we
actively decided to not include them in the study at this time.

Many of the masses and distances are unknown or not very well
constrained.  Where they are unknown we have assumed values of
$10M_\odot$ and $5\kpc$ respectively.  These uncertainties effect the
calculation of the Eddington Luminosities ($L_{\rm Edd}$) for these
BHXRBs which are used extensively throughout this analysis to scale
the BHXRBs to one another.  Until the distances and masses are well
determined, there will always be some uncertainty when comparing
between sources.

\begin{table*}
\centering
\caption{\label{tab:obj} {\sc X-ray Binary Parameters}}
\begin{tabular}{lrlrlrlrlrlrlrl}
\hline
\hline
Object & \multicolumn{2}{c}{$M_{\rm BH}$} & \multicolumn{2}{c}{$D$} & \multicolumn{2}{c}{$N_{\rm H}$} & \multicolumn{2}{c}{$P_{\rm orb}$}& \multicolumn{2}{c}{$M_*$}& \multicolumn{2}{c}{Inclination}\\
&\multicolumn{2}{c}{$(M_{\odot})$}&\multicolumn{2}{c}{$(\kpc)$}&\multicolumn{2}{c}{$(\times 10^{22}\pcm2)$}& \multicolumn{2}{c}{$(h)$}&\multicolumn{2}{c}{$(M_\odot)$}&\multicolumn{2}{c}{$(^\circ)$}\\
\hline
4U 1543-47    	&$9.4\pm2.0$&(1,2)&$7.5\pm0.5$&(3,4)&$0.43$&(2,4)&$26.8$&(4)&$2.45$&(1)&$21$&(2)\\
4U 1630-47	&$[10]$&&$10.0\pm5.0$&(5)&$>6$&(6)&$-$&&$-$&&$-$\\
4U 1957+115	&$[10]$&&$[5]$&&$0.15$&(7)&$9.3$&(8)&$1.0$&(9)&$-$\\
GRO J1655-40	&$7.0\pm0.2$&(10,11)&$3.2\pm0.2$&(4,12)&$0.8$&(13)&$62.9$&(4)&$2.35$&(10)&$70$&(48)\\
GRS 1737-31	&$[10]$&&$[5]$&&$6.0$&(14)&$-$&&$-$&&$-$\\
GRS 1739-278	&$[10]$&&$8.5\pm2.5$&(15)&$2$&(15)&$-$&&$-$&&$-$\\
GRS 1758-258	&$[10]$&&$[5]$&&$1.50$&(16)&$18.5$&(17)&$-$&&$-$\\
GS 1354-644	&$>7.8=10.0\pm2.0$&(1)&$>27=33\pm6$&(18)&$3.72$&(18,19)&$61.1$&(18)&$1.02$&(1)&$-$\\
GS 2023+338	&$10\pm2$&(1)&$4.0\pm2.0$&(4)&$0.7$&(4)&$155.3$&(4)&$0.65$&(1)&$-$\\
GX 339-4	&$5.8\pm0.5$&(20)&$8.0\pm4.0$&(21)&$0.4$&(22)&$42.1$&(4)&$0.52$&(20)&40\\
H 1743-322	&$[10]$&&$[5]$&&$2.4$&(23)&$-$&&$-$&&$-$\\
XTE J1118+480	&$6.8\pm0.4$&(1,24)&$1.7\pm0.05$&(25,26)&$0.01$&(25)&$4.08$&(4)&$0.28$&(1)&$68$&(26)\\
XTE J1550-564	&$10.6\pm1.0$&(3)&$5.3\pm2.3$&(4)&$0.65$&(27)&$37.0$&(4)&$1.30$&(3)&$72$&(3)\\
XTE J1650-500	&$<7.3=6\pm3$&(28)&$2.6\pm0.7$&(29)&$0.7$&(30)&$7.7$&(28)&$-$&&$30$&(49)\\
XTE J1720-318	&$[10]$&(31)&$>8=8\pm6$&(31)&$1.24$&(31)&$-$&&$-$&&$-$\\
XTE J1748-288	&$[10]$&&$>8=10\pm2$&(32)&$7.5$&(33)&$-$&&$-$&&$-$\\
XTE J1755-324	&$[10]$&&$[5]$&&$0.37$&(34)&$-$&&$-$&&$-$\\
XTE J1817-330	&$<6=4\pm2$&(35)&$>1=[10]$&(35)&$0.15$&(35)&$-$&&$-$&&$-$\\
XTE J1859+226	&$10\pm5$&(36)&$6.3\pm1.7$&(4)&$0.34$&(36)&$9.17$&(4)&$0.9$&(36)&$-$\\
XTE J2012+381	&$[10]$&&$[5]$&&$1.3$&(37)&$-$&&$-$&&$-$\\
LMC X-1		&$10\pm5$&(38)&$52\pm1.0$&(39)&$0.5$&(13)&$93.8$&(40)&$-$&&$45$&(38)\\
LMC X-3  	&$10\pm2$&(41)&$52\pm1.0$&(39)&$0.06$&(42)&$40.8$&(43)&$6$&(41)&$60$&(41)\\
SAX 1711.6-3808	&$[10]$&&$[5]$&&$2.8$&(44)&$-$&&$-$&&$-$\\
SAX 1819.3-2525	&$10\pm2$&(46)&$10\pm3$&(46)&$0.1$&(47)&$67.6$&(46)&$-$&&$65$&(46)\\
SLX 1746-331	&$[10]$&&$[5]$&&$0.4$&(45)&$-$&&$-$&&$-$\\
\hline
\end{tabular}
\begin{quote}
Many of the objects do not have well determined distances or masses.
In this case we have taken the distances to be $5\kpc$ and the masses
$10~M_{\odot}$.  A recent critical look at the distance estimates for
GRO~J1655-40 by \citet{Foellmi08} indicates a revised estimate of the
distance of $<2.0\kpc$.  References: 

(1) \citet{Ritter03}
, (2) \citet{Park04}
, (3) \citet{Orosz02}
, (4) \citet{Jonker04}
, (5) \citet{Augusteijn01}
, (6) \citet{Tomsick05}
, (7) \citet{Nowak08}
, (8) \citet{Thorstensen87}
, (9) \citet{Shahbaz96}
, (10) \citet{Hynes98}
, (11) \citet{Shahbaz99}
, (12) \citet{Hjellming95}
, (13) \citet{Gierlinski01}
, (14) \citet{Cui97}
, (15) \citet{Greiner96}
, (16) \citet{Pottschmidt06}
, (17) \citet{Smith02}
, (18) \citet{Casares04}
, (19) \citet{Kitamoto90}
, (20) \citet{Hynes03}
, (21) \citet{Zdziarski04}
, (22) \citet{Miller04}
, (23) \citet{Capitanio05}
, (24) \citet{Wager01}
, (25) \citet{Chaty03}
, (26) \citet{Gelino06}
, (27) \citet{Gierlinski03}
, (28) \citet{Orosz04}
, (29) \citet{Homan06}
, (30) \citet{Miniutti04}
, (31) \citet{CadolleBel04}
, (32) \citet{Hjellming98}
, (33) \citet{Kotani00}
, (34) \citet{Revnivtsev98}
, (35) \citet{Sala07}
, (36) \citet{Hynes02}
, (37) \citet{Campana02}
, (38) \citet{Hutchings87}
, (39) \citet{DiBenedetto97}
, (40) \citet{Orosz08}
, (41) \citet{Cowley83}
, (42) \citet{Haardt01}
, (43) \citet{Hutchings03}
, (44) \citet{IntZand02}
, (45) \citet{Wilson03}
, (46) \citet{Orosz01}
, (47) \citet{IntZand00}
, (48) \citet{VanDerHooft98}
, (49) \citet{SanchezFernandez02}

\end{quote}
\end{table*}

The investigation presented in this work concentrates on the variation
of the disc characteristics during the outburst as the changes in the
disc parameters are the most prominent changes in the spectrum
during a BHXRB outburst.  Our
results are therefore dominated by those objects which have had
outbursts well monitored by \rxte.  Roughly this ``removes'' all the
BHXRBs from our study which have only had a few \rxte\ observations.
Some objects which have had a comparatively large number of
observations are not observed to undergo the canonical outburst
structure outlined in Section \ref{sec:intro}.  These sources are
less able to show what changes disc undergoes during a
complete outburst, but are still useful for the hard/powerlaw
dominated states.

\section{Disc Parameters}\label{sec:DiscParams}

\begin{table}
\centering
\caption{\label{tab:obstimes} {\sc Observation Numbers, Times and Disc
Detections}}
\begin{tabular}{llll}
\hline
\hline

Object & Selected Obs & Exposure & Disc Detections\\
&&$\Ms$\\
\hline
4U 1543-47    	&$61   $&$0.147$&36\\
4U 1630-47	&$704  $&$1.371$&491\\
4U 1957+115	&$59   $&$0.260$&25\\
GRO J1655-40	&$484  $&$1.829$&368\\
GRS 1737-31	&$5    $&$0.045$&2\\
GRS 1739-278	&$6    $&$0.017$&6\\
GRS 1758-258	&$9    $&$0.007$&9\\
GS 1354-644	&$8    $&$0.049$&1\\
GS 2023+338	&$0    $&$0.000$&0\\
GX 339-4	&$709  $&$1.682$&284\\
H 1743-322	&$346  $&$0.998$&224\\
XTE J1118+480	&$81   $&$0.170$&1\\
XTE J1550-564	&$365  $&$0.833$&168\\
XTE J1650-500	&$108  $&$0.191$&37\\
XTE J1720-318	&$63   $&$0.125$&33\\
XTE J1748-288	&$21   $&$0.074$&12\\
XTE J1755-324	&$2    $&$0.006$&1\\
XTE J1817-330	&$123  $&$0.329$&100\\
XTE J1859+226	&$121  $&$0.292$&101\\
XTE J2012+381	&$15   $&$0.036$&15\\
LMC X-1		&$69   $&$0.349$&64\\
LMC X-3  	&$471  $&$1.048$&173\\
SAX 1711.6-3808	&$13   $&$0.029$&5\\
SAX 1819.3-2525	&$48   $&$0.114$&2\\
SLX 1746-331	&$28   $&$0.091$&14\\
\hline
Totals		&3919	&10.09&2172\\
\hline
\hline
\end{tabular}
\begin{quote}
\end{quote}
\end{table}

The discs around black holes are thought to be optically thick and
geometrically thin \citep{Shakura73}.  The spectrum expected from this kind of disc around a
non-rotating black hole is easily calculated.  It is the sum of a set
of blackbody spectra, one for each radius, $R$, with a characteristic
temperature $T_{\rm eff}(R)$.  The total spectrum resulting from this
sum is then a multicolour disc blackbody, with a peak temperature
$T_{\rm eff,max}$ coming from close to the innermost stable orbit.
However, this spectrum is effected by the opacity of the disc, which
results in a colour temperature correction factor, $f_{\rm col}$
\citep{Shimura95,Merloni00, Davis06}.  This factor was shown by
\citet{Shimura95} to be $\sim 1.8$ for almost all black hole masses
and emission luminosities and is discussed further in Section
\ref{sec:spurs}.  Of course, the description of the disc may not be
quite as simple as envisaged by \citet{Shakura73} and radiatively
inefficient flows (\eg Advection Dominated Accretion Flows,
\citealp{Ichimaru77, Narayan94}) or slim discs (\eg
\citealp{Abramowicz88}) may exist.  However, we concentrate on the
disc model proposed by \citet{Shakura73} in this study.

We use the physical description in \citet{Gierlinski04, Gierlinski99}
to calculate the relation between the disc luminosity and the
temperature as for a Schwarzschild black hole,

\begin{equation}
\frac{L_{\rm Disc}}{L_{\rm Edd}} \approx 0.583 \left(\frac{1.8}{f_{\rm
    col}}\right)^4   \left(\frac{M}{10 {\rm M_\odot}}\right)
\left(\frac{kT_{\rm max}}{1\kev}\right)^4, \label{eq:LT}
\end{equation}

\noindent which assumes a constant inner disc radius, $R_{\rm
  in}$.  We investigate the effects of the black hole spin in Section
\ref{sec:Dists}.  We include the adjustments to the observed disc temperature,
$T_{\rm obs}$, for relativistic effects close to
the black hole \citep{Gierlinski04}. We add a 4 per cent
temperature shift to account for the stress-free boundary layer, and
also the adjustment from \citet{Zhang97} which accounts for the strong
gravitational potential.
\[
T_{\rm max}=T_{\rm obs}/f_{\rm GR}(\theta,a^*)\xi,
\]
where $\xi=1.04$ is for the stress-free boundary layer, $\theta$ is
the inclination angle and $a^*$ the dimensionless spin parameter.    Out of the $\sim
3900$ observations $\sim 2200$ have disc 
detections (see Table \ref{tab:obstimes}).  Although the number of
observations in which a disc is well determined 
depends on the state of the BHXRB at the time it was observed, we show
the number of disc detections so that it is clear that a few BHXRBs
have many more detections than most of the others.  Therefore our
conclusions are depend more on the results from these BHXRBs.

In the data
reduction routine and best fitting model selection proceedure we have
been conservative in determining which observations have discs (see
Section \ref{sec:model}).  Initially we
investigate the degree to which the BHXRB discs follow the expected $L-T$
relation (Equation \ref{eq:LT}), using an $f_{\rm col}=1.8$ and assuming that the inner
radius of the disc is constant.  In later Sections we relax these assumptions.

In the following, we define the Powerlaw Fraction (PLF) and the Disc Fraction (DF) as
\begin{eqnarray}
{\rm PLF}=\frac{L_{1-100\kev,\ {\rm
      PL}}}{L_{0.001-100\kev,\ {\rm Disc}}+L_{1-100\kev,\ {\rm PL}}}
\nonumber \\
{\rm DF}=\frac{L_{0.001-100\kev,\ {\rm
      Disc}}}{L_{0.001-100\kev,\ {\rm Disc}}+L_{1-100\kev,\ {\rm
      PL}}}, \nonumber
\end{eqnarray}
following \citet{Dunn10} as well as \citet{Dunn08, Koerding06}.  These
two quantities, used when creating Disc Fraction Luminosity Diagrams,
allow the natural separation of the outburst into two states -
powerlaw and disc dominated.  These correspond roughly to the hard and
soft states more commonly used in BHXRB studies.  For an in depth study
of the relation between these state conventions see \citet{Dunn10}.

\section{Disc Temperature and Luminosity}\label{sec:L-T}

We show the variation of the disc temperature with unabsorbed disc
luminosity for 
each BHXRB individually in Appendix Fig. \ref{fig:DL_DT_objs}.  The
errorbars are only from the uncertainties arising in the spectral
fitting.  We do not include the uncertainties in physical parameters
of the BHXRB system (\eg mass and distance), as in many cases the
physical parameters are 
unknown, and would further complicate the diagram.  We also show the
theoretically expected $L-T$ relation for $f_{\rm col}=1.8$ for each
BHXRB on each diagram in Appendix Fig. \ref{fig:DL_DT_objs} as the
dashed black line, using the masses as shown in Table 
\ref{tab:obj} (again without including the uncertainties).  Also shown
in Appendix Fig. \ref{fig:DL_DT_objs} is a schematic showing the
motion of the BHXRB through the $L-T$ plane as an outburst
progresses.  For
clarity, for the remainder of this section the theoretical
$L-T$ relation is that from Eq. \ref{eq:LT} under the assumption of a
constant inner disc radius and colour correction factor.

\begin{figure}
\centering
\includegraphics[width=1.0\columnwidth]{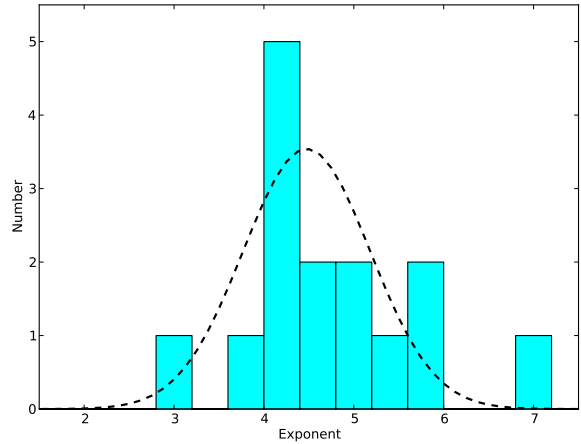}
\caption{\label{fig:LTExponents} The distribution of the best fitting
  slopes in the most disc dominated states.  The indicated Gaussian is
a least-squares fit to the histogram, and peaks at $4.48$ with a width
of $0.33$. GRS~1739-278, XTE~J1650-500 and LMC~X-1 are beyond the
edges of the plot (see Table \ref{tab:DiscT}).}
\end{figure}

In Appendix Fig. \ref{fig:DL_DT_objs} it is clear that most of the BHXRB's
discs do closely follow the theoretically expected $L-T$ relation.  We fit the
most disc dominated points (DF $>0.8$) of each BHXRB with a powerlaw
in the $\log_{10}T-\log_{10}L_{\rm Disc}$ plane, where for a constant
size black body a slope of four is expected.  We show the resulting slopes in 
Table \ref{tab:DiscT} and a histogram of their distribution in
Fig. \ref{fig:LTExponents}.  The best-fit Gaussian distribution to the
histogram peaks at a slope of $4.48$.  We select the most disc dominated observations in
order to focus on those where the disc parameters (temperature and
disc normalisation in {\scshape
  xspec}) were very well determined and also to exclude points close
to the intermediate state, where the relation 
may not apply (similar to the selection performed in
\citealp{Gierlinski04}).  The behaviour of the disc 
temperature at smaller disc fractions, corresponding 
to states closer to the intermediate states, and are discussed in
Section \ref{sec:spurs}.  

Of the ten BHXRBs studied by \citet{Gierlinski04}, nine are included
in our study. For four of the BHXRBs, their distribution of observations match
between their study and those in Appendix Fig \ref{fig:DL_DT_objs}
(GRS~1739-278, XTE~J2012+381, LMC~X-1 and LMC~X-3).  There are many more
observations of GX~339-4 presented in our study, and so it is
difficult to determine any differences between the two studies.  Of the remaining four BHXRBs
(GRO~J1655-40, XTE~J1550-564, XTE~J1650-500 and XTE~J1859+226), the
trends observed in \citet{Gierlinski04} show very clear and linear
$\sim T^4$ relations.  However, in Appendix Fig \ref{fig:DL_DT_objs}
we find that, although the most disc dominated observations do on the
whole follow the expected $L-T$ relation, there are a large
number of points at low disc fractions which fall ``below'' the
expected $L-T$ relation (see also Section \ref{sec:spurs}).  In this
study, we include and
show all observations in which a disc+powerlaw model was the best fit, whereas
those in \citet{Gierlinski04} select ``disc dominated spectra'' where
up to 15 per cent of the total bolometric emission can be present in a
Comptonized tail.  In Appendix Fig \ref{fig:DL_DT_objs} we only fit
those observations for which DF$>0.8$ and therefore the
apparent observed differences are large because of the plotted low
disc fraction points.

In a number of the BHXRBs, the statistically best fitting line is not
similar to the expected $L-T$ relation.  This mismatch in between the
slopes of the expected $L-T$ relation and 
that of the best fit to the most disc dominated states could arise
from the limitations on the range of disc temperatures probed and the
spectral response of 
\rxte.  If only a few observations have a detected disc, then the
variation in the disc temperature and luminosity may be small, which
could mask a $T^4$ trend if the scatter is naturally high, for example
GRS~1739-278 and LMC~X-1.  

In a large number of cases, although
the shape of the relation is close to that of the $L-T$ relation, the
normalisations are not always correct (see Equation \ref{eq:LT}).  Given the uncertainties in the
masses and distances an offset between the observed and expected
behaviours is not unexpected.  We also
note that the expected $L-T$ relation assumes that the black hole is not
spinning.  We discuss the effects of these unknown system parameters
further in Section \ref{sec:Dists}. 

Therefore it is clear that under the assumption of a constant $f_{\rm
  col}$ and a constant inner disc radius, $R_{\rm in}$, the majority
of the BHXRBs show a variation of the disc luminosity, $L_{\rm Disc}
\,\tilde\propto\, T^4$ in the most 
disc dominated states over an
order of magnitude in luminosity and a factor of two in disc
temperature.  This has been found by earlier studies (\eg
\citealp{Gierlinski04, Davis06,Dunn08}), though in some cases the
analysis as restricted to
the most disc dominated (soft) states.  Specifically, the ratio of $R_{\rm in}:f_{\rm col}$
is constant in these states.  However, it would be very strange if the
dynamics of the accretion disc and the radiation transfer processes
conspired to keep this ratio constant, and so it is likely that each
quantity is itself, constant, in the disc dominated state.  However, what of the behaviour in the less
disc dominated states?  As noted above, many of the observations at
low disc fractions fall ``below'' the expected ``L-T'' relation.  We
investigated whether this could result from slightly different $L-T$
relations from different outbursts of the same BHXRB, however no clear
or variation was found.  A number of the low disc fraction
observations are seen as almost perpendicular deviations from 
the expected $L-T$ relation.  The trend for these ``spurs'' is for the
disc luminosity to decrease as the disc temperature increases and
are observed both at the beginning and the end of the outbursts (see
e.g. 4U~1543-47 and GRO~J1655-40).  We discuss
these deviations further in Section \ref{sec:spurs}.  

There are a few cases where the best fit line to the most disc
dominated observations is radically different
to the expected $L-T$ relation even though there are a large number of
observations, e.g. XTE~J1650-500, LMC~X-1.  It is almost
as if whatever is causing the ``spurs'' dominates the variation of the
disc temperature and luminosity in these objects.
In these two BHXRBs there are two effects which conspire to give
best-fit relations different to the expected $L\propto T^4$.  Firstly, in
both of these sources the majority (if not all) of observations have
the same luminosity.  In XTE~J1650-500 those observations occur just after the
disc fraction reaches $0.8$, rather than in the most disc dominated
state of the outburst (DF$>0.9$).  If we select those observations
with DF$>0.9$ as opposed to DF$>0.8$ then the slope of the best
fitting line is 9.52.  Whereas in LMC X-1 the only a few observations 
have a sufficiently high disc fraction to be fitted, with almost no
variability in the total luminosity\footnote{We note that the not all
  of the best fitting models appear appropriate for LMC X-1 in
  \citet{Dunn10}, as all the observations have similar X-ray
  colours, yet some have no disc component as seen in the Disc
  Fraction Luminosity Diagram (see \citet{Koerding06, Dunn08}  and
  \citet{Dunn10} for more details).  This
  is likely to be the result of the similarity between a broken
  powerlaw model and a disc + powerlaw model when the disc does not
  dominate the spectrum.  This is exacerbated in LMC X-1 as the hard
  X-rays are also faint as a result of the large distance to the
  BHXRB, which make fitting the powerlaw difficult.}.
Secondly, as seen in a number of the other BHXRBs, observations at
intermediate states, or those which do not have overly strong disc
fractions show ``spurs'' running approximately perpendicular to the
$L-T$ relation.  Both of these effects are combined with the
scatter observed in the the relations exhibited by other BHXRBs.

Therefore, if there are few points in the most disc
dominated states (XTE~J1650-500) or there is a small variation in the
luminosity (LMC X-1), it is possible that the best fit relation will
not be close to the theoretically expected one.  However, in
XTE~J1650-500, the best fit relation to the observations with DF$>0.9$
is much closer to the expected $L-T$ relation.  In a
study of the discs in BHXRBs \citet{Gierlinski04} also find that
LMC~X-1 does not appear to follow the relation.  However the range in
disc temperatures and luminosities in their study, like the observations
presented here, is comparatively small.  XTE~J1650-500 is also
included in their study, and although they find departures from the
$L_{\rm Disc}\propto T^4$ law, these observations have large error bars and so are
consistent still with it. 

\begin{table}
\centering
\caption{\label{tab:DiscT} {\sc Disc Temperature Fits}}
\begin{tabular}{lrll}
\hline
\hline

Object & Exponent& Points& Note\\

\hline
4U 1543-47    	&$5.76\pm0.76$&26\\
4U 1630-47	&$5.67\pm0.23$&245\\
4U 1957+115	&$4.10\pm0.24$&16\\
GRO J1655-40	&$4.16\pm0.14$&209\\
GRS 1737-31	&$-    $&\\
GRS 1739-278	&$-0.59\pm0.02$&5&1\\
GRS 1758-258	&$4.80\pm0.42$&2\\
GS 1354-644	&$-    $&\\
GS 2023+338	&$-    $&\\
GX 339-4	&$7.08\pm0.37$&197\\
H 1743-322	&$3.70\pm0.14$&99\\
XTE J1118+480	&$-    $&\\
XTE J1550-564	&$4.26\pm0.09$&109\\
XTE J1650-500	&$24.64\pm7.79$&28&2\\
XTE J1720-318	&$5.21\pm0.19$&29\\
XTE J1748-288	&$3.19\pm0.16$&5\\
XTE J1755-324	&$-    $&\\
XTE J1817-330	&$4.68\pm0.23$&88\\
XTE J1859+226	&$4.11\pm0.34$&52\\
XTE J2012+381	&$4.28\pm0.55$&15\\
LMC X-1		&$-1.88\pm0.88$&14&3\\
LMC X-3  	&$4.64\pm0.11$&109\\
SAX 1711.6-3808	&$-    $&\\
SAX 1819.3-2525	&$-    $&\\
SLX 1746-331	&$4.88\pm0.40$&12\\

\hline
\end{tabular}
\begin{quote}
The number of points are those used in the fitting, so with a Disc
Fraction $>0.8$.  Notes: 1 -- GRS~1739-278 has very few points for the
fitting.  2 -- XTE~J1650-500 has a strange distribution of
observations along the outburst, causing the erroneous fit.  3 --
LMC~X-1: small range in disc temperatures and luminosities and the
slope is not well defined.  
\end{quote}
\end{table}

The theoretical $L-T$ relation depends on the mass of the BHXRB - which can
reasonably be assumed to be constant during the outburst - the $f_{\rm
  col}$, the colour temperature correction factor, the spin and also the inner
radius of the disc, $R_{\rm in}$.  The behaviour of the disc
temperature and luminosity seen in Appendix Fig \ref{fig:DL_DT_objs} for the
majority of the observations of the BHXRBs, indicates that over a
large range in disc temperature and luminosity, both the $f_{\rm col}$ and
$R_{\rm in}$ are relatively constant, at least in the most disc
dominated states.  There is
some scatter around the best fitting 
relation which on the whole appears random with no clear secondary
trend (see Fig. \ref{fig:LTExponents}).  However, in the ``spurs'' at the beginning and ends of the
outbursts, the deviation from the $L-T$ relation is large, and we now
investigate whether coherent variations either $R_{\rm in}$ or $f_{\rm
  col}$ could cause them.

\section{Disc Radius}\label{sec:discinner}

Only when the inner radius of the disc, $R_{\rm in}$, is constant will
the disc luminosity and temperature follow the expected $L \propto
T^4$ relation assuming that $f_{\rm col}$ is also constant.  As we
have shown in Section \ref{sec:L-T} that for the
majority of the observations the expected $L-T$ relation is a good
description of the behaviour, then we expect that the disc radius
remains constant for the most disc dominated observations.

To calculate the disc radius, we use the
normalisation of the {\scshape diskbb} model, $\mathcal N$, from
{\scshape xspec} as this explicity includes the
inner radius of the disc, $R_{\rm in}$. 

\begin{equation}
{\mathcal N}=\left[\frac{R_{\rm in}}{D_{\rm 10\kpc}}\right]^2 \cos
\theta, \label{Eq:DiskBB}
\end{equation}

\noindent where $D_{\rm 10\kpc}$ is the distance in units if $10\kpc$ and
$\theta$ is the inclination of the system.  Where the inclination of
the system is not known (Table \ref{tab:obj}) we use
$\theta=60^\circ$.  It is possible that the discs in some of these
BHXRBs are misaligned with respect to the binary's inclination
\citep{Maccarone02, Fragos10}.  In Appendix Fig
\ref{fig:DL_DT_objs} we show the inner radius against the disc
temperature for each BHXRB.  We do not include the effect of the uncertainties
in the distance and inclination in the error bars shown.  We also show
in Appendix Fig \ref{fig:DL_DT_objs} a schematic diagram demonstrating
the motion of the BHXRB through the
$R_{\rm in}-T$ plane as the outburst progresses.  As, on the
whole, the disc temperature decays during the most disc dominated
stages of the outburst, the track of the BHXRB through the $R_{\rm
  in}-T$ plane should be clear.

As was expected from the behaviour of the disc luminosity and
temperature, and from the study of GX~339-4 by \citet{Dunn08}, the
majority of points are at a relatively constant inner disc 
radius.  We have chosen to plot the disc radius in kilometres on the
primary $x$-axis rather
than $R$ as fractions of the gravitational radius, $R_{\rm g}$, as in
many of the BHXRBs the masses 
are not accurately known.  The secondary $x$-axis shows the radii as a
fraction of $R_{\rm g}$ for comparison.  We show in Table
\ref{Eq:DiskBB} the best fit inner 
radii for the observations with a Disc Fraction $>0.8$. 

In some BHXRBs (4U~1957+115, GRS~1758-258 and
SLX~1746-331) the disc radii are very small, less than $10\km$ or
below $1R_{\rm g}$.  The innermost stable circular orbit (ISCO) for
a black hole is $6R_{\rm g}$ for a non-rotating black hole, where
$R_{\rm g}=GM/c^2=$ is the Schwarzschild radius.  For a maximally
rotating black hole, the ISCO can go down to $R_{\rm g}$.  In these
three BHXRBs the distances, masses and inclinations are not
known, and so these effect the estimates on the inner disc radius.
From the normalisation, ${\mathcal N}$, the distance is directly
proportional to the $R_{\rm in}$, the mass inversely proportional
(when measured in units if $R_{\rm g}$) and
the inclination has a $(\cos\theta)^{-1/2}$ dependence.  The most
change would arise if the inclination would increase, though
increasing the distance or decreasing the mass would also increase the
inner disc radius.  However, not knowing the true values of the inner disc
radius, it is difficult to determine which of these parameters should change
and by how much.  Therefore,
although the estimates on the inner disc radius are smaller than
physically sensible, it is likely to be the result of our incomplete
knowledge of the system parameters.

The ``spurs'' which were mentioned in Section \ref{sec:L-T} are also
seen in the $R_{\rm in}-T$ plots.  If taken at face value, then they indicate that
at the end of the outburst the inner disc radius decreases as the disc
temperature rises, and vice versa at the beginning of the outburst.
This behaviour does not appear to be physically meaningful, as the
minimum radius measured for some of the observations fall well
within the smallest stable orbit for a $10 M_{\odot}$ black hole.  We
discuss these spurs and their possible causes further in the following section.

\begin{table}
\centering
\caption{\label{tab:DiscR} {\sc Disc Inner Radius Fits}}
\begin{tabular}{lll}
\hline
\hline

Object & $R_{\rm in} (\km)$& $R_{\rm in}/R_{\rm g}$\\

\hline
4U 1543-47    	&$  53.6 ^{+ 10.4} _{\quad\quad -  8.7} $ & $   3.86 ^{+  0.75} _{\quad\quad -  0.62} $\\
4U 1630-47	&$  20.9 ^{+  3.6} _{\quad\quad -  3.1} $ & $    1.42^{+  0.24} _{\quad\quad -  0.21} $\\
4U 1957+115	&$  2.43 ^{+ 0.07} _{\quad\quad -  0.07}$ & $   0.16 ^{+  0.005} _{\quad\quad -  0.005} $\\
GRO J1655-40	&$  21.8 ^{+  2.1} _{\quad\quad -  1.9} $ & $   2.11 ^{+  0.21} _{\quad\quad -  0.19} $\\
GRS 1739-278	&$  29.3 ^{+  0.6} _{\quad\quad -  0.6} $ & $   1.98 ^{+  0.04} _{\quad\quad -  0.04} $\\
GRS 1758-258	&$  4.90 ^{+ 0.01} _{\quad\quad -  0.01}$ & $   0.33 ^{+  0.0009} _{\quad\quad -  0.0009} $\\
GX 339-4	&$  42.0 ^{+  8.6} _{\quad\quad -  7.1} $ & $   4.90 ^{+  1.00} _{\quad\quad -  0.83} $\\
H 1743-322	&$  22.4 ^{+  2.2} _{\quad\quad -  2.0} $ & $   1.52 ^{+  0.15} _{\quad\quad -  0.14} $\\
XTE J1550-564	&$  49.4 ^{+  5.4} _{\quad\quad -  4.8} $ & $   3.15 ^{+  0.34} _{\quad\quad -  0.31} $\\
XTE J1650-500	&$  18.9 ^{+  4.3} _{\quad\quad -  3.5} $ & $   2.14 ^{+  0.49} _{\quad\quad -  0.40} $\\
XTE J1720-318	&$  69.5 ^{+ 16.7} _{\quad\quad - 13.4} $ & $   4.70 ^{+  1.13} _{\quad\quad -  0.91} $\\
XTE J1748-288	&$  28.2 ^{+  1.4} _{\quad\quad -  1.3} $ & $   1.91 ^{+  0.09} _{\quad\quad -  0.09} $\\
XTE J1817-330	&$  66.2 ^{+ 12.9} _{\quad\quad - 10.8} $ & $  11.12 ^{+  2.18} _{\quad\quad -  1.82} $\\
XTE J1859+226	&$  35.5 ^{+  6.5} _{\quad\quad -  5.5} $ & $   2.40 ^{+  0.44} _{\quad\quad -  0.37} $\\
XTE J2012+381	&$  27.2 ^{+  1.4} _{\quad\quad -  1.4} $ & $   1.84 ^{+  0.10} _{\quad\quad -  0.09} $\\
LMC X-1		&$  38.2 ^{+  5.6} _{\quad\quad -  4.9} $ & $   2.59 ^{+  0.38} _{\quad\quad -  0.33} $\\
LMC X-3  	&$  36.1 ^{+  2.1} _{\quad\quad -  2.0} $ & $   2.44 ^{+  0.14} _{\quad\quad -  0.14} $\\
SLX 1746-331	&$  6.7  ^{+  0.7} _{\quad\quad -  0.7} $ & $   0.45 ^{+  0.05} _{\quad\quad -  0.04} $\\
\hline
\end{tabular}
\begin{quote}
The inner radius estimates do not include the uncertainties on the
values or estimates on the distance or mass of the BHXRB (see Equation
\ref{Eq:DiskBB}).  As a reminder, the inner radius of the disc is
$6R_{\rm g}$ for a non-rotating black hole, and $1R_{\rm g}$ for a
maximally rotating one.
\end{quote}
\end{table}

\section{The Spurs - an $f_{\rm col}$ connection?}\label{sec:spurs}

As has been alluded to in the above sections, apart from the
theoretically expected $L-T$ relation at constant disc radius $R_{\rm in}$, the
other main trend is perpendicular to the $L_{\rm Disc} \propto T^4$ relation,
leading to an apparent decrease in the disc radius at the beginnings
and ends of the outburst.  These spurs are seen in most of the BHXRBs,
and were also seen in GX~339-4 in 
\citet{Dunn08}, XTE~J1650-500 in
\citet{Gierlinski04} and GRO~J1655-40 in \citet{Done07}.  

The plots of $R_{\rm in}-T$ in Appendix Fig. \ref{fig:DL_DT_objs} show
that the disc radius decreases to 
very small values at the beginnings and ends of the outbursts.  It is
the extent of the decreases, down to values well below $1R_{\rm g}$,
that lead us to investigate whether changes in $f_{\rm col}$ could be
responsible.  Although changing the system parameters can shift the
location of the observations in the $R_{\rm in}-T$ plane, they are
unlikely to move the observations with the smallest
calculated $R_{\rm in}$ sufficiently far.  We note that there are 
also points which occur at quite a distance from the main $L-T$
relation, and also have very low disc fractions ($\lesssim 0.5$) and
also have large uncertainties in the disc temperature and luminosity.
These are unlikely to be explained by a variation in $f_{\rm col}$,
and are discussed further in Section \ref{sec:bpl}.

The limited low energy response of the \rxte\ \pca\ instrument may
restrict the accurate fitting of a disc components when it does not
dramatically dominate over the remainder of the continuum.  The
calibrated range of the \pca\ starts at around $3\kev$ whereas the
disc temperatures peak at around $1-2\kev$.  As we are therefore
fitting only one side of the disc component, as the temperature 
and the luminosity fade, the slight curvature could be difficult to
accurately fit especially in short observations or ones which have a
small number of signal counts.  Although some of the spurs could arise from
weaknesses in our data analysis proceedure, as they have also been
seen in other studies (\eg \citealp{Gierlinski04,Done07}) it is
likely that these weaknesses are not the full explanation.  However
\citet{Gierlinski04} only use the most disc dominated observations in
their analysis.  We have emulated this approach in this study when fitting lines or
finding averages by selecting those observations with a very high disc
fraction ($>0.8$).  However, the plots show all the observations which
have a detected disc component.  Therefore, our plots show these spurs, which may not
have been shown in other studies where only the selected observations
were plotted.

We also note that in a study of LMC~X-3 by \citet{Steiner10}, the
observations from \rxte\ \pca\ PCU-2 gave a very consistent value for
the inner radius, a value which was also consistent with those obtained
from other detectors (\eg\ {\it Suzaku}, {\it Swift} and {\it
  XMM-Newton}).  In our study we also find no evidence for the spurs
seen in other BHXRBs in LMC~X-3, and so their study does not help in
clarifying whether the spurs come from instrumental effects.  However the close correspondence of
the $R_{\rm in}$ between \rxte\ and the other detectors does indicate
that in the disc dominated state, the spectral coverage of the
\pca\ is sufficient to be able to accurately determine the disc parameters.

In the theoretical $L-T$ expectation, we have used a constant value
for the colour correction, $f_{\rm col}=1.8$ \citep{Shimura95}.  The departures from the
$L\propto T^4$ law could be the result of the variation of the value
of $f_{\rm col}$.  The colour correction factor accounts
for the change in the dominant emission process in the inner disc, and
therefore is a function of $L_{\rm Disc}$.   \citet{Gierlinski04}
showed that the $f_{\rm col}$ is approximately constant throughout the
outburst (see also \citealp{Shimura95, Merloni00, Davis06}).  However
these were only for the most disc dominated 
observations.  Therefore it is possible that $f_{\rm col}$ variations
could have occurred in the BHXRBs presented in these studies, but not
be shown in the figures.  If $f_{\rm col}$ is constant 
during the most disc dominated phases of the outburst, and only those
phases are shown, then any variation would not be detected.

To account for these spurs, $f_{\rm col}$ would decrease at the
beginning of the outburst, and then increase on the exit of the
outburst as the BHXRB goes through the intermediate states (see the
schematic in Appendix Fig. \ref{fig:DL_DT_objs}).  The motion off the $T^4$ 
relation is approximately perpendicular.  Therefore a simple change in
$f_{\rm col}$ would explain the deviations, without needing any further variation (of
inner disc radius, for example).  It is of course possible that the
disc radius is not constant at the very beginnings and ends of the
outburst.  However, if the $f_{\rm col}$ is not constant then it will
be difficult to determine what the true $R_{\rm in}$ is in these non
dominant discs observed by \rxte.  

We show on the $L-T$ plane in Appendix Fig. \ref{fig:DL_DT_objs} the theoretically
expected relation for $f_{\rm col}=1.8$ but also for a range of values
for $f_{\rm col}$ between 1.6 and 2.6.  The lower limit arises from
the initial investigation into $f_{\rm col}$ by \citet{Shimura95},
whereas the upper comes from the best characterised BHXRB in
\citet{Gierlinski04}.  In cases where the values of 
the distance and mass used are such so that the theoretical $L-T$
relation is a good match to the observed $L-T$ relation, then the
spurs, should they be present, mostly fall within this $1.6<f_{\rm
  col}<2.6$ range.   In Appendix Fig. \ref{fig:DL_DT_objs} we also
show a schematic which indicates the route taken by a BHXRB in this diagram.

If these spurs are purely the result of changes in $f_{\rm col}$, we
can calculate the change in $f_{\rm col}$ required, $\delta f_{\rm col}$, for the spurs to
be part of the expected $L-R_{\rm in}^2T^4$ relation, under the
assumption that the disc radius is constant.  We assume that $f_{\rm
  col}=1.8$ when the $T^4$ relation is followed, and so adjust the
normalisation of Equation \ref{eq:LT} so that the expected relation falls
under the observed points at $kT=1\kev$.  As this normalisation is
affected by the distance, mass, and inclination, which in many BHXRBs
are only estimates, this simplifies our approach, without affecting
our conclusions on the variation of $f_{\rm col}$.  We also note that the spin of a black hole can
affect the normalisation.  We are currently assuming that the black
hole is not rotating, but discuss spinning black holes in see Section
\ref{sec:Dists}.  In
Appendix Fig. \ref{fig:DL_DT_objs} we show for each BHXRB the excess 
$f_{\rm col}$ required for the observation to lie on the 
$T^4$ relation, $\delta f_{\rm col}$,  against the powerlaw and disc fractions of the
observation.  In most cases this centred on $\delta f_{\rm col}=0.0$, which is
by design, though where the $L-T$ relation slope is very different from
$4$ (Section \ref{sec:L-T}), then the position along the $x$-axis can vary.

There appear to be three regions in the diagrams.  The
observations with the largest disc fractions cluster around $\delta f_{\rm
  col}=0.0$, as defined by the normalisation adjustment mentioned
above.  The $\delta f_{\rm col}$ remains almost constant at zero over around an order of
magnitude change in the disc fraction.  These are the observations
which scatter around the theoretically expected $T^4$ relation, as the
$f_{\rm col}$ is constant.  

As the disc fraction decreases the
trend is for the observations to move gradually towards progressively
higher values of $f_{\rm col}$, $\delta f_{\rm col}$ increases.  These are the beginnings
of the spurs, but are also visible as lopsidedness in the scatter around
the most disc dominated observations in the plots of $L_{\rm Disc}$
versus $T$.

At around $\delta f_{\rm col}=0.5$ the trend in the observations flattens off, as
the disc fraction approaches zero, resulting in large changes in
$f_{\rm col}$ over small changes in the disc fraction.  These observations are the ones
from the spurs and extend up to $\delta f_{\rm col} \sim 1.0$.  The value of
$f_{\rm col}$ required for these observations to lie on the $T^4$
relation become larger with very little change in the disc fraction.
The $x$-axes of the figures has been truncated as the observations 
with very low disc fractions which lie well below the main cluster of
points have up to $\delta f_{\rm col} \sim 20$ (see Section \ref{sec:bpl}).  These observations are
unlikely to be explained by a varying $f_{\rm col}$ and hence we do
not show them in the figure.

The most recent investigation into $f_{\rm col}$ by \citet{Done08}
shows that there is a positive correlation between the $f_{\rm col}$
and the mass accretion rate.  The effect is stronger for a
proportional counter array (\eg\ the \rxte\ \pca) than for a
Charge-Coupled Device (CCD) and shows that for an alpha disc, with
$\alpha=0.1$, $f_{\rm col}$ can reach values of around $2.1$ for
accretion rates of $10^{19}\thinspace{\rm g/s}$ (for a {\scshape
  kerrbb} disc model).  This is lower than
the $f_{\rm col}$ increase inferred in Appendix
Fig. \ref{fig:DL_DT_objs}, but links the accretion rate to the $f_{\rm
  col}$.  Although the evolution of the $f_{\rm col}$ over time indicated by
\citet{Done08} is different, the clear link between the deviations
from $L-T^4$ and $f_{\rm col}$ suggests a link to the accretion rate.
However, it must be noted that the color correction fraction and, in
general, the observed properties of the high-energy tail of the disc
emission are quite sensitive to the vertical structure of the disc. In
particular, as discussed in more detail in \citet{Davis05} and
\citet{Done08}, the vertical dissipation profile 
may be very different in spectral states where a non-thermal
(power-law) component is significantly detected, as a larger fraction
of the total accretion power has to be released near or above the disc
surface, leading to a possible increase in the estimated color
correction factors. 

Both a constant $f_{\rm col}$ and a constant $R_{\rm in}$
are observed when the disc fraction is high ($\gtrsim 0.8$), i.e. in
those observations where the
disc emission dominates over the powerlaw emission, and the disc
parameters have been well determined.  However, under the assumption
of a constant $f_{\rm col}$, as the disc fraction
reduces the inferred $R_{\rm in}$ decreases, which has a knock-on effect on the
behaviour of the disc temperature with the luminosity.  However, the
drastic nature of the decrease in the $R_{\rm in}$ is such, that a
increase in $f_{\rm col}$ may be a more reasonable explanation.  If
the true underlying behaviour of the BHXRB is that the inner radius 
remains constant, then an increase of the $f_{\rm col}$ could
account for the majority of the spurs observed.

\section{Distances, Masses and Spins from Disc
  Parameters}\label{sec:Dists} 

During the most disc dominated parts of the outburst, the inner radius
of the disc and the $f_{\rm col}$ are approximately constant.  The estimated disc parameters
(temperature and luminosity) depend on the physical parameters of the
BHXRB system - the distance, mass and spin of the black hole.
However, these parameters are constant for a particular BHXRB.
Therefore, the {\it shape} that the observations make in the $L-T$
plane are fixed, but their {\it location} within the plane could vary,
depending on these parameters.
   
Using the theoretical relation (Equation \ref{eq:LT}) and the data we are able to place
limits on the distances and masses of the BHXRBs, in the case of a
Schwarzschild black hole.  From Equation \ref{eq:LT}, the
theoretically expected $L-T$ relation is of the form of
\begin{equation}
\frac{L_{\rm Disc}}{L_{\rm Edd}}=\mathcal{A} M T^4,
\end{equation}
\noindent where $M$ is the mass of the black hole.  When fitting the trend in the $L-T$ plane for the most disc
dominated states, the form is (for constant $R_{\rm in}$ and $f_{\rm col}$)
\begin{equation}
\frac{L_{\rm Disc}}{L_{\rm Edd}}=\mathcal{B} \frac{D^2}{M} T^{\rm C},
\end{equation}
\noindent where $T$ is the observed temperature, $B$ is derived from the
normalisation of the {\scshape diskbb} component in {\scshape xspec}
and $D$ is the distance of the BHXRB.  The dependence on $M$ and $D$
in these two equations are different.  
As these two Equations should be equal, assuming the fitted value of $C=4$, then
\begin{equation}
\mathcal{A} M =\mathcal{B} \frac{D^2}{M}.
\end{equation}
However, the slope of the fits to the $L-T$ relation do not always end
up with ${\rm C}\sim 4$.  Therefore, in order to remove this dependence we
calculate the match at $kT=1\kev$, which is close to the temperatures
of the observed discs.  

Therefore, under the assumption that $R_{\rm in}$ and $f_{\rm col}$
are constant, we can determine the distances and masses which are required
in order that the location of the expected $L-T$ relation matches those
which are observed.  In cases where limits have been placed on either the
distance or mass, then we are able to constrain the acceptable values
for the mass or distance respectively.  These loci of points in the distance-mass
diagram are shown in the Appendix, Fig. \ref{fig:DL_DT_objs} by the
blue line, and we
also show the current best estimates on the distances and masses and
their uncertainties, where they exist.  In some cases (e.g. GRS~J1739-278,
XTE~J1650-500 LMC~X-1) the fitted slope is very
different to $4$, and so there in these cases, these loci are not reliable.

In some cases the current best observational estimates on the
distances and masses 
do coincide with the estimates from this work (\eg 4U~1543-47,
GX~339-4, XTE~J1720-318).  However in many cases there is no overlap
between the observed estimates on the distance and mass and those
calculated here (\eg GRO~1655-40, XTE~J1550-564).  Although
at face value, cases where there is no overlap would allow the
distance and mass estimates to be refined, it is not quite that
simple.  These constraints are for a non-rotating black hole, and
there is significant evidence that at least some black holes have
significant spin (see \eg \citealp{Miller09,McClintock06,Middleton06}
and also \citealp{Fender10} and references therein).  

The normalisation of the theoretically expected relation between
$L_{\rm Disc}$ and $T$ changes when the black hole is maximally
spinning $(a^*=0.998)$ (see \eg \citealp{Gierlinski04}).  We use a very simple
parameterisation from \citet{Makishima00} who include first order
effects of the black hole spin on the theoretically expected relation
in terms of
\[
\alpha=R_{\rm in}/3R_{\rm S},
\]
the ratio of the inner disc radius to the Schwarzschild radius
($\alpha=1$ and $\alpha=1/6$ for a non rotating and maximally rotating black hole respectively).  This
appears as an $\alpha^2$ term in their version of Equation \ref{eq:LT}.  We
add this correction factor into Equation \ref{eq:LT}, and also take
into account the changes in the general relativistic correction
factors from \citet{Zhang97} for rotating black holes. 

Therefore, we show in Appendix, Fig \ref{fig:DL_DT_objs}, as well as the range of
distances and masses allowed for a non-rotating black hole, we also
show those for a maximally rotating black hole ($a^*=0.998$, red line) and one for
$a^*=0.5$ (green line).  This results in an
area in the $D-M$ plane in which both values are allowed.  We note,
that, a more accurate investigation would start 
with a more appropriate model for the disc
emission including the relativistic effects of the black hole spin
(\eg {\scshape kerrbb}).
We have, as yet, not re-fitted all our results with such a model.  
These plots show that many of the BHXRBs whose distances and masses did not
match the loci for a non-rotating black hole, do match if
the black hole is rotating (e.g. GRO~J1655-40, XTE~J1550-564, XTE~J1650-500,
LMC~X-3).  The eight BHXRBs which have both mass and
distance estimates, are all consistent with spin values $0<a^*<0.998$,
or, alternatively the inner radii are consistent with $R_{\rm
  g}<R_{\rm in}<6R_{\rm g}$.  In fact, none of the estimates on the
spin of the black hole require high spin values ($a\sim 1$).  This
also indicates that there are no counter-rotating discs in these
BHXRBs.  As the estimated spins fall in the range expected for Kerr
black holes, this indicates that the position of $R_{\rm in}$ in the
disc dominated states is mainly determined by strong gravitational effects.

For many of the BHXRBs in this sample, either the mass or the distance
or both are unknown.  If only one is known, then limits on the other
can be placed from the range allowed by the black holes spin.
However, in many cases these limits are not very constraining (\eg
GRS~J1739-278, XTE~J1720-318).  If both are unknown, then the best that can be
obtained is a lower limit on the distance (assuming a reasonable
lower bound for the black hole mass).

\section{Broken Powerlaw - Disc model degeneracy}\label{sec:bpl}

\begin{figure}
\centering
\includegraphics[width=1.0\columnwidth]{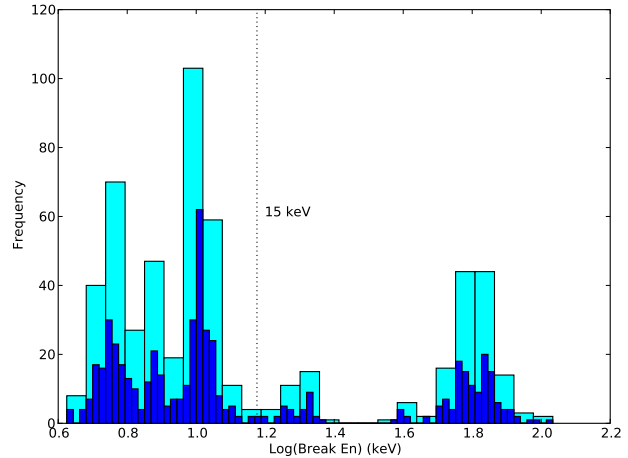}
\caption{\label{fig:BknEn} The distribution of the break energy for
  the broken powerlaw fits using two different binning levels.  The
  gap between $1.4<\log_{10} E_{\rm Break} <1.6$ arises from the cross over of
  the \pca\ and \hexte\ instruments.  }
\end{figure}

The limitations of the spectral response of the \rxte\ \pca\ make the
unequivocal detection of a non-dominant disc very difficult.  When the
disc's spectral component begins to rise above the powerlaw, the
difference in the $\chi^2_\nu$ between the broken powerlaw and disc +
powerlaw model is very small.  It is therefore difficult to say {\it ab
  initio} which of the two models is the most appropriate to
select for the observation (see \citealp{Dunn10}).  In our selection procedure we select
merely on the lowest reduced $\chi^2$, with some restrictions on the
model parameters.  We now investigate how adapting the model selection proceedure
affects the numbers and parameters of the discs detected.  

As the broken powerlaw model could mimic a disc model, especially when
the break energy is low ($\lesssim 10 \kev$).  We show in Fig. \ref{fig:BknEn} the
distribution of the break energies in the observations best fit by the
broken powerlaw model, using two different binning schemes.  The gap
around $\log_{10} E_{\rm break}=1.4$ is the result of the crossover between the \pca\ and the
\hexte\ instruments at $25\kev$.  

The main cluster of points occur at low energies, with another smaller
cluster at around $\log_{10} E_{\rm break}=1.8$.  These observations
with high break energies ($E_{\rm break}\sim 60 \kev$) are likely to
be reliable fits to a true break in the spectrum.  However, those
which fall below around $E_{\rm break}\sim 10 \kev$ could be the
result of a non-dominant disc mimicking a broken powerlaw

We redo the model selection proceedure but penalise the broken
powerlaw models where the break energy is lower than $15\kev$.  In
some cases only one of the two powerlaw fits (with and without line)
has a break energy $<15\kev$, and then the remaining broken powerlaw
fit is still allowed.  However, this only occurs in rare cases.
The other two remaining options are then single 
powerlaw or disc+powerlaw models (each with and without a line).   We
expect that this will have most effect in the intermediate states,
where non-dominant discs are expected, rather than in the most disc
dominated states where $L \propto T^4$.

\begin{figure}
\centering
\includegraphics[width=1.0\columnwidth]{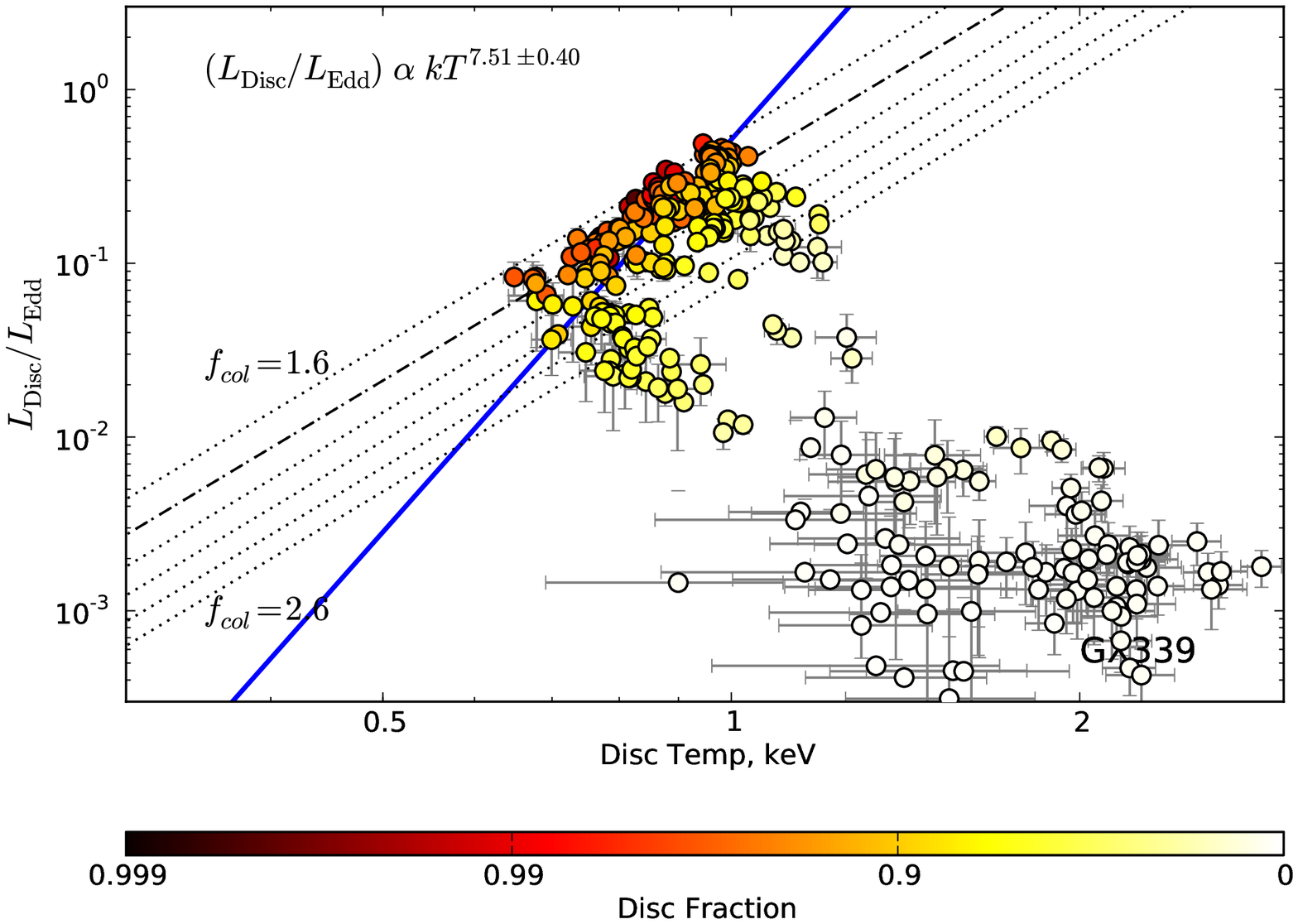}
\includegraphics[width=1.0\columnwidth]{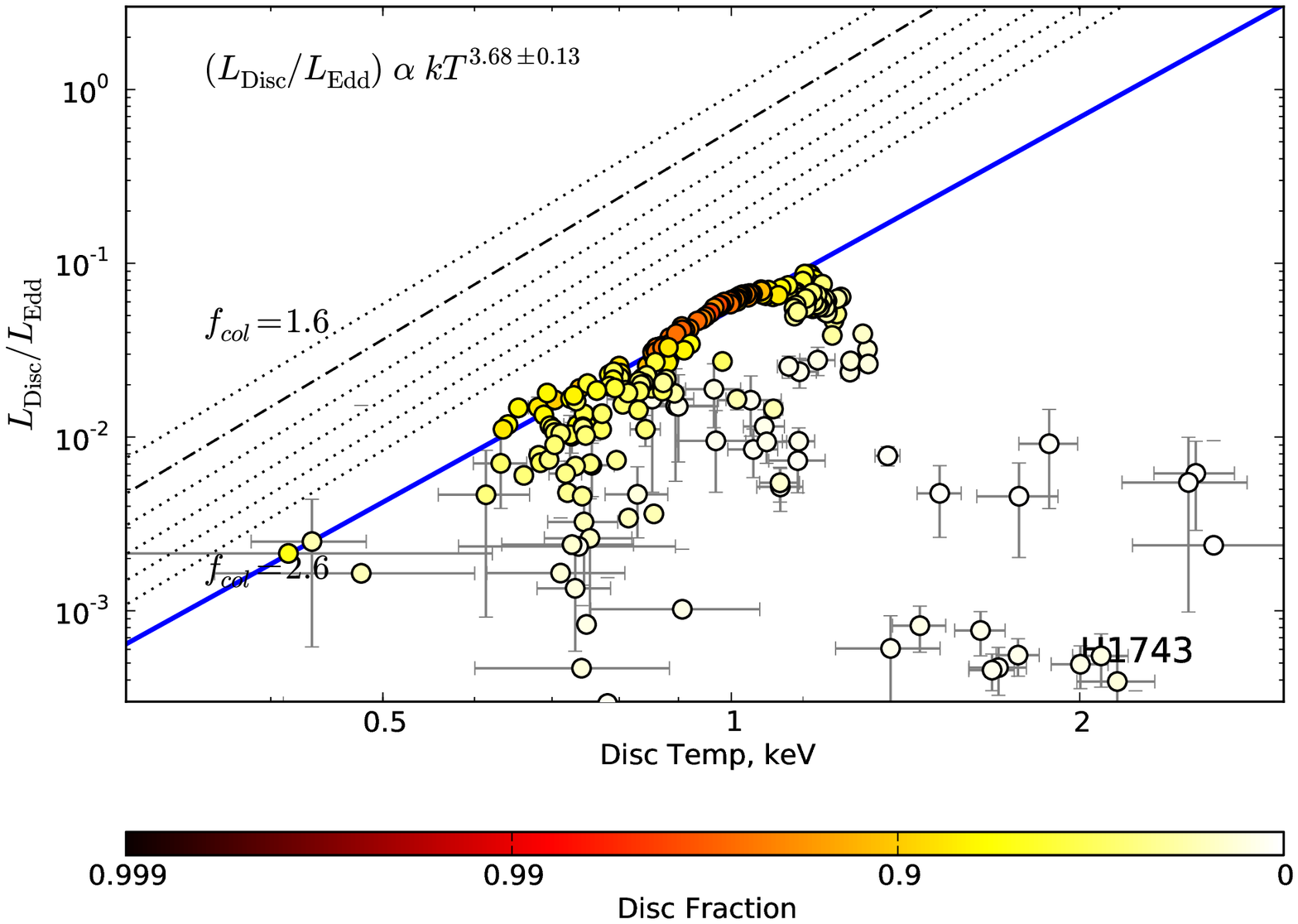}
\caption{\label{fig:LT-bpl} The variation of the disc luminosity with
  the disc temperature for GX~339-4 and H~1743-322 when penalising the
$\chi^2$ of broken powerlaw models which have a break energy
  $<15\kev$.  For comparison plots see Appendix Fig. \ref{fig:DL_DT_objs}. }
\end{figure}

In Fig. \ref{fig:LT-bpl} we show the $L-T$ plane for GX~339-4 and H~1743-322 as an
example to show the increase in the number of observations which have
a low disc fraction.  These extra observations appear below the $T^4$
relation, and the spurs appear to merge in with them.  Most of these
observations have large uncertainties in both the disc temperature and
luminosity.  

Doing the converse on the selection procedure - favouring the broken
powerlaw models in these situations - removes most of the observations
with a disc detection at very low disc fractions (lower right in the
$L-T$ plane) in Appendix
Fig. \ref{fig:DL_DT_objs}.  Although this is may be a more
conservative selection procedure (no discs detected when they may be
uncertain), it is less informative on the behaviour of the BHXRBs on
the transitions between the different states.

It is therefore difficult to determine what the appropriate selection
procedure is when broken powerlaw and disc models are both good fits
to the data.  The spectral resolution and low energy range of the
\rxte\ \pca\ are very limiting in this case.  Future investigations or
instruments may aid in pin-pointing the disc behaviour in the
transition regions

\section{Powerlaw Evolution}\label{sec:PLevol}

Using the Disc Fraction Luminosity Diagram (DFLD) to investigate the behaviour of a BHXRB during an
outburst restricts what information can be extracted about the
variation of the powerlaw component, especially in the
hard/powerlaw-dominated state.  In this state, the variation in the
powerlaw slope or break energy do not effect the disc fraction, and so
all observations fall on a single line.  

In Appendix Fig. \ref{fig:DFLD_PLpanel} we show the DFLDs for the X-ray
binary in question, where the colourscale shows the variation of the
powerlaw slope (below the break if it is a broken powerlaw).  We show
in the neighbouring panel, the powerlaw slope against the total
luminosity for those observations with a disc fraction of $<0.2$
(powerlaw fraction $>0.8$).  This allows the change in the powerlaw
slope to be tracked in the powerlaw dominated state.  In a number of
binaries there are insufficient observations in the hard state to
determine any trend with time.  Also, 4U~1630-47, the variation of the 
powerlaw slope appears complex, with no easily discernable global
trend.  However, the outburst structure in this BHXRB is also complex
and so this variation is expected (see \citealp{Dunn10}).

In the majority of BHXRBs, in the low luminosity ``stalk'', the
powerlaw slope increases as the luminosity falls -- the spectrum
softens.  This has been seen in the 
HIDs of the BHXRBs before, as a change in the X-ray colour.  The
re-emergence of the disc at very low luminosities has been observed in
deep pointed observations and may also play a role in the softening of
the spectrum at low luminosities (see \citealp{Cabanac09}).  However, at these low luminosities the effects of the
Galactic Ridge Emission (GRE) play a role.  None of the BHXRBs were
fitted with a model which takes into account the effects of the
GRE in their vicinity \citep{Dunn10}.  At low luminosities, the GRE
can have an appreciable effect on the shape and flux of the spectrum.
However, the curvature in the stalk was seen in the study of GX~339-4
by \citet{Dunn08}, where the GRE was added to the model spectrum as a
fixed component.  Therefore only part of the softening at low
luminosities can be explained by the GRE.

However, at the top of the powerlaw dominated state in some BHXRBs, there is an
increase in the powerlaw slope.  The increase in $\Gamma$ has been
observed as the BHXRB enters the soft or disc-dominated state.
However, in these Figures, the $\Gamma$ increases far beyond what has
been observed in other studies of these BHXRBs, using the same data
(\eg \citealp{Motta09}).  The likeliest explanation is that the broken
powerlaw is accounting for a rising disc, which is not being well fit
by a disc model.  What can also be seen is that these softer
powerlaw slopes are from broken powerlaws, and they have a comparatively
low break energy.  As noted in \citet{Dunn10} and Section \ref{sec:bpl} there is a
possibility for the broken 
powerlaw to mimic the disc (and powerlaw) model.  It is probable that as
the disc rises in luminosity the limited spectral range of the
\rxte\ \pca\ means that the curvature of the disc cannot be
determined, and the broken powerlaw resulted being a better fit.
Restricting the powerlaw break to being above the peak mentioned in
Section \ref{sec:bpl} would prevent this occurring.  However, the
accuracy of the fitted disc parameters is not clear.  Therefore, the
softening of the powerlaw on the transitions to the disc-dominated
state is likely to be the result of the limitations of the
\rxte\ \pca.

\section{Summary}

We have investigated the behaviour of the disc and powerlaw components
in the 25 BHXRBs presented in \citet{Dunn10}.  In the majority of
BHRXBs in which at least most of an outburst has been observed, the
disc luminosity scales close to $T^4$ in the most disc dominated
observations.  This behaviour had been seen in other
studies (\eg \citealp{Gierlinski04}).  The scaling of $T^4$ implies
that both the disc's inner radius, $R_{\rm in}$ and the colour
correction factor, $f_{\rm col}$ are relatively constant in the most
disc dominated states.  A number of BHXRBs do not show a clear $T^4$
relation, but these could be the result of the limitations of the
model fitting routine or the frequency of observations.

However, in observations where the disc is no longer overly dominant,
there are deviations from the $T^4$ law.  If interpreted as changes in
the disc's inner radius, these deviations imply that the disc is
moving inwards at the end of an outburst, and outwards at the
beginning of the outburst.  Although we do not rule this behaviour
out, it seems an unlikely scenario.  If these deviations are
attributed to changes in the colour correction factor, then $f_{\rm
  col}$ rises as the disc fraction decreases.

There are a number of observations in which the disc parameters
determined are unlikely to be explained by reasonable values for the
disc radius or the $f_{\rm col}$.  The spectral fits for these
observations tend to have $\chi^2$ values which are very similar to
those for the broken-powerlaw model, which makes selecting the most
appropriate model difficult.  This also makes determining the true
behaviour of the disc temperature, radius and $f_{\rm col}$ in these
intermediate states difficult.

The luminosity of the disc in Eddington units can be calculated from
theoretical arguments from the BHXRB parameters and the disc
temperature.  When calculating the observed luminosity the distance of
the BHXRB system also enters the calculation.  The combination of
these two calculations allows the ratio of $D/M$ to be estimated from
the $L-T$ relation of the BHXRB disc.  We have therefore placed limits
on the range values of $D$ and $M$ values allowed for different values
of the spin of the black hole.

\section*{Acknowledgements}

We thank the referee for a prompt and useful report which helped
improve the manuscript.  RJHD acknowledges support from the Alexander
von Humboldt Foundation.  EK acknowledges financial support by a Marie
Curie Reintegration Grant under contract number PERG05-GA-2009-243869.
TMB acknowledges support ASI via 
contract I/088/06/0  and thanks the International Space Science
Institute (ISSI).  The research leading to these results has received
funding from the European Community’s Seventh Framework Programme
(FP7/2007-2013) under grant agreement number ITN 215212 “Black Hole
Universe”.  This research was supported
by the DFG cluster of excellence ‘Origin and Structure of the
Universe’ (www.universe-cluster.de).   

\bibliographystyle{mn2e} 
\bibliography{mn-jour,dunn}

\section*{Appendix}

\renewcommand{\thefigure}{A.\arabic{figure}}
\setcounter{figure}{0}
\begin{figure*}
\centering
\includegraphics[width=0.3\textwidth]{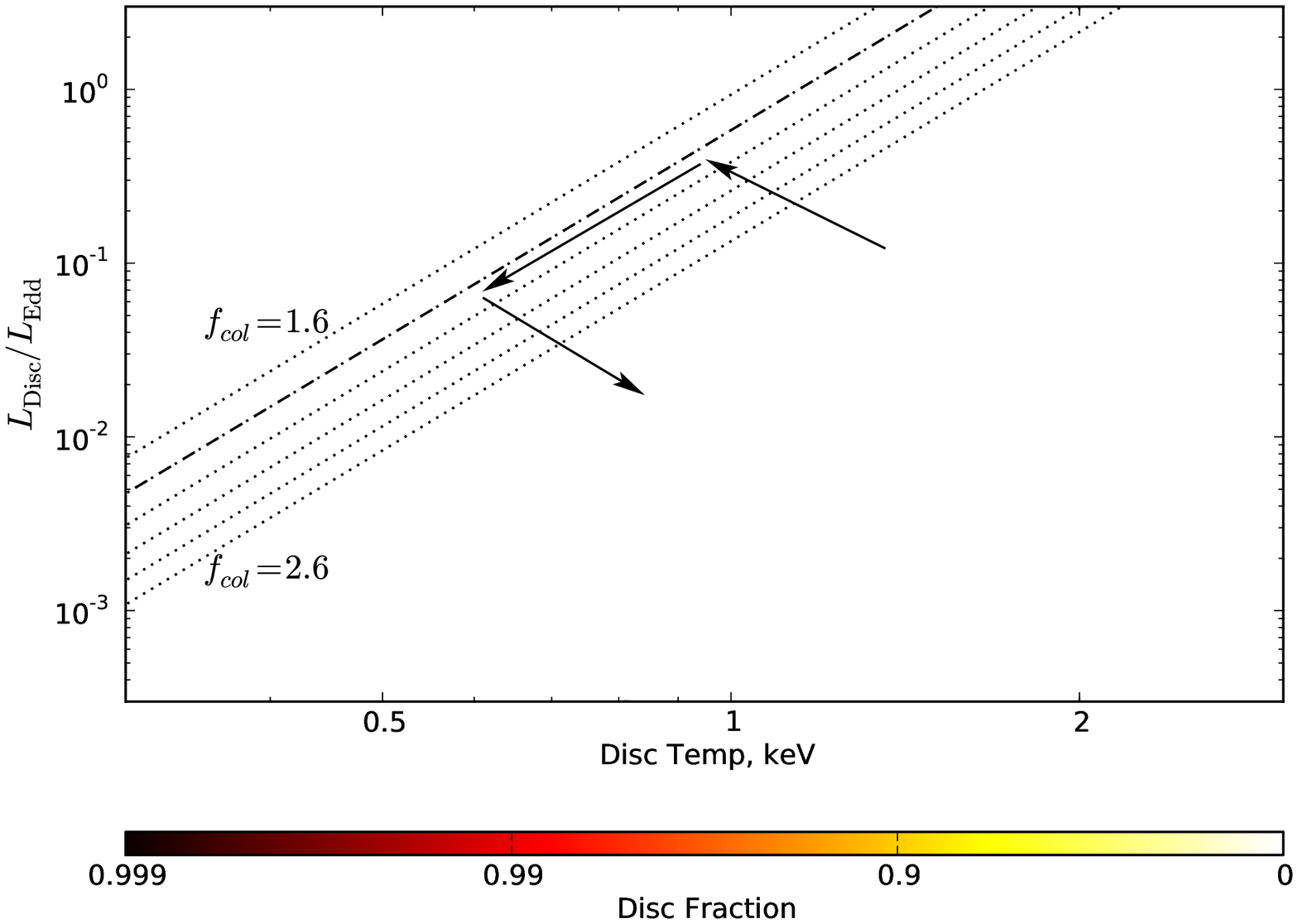}
\includegraphics[width=0.3\textwidth]{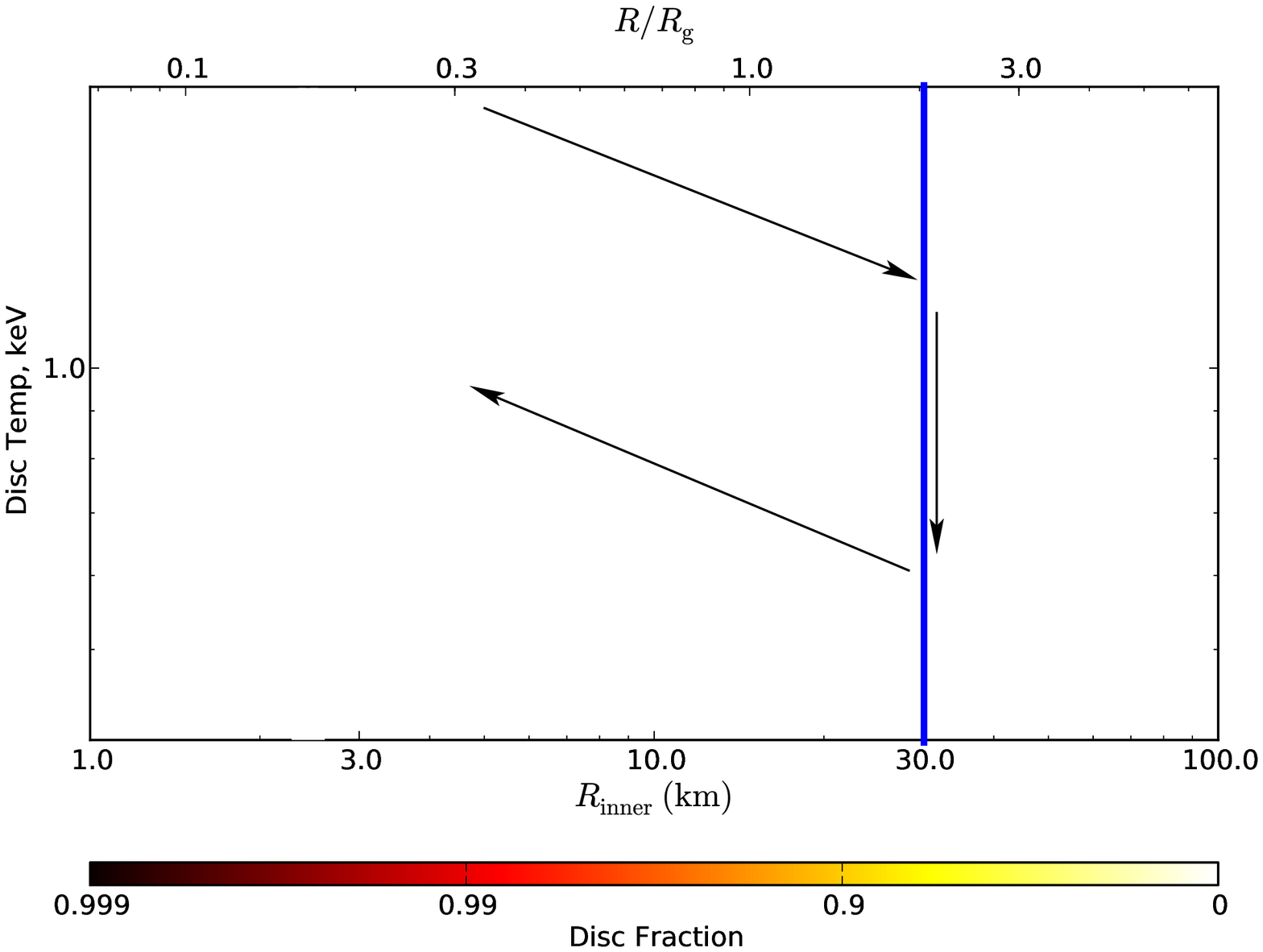}
\includegraphics[width=0.3\textwidth]{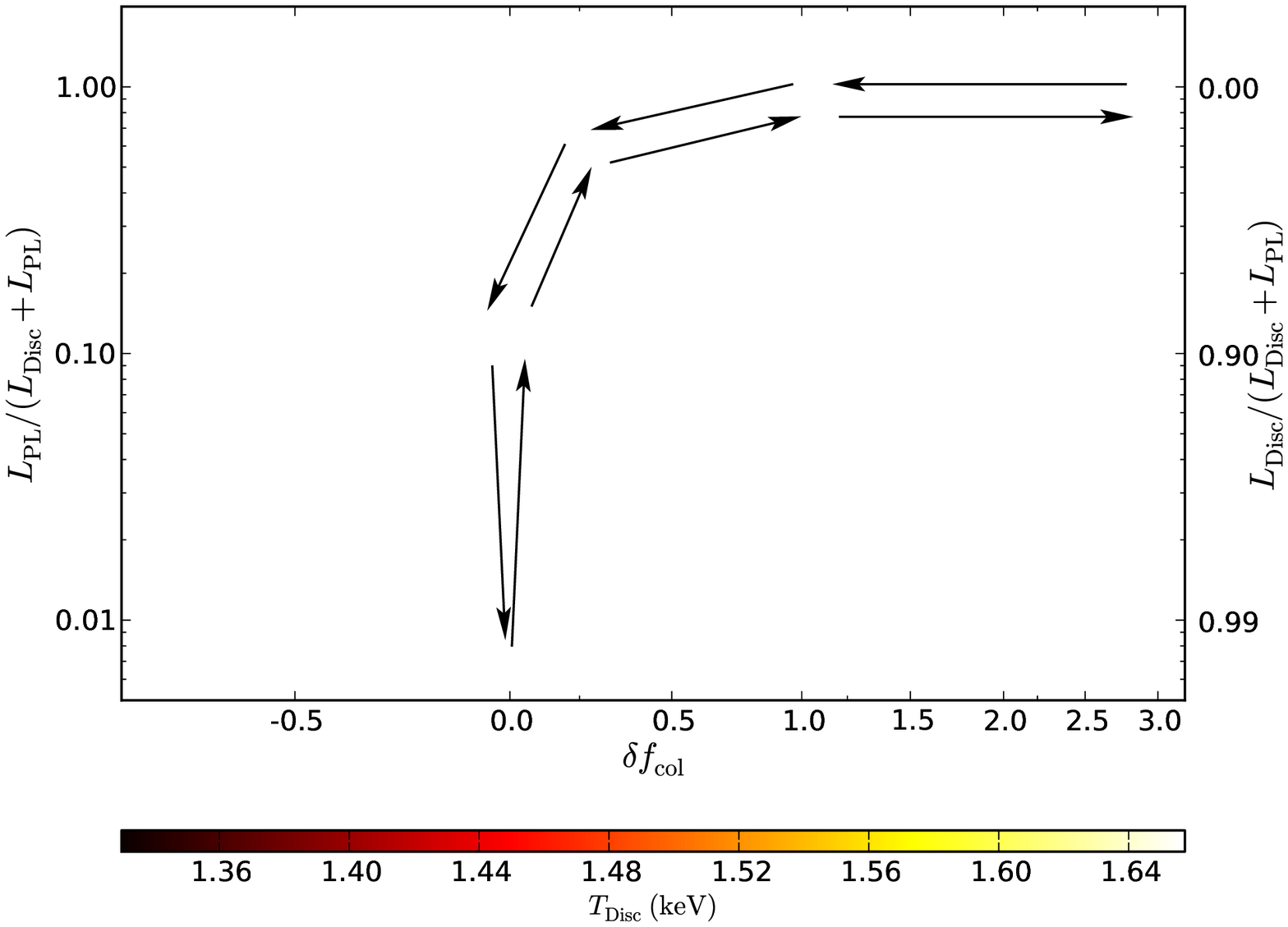}
\caption{\label{fig:DL_DT_objs} Schematic figures of {\scshape left}: Disc luminosity against
disc temperature, {\scshape middle}: Disc radius against disc
temperature and {\scshape right}: the excess $\delta f_{\rm col}$
required against the disc fraction.  The arrows show the motion
through the diagram.  In the right-hand figure, the offset between the
inward and outward tracks are for clarity only.}
\end{figure*}
\addtocounter{figure}{-1}
\begin{figure*}
\centering
\includegraphics[width=0.4\textwidth]{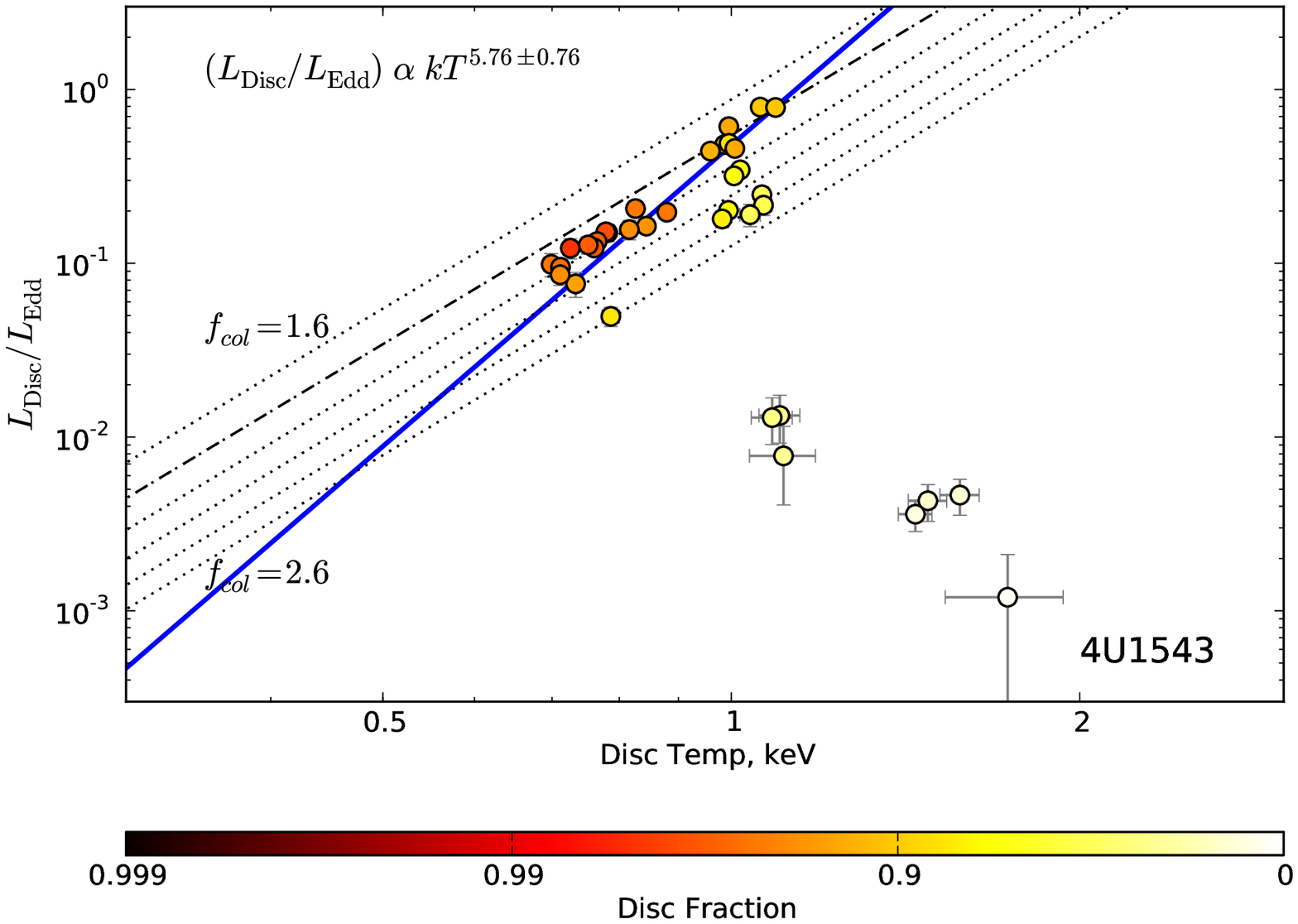}
\includegraphics[width=0.4\textwidth]{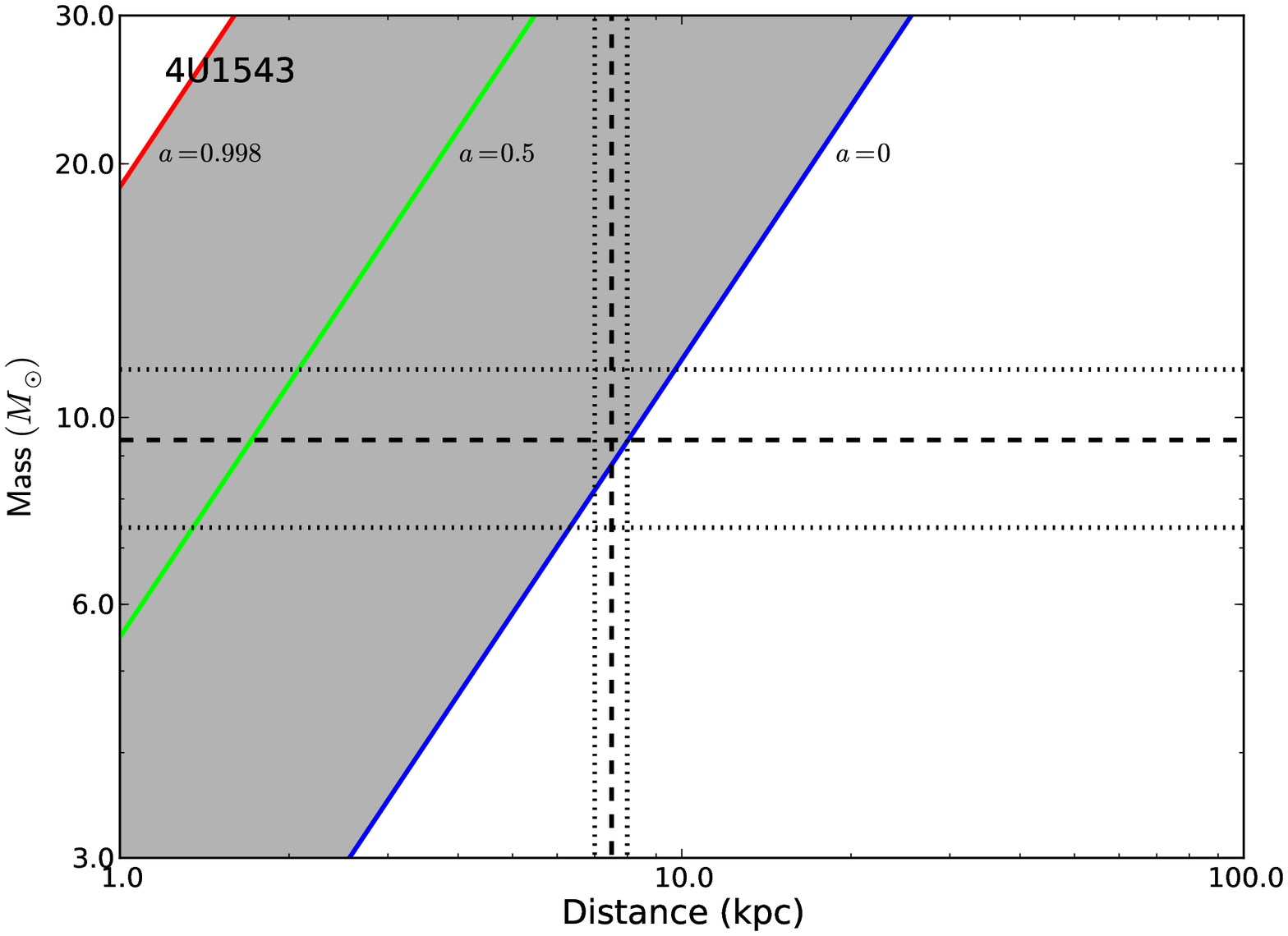}
\includegraphics[width=0.4\textwidth]{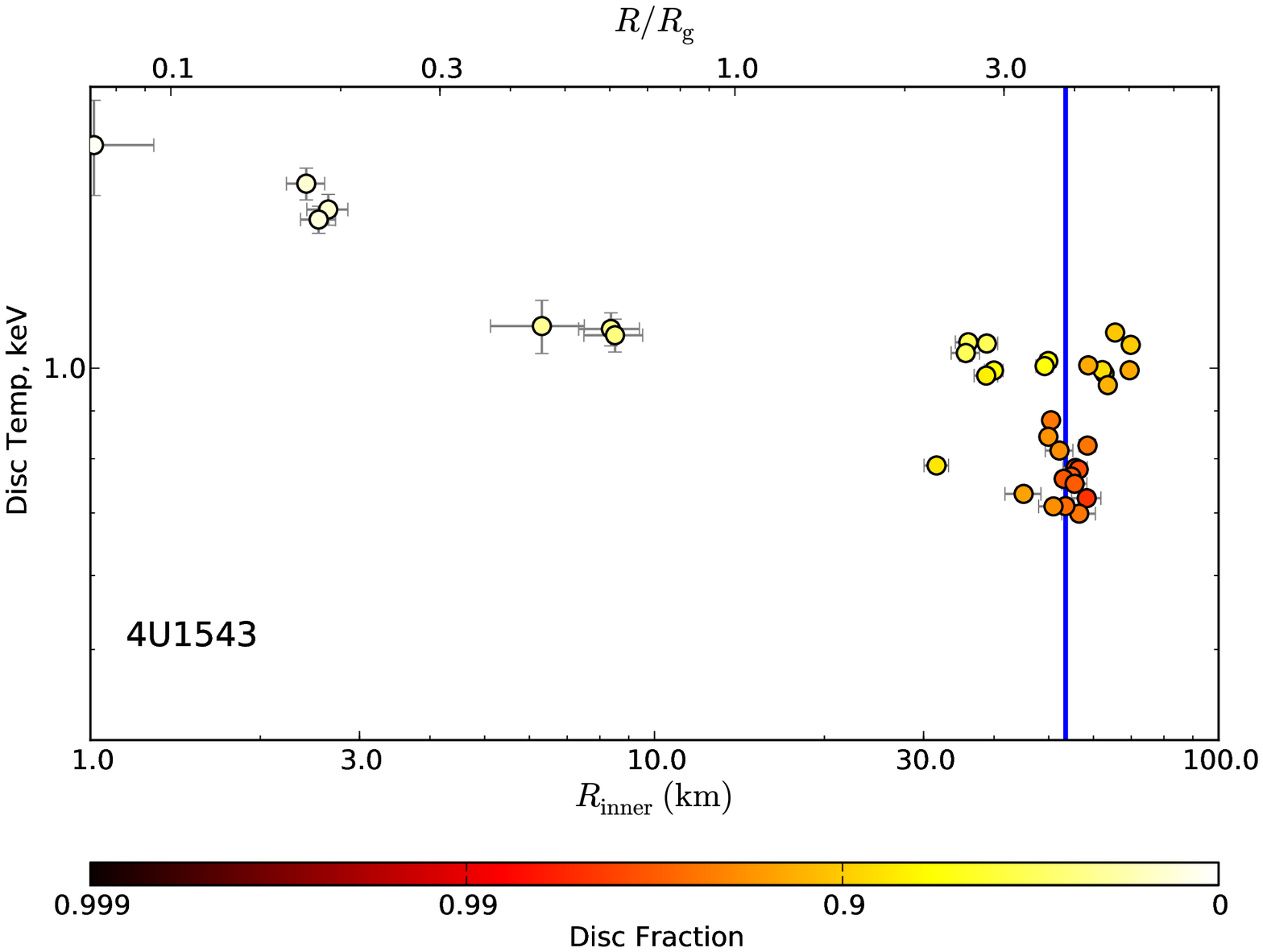}
\includegraphics[width=0.4\textwidth]{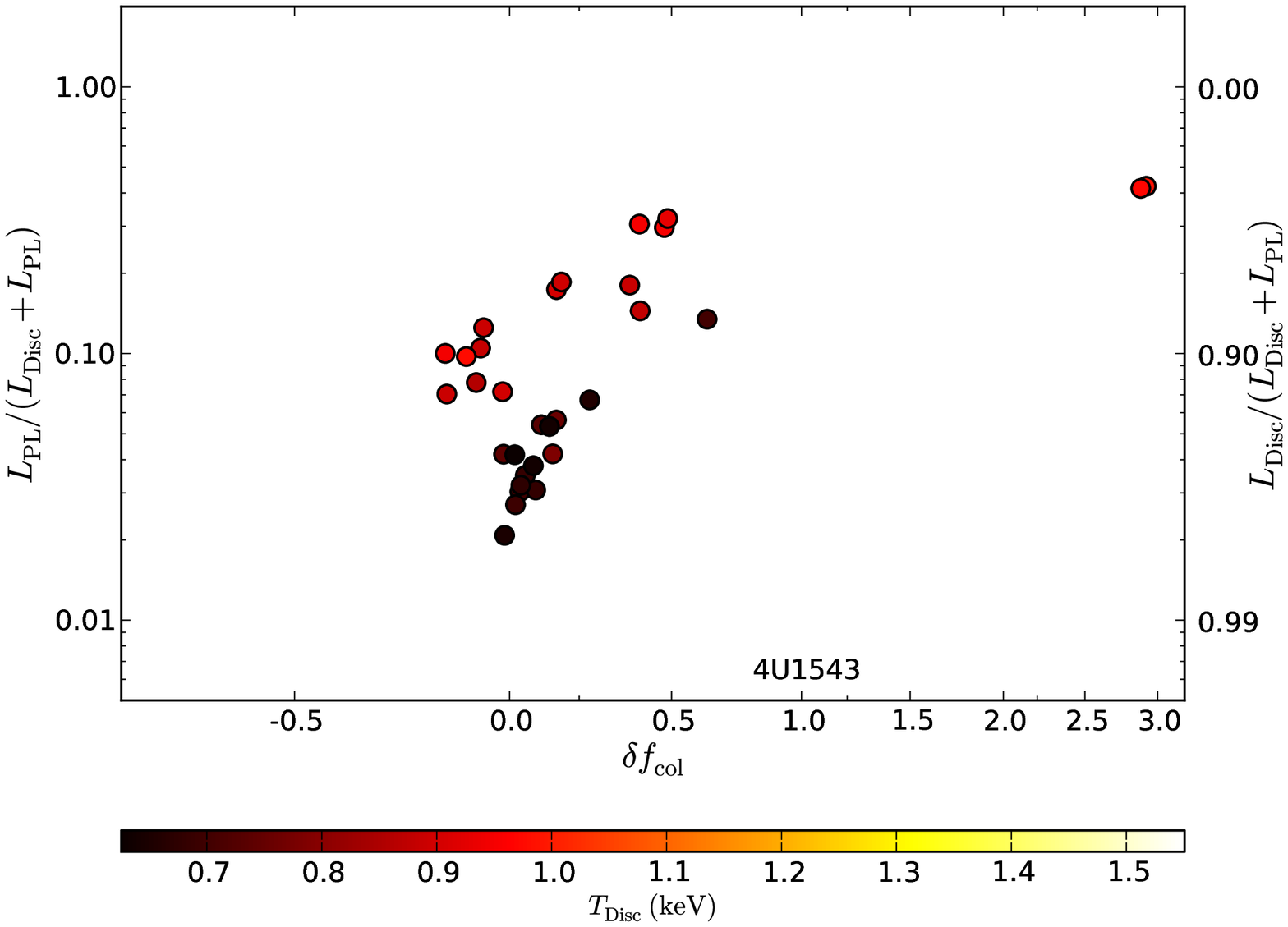}
\caption{(cont) 4U~1543-47 {\scshape top left} We show the Luminosity
of the disc as a function of the temperature of the {\scshape discbb}
component in {\scshape xspec}.  The colour scale is the disc
fraction of the observation.  The theoretical $L-T$ relations are shown
by the sets of dotted
line for a number of values of $f_{\rm col}$.  The solid line shows the fit to the most disc dominated
states (Disc Fraction $>0.8$).  {\scshape top right}  We show the range of
distances and masses allowed, if the observations are to match the
theoretical relation.  The blue line is for a non-rotating black hole,
and the direction for increasing $a$ is shown.  The dashed and
dotted lines show the current values and uncertainties on the masses
and distances where available.  {\scshape bottom left} We show the
variation of the inner disc radius ($R_{\rm in}$) with the disc
temperature.  The colour scale is the disc
fraction of the observation.  The solid line shows the average $R_{\rm
  in}$ of the most disc dominated states (Disc Fraction $>0.8$).  {\scshape
  bottom right} The excess the colour correction required, $\delta f_{\rm
  col}$, with the disc fraction.  The disc temperature is the colour scale.}
\end{figure*}
\addtocounter{figure}{-1}

For each BHXRB in our sample we show in Fig. \ref{fig:DL_DT_objs}
{\scshape top left} the disc temperature as a
function of the disc luminosity, along with the theoretically expected
relation for the case that the BH is not rotating for a variety of
values for $f_{\rm col}$.  The $f_{\rm col}=1.8$ line is depicted
thicker than the rest, which increment in $0.2$ intervals.  The line which
best fits the most disc dominated spectra is shown, as well as its
slope.  We show {\scshape top right} the region in the distance-mass plane which is
allowed by the theoretical expectation of the luminosity-temperature
relation and the observed data points.  The best determined values for
the distance and mass along with their uncertainties are also shown
where available.

In the {\scshape bottom left} we show the variation of the disc radius
($R_{\rm in}$) with the disc temperature.  The colourscale is the disc
fraction.  The line is the average of the disc radii for the
observations which have a disc fraction $>0.8$.  The variation of the
$f_{\rm col}$ with the disc fraction is show in 
the {\scshape bottom right}.  The disc temperature is the colour
scale.  

By comparing all three of the scatter plot figures, the behaviour of
the disc in the BHXRB in the outburst becomes clearer.

\begin{figure*}
\centering
\includegraphics[width=0.4\textwidth]{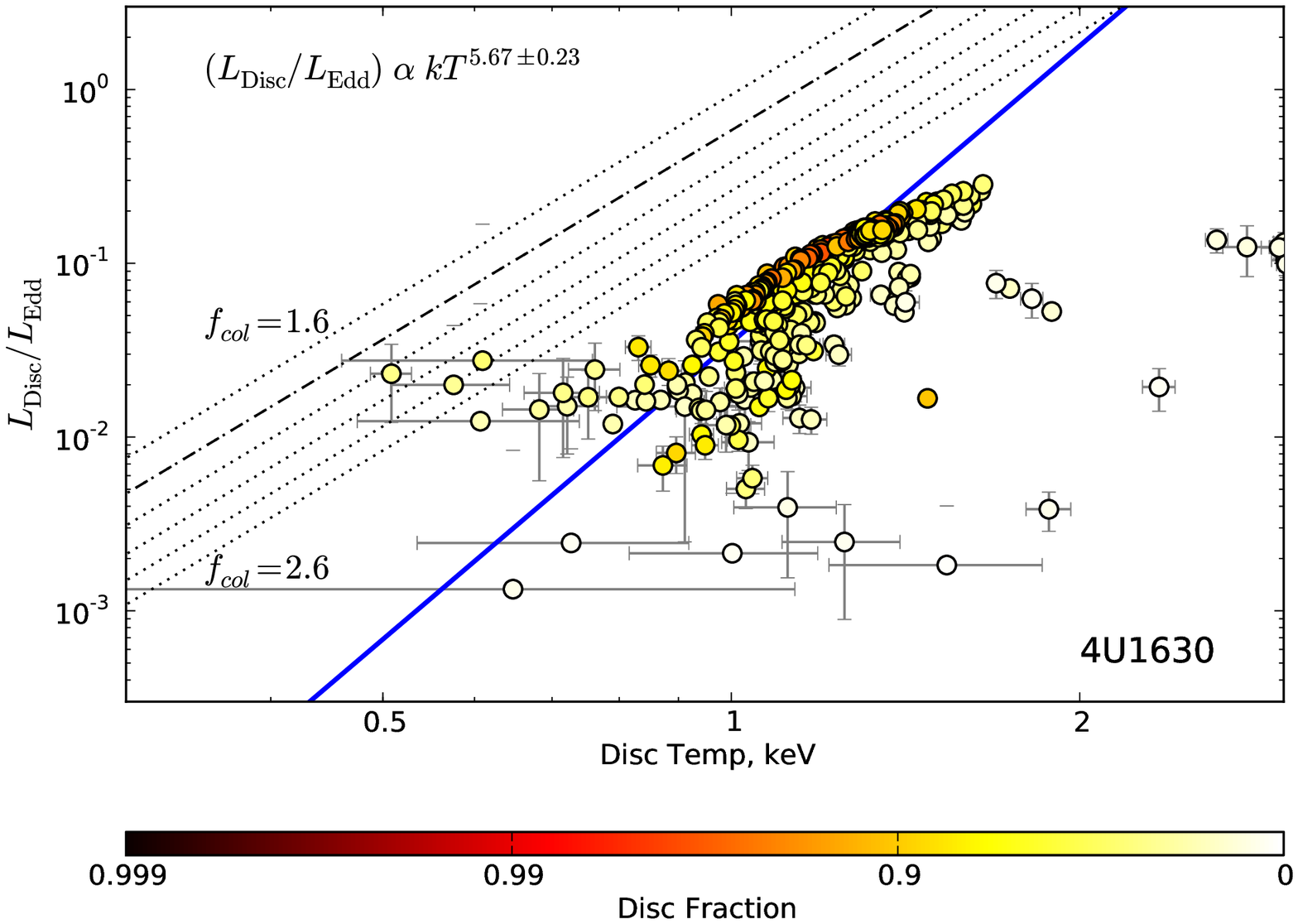}
\includegraphics[width=0.4\textwidth]{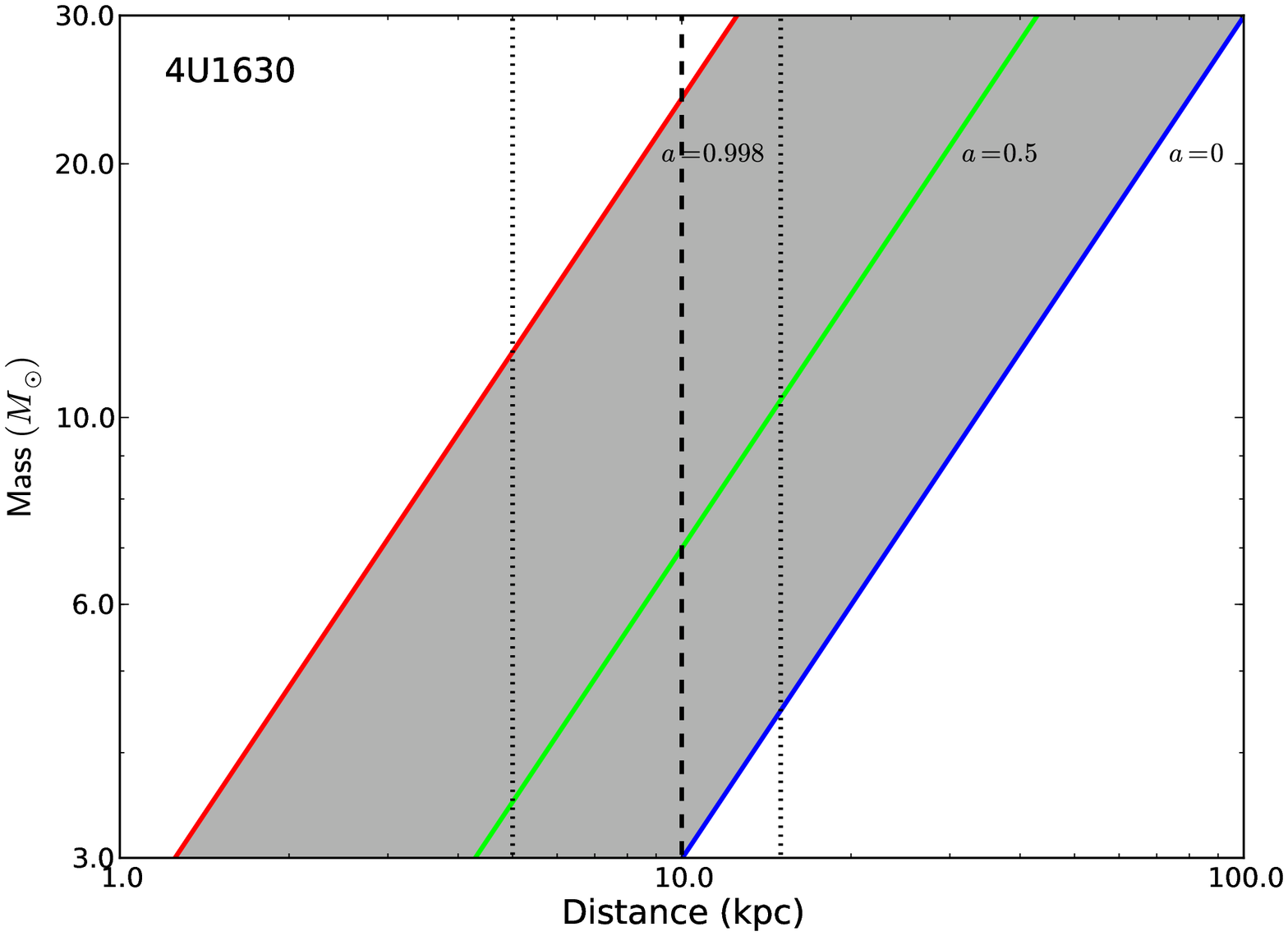}
\includegraphics[width=0.4\textwidth]{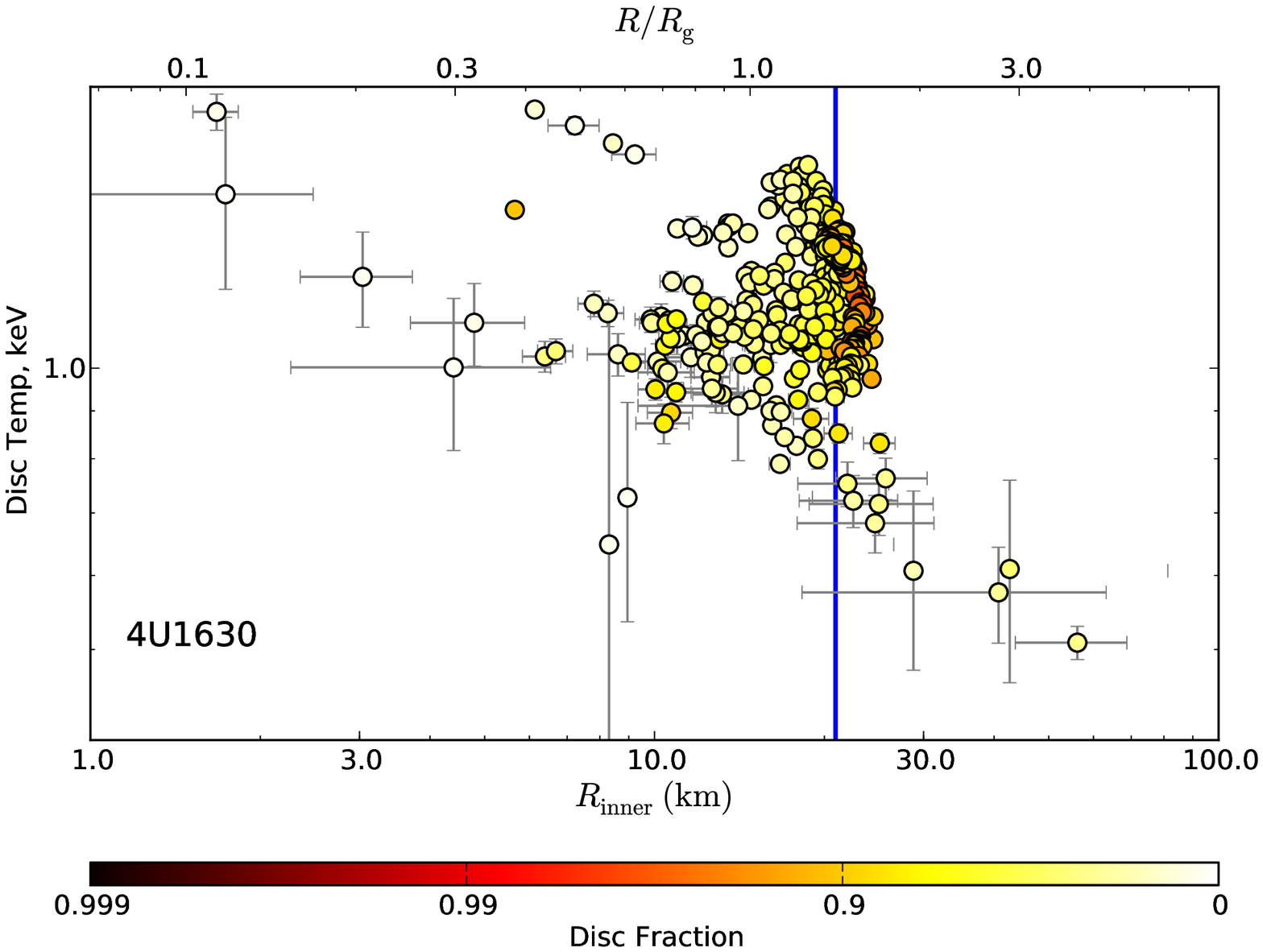}
\includegraphics[width=0.4\textwidth]{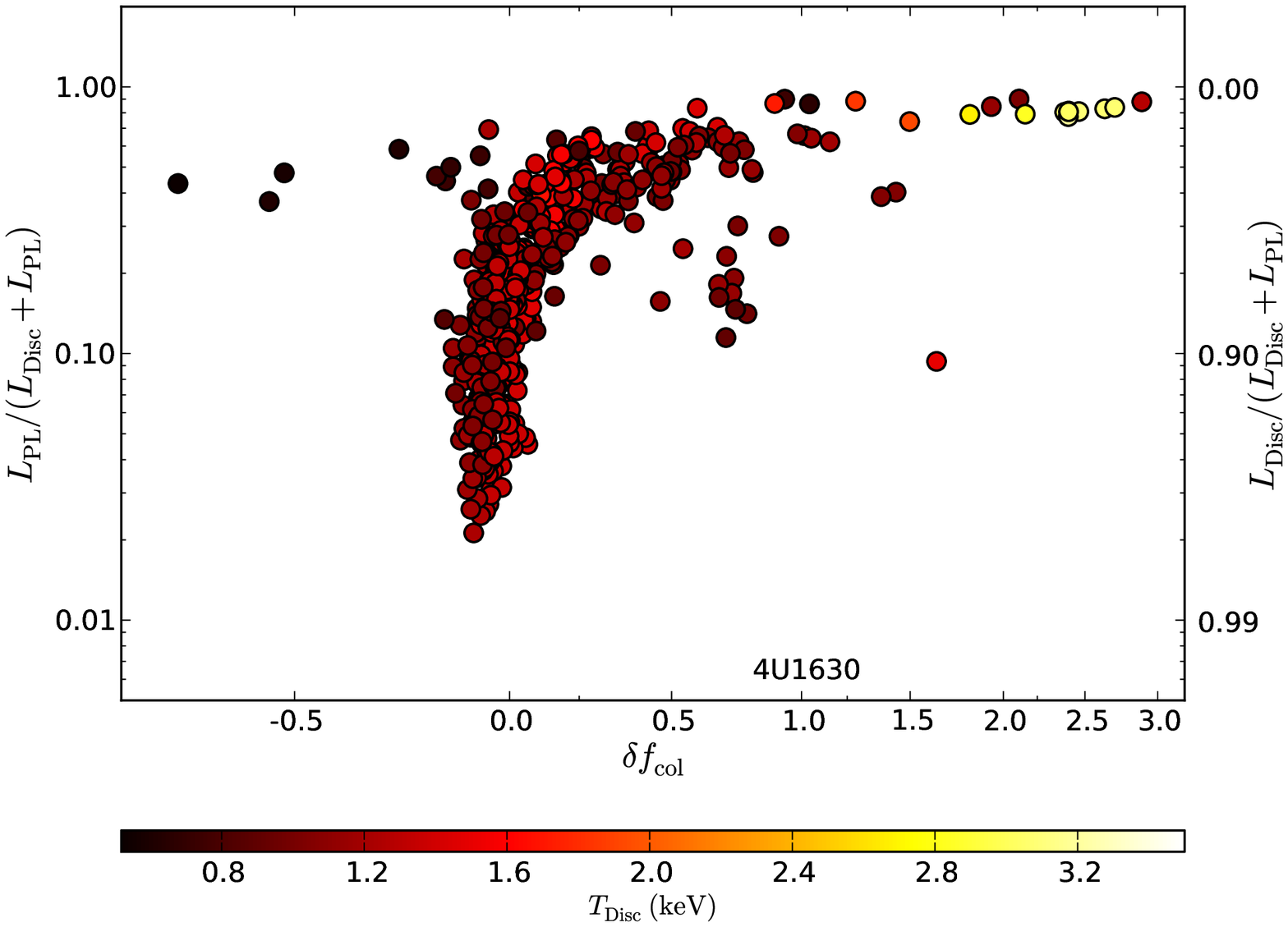}
\caption{(cont) 4U~1630-47 }
\end{figure*}
\addtocounter{figure}{-1}
\begin{figure*}
\centering
\includegraphics[width=0.4\textwidth]{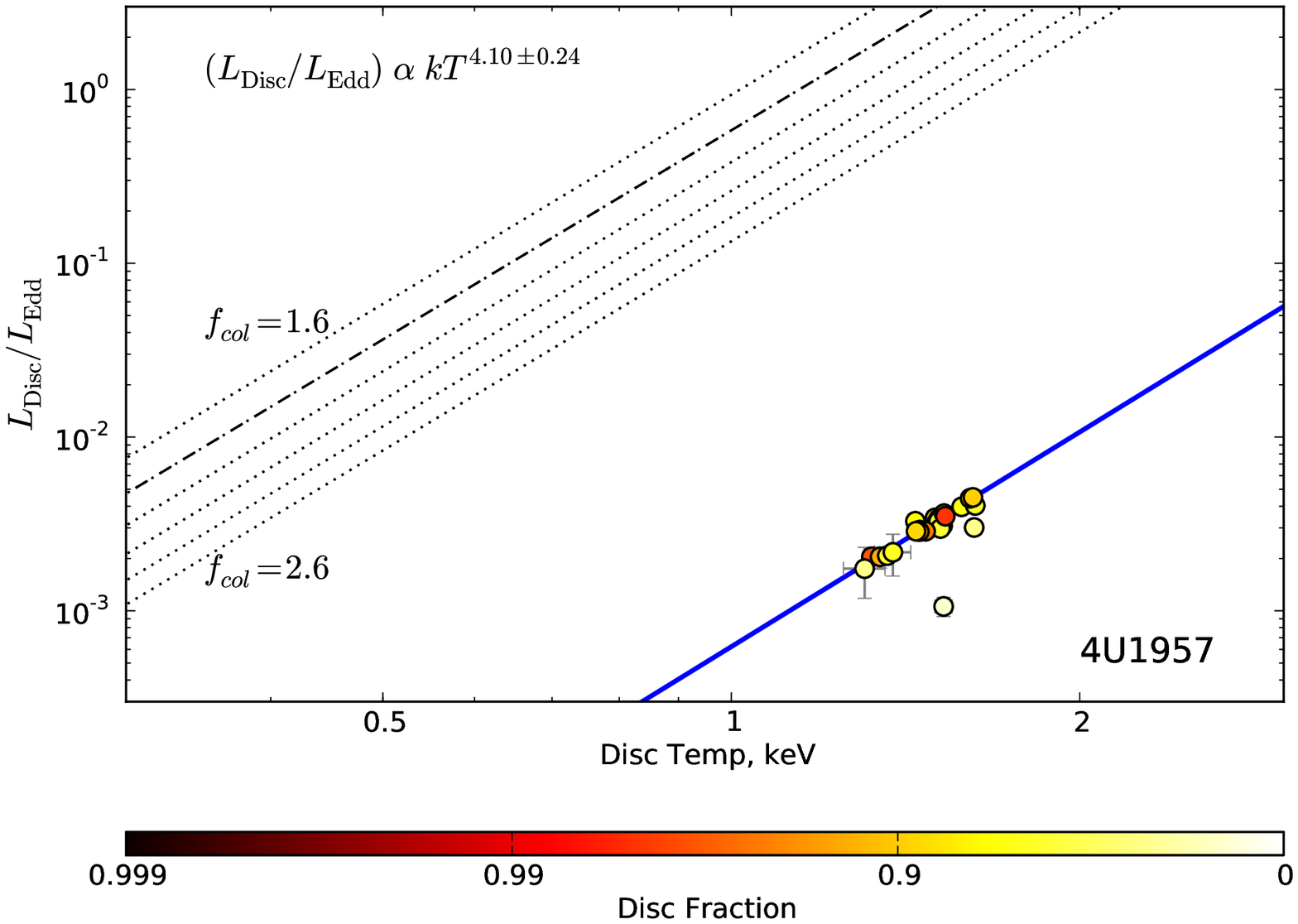}
\includegraphics[width=0.4\textwidth]{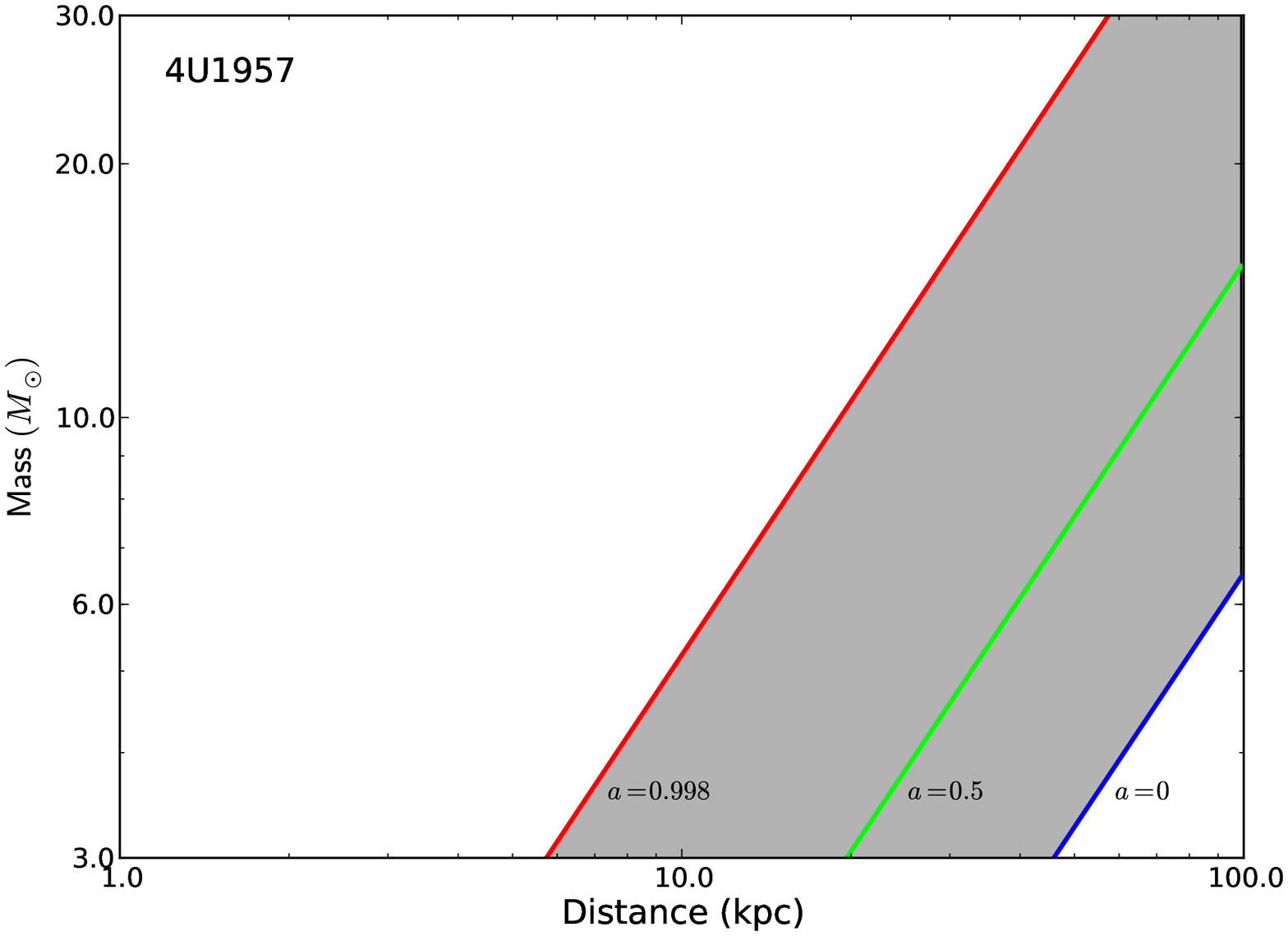}
\includegraphics[width=0.4\textwidth]{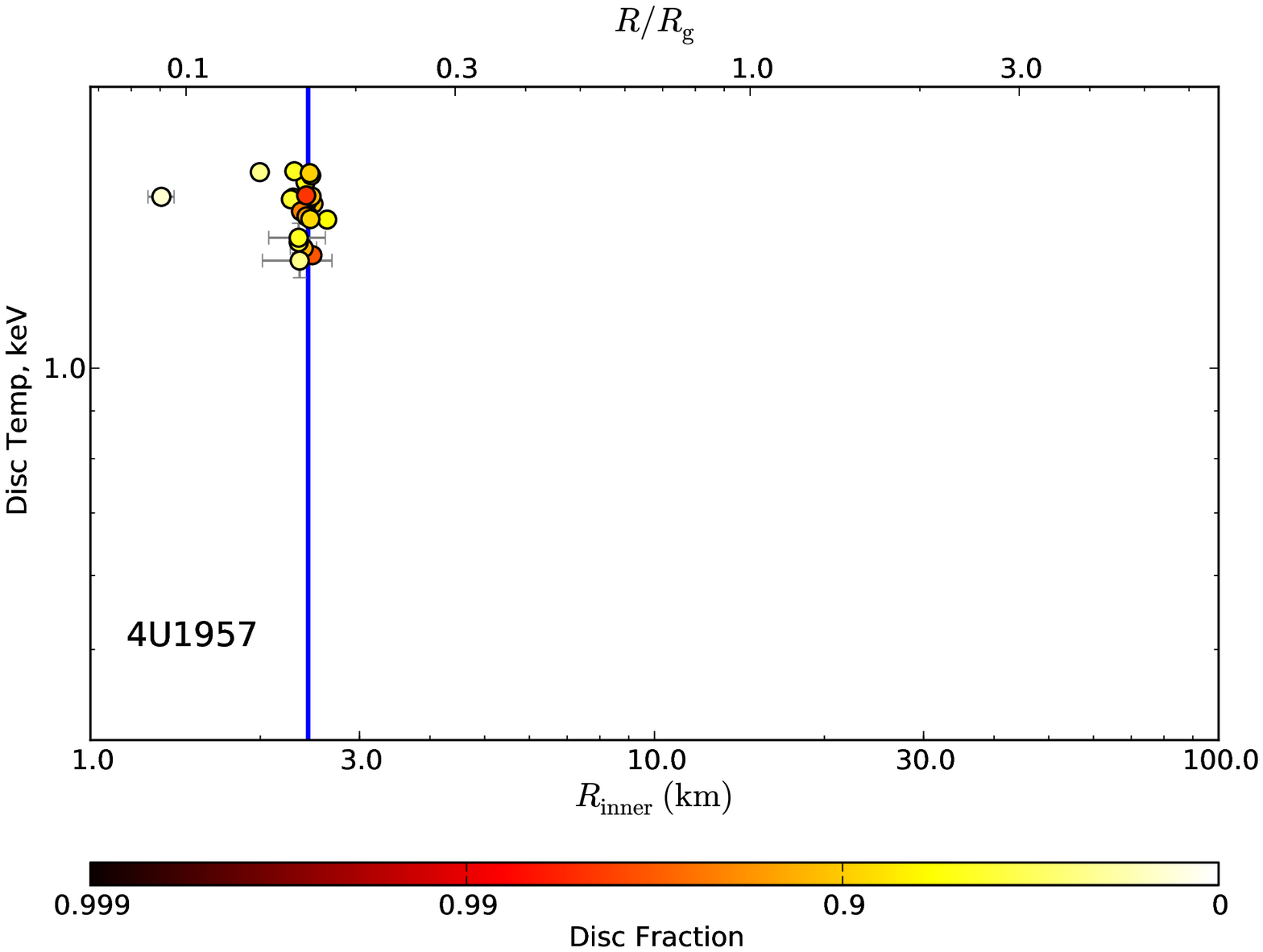}
\includegraphics[width=0.4\textwidth]{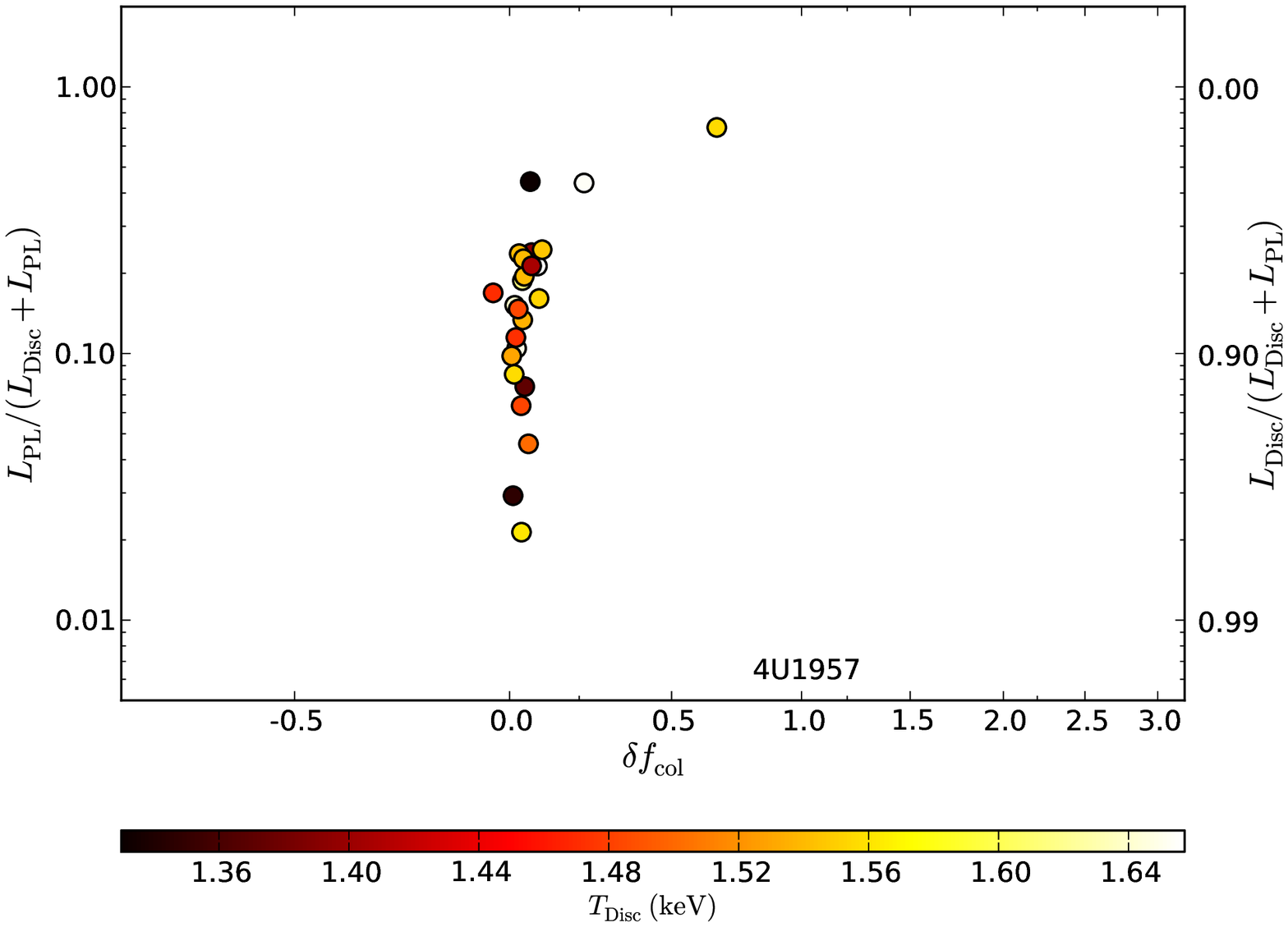}
\caption{(cont) 4U~1957+115}
\end{figure*}
\clearpage
\addtocounter{figure}{-1}
\begin{figure*}
\centering
\includegraphics[width=0.4\textwidth]{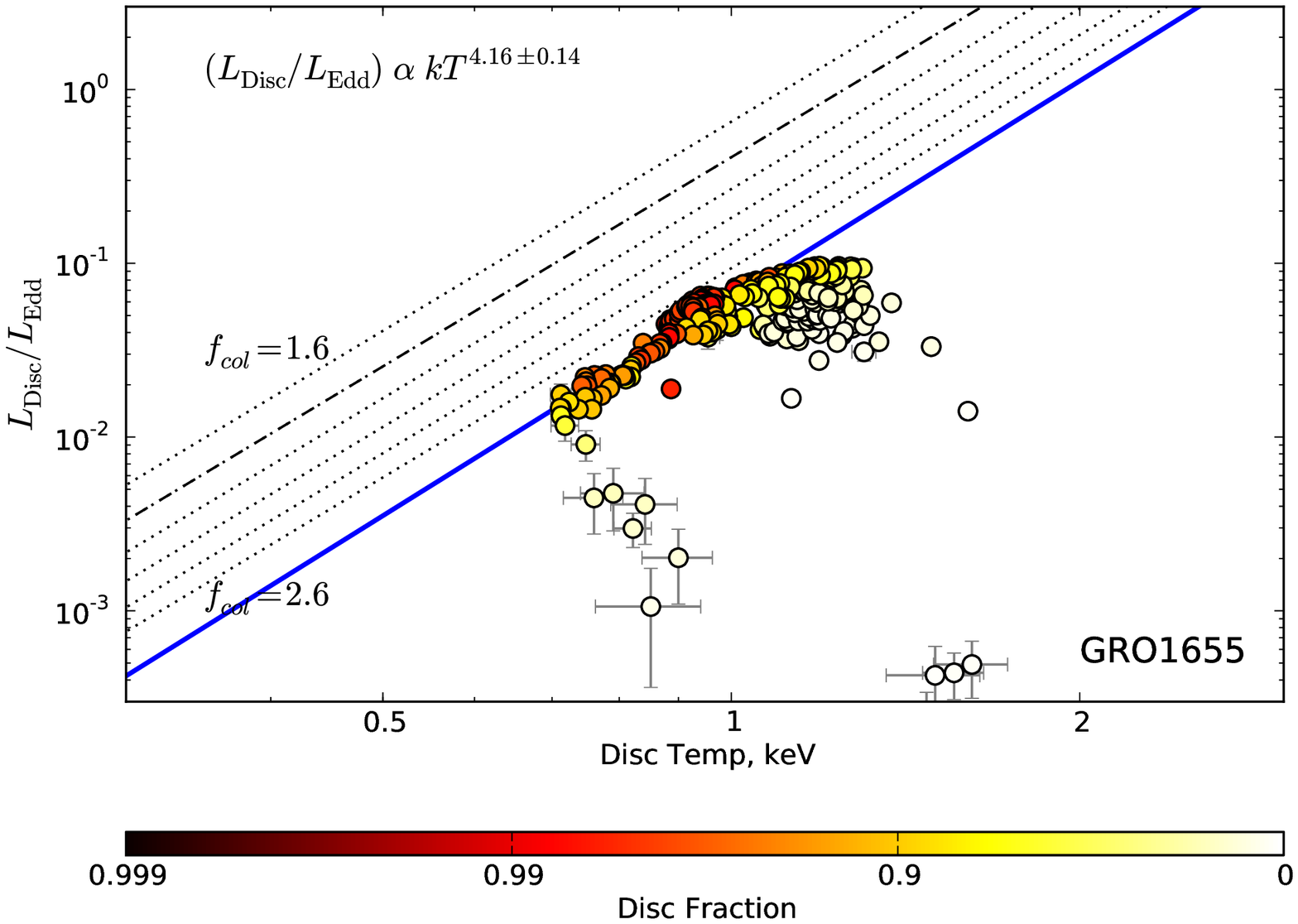}
\includegraphics[width=0.4\textwidth]{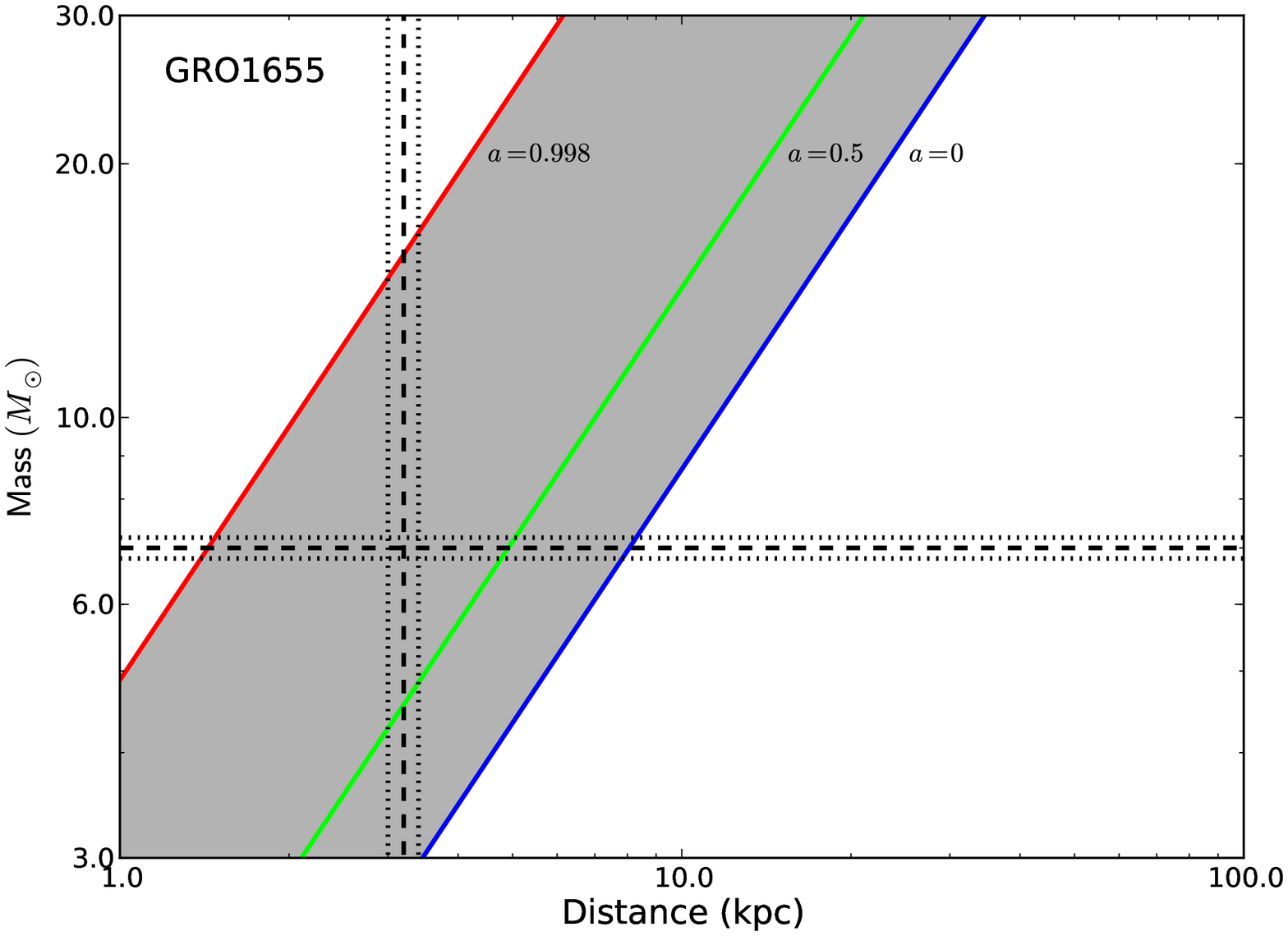}
\includegraphics[width=0.4\textwidth]{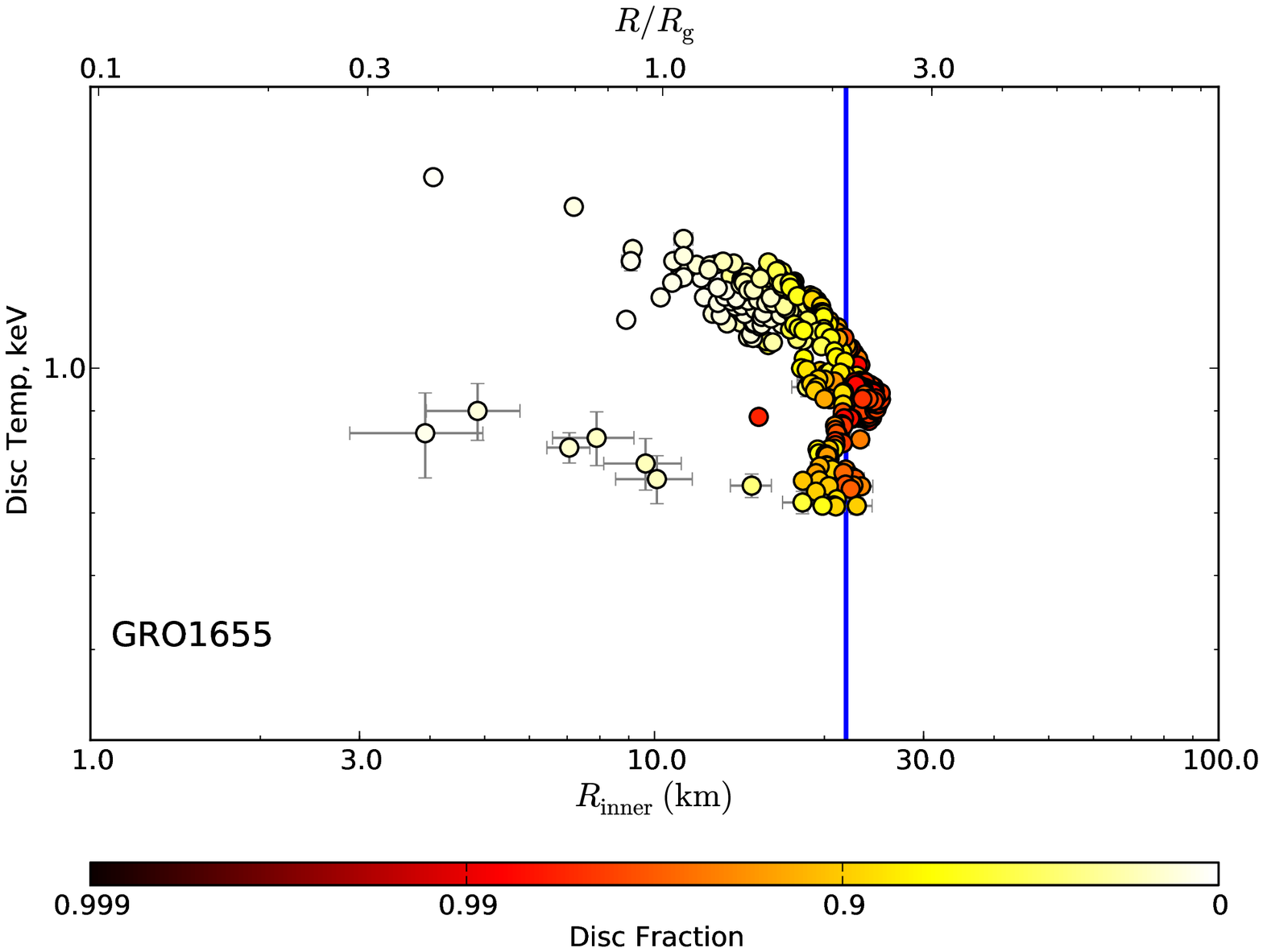}
\includegraphics[width=0.4\textwidth]{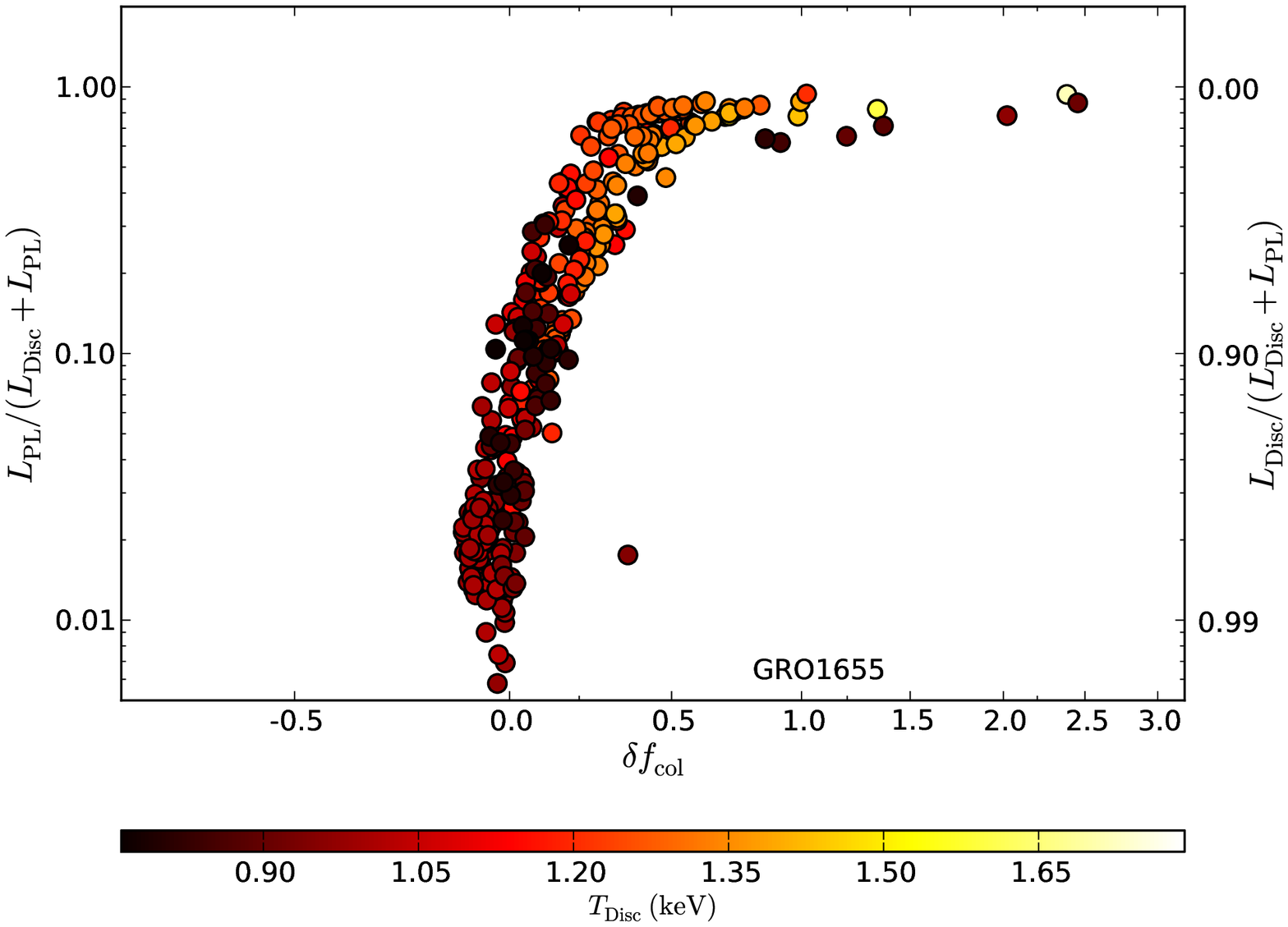}
\caption{(cont) GRO~J1655-40}
\end{figure*}
\addtocounter{figure}{-1}
\begin{figure*}
\centering
\includegraphics[width=0.4\textwidth]{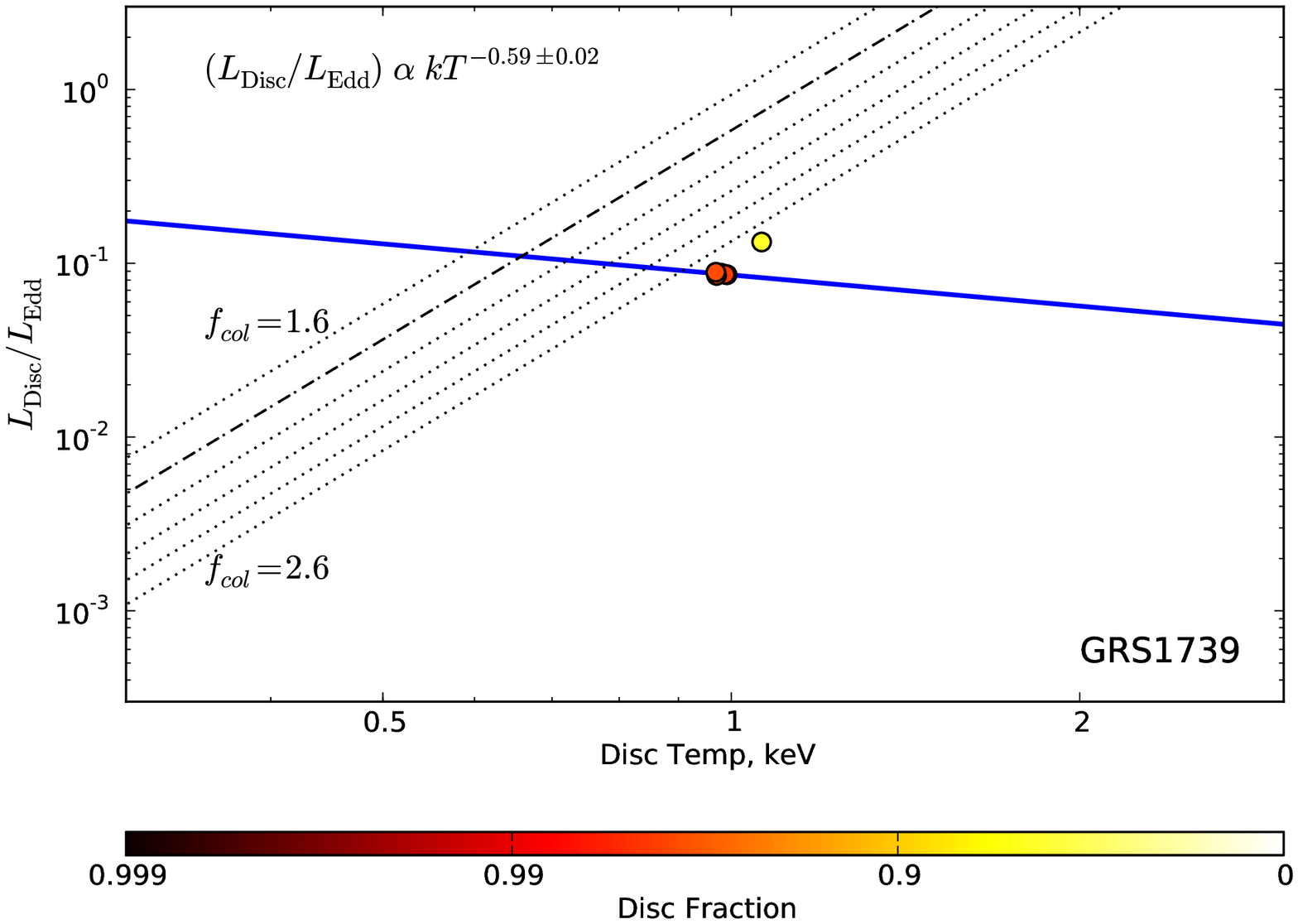}
\includegraphics[width=0.4\textwidth]{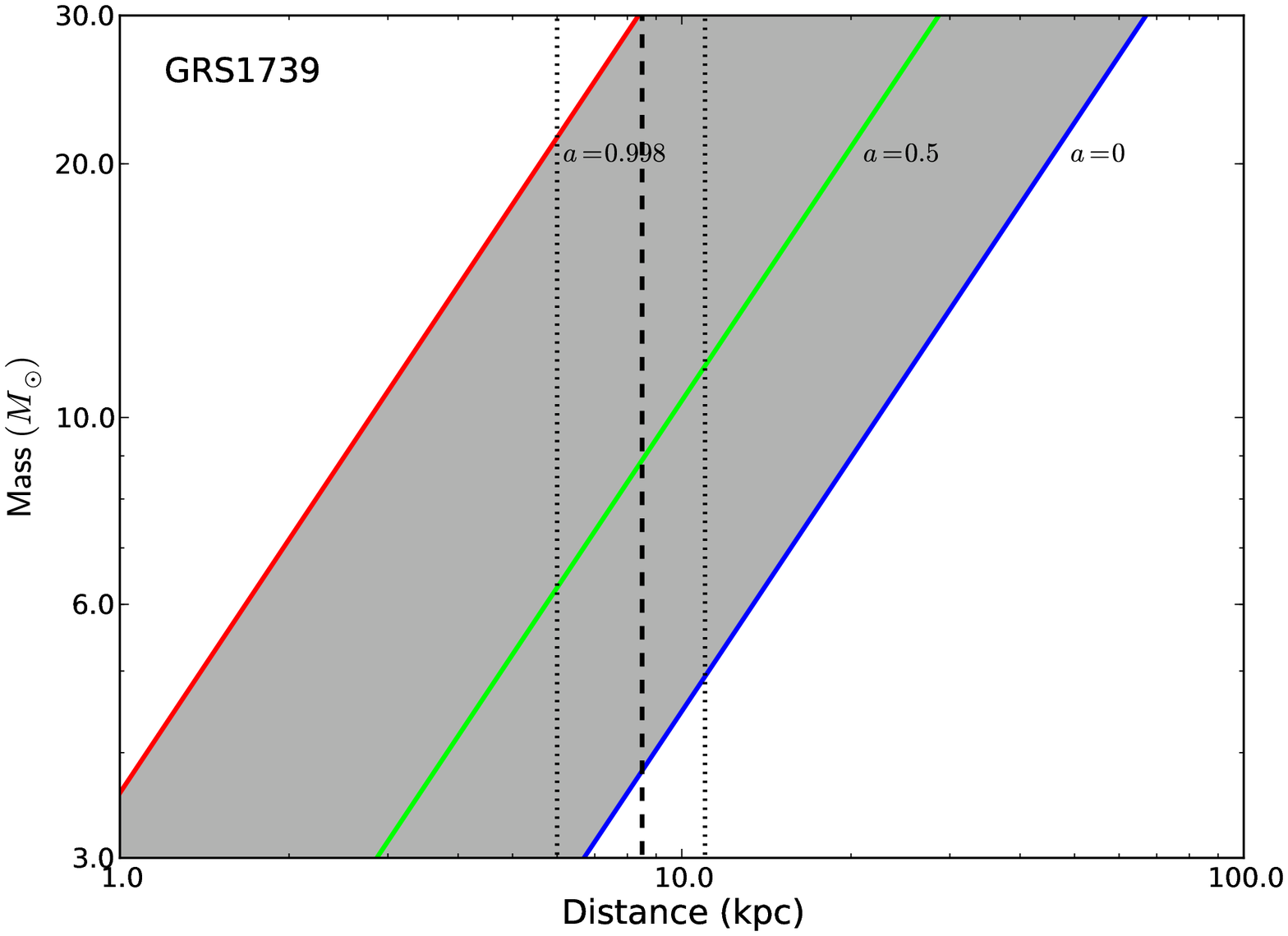}
\includegraphics[width=0.4\textwidth]{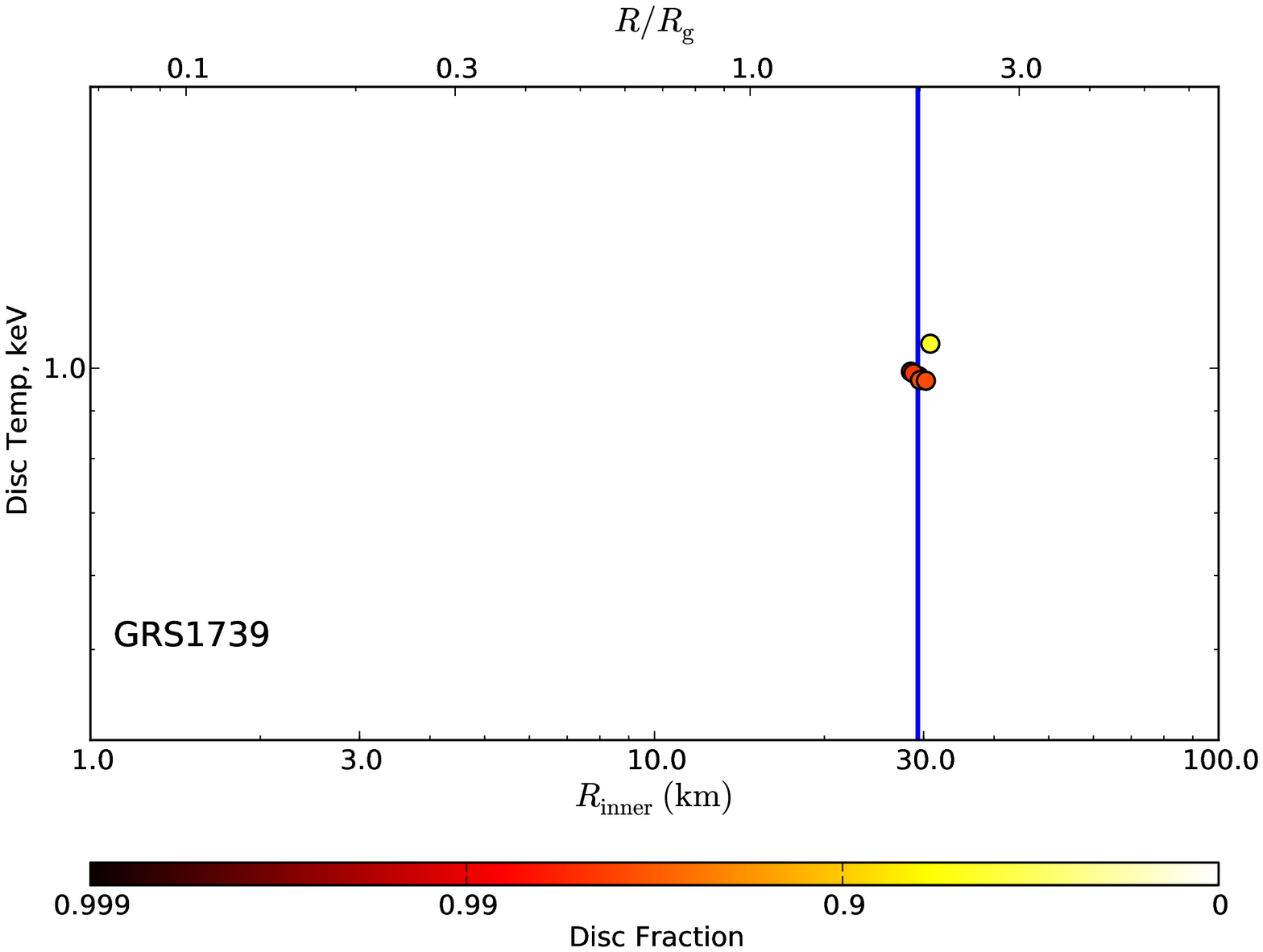}
\includegraphics[width=0.4\textwidth]{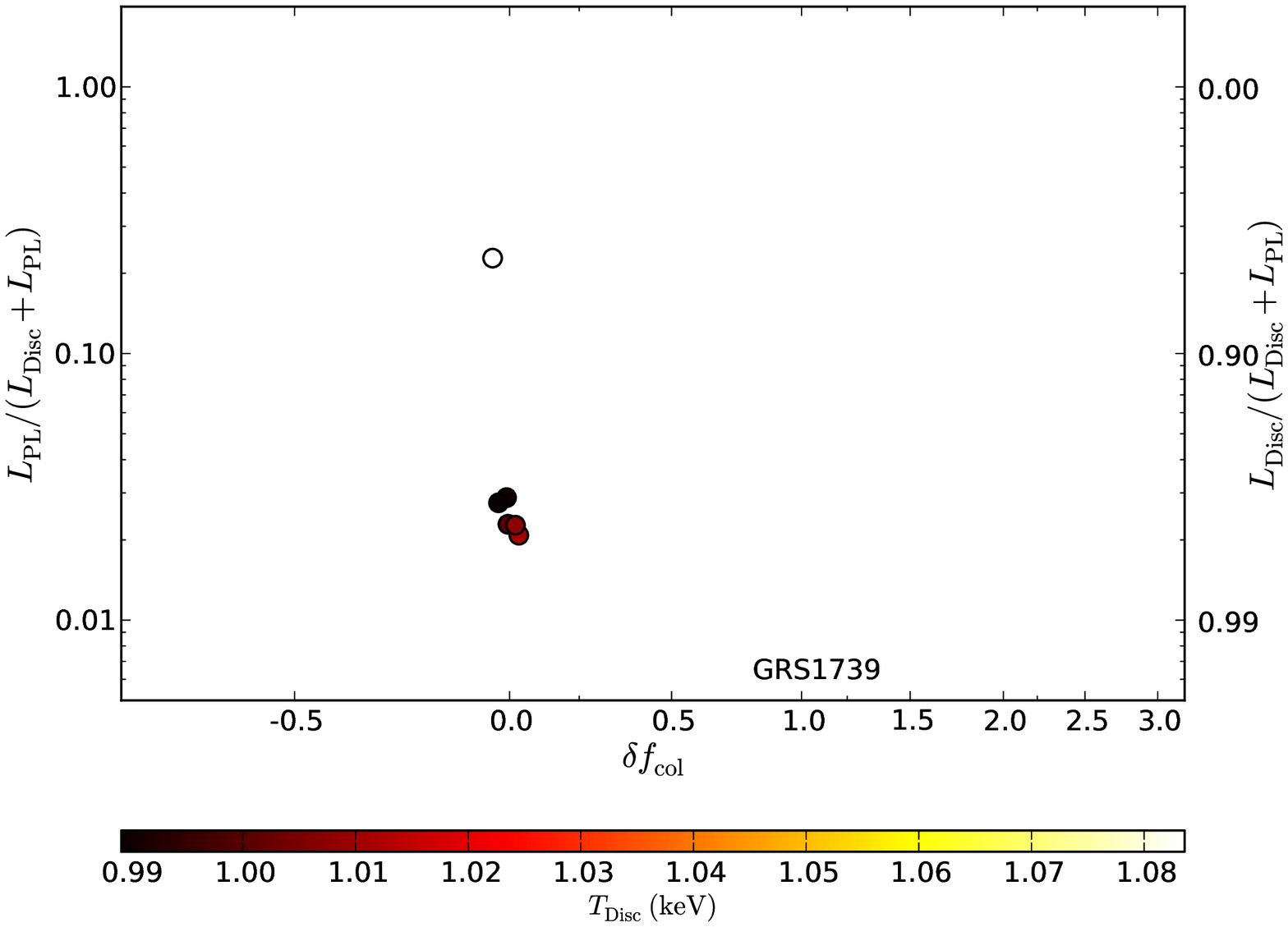}
\caption{(cont) GRS~1737-31}
\end{figure*}
\addtocounter{figure}{-1}
\begin{figure*}
\centering
\includegraphics[width=0.4\textwidth]{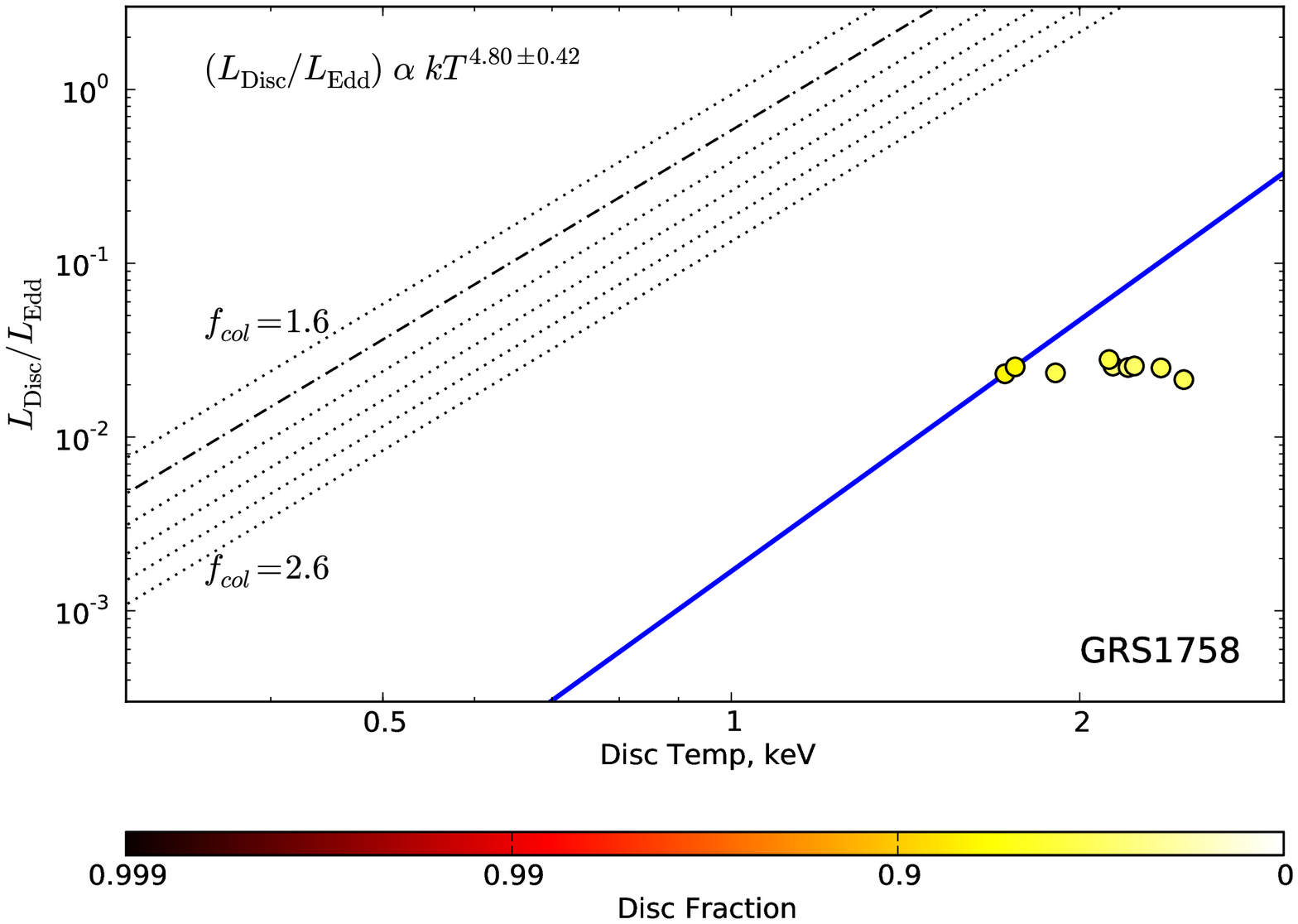}
\includegraphics[width=0.4\textwidth]{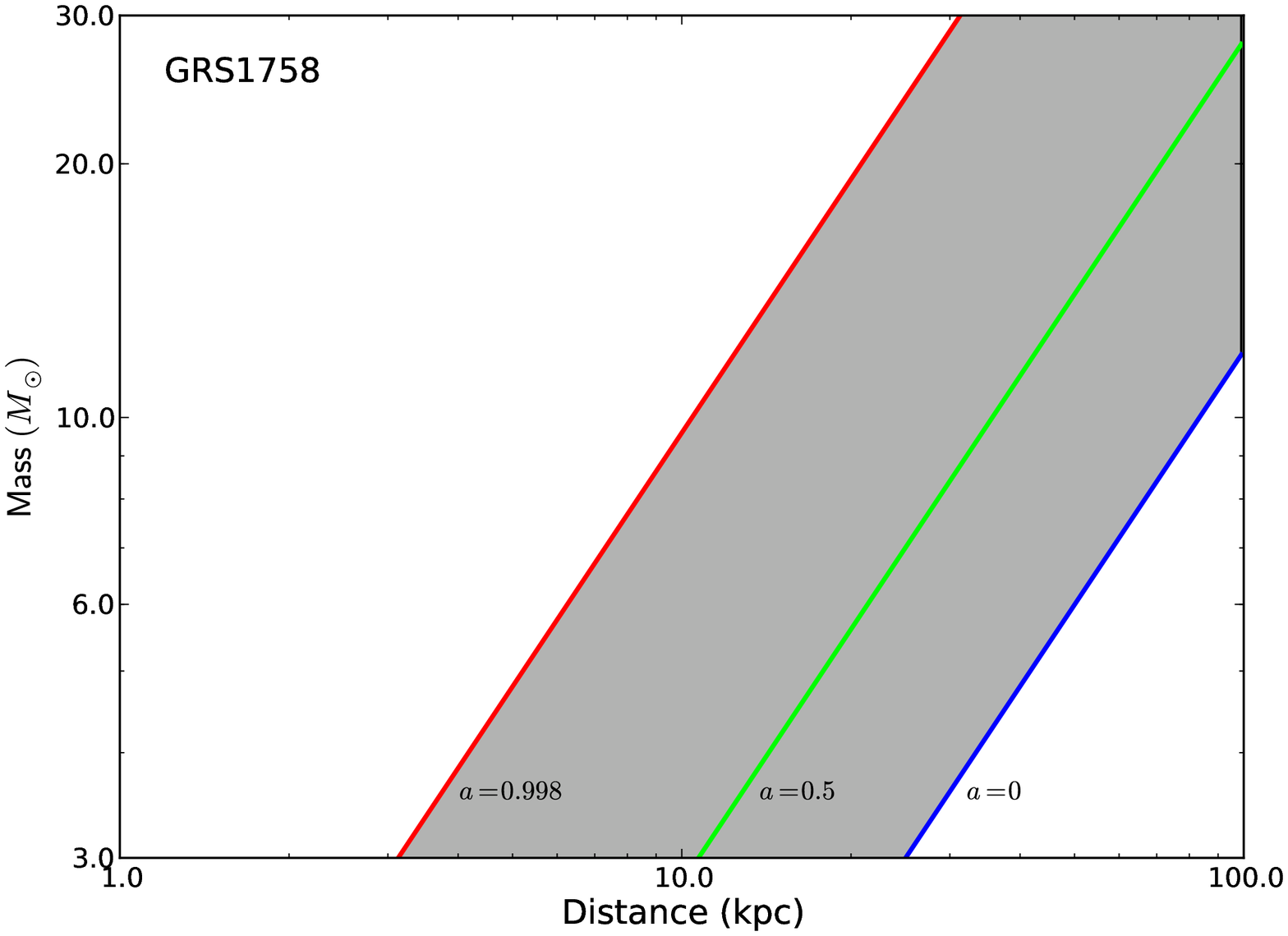}
\includegraphics[width=0.4\textwidth]{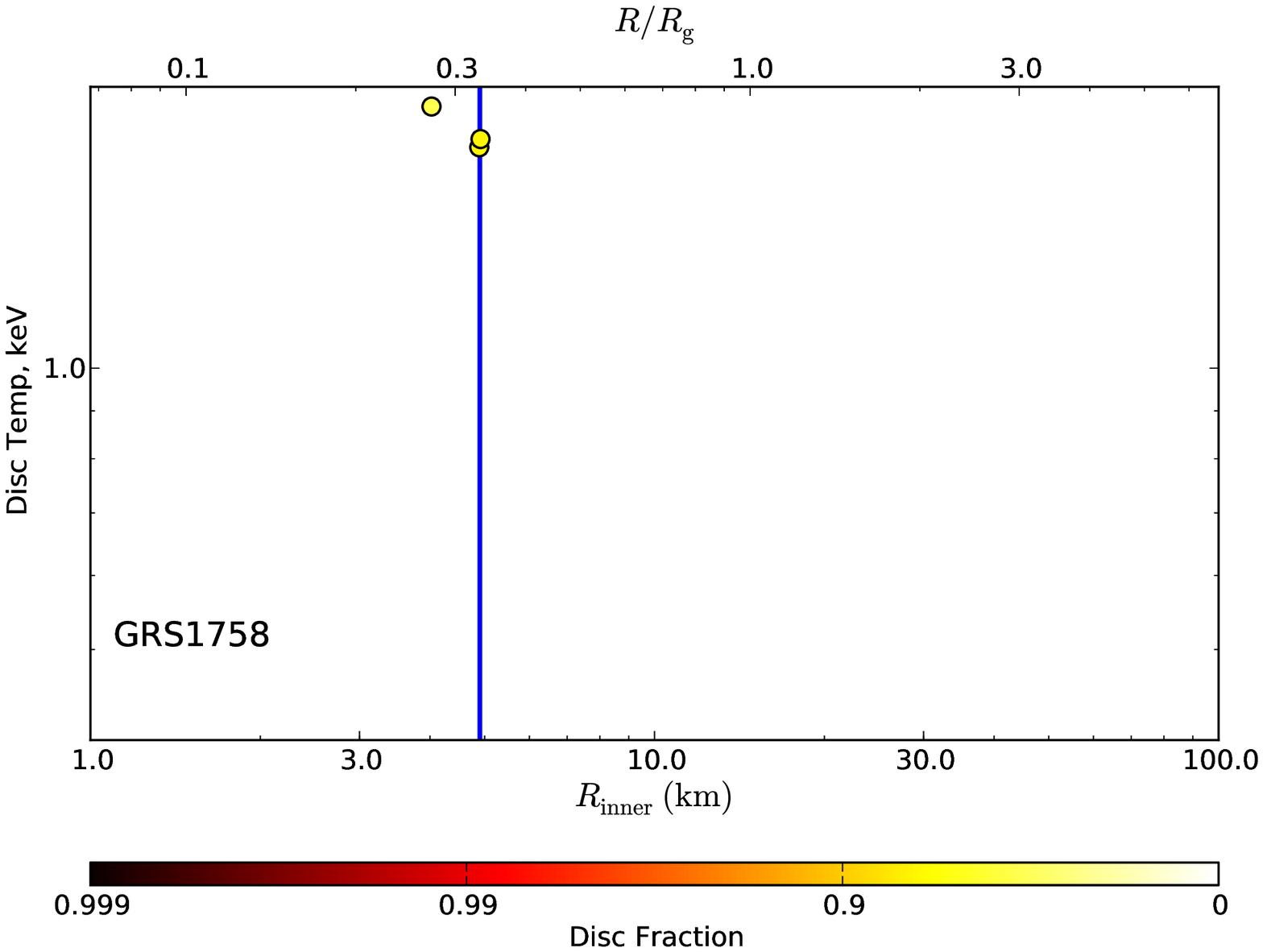}
\includegraphics[width=0.4\textwidth]{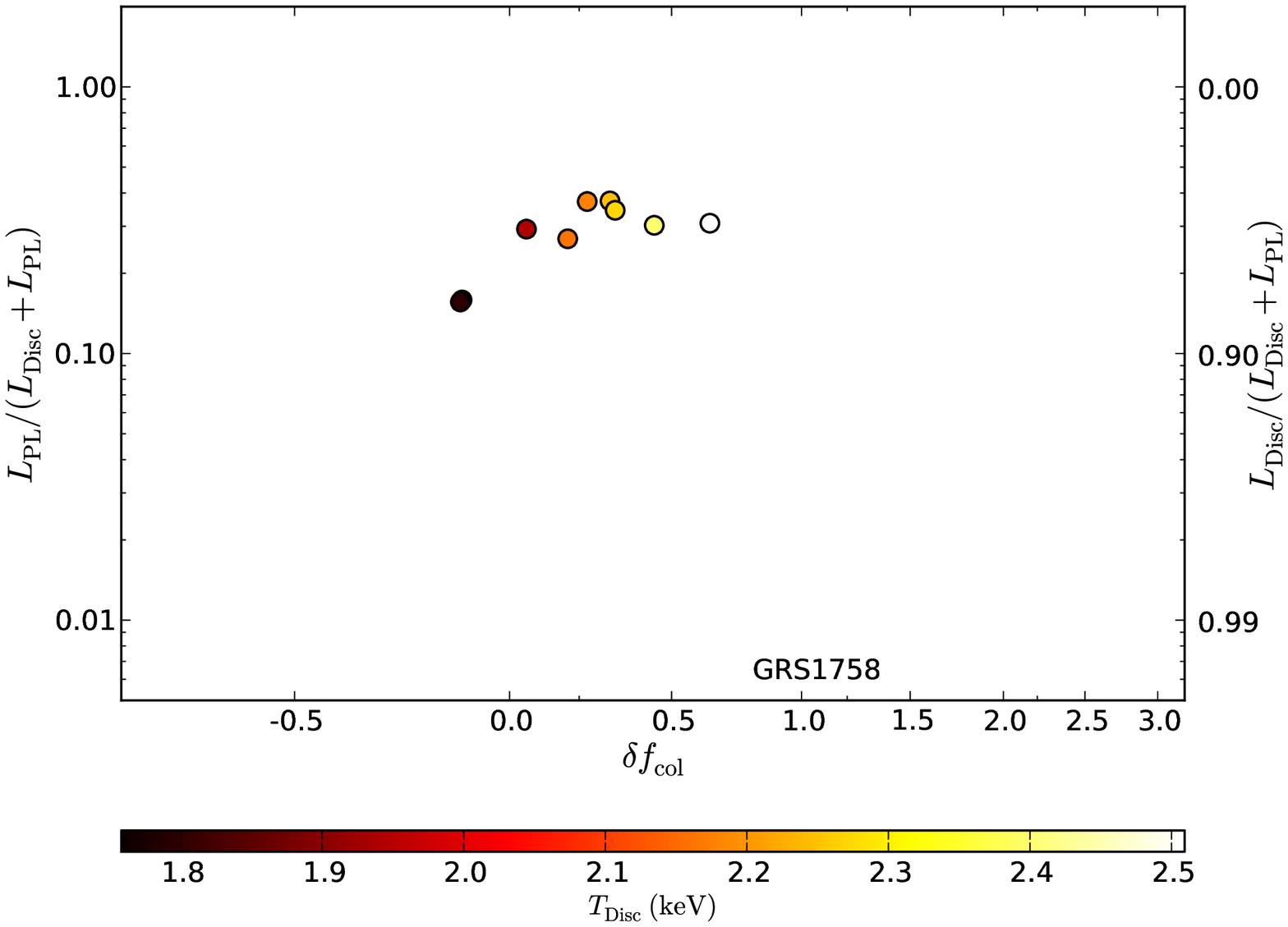}
\caption{(cont) GRS~1758-258}
\end{figure*}
\addtocounter{figure}{-1}
\begin{figure*}
\centering
\includegraphics[width=0.4\textwidth]{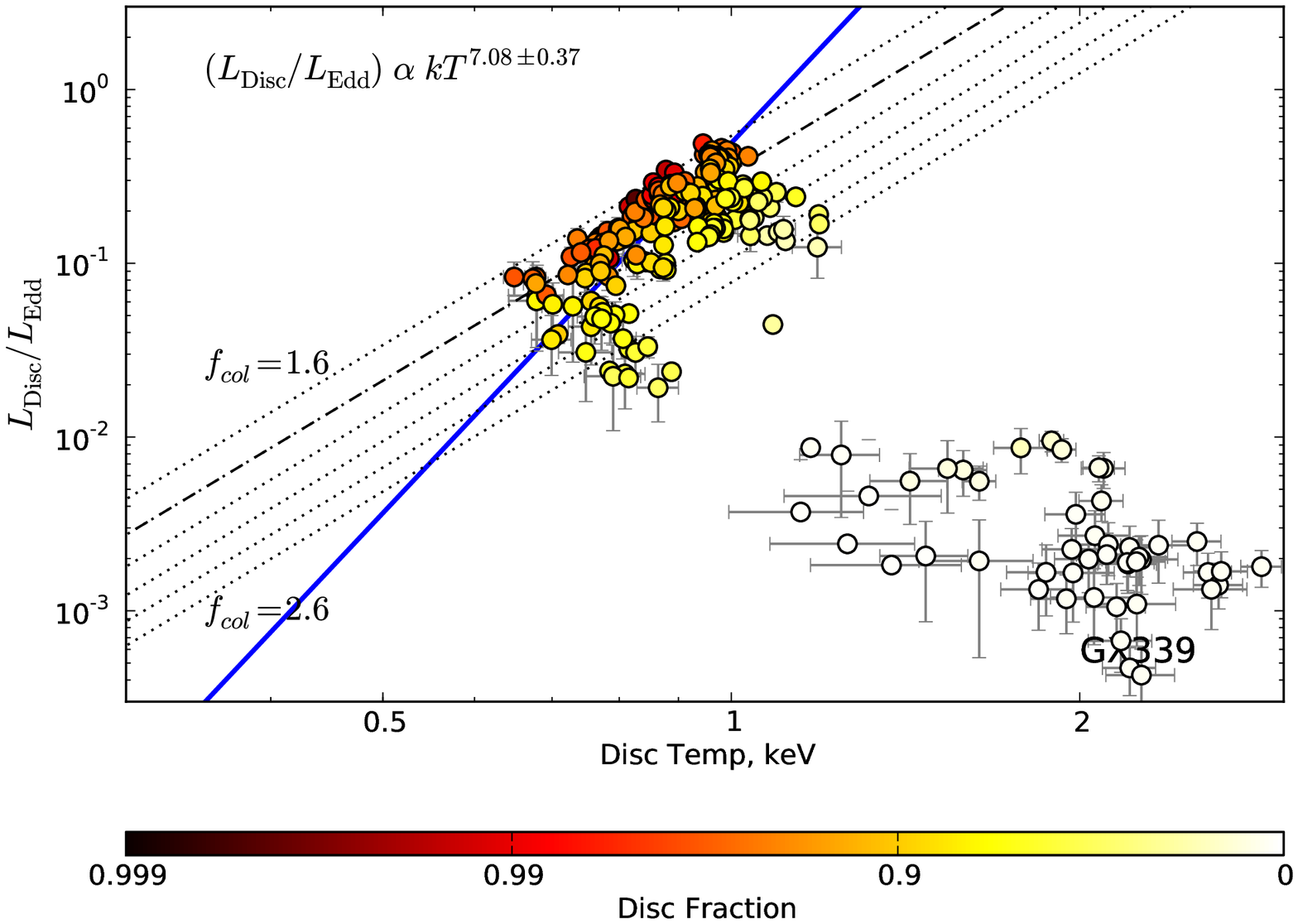}
\includegraphics[width=0.4\textwidth]{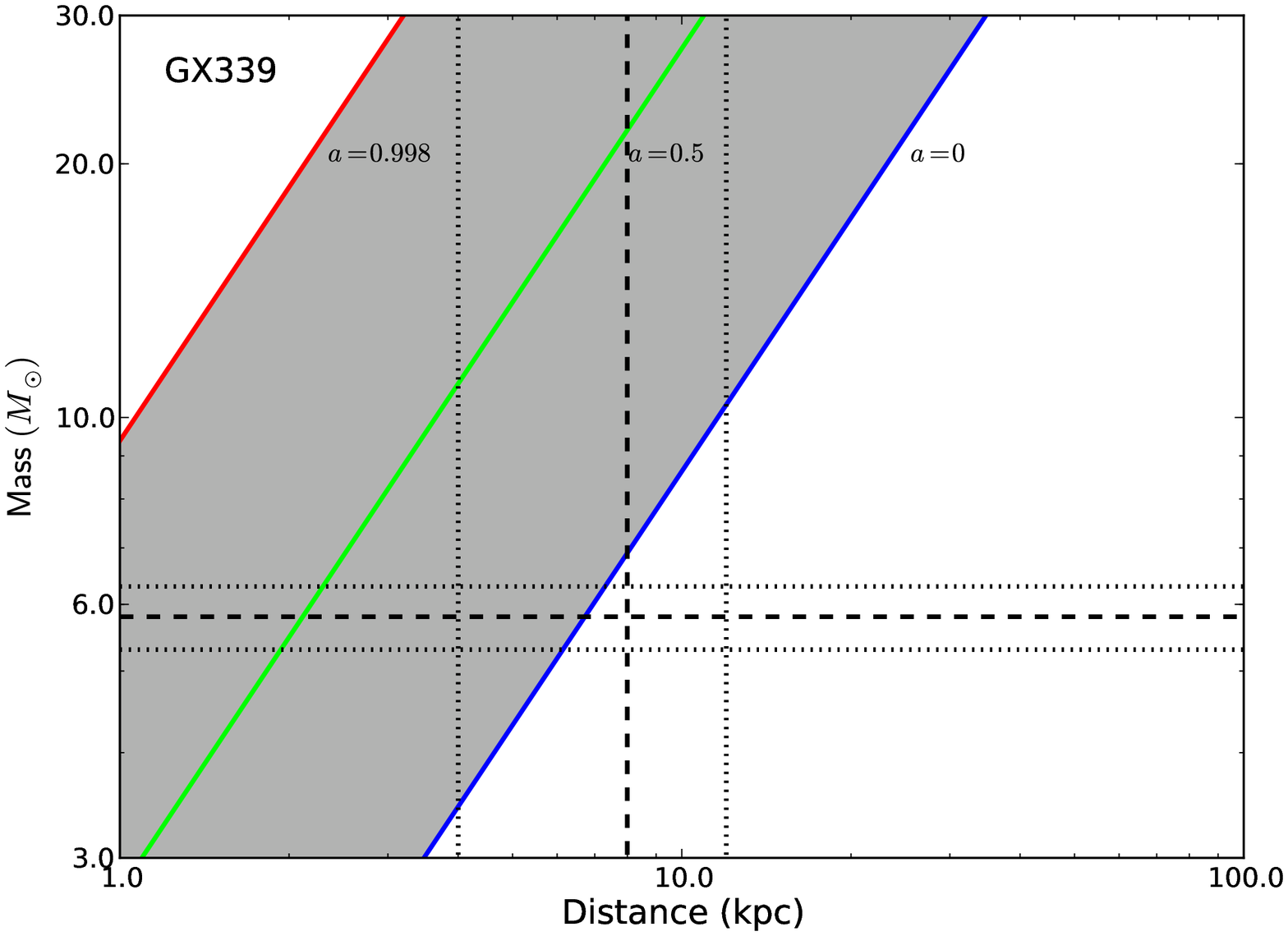}
\includegraphics[width=0.4\textwidth]{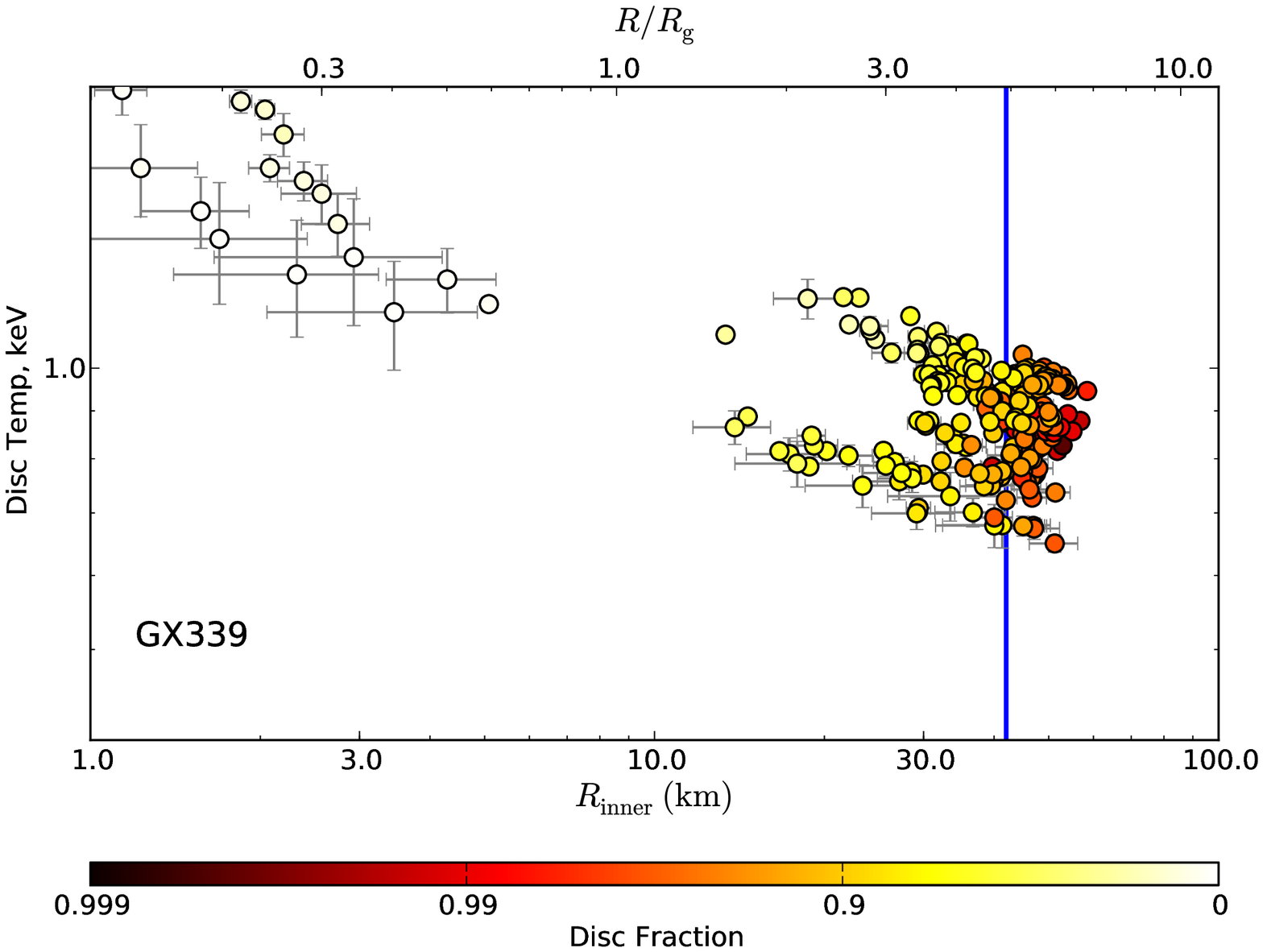}
\includegraphics[width=0.4\textwidth]{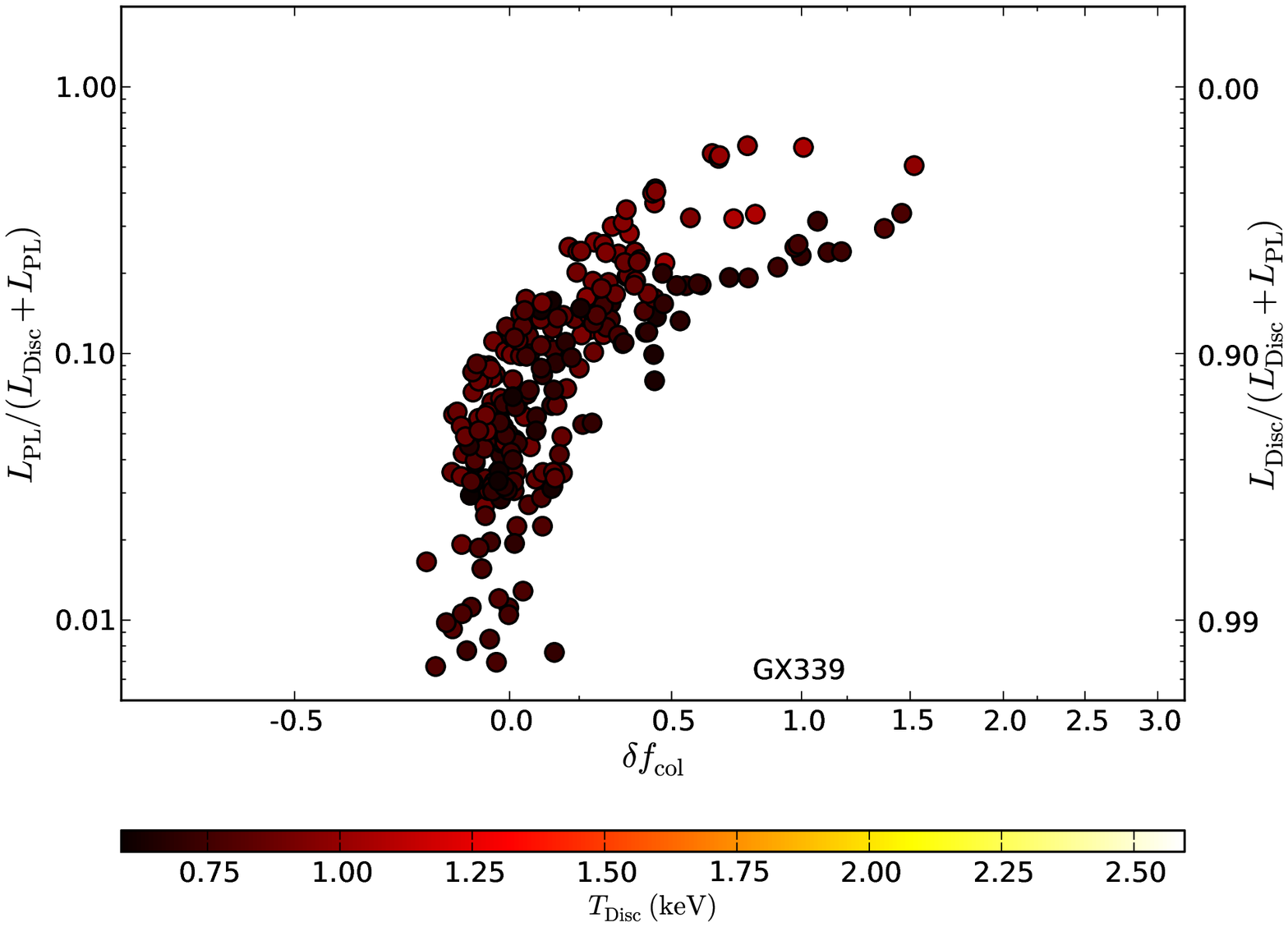}
\caption{(cont) GX~339-4}
\end{figure*}
\addtocounter{figure}{-1}
\begin{figure*}
\centering
\includegraphics[width=0.4\textwidth]{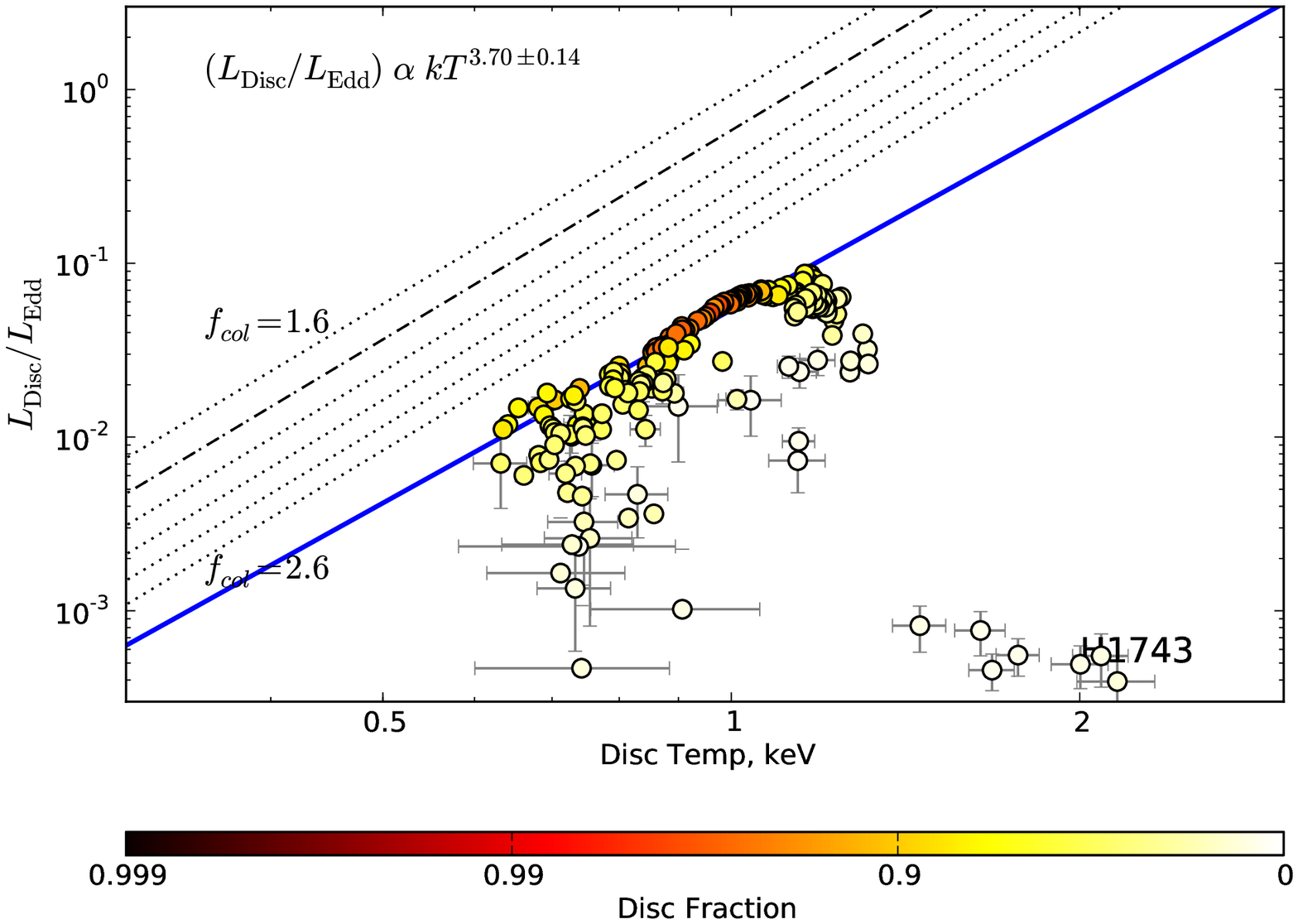}
\includegraphics[width=0.4\textwidth]{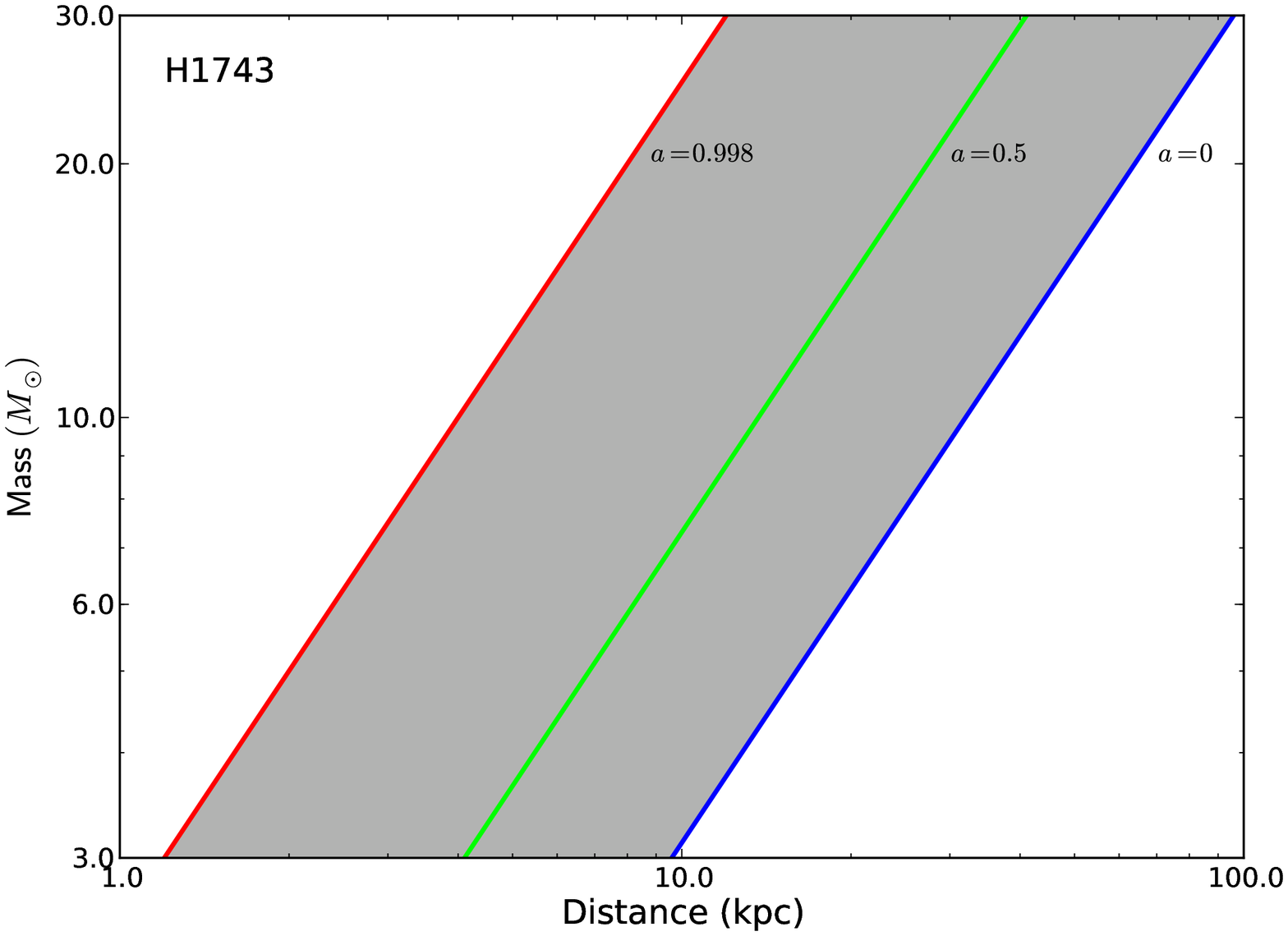}
\includegraphics[width=0.4\textwidth]{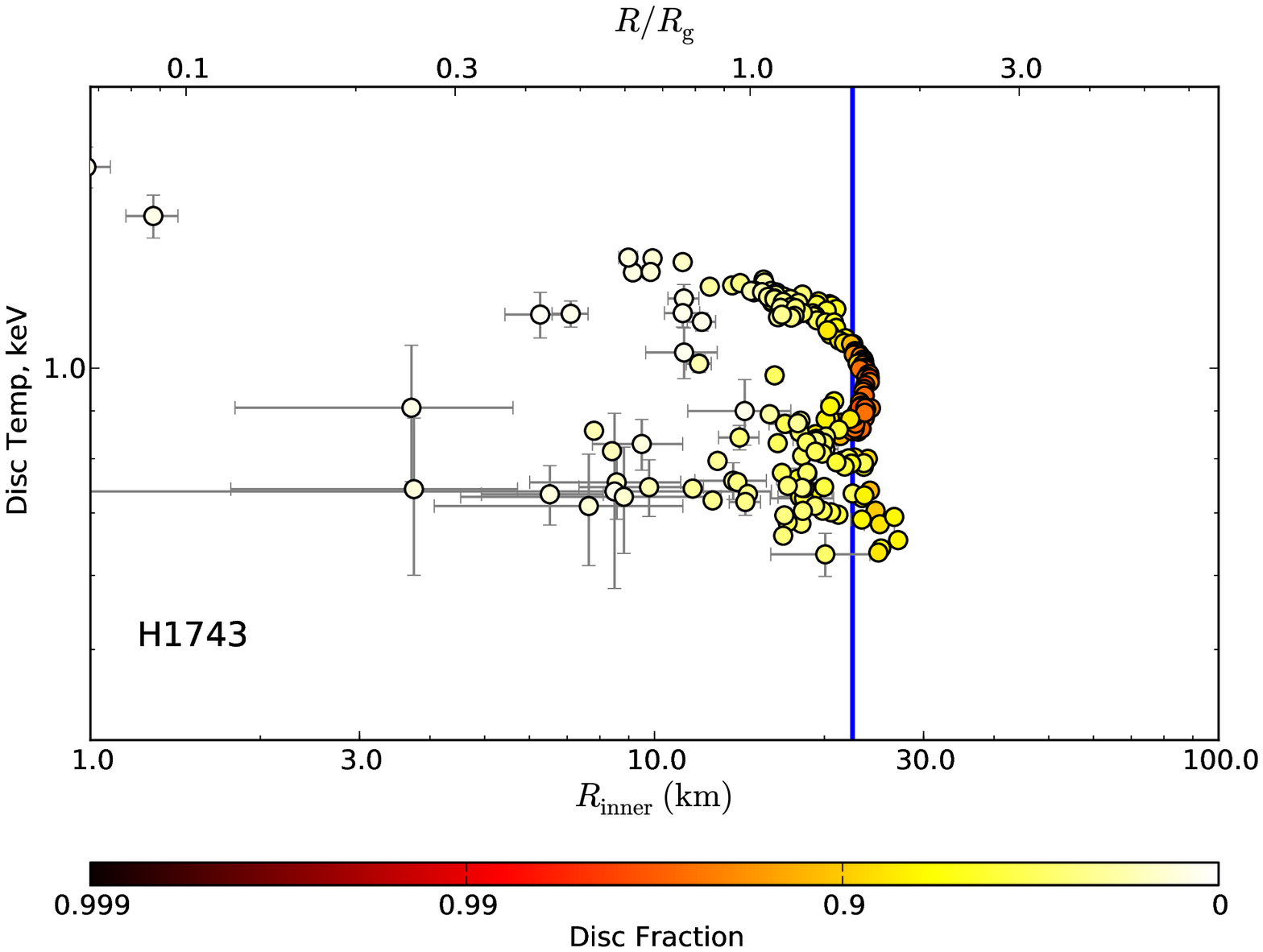}
\includegraphics[width=0.4\textwidth]{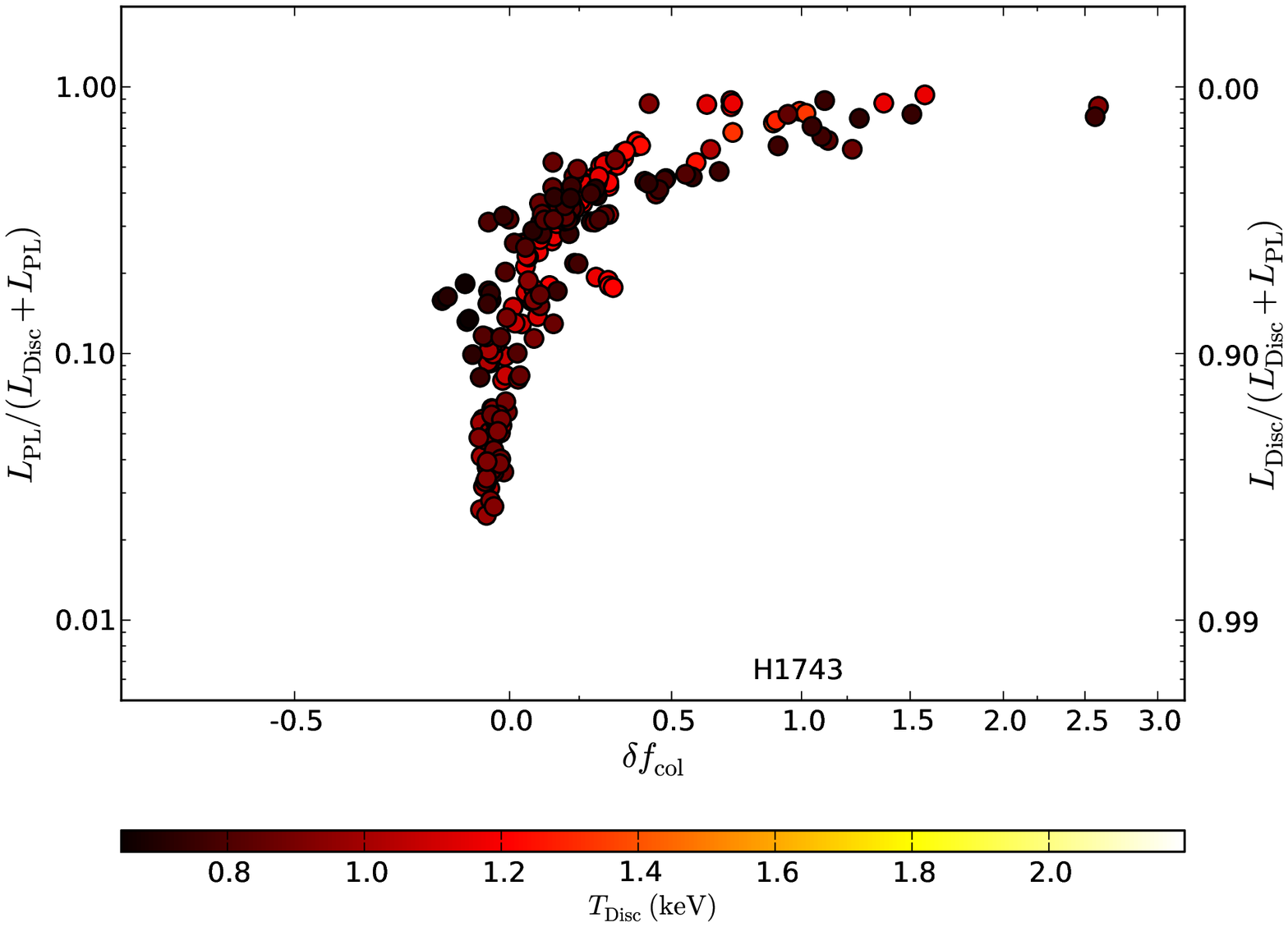}
\caption{(cont) H 1743-322}
\end{figure*}
\addtocounter{figure}{-1}
\begin{figure*}
\centering
\includegraphics[width=0.4\textwidth]{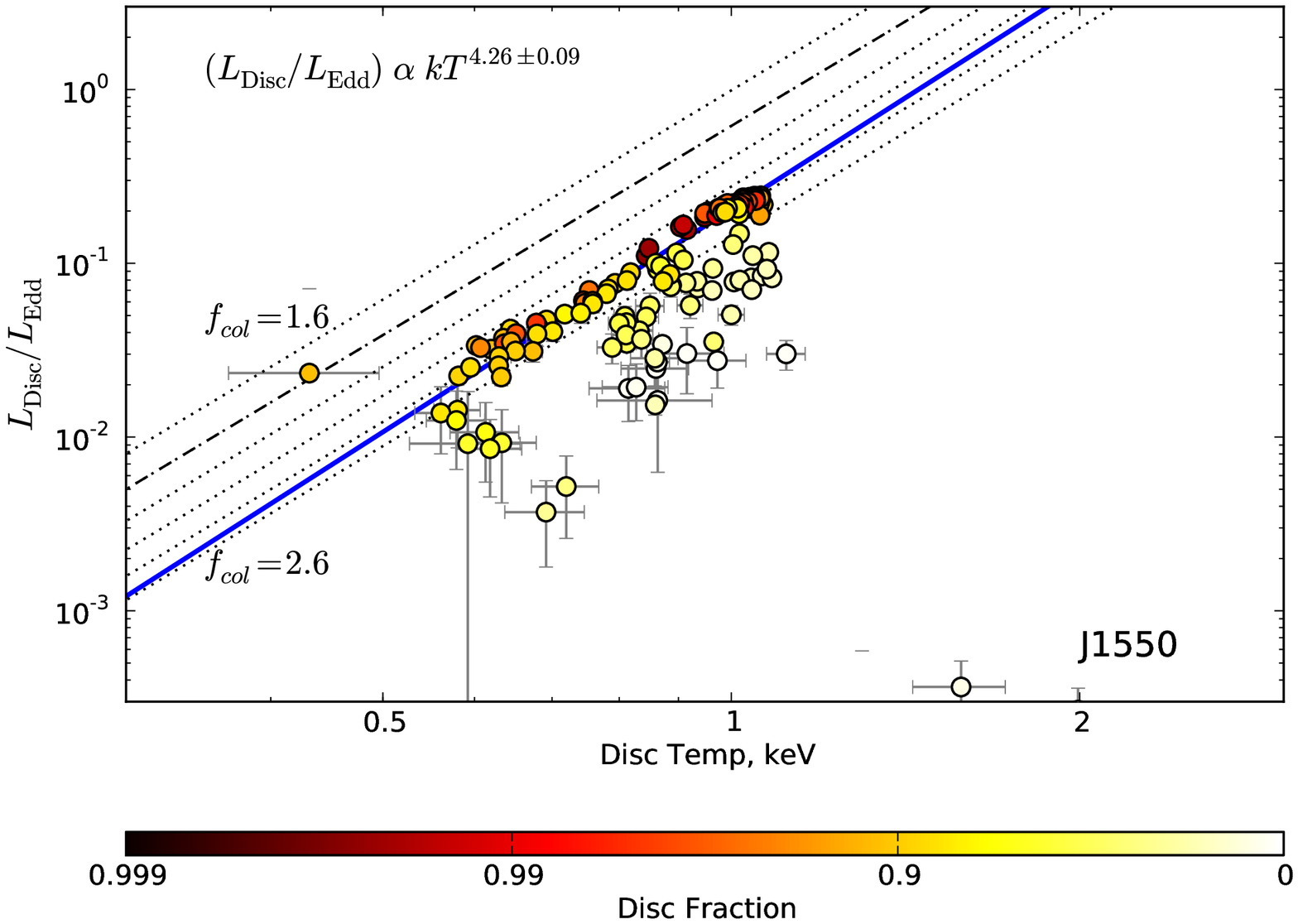}
\includegraphics[width=0.4\textwidth]{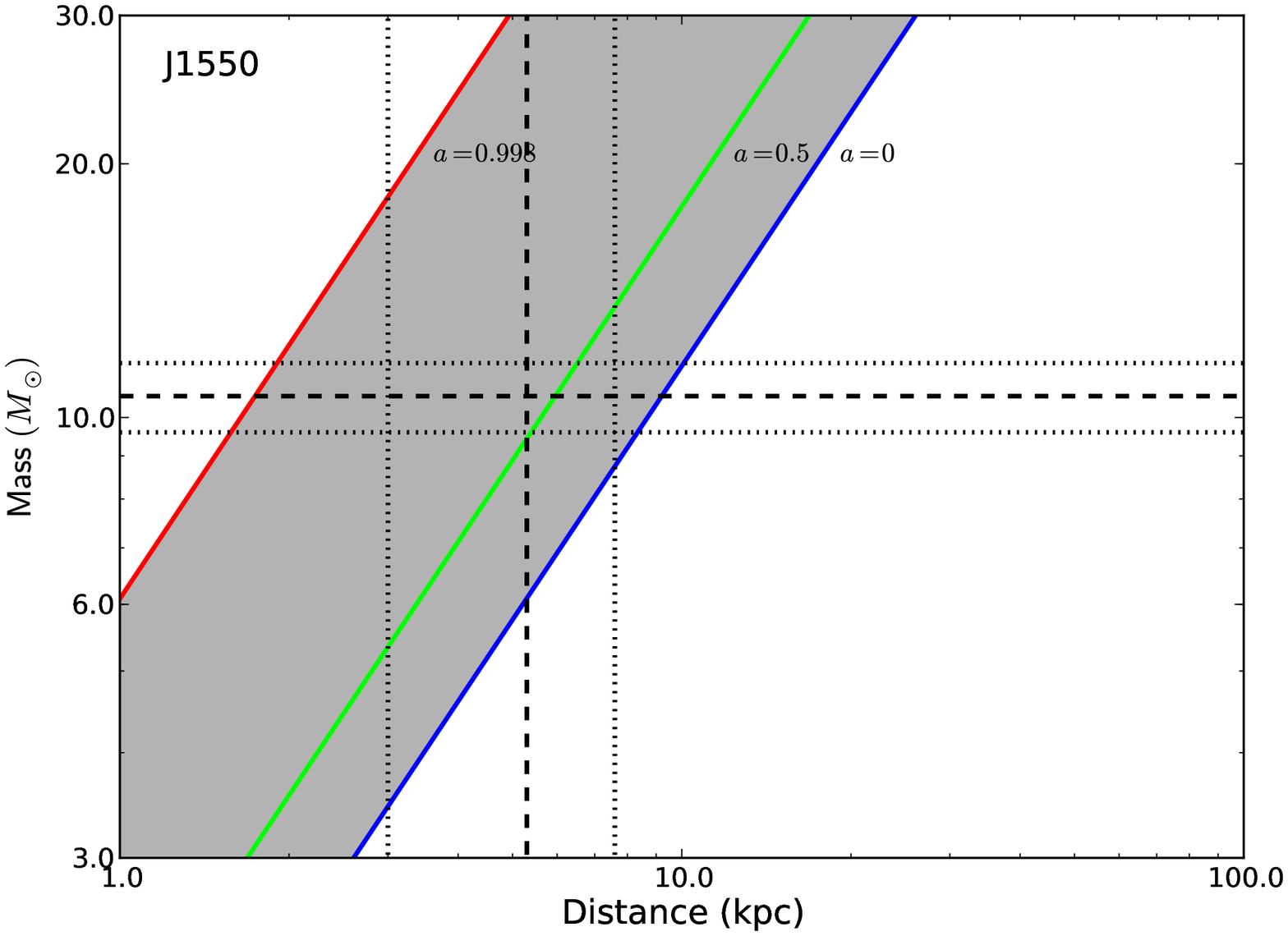}
\includegraphics[width=0.4\textwidth]{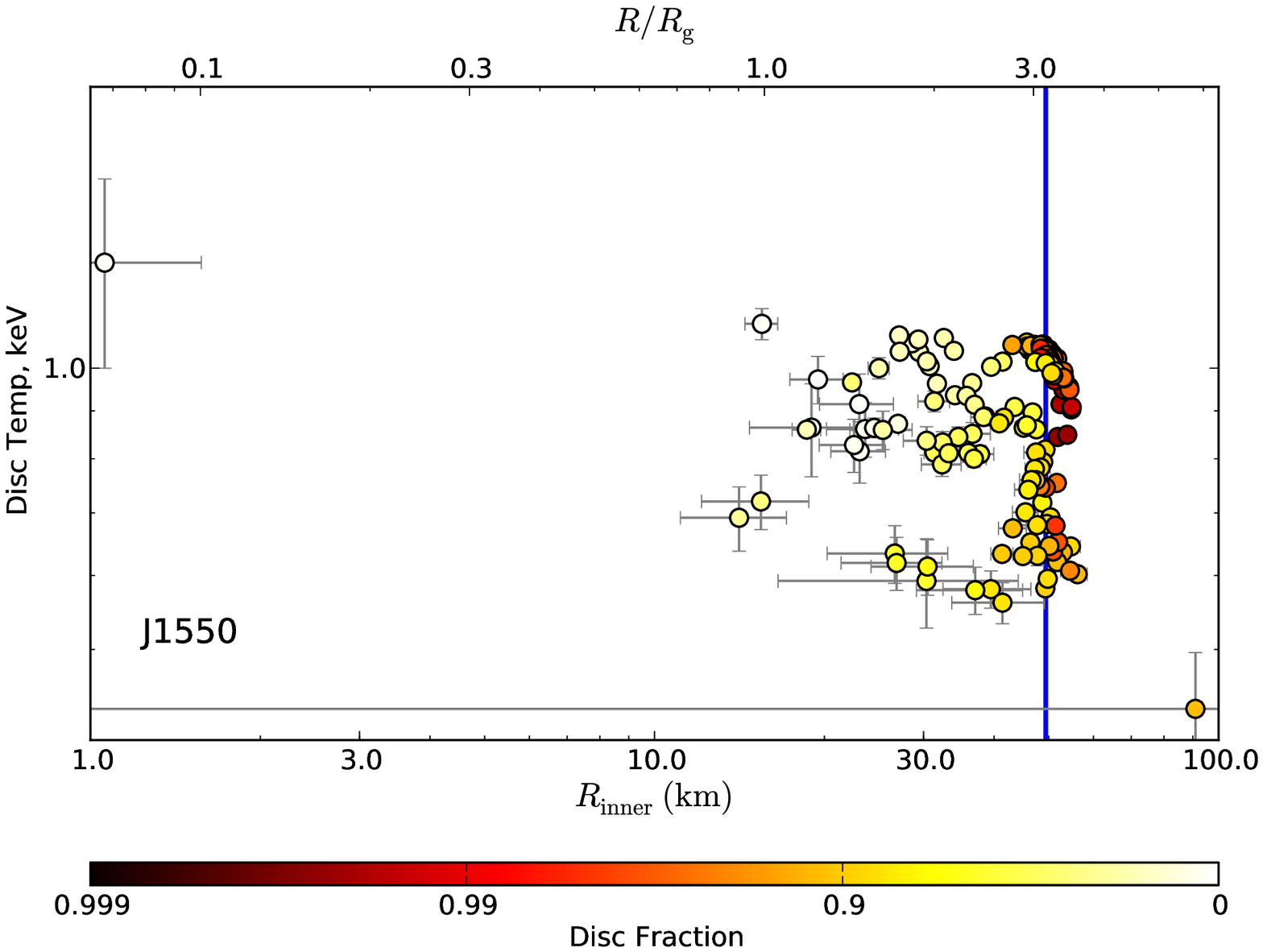}
\includegraphics[width=0.4\textwidth]{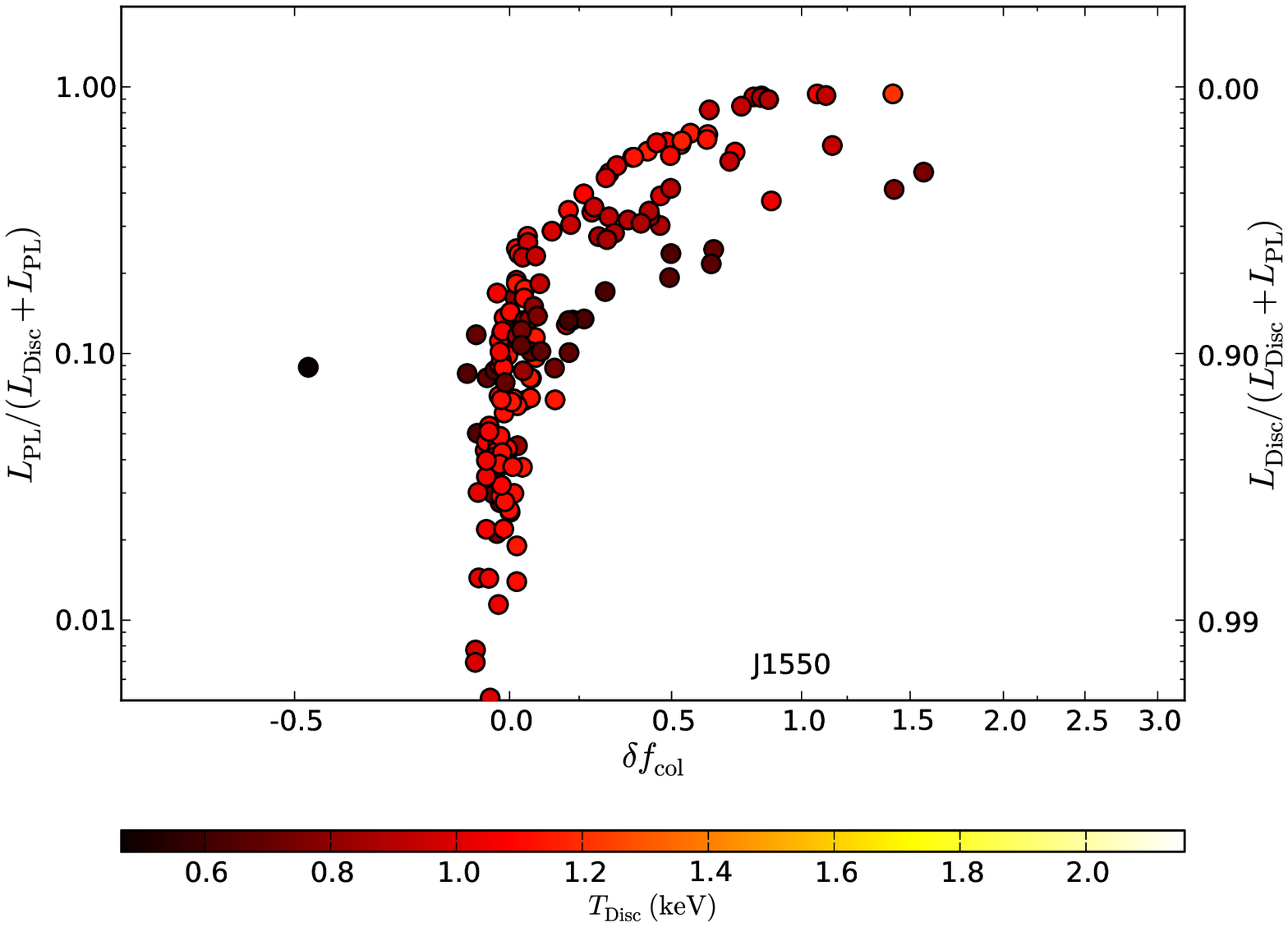}
\caption{(cont) XTE~J1550-564}
\end{figure*}
\addtocounter{figure}{-1}
\begin{figure*}
\centering
\includegraphics[width=0.4\textwidth]{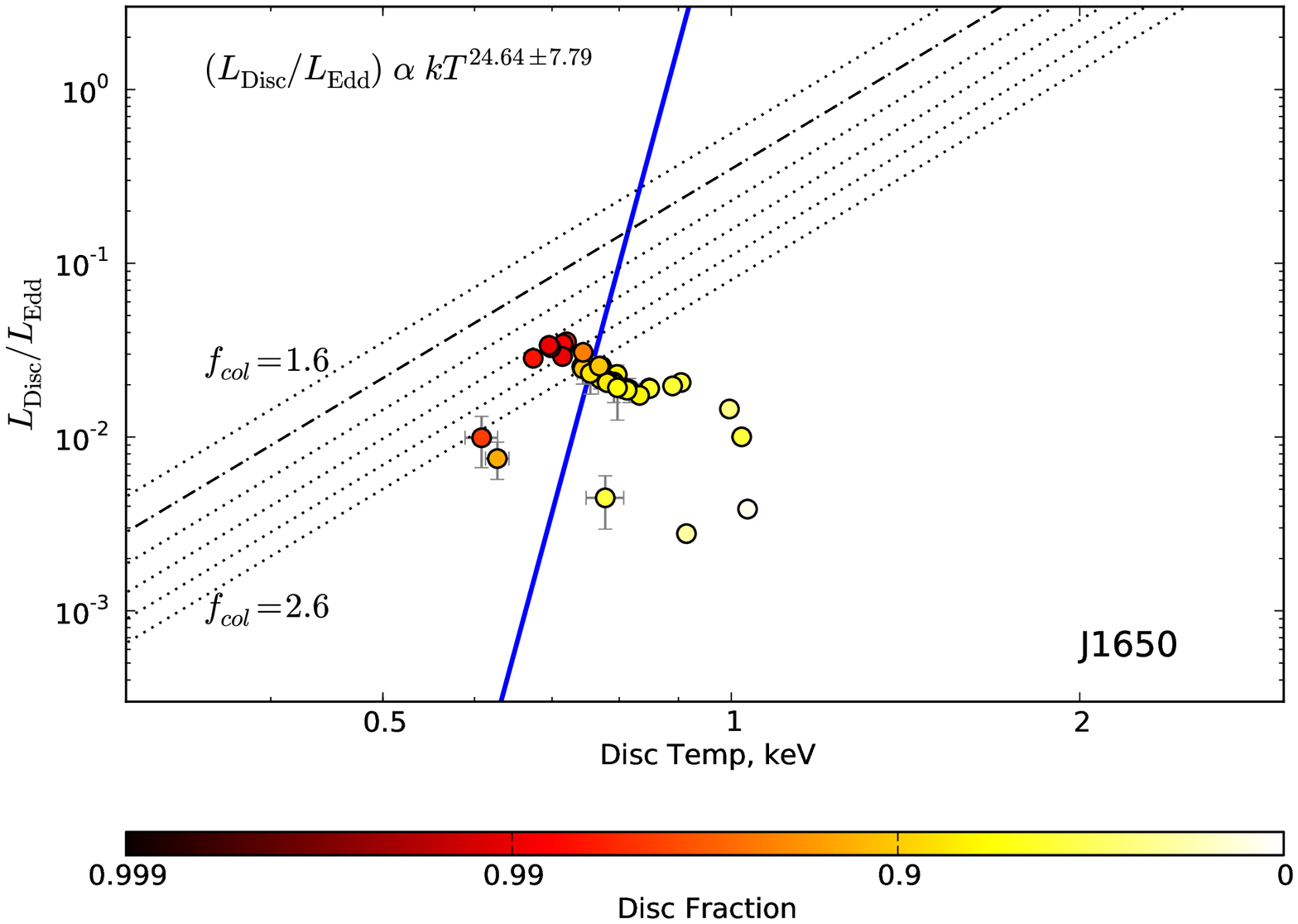}
\includegraphics[width=0.4\textwidth]{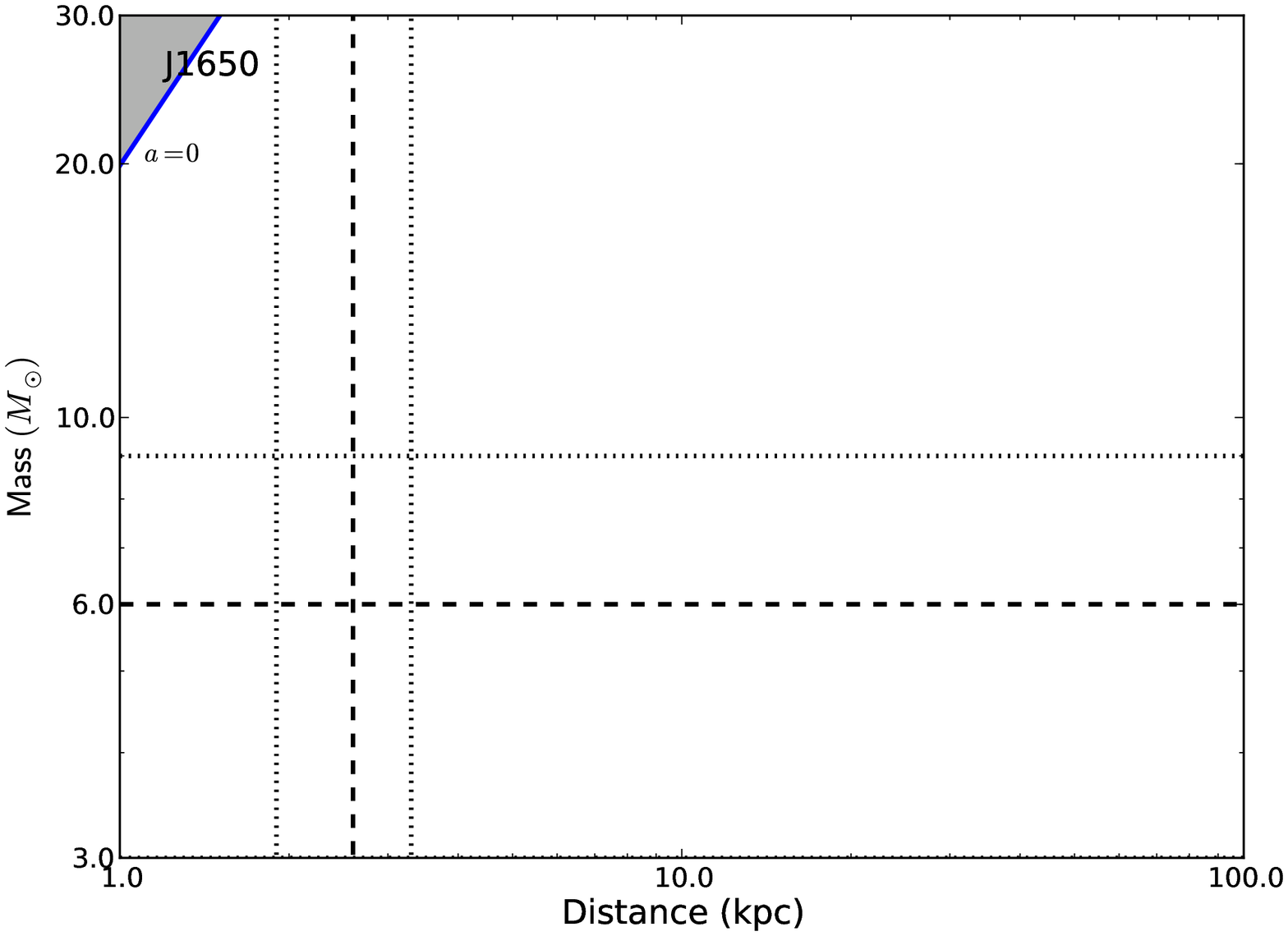}
\includegraphics[width=0.4\textwidth]{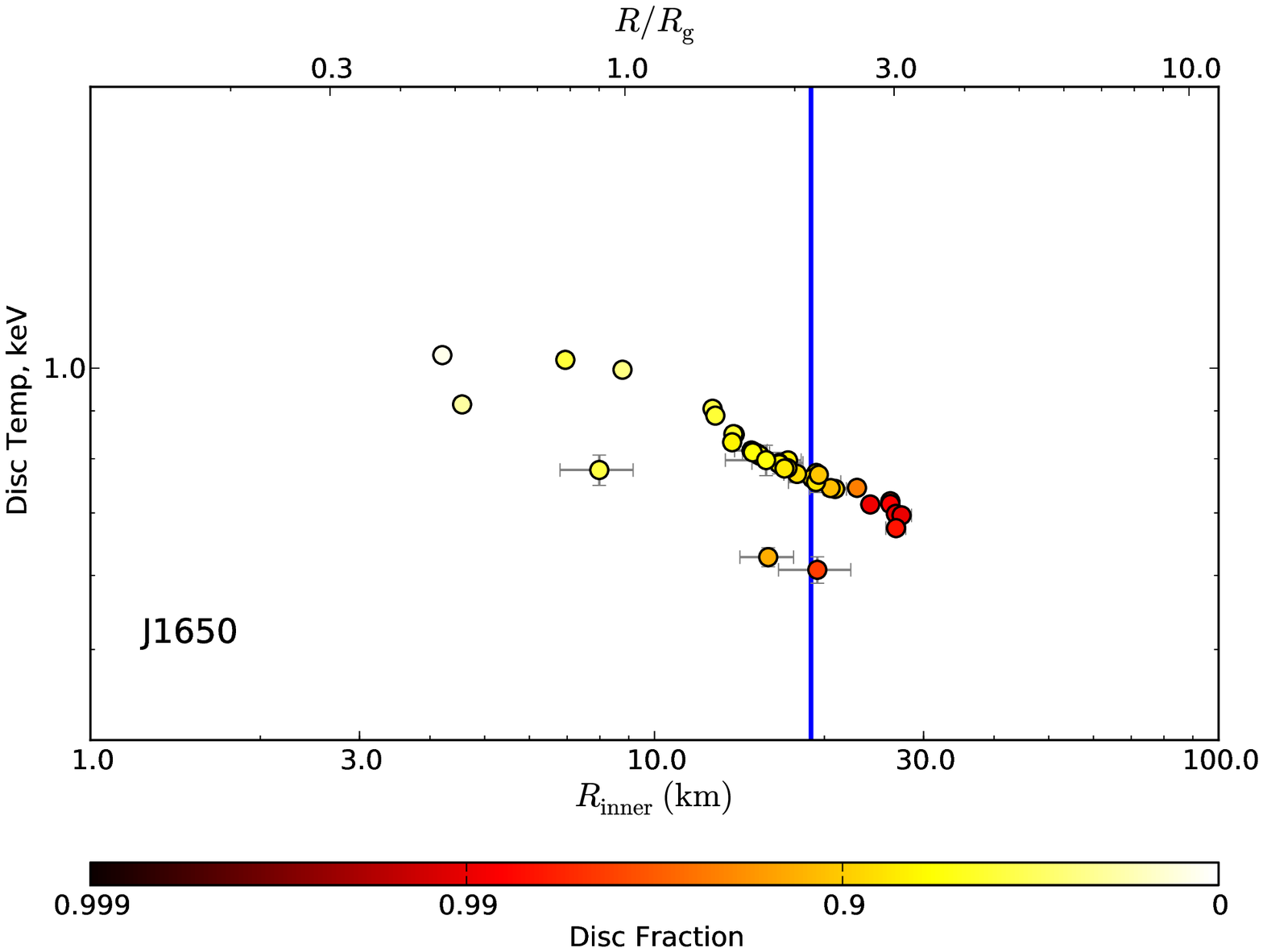}
\includegraphics[width=0.4\textwidth]{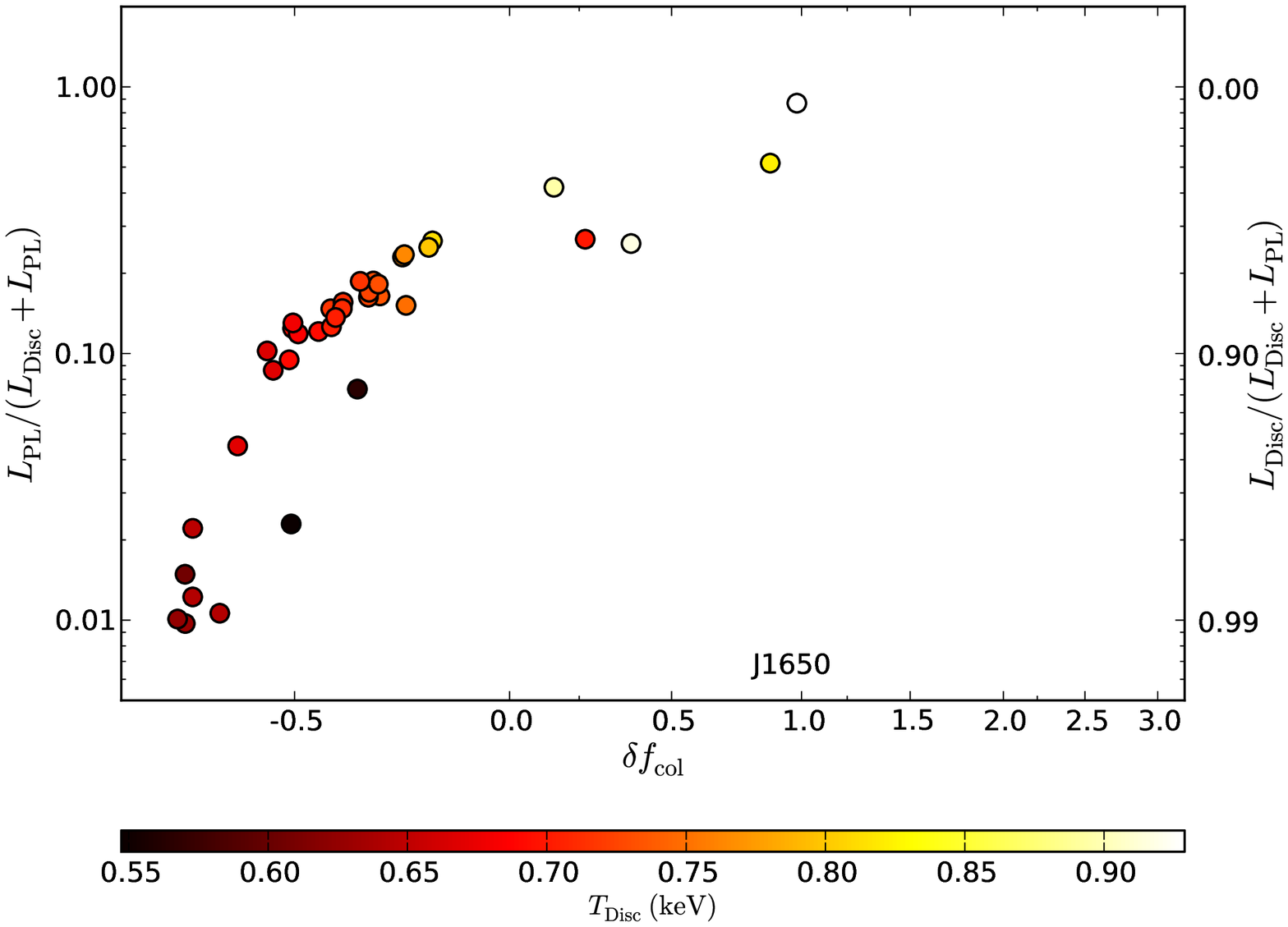}
\caption{(cont) XTE~J1650-500}
\end{figure*}
\addtocounter{figure}{-1}
\begin{figure*}
\centering
\includegraphics[width=0.4\textwidth]{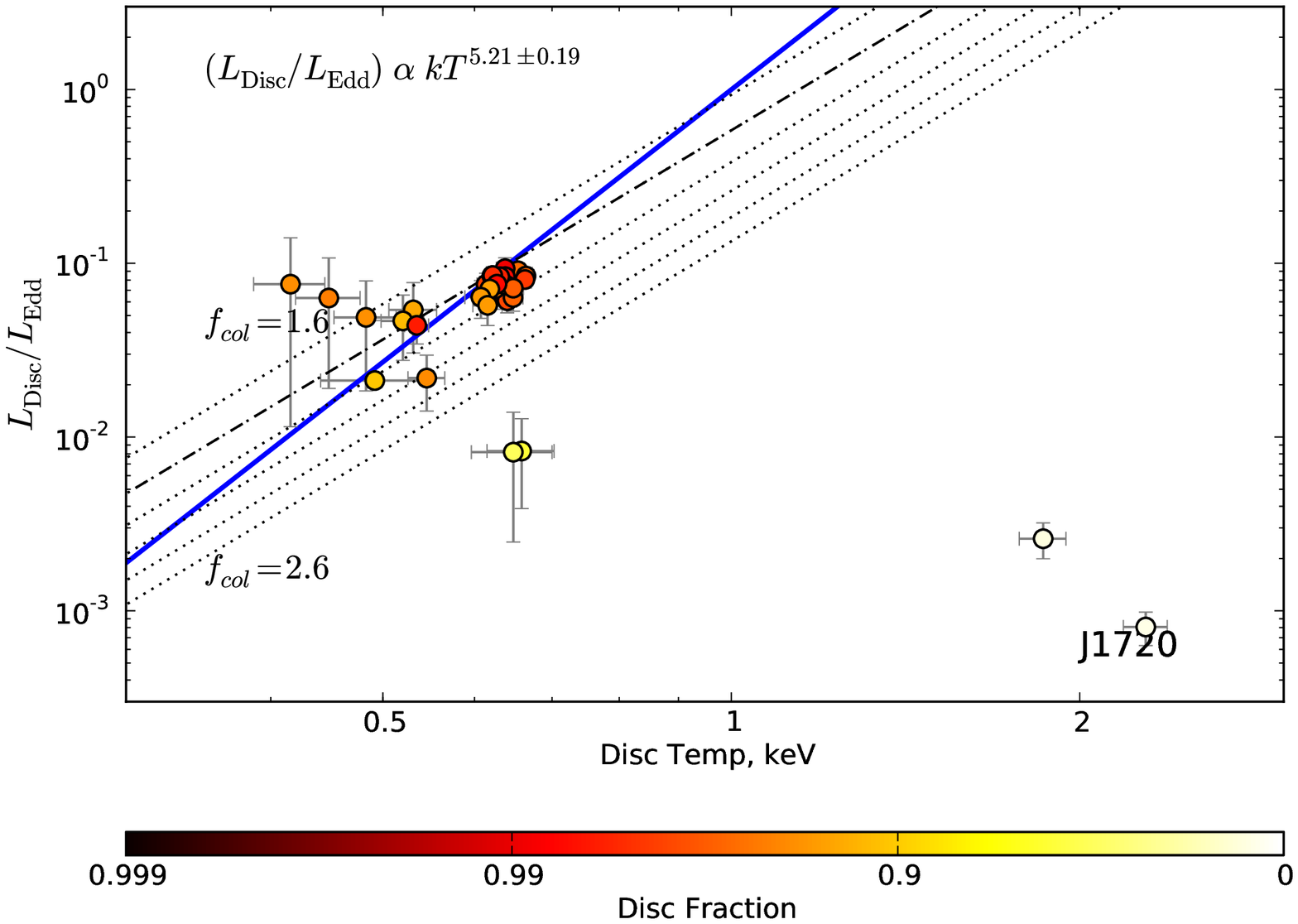}
\includegraphics[width=0.4\textwidth]{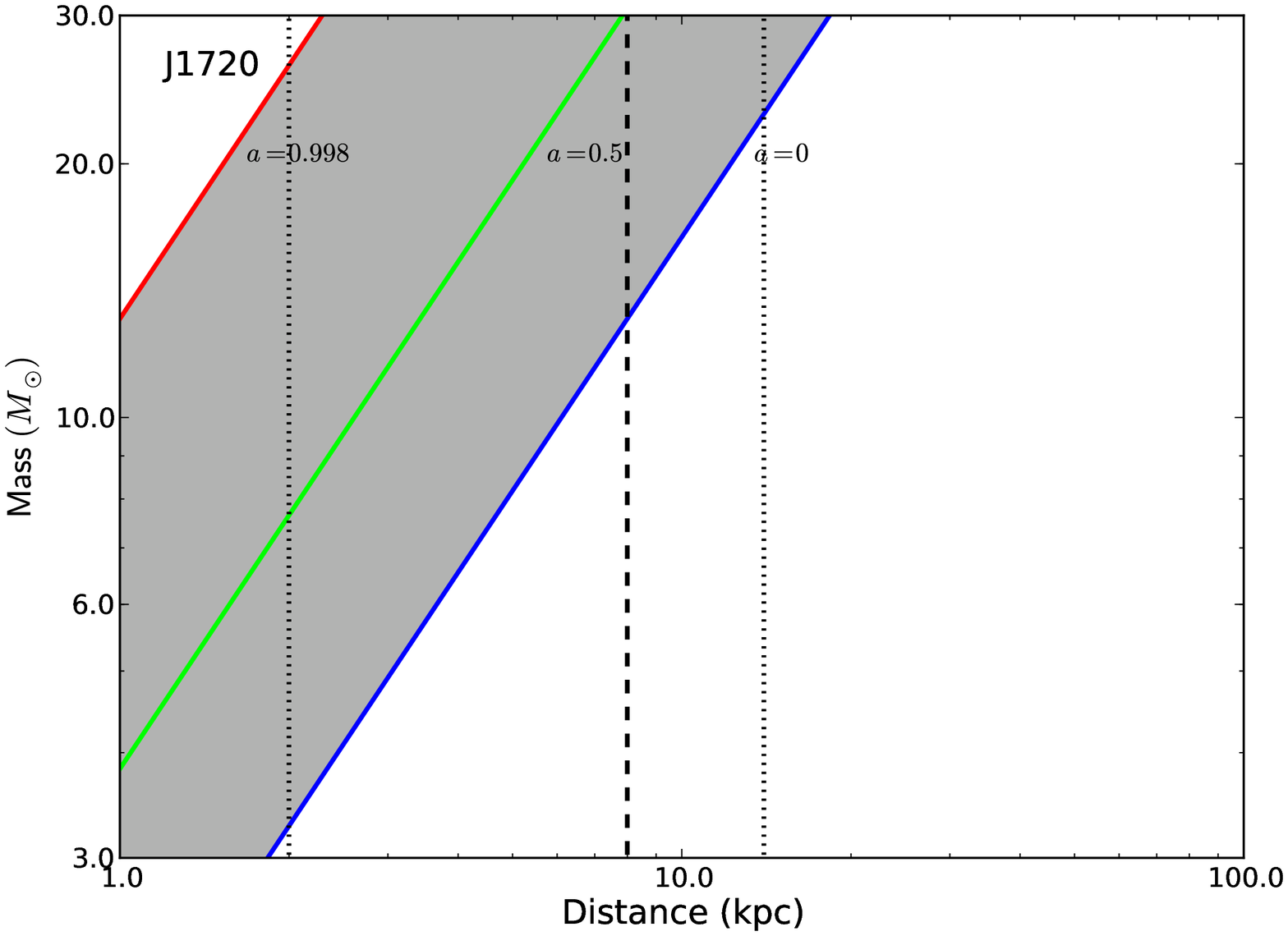}
\includegraphics[width=0.4\textwidth]{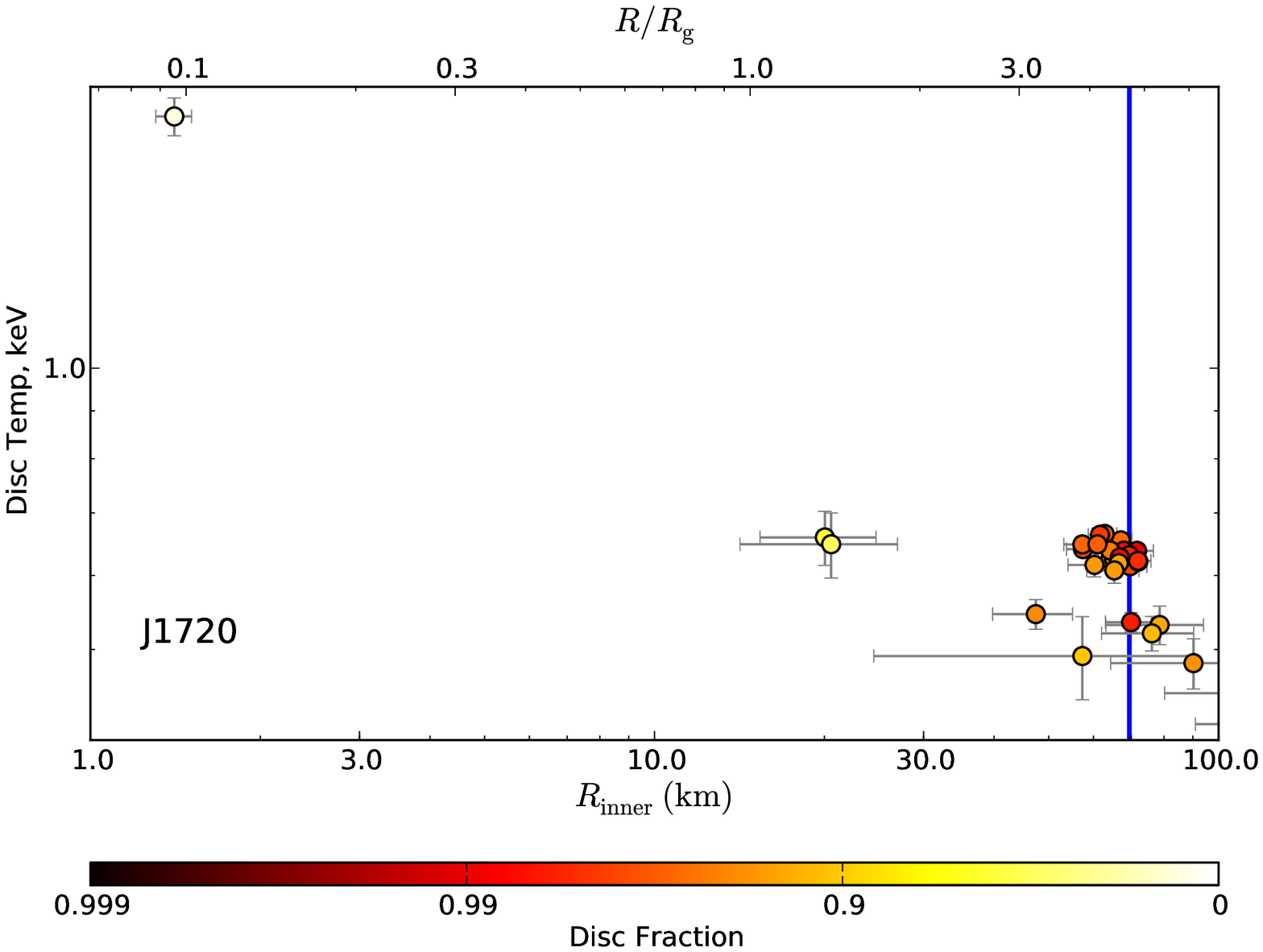}
\includegraphics[width=0.4\textwidth]{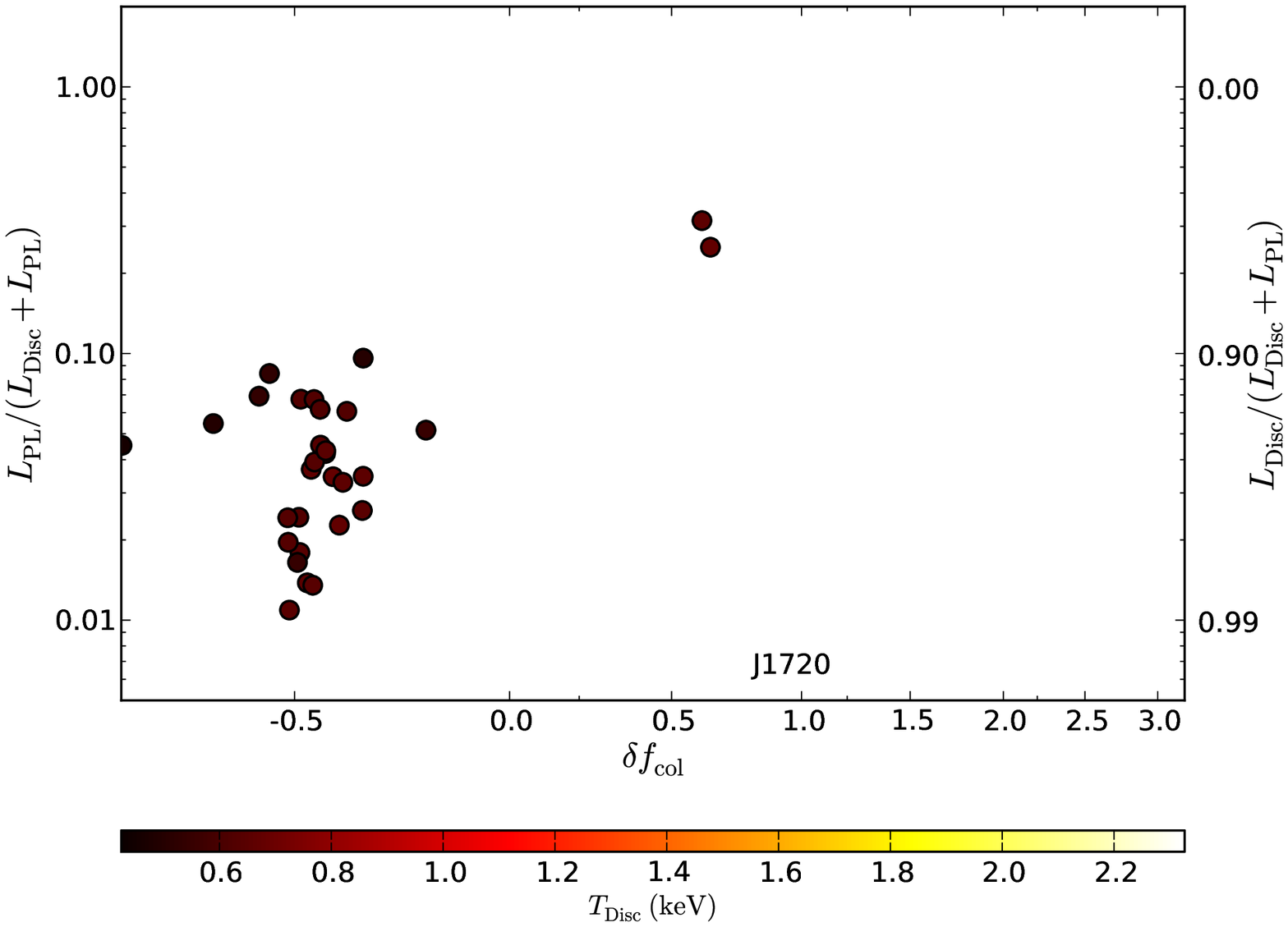}
\caption{(cont) XTE~J1720-318}
\end{figure*}
\addtocounter{figure}{-1}
\begin{figure*}
\centering
\includegraphics[width=0.4\textwidth]{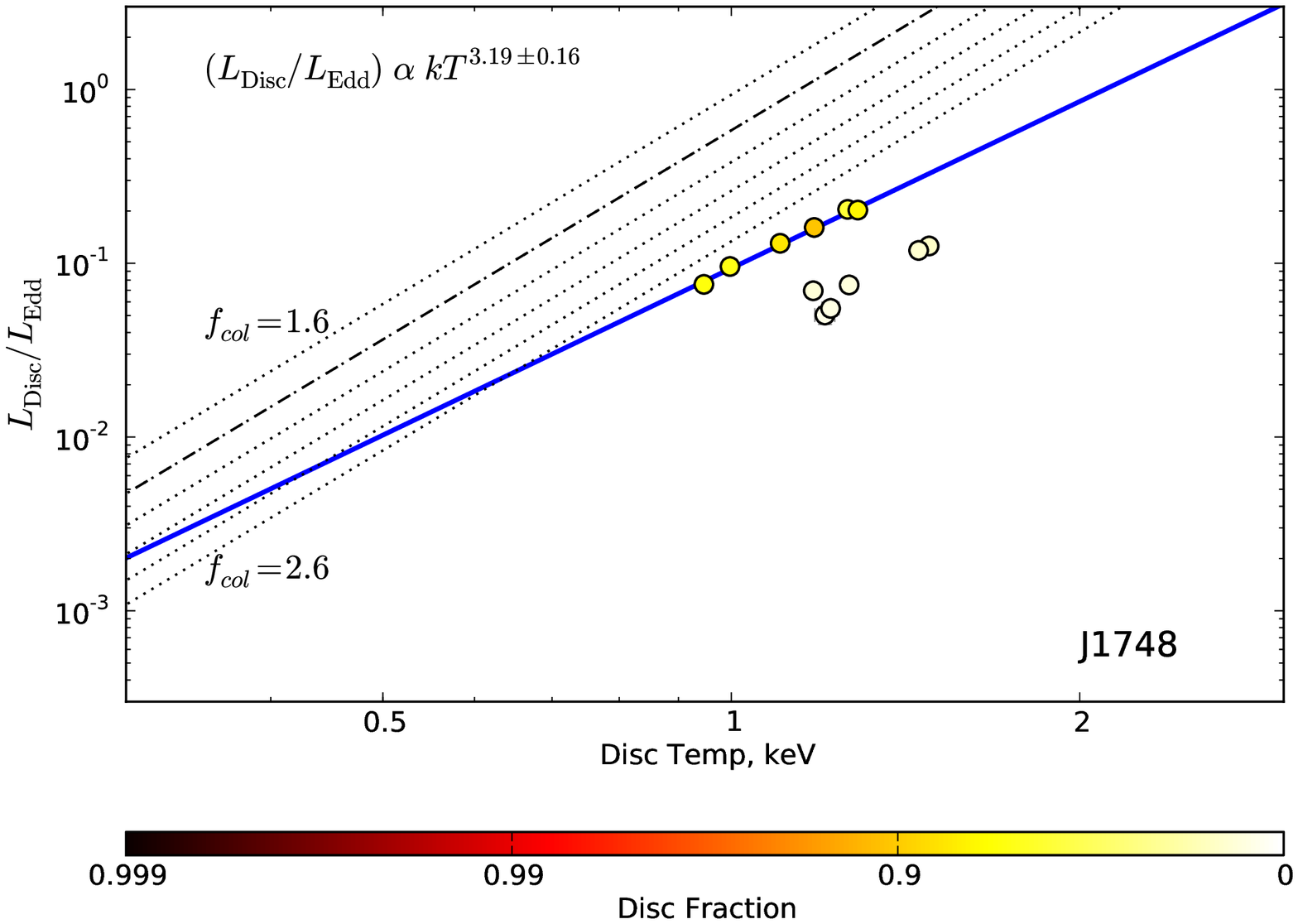}
\includegraphics[width=0.4\textwidth]{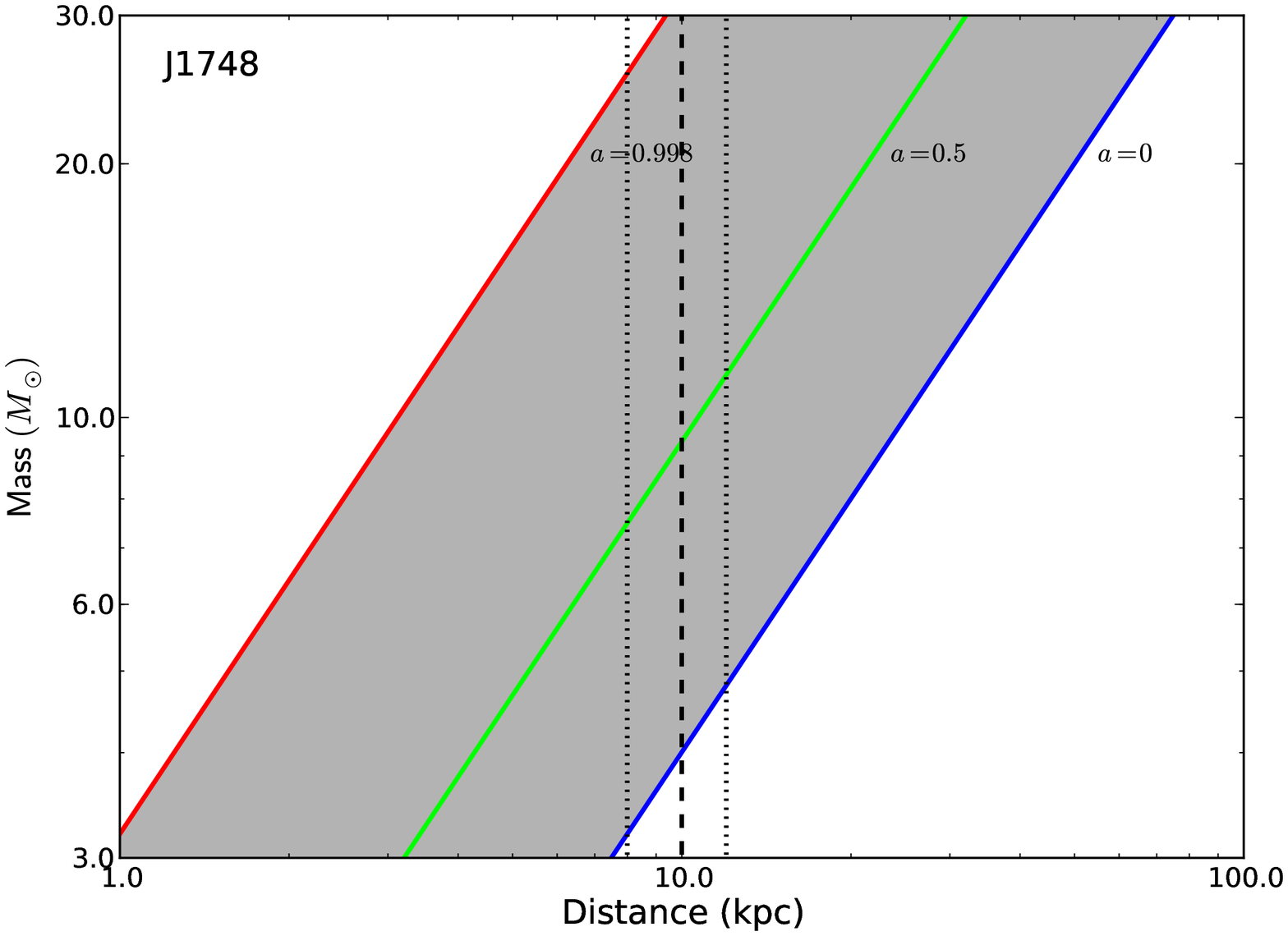}
\includegraphics[width=0.4\textwidth]{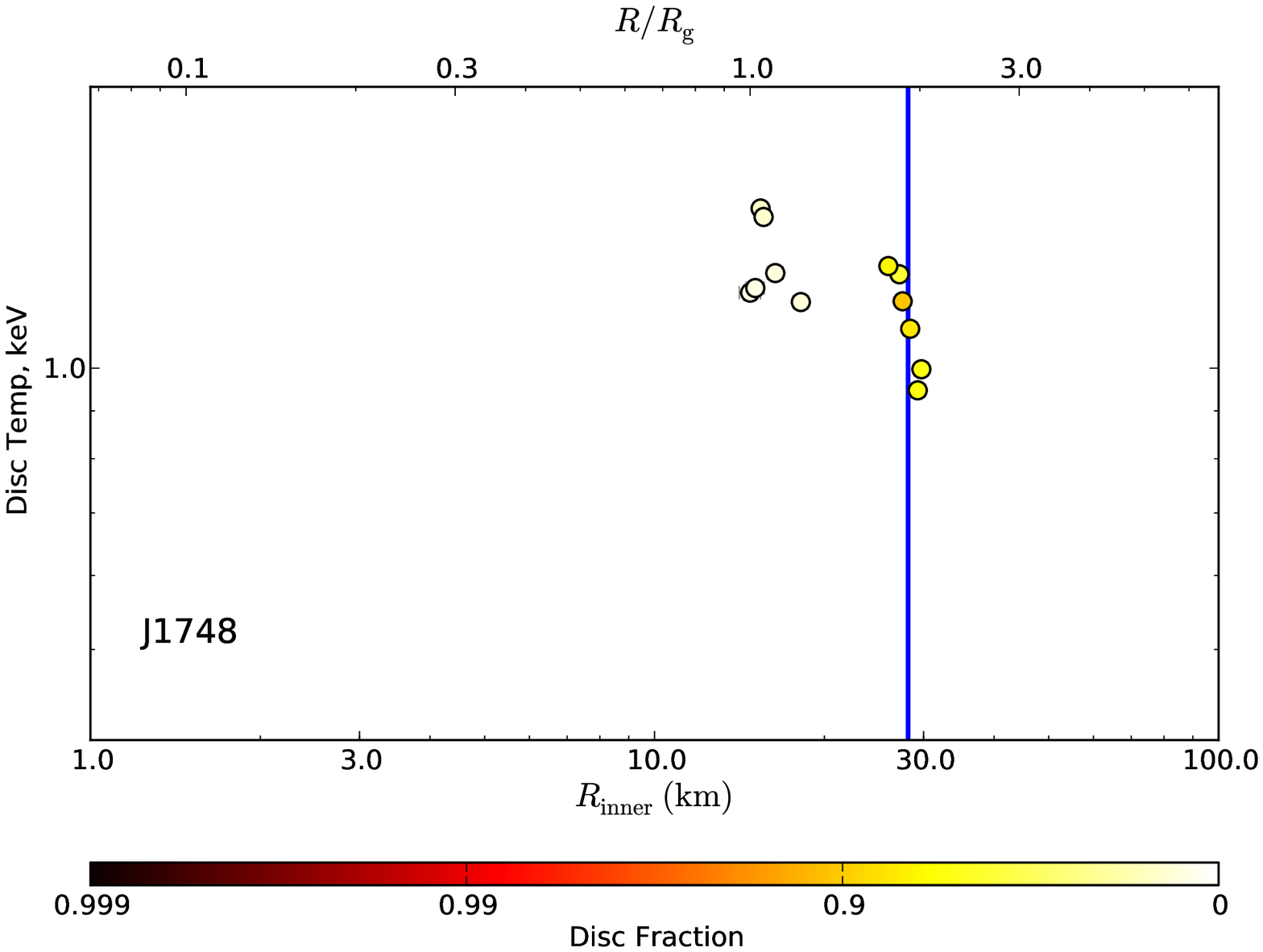}
\includegraphics[width=0.4\textwidth]{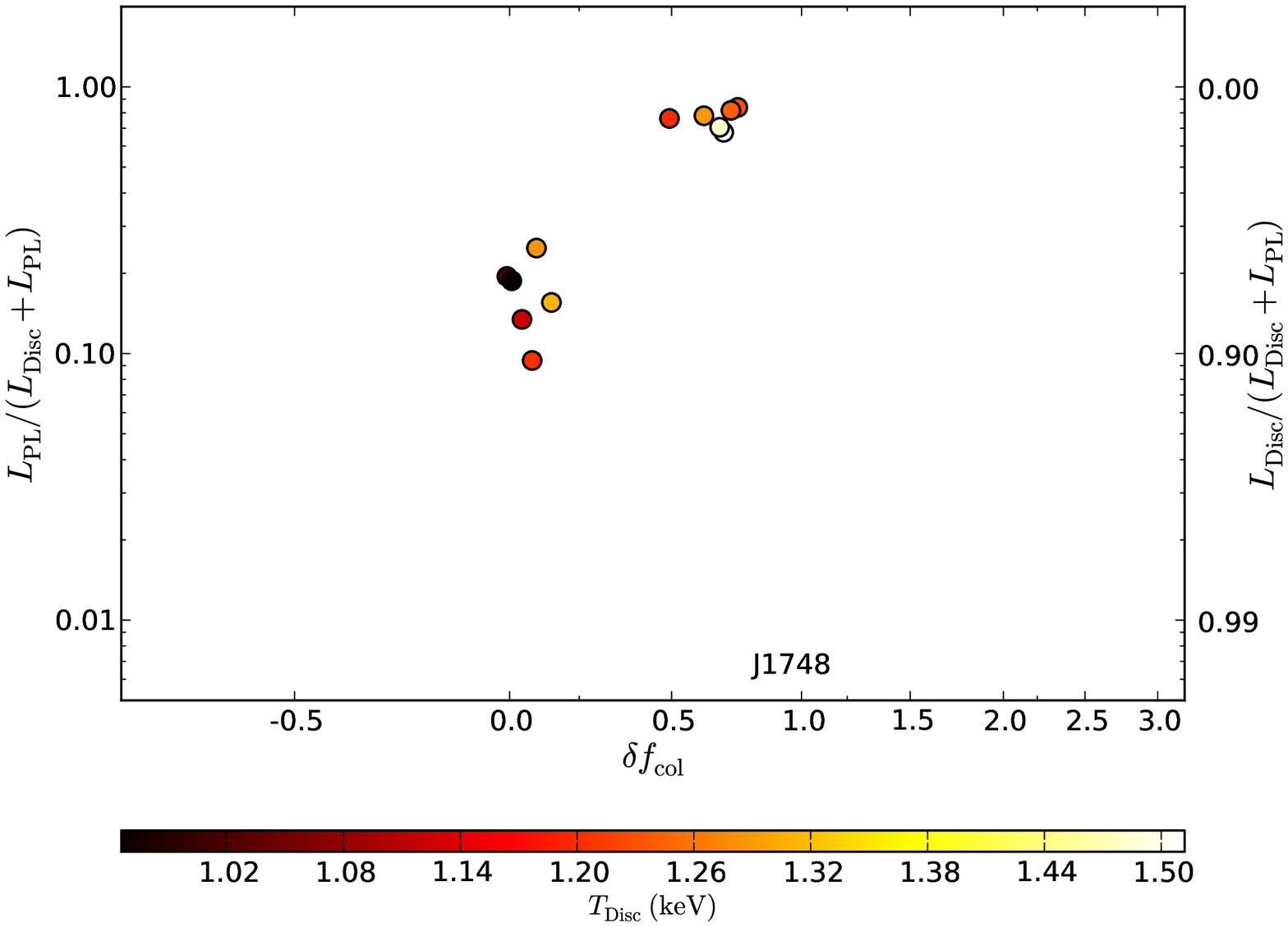}
\caption{(cont) XT~J1748-288}
\end{figure*}
\addtocounter{figure}{-1}
\begin{figure*}
\centering
\includegraphics[width=0.4\textwidth]{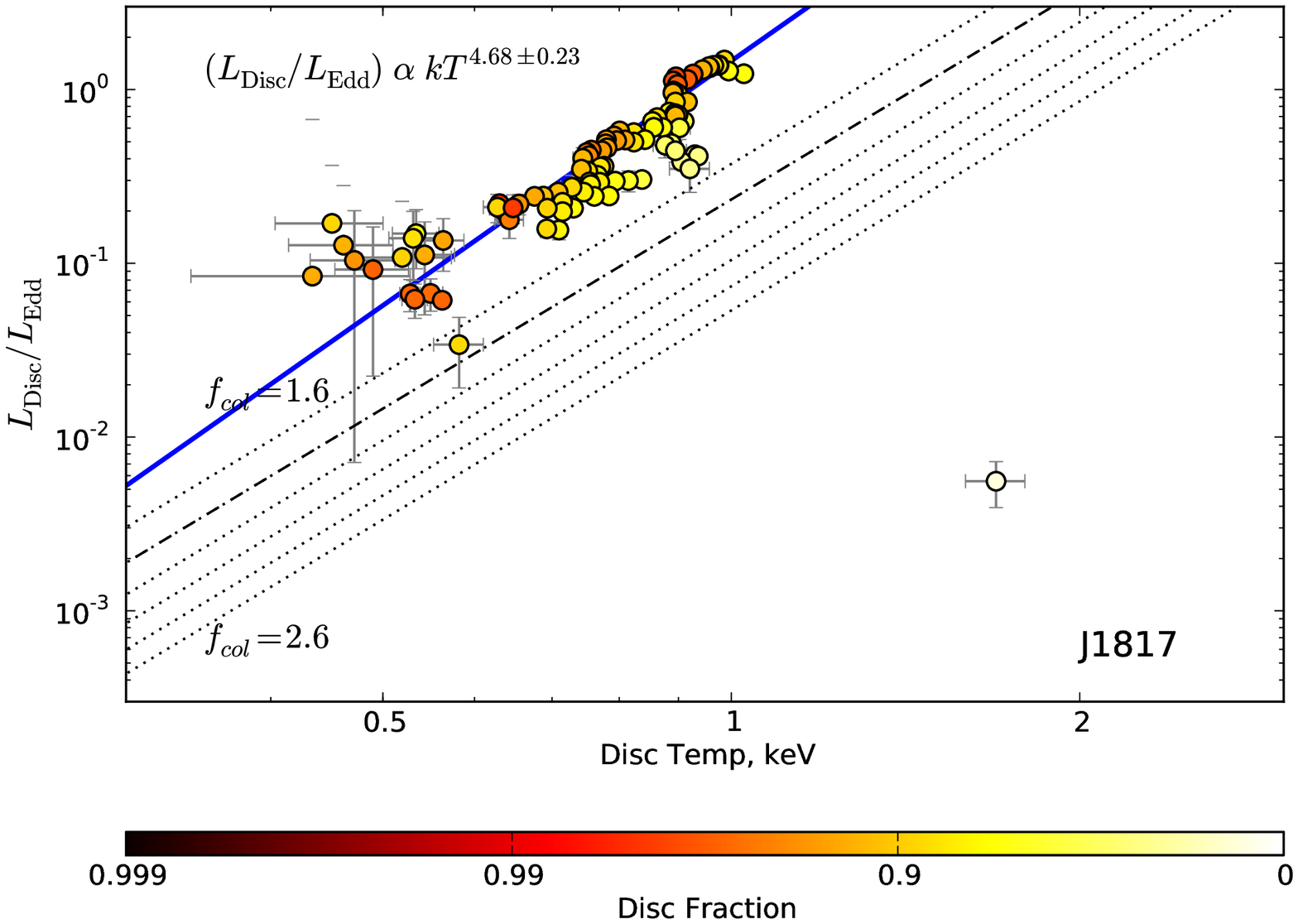}
\includegraphics[width=0.4\textwidth]{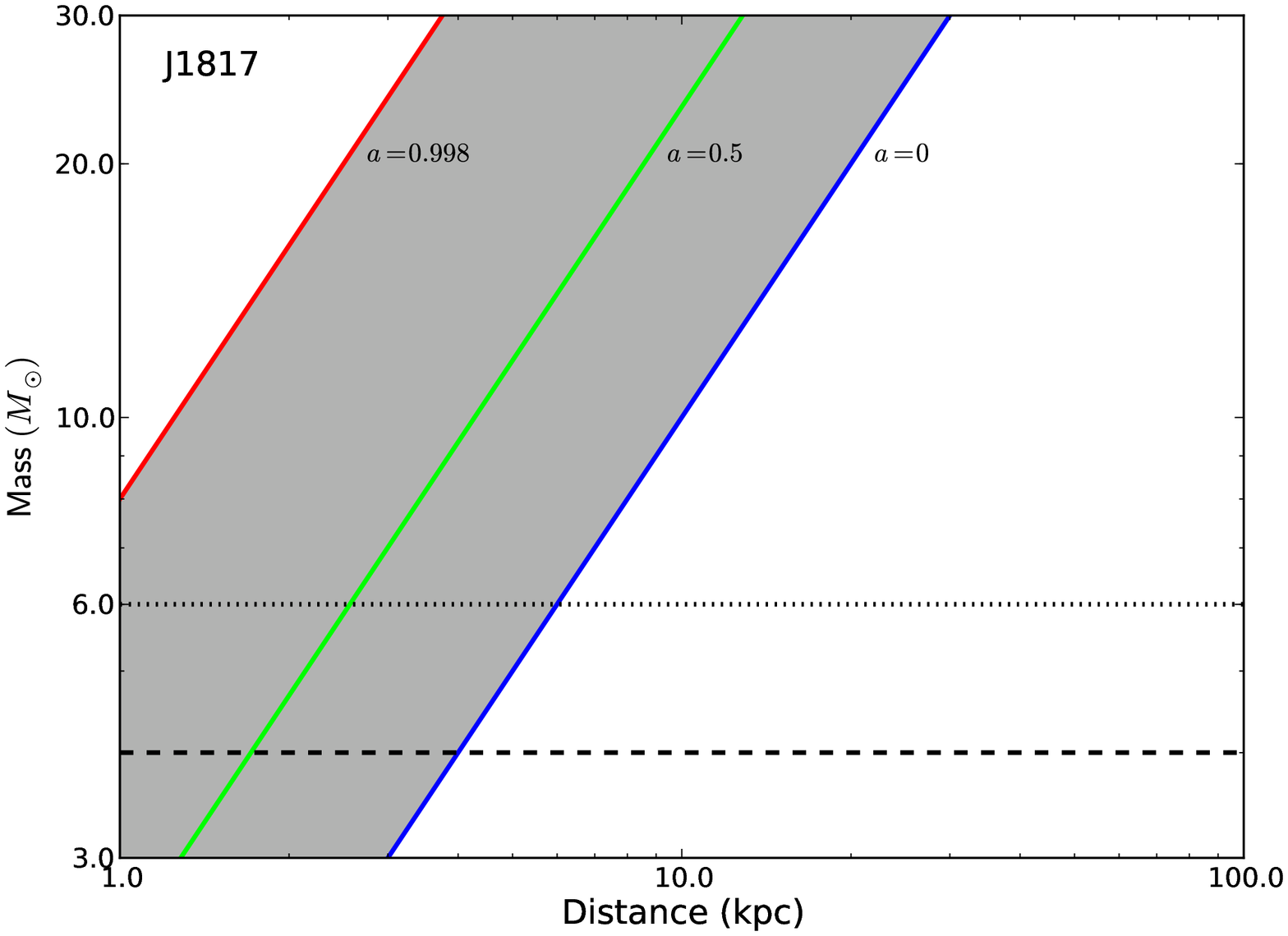}
\includegraphics[width=0.4\textwidth]{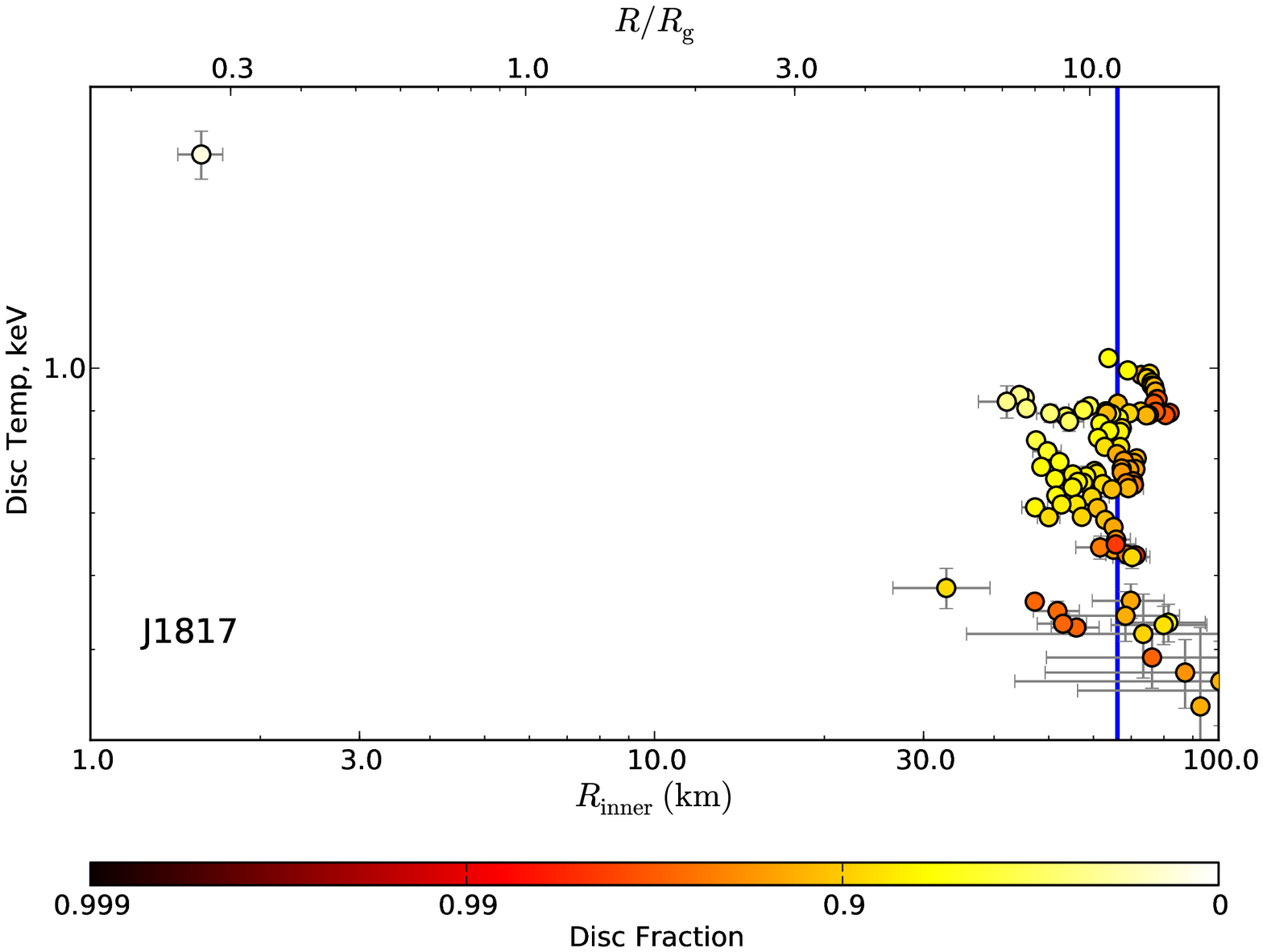}
\includegraphics[width=0.4\textwidth]{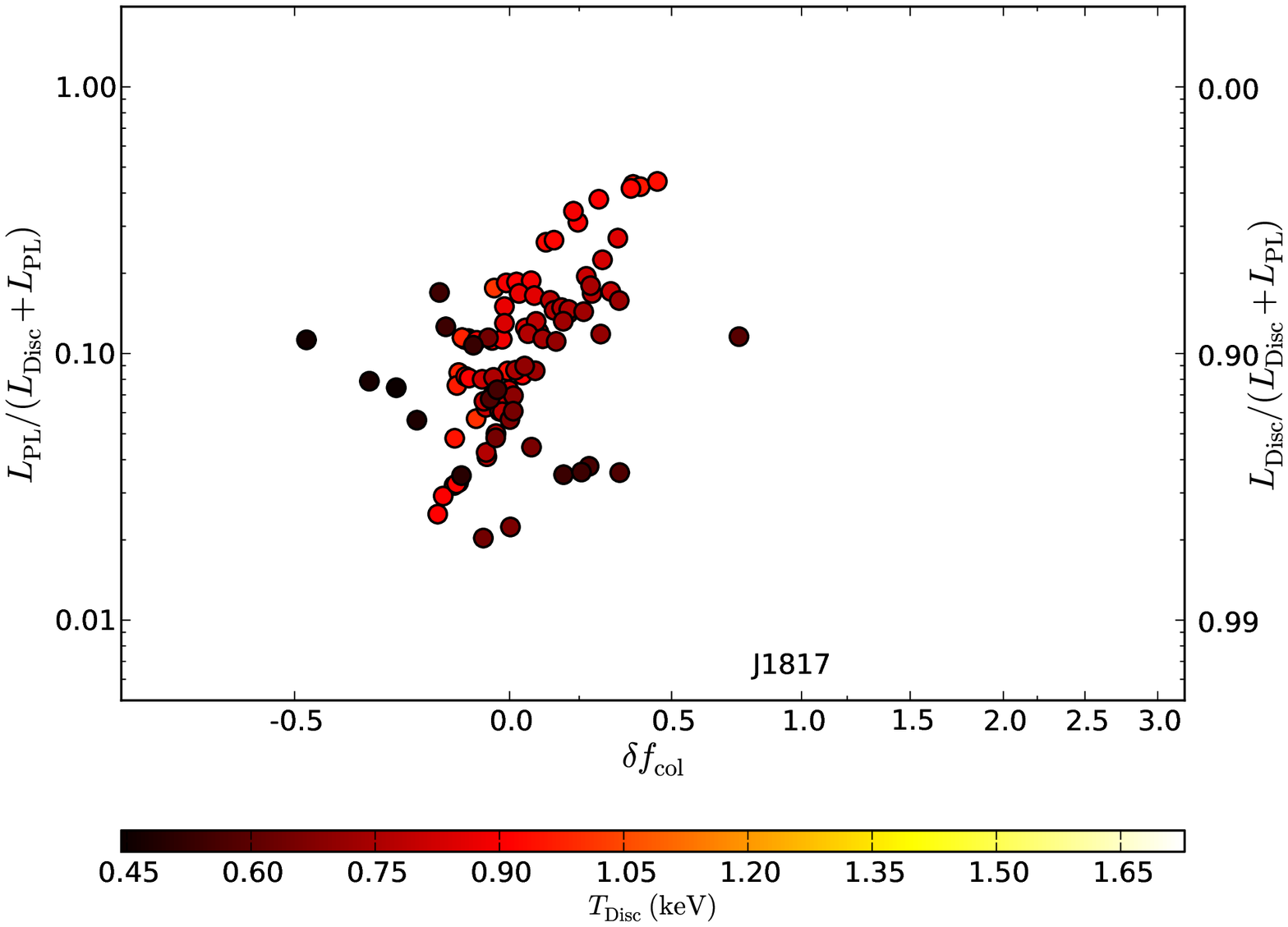}
\caption{(cont) XTE~J1817-330}
\end{figure*}
\addtocounter{figure}{-1}
\begin{figure*}
\centering
\includegraphics[width=0.4\textwidth]{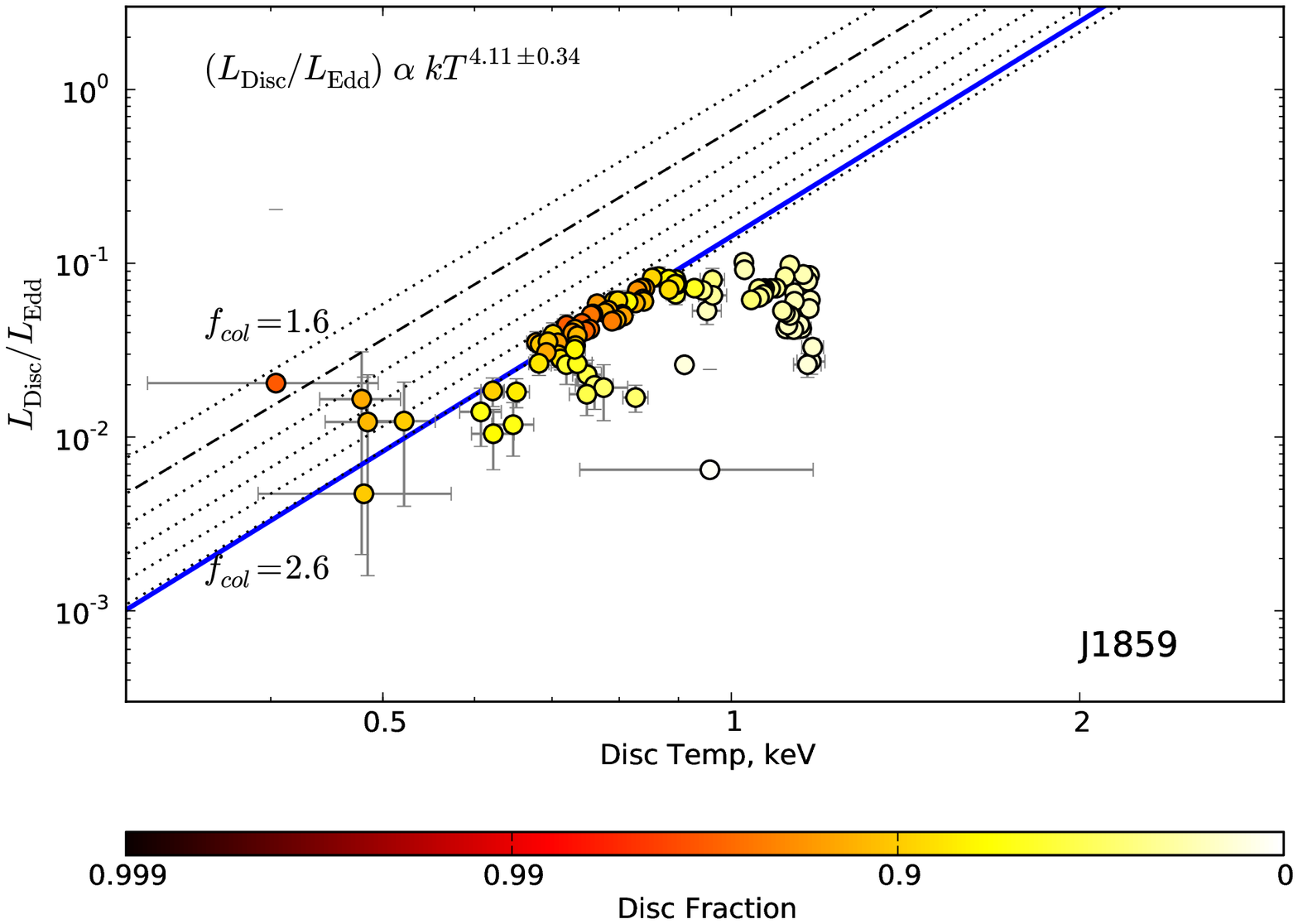}
\includegraphics[width=0.4\textwidth]{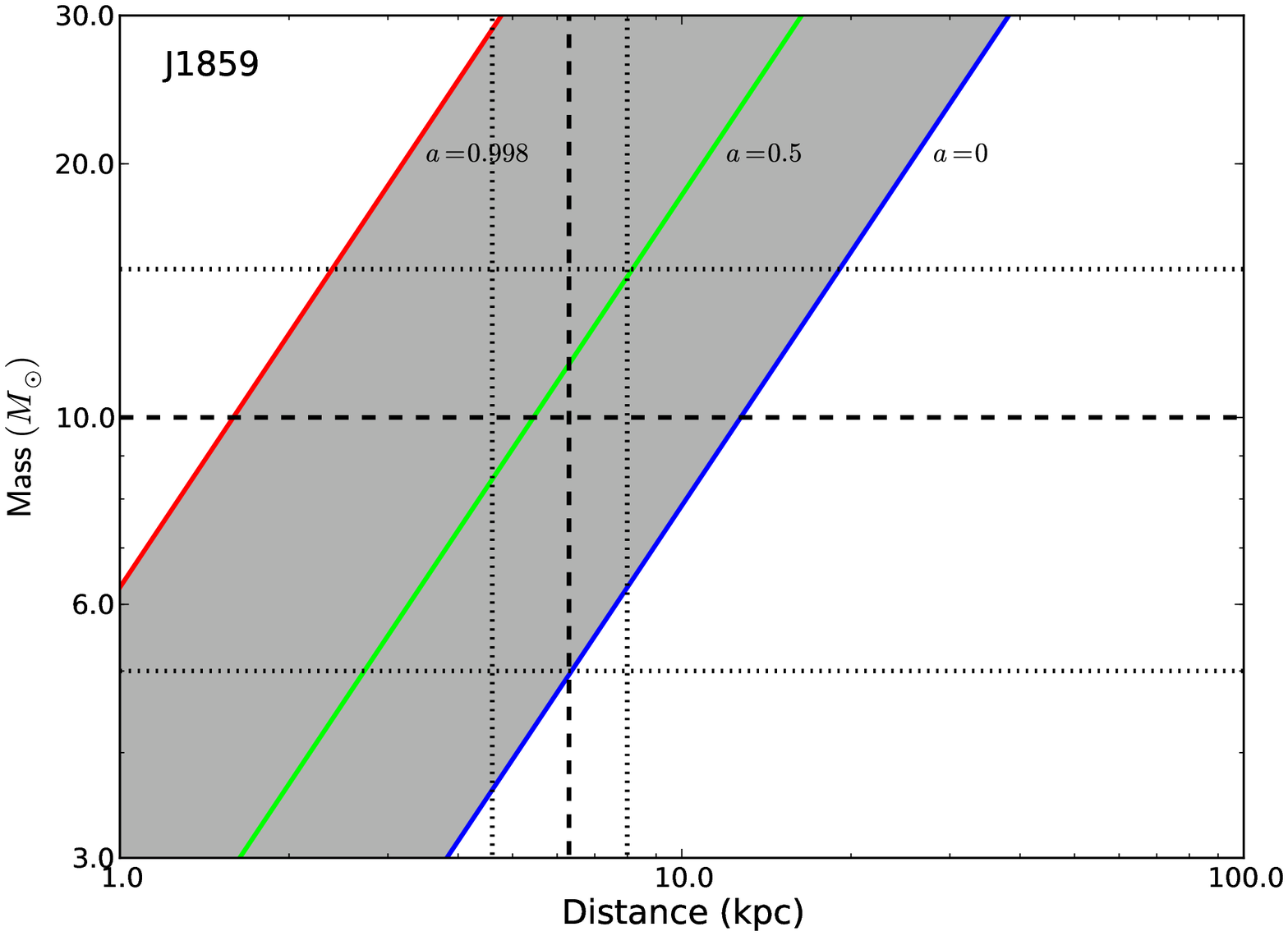}
\includegraphics[width=0.4\textwidth]{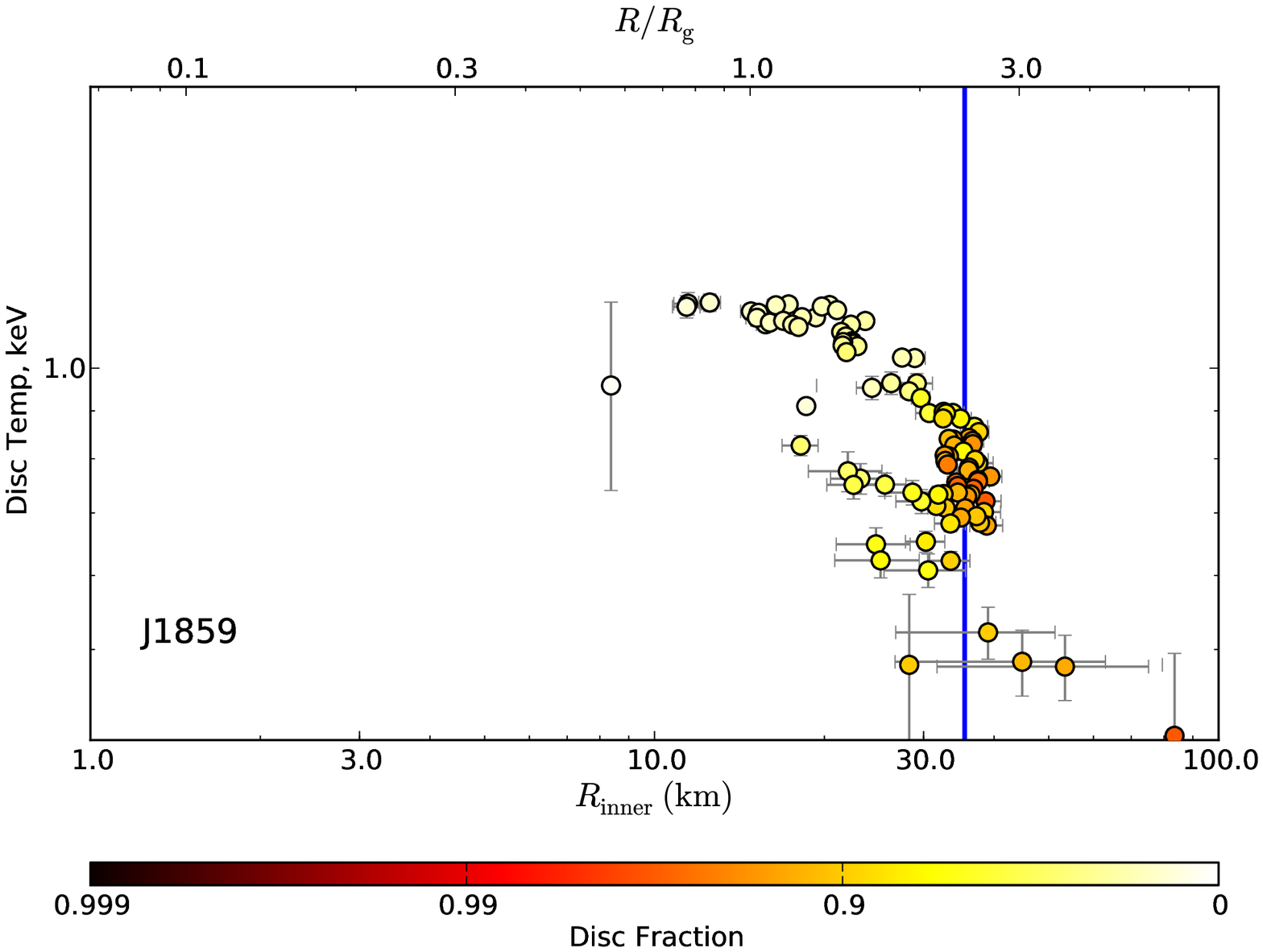}
\includegraphics[width=0.4\textwidth]{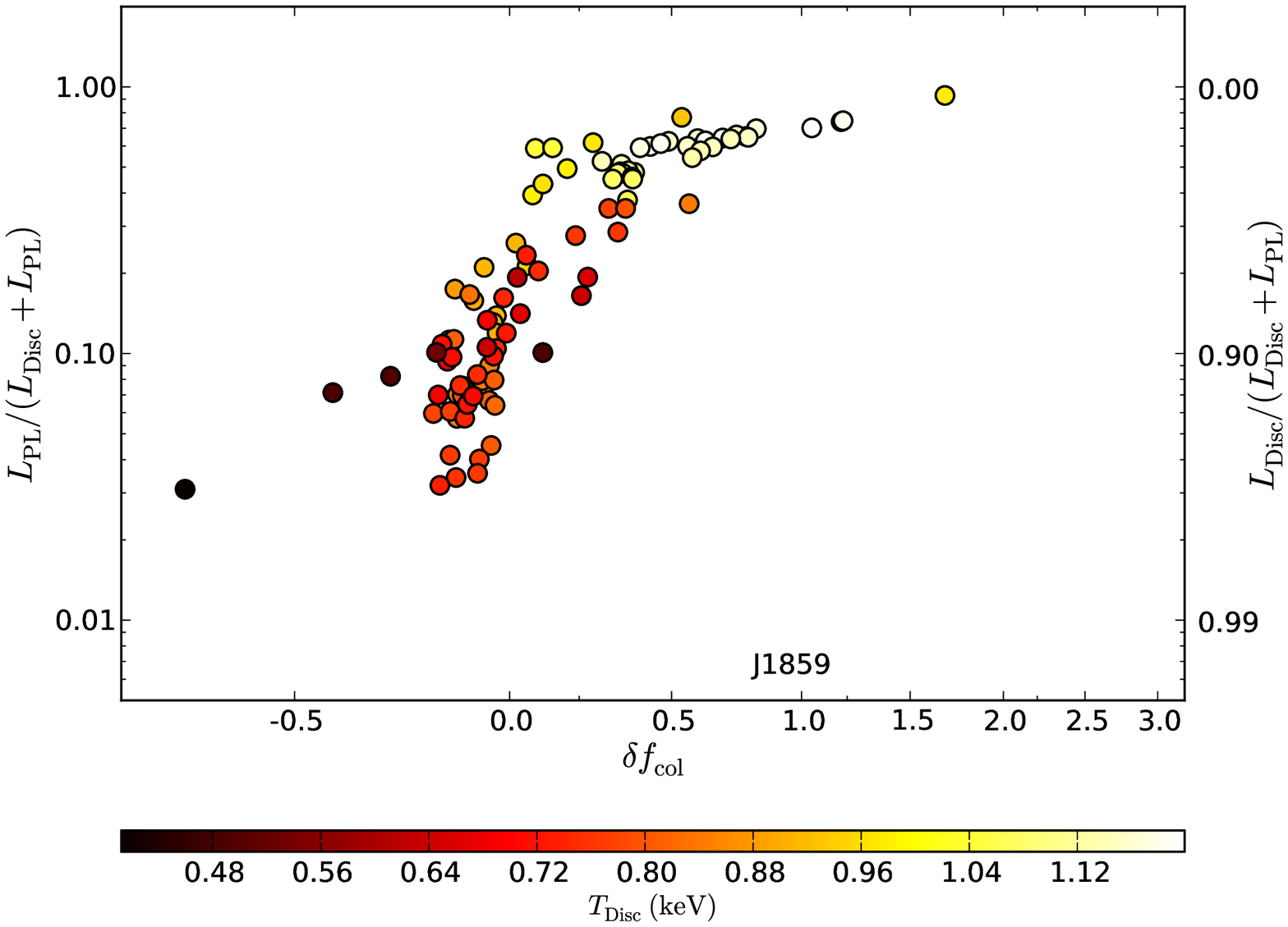}
\caption{(cont) XTE~J1859+226}
\end{figure*}
\addtocounter{figure}{-1}
\begin{figure*}
\centering
\includegraphics[width=0.4\textwidth]{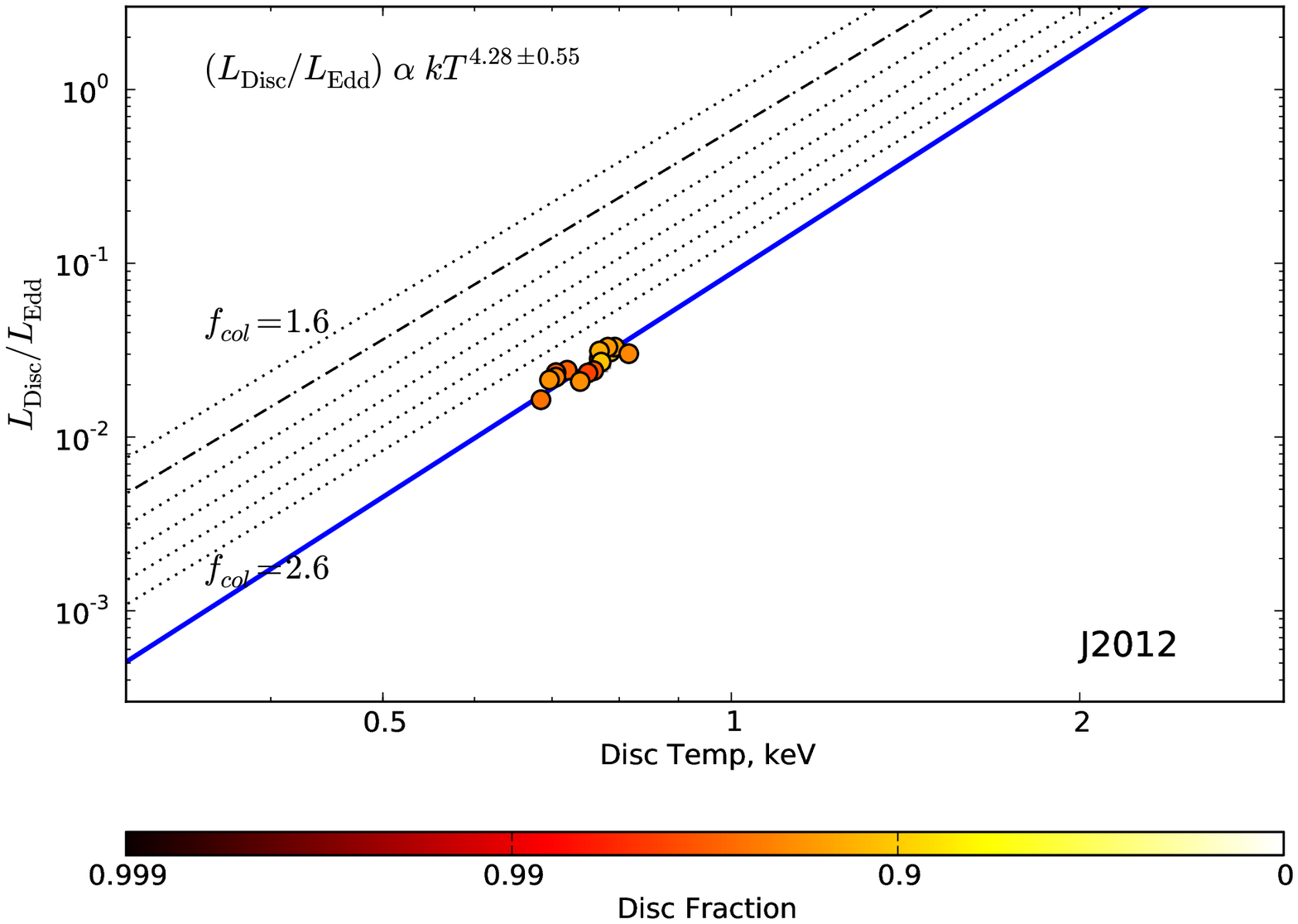}
\includegraphics[width=0.4\textwidth]{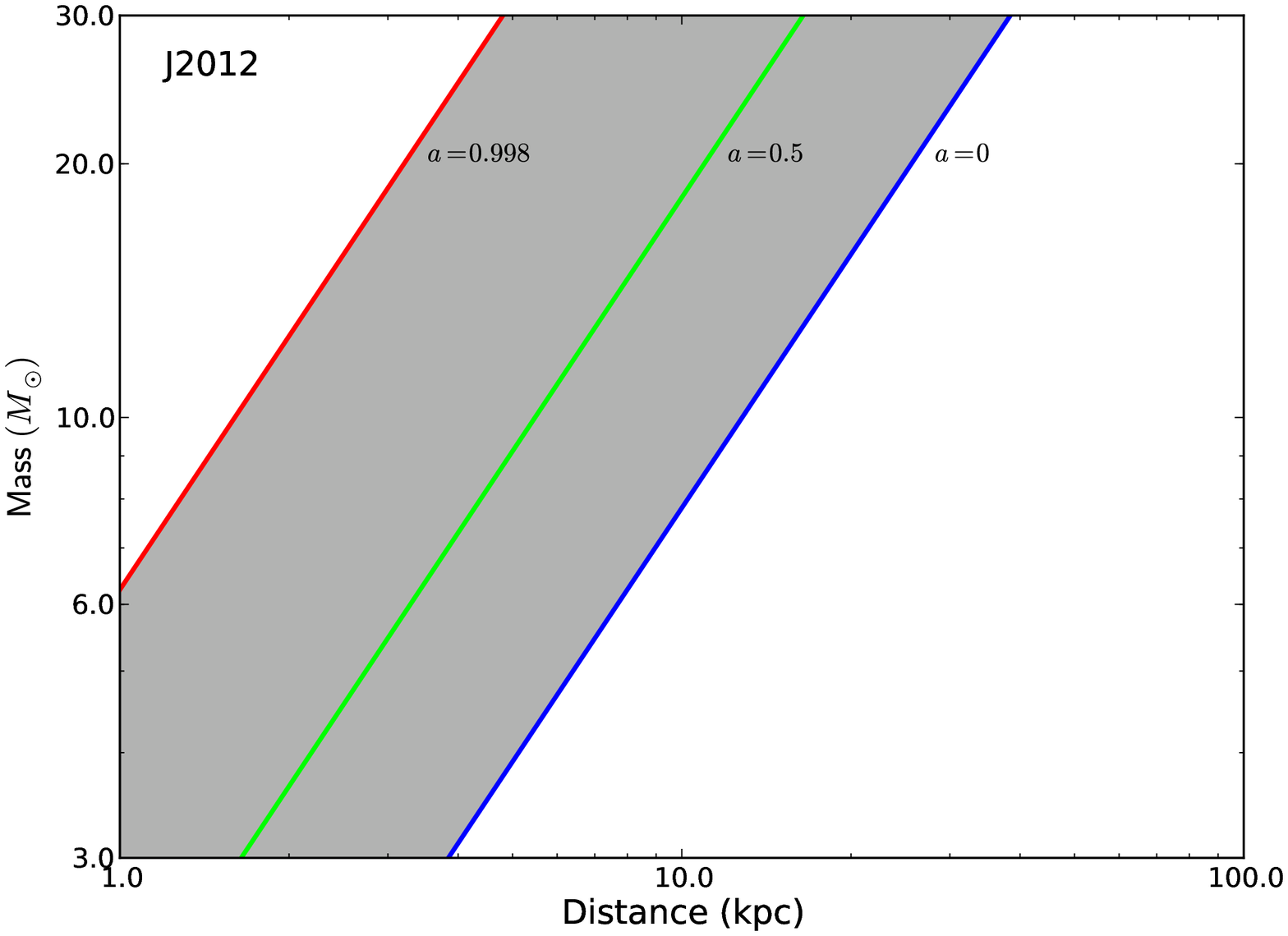}
\includegraphics[width=0.4\textwidth]{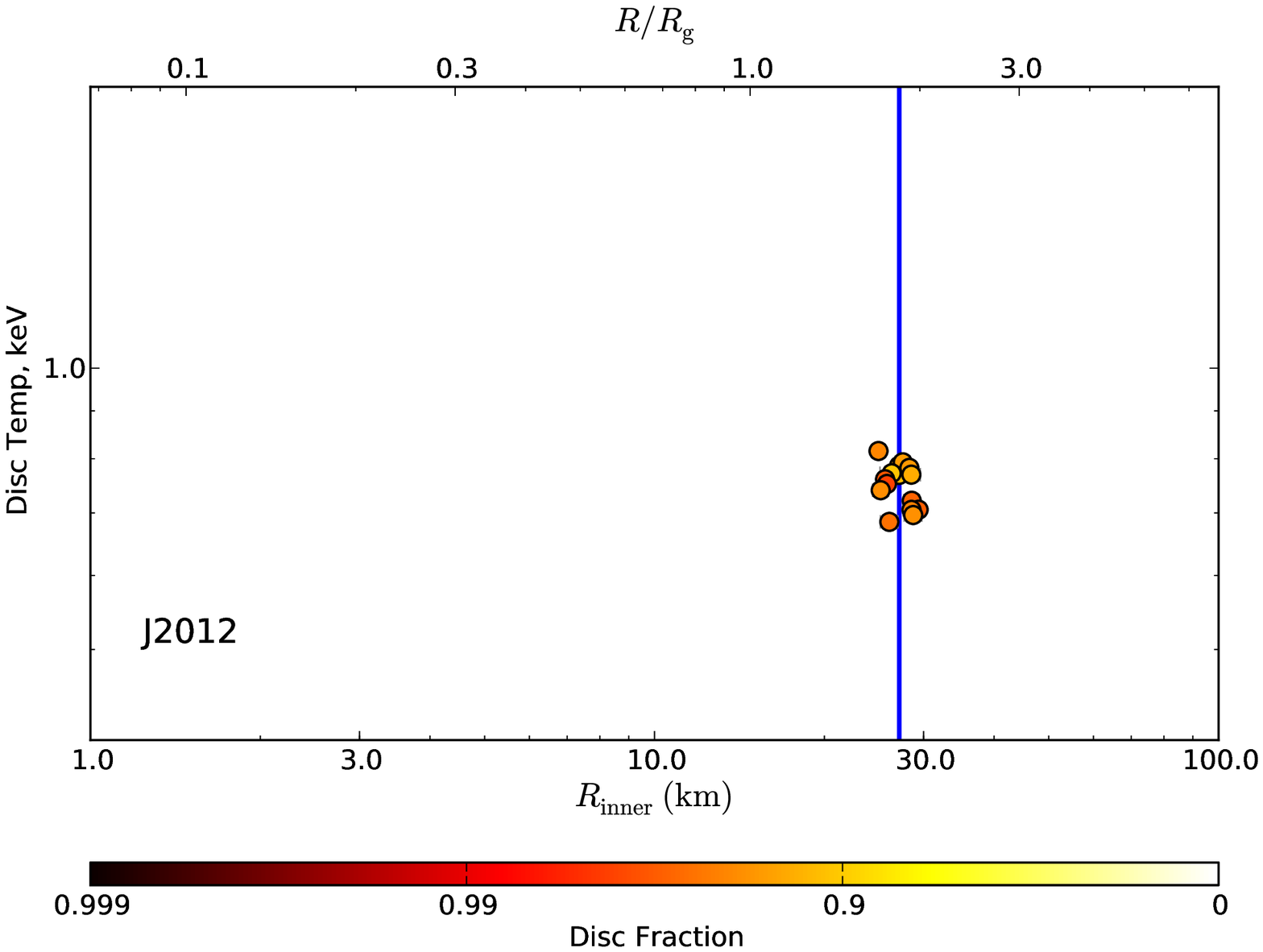}
\includegraphics[width=0.4\textwidth]{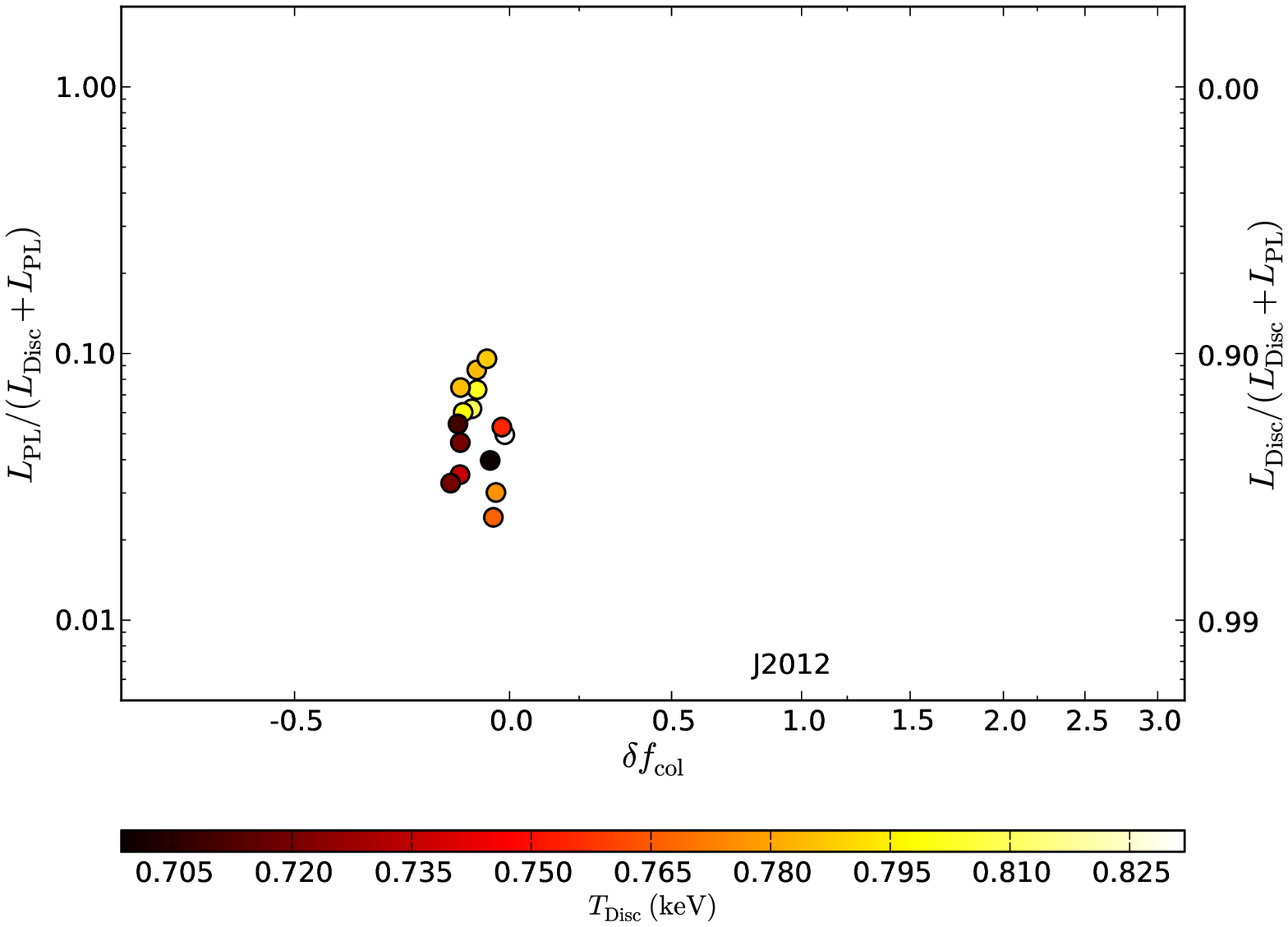}
\caption{(cont) XTE~J2012+381}
\end{figure*}
\addtocounter{figure}{-1}
\begin{figure*}
\centering
\includegraphics[width=0.4\textwidth]{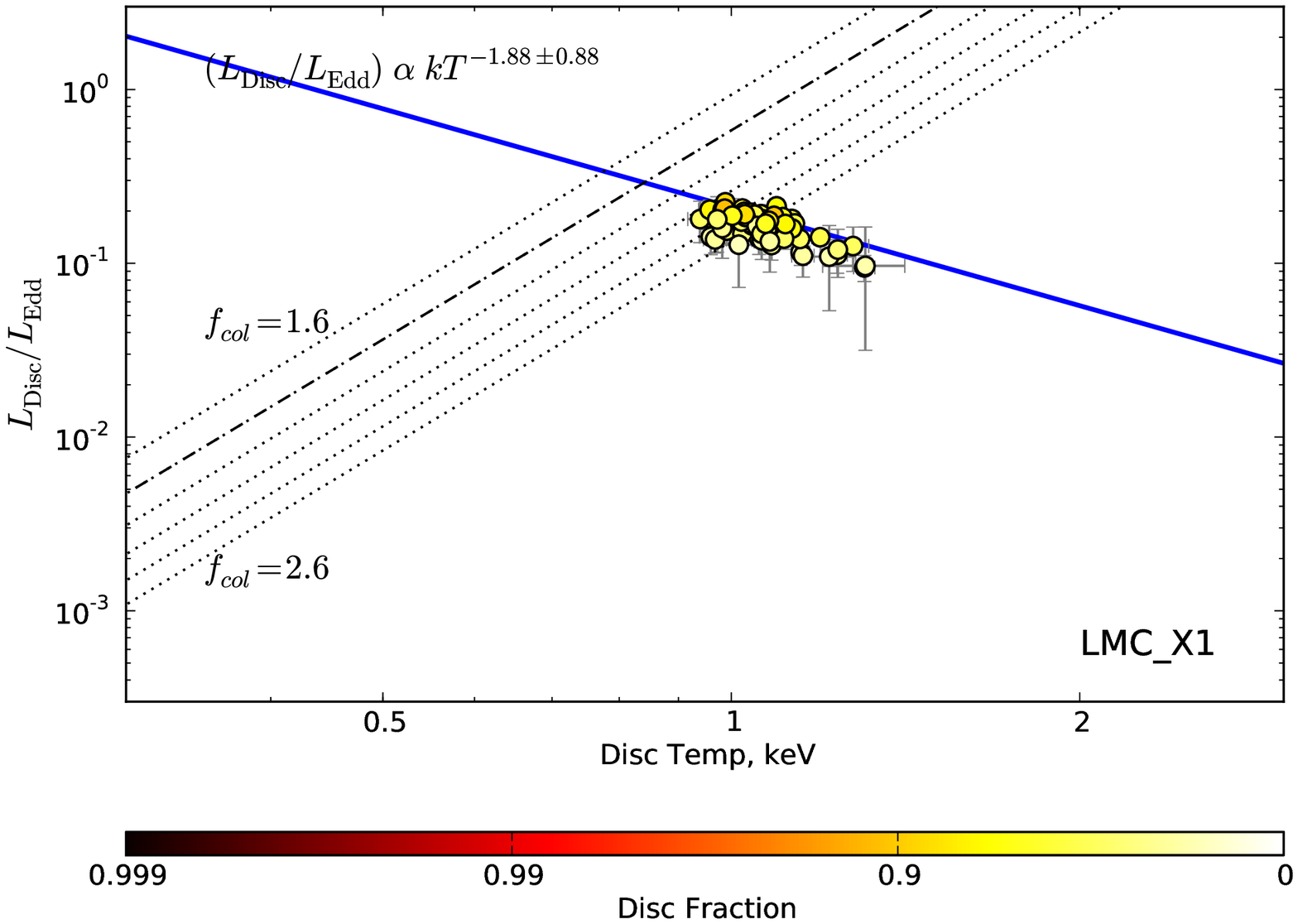}
\includegraphics[width=0.4\textwidth]{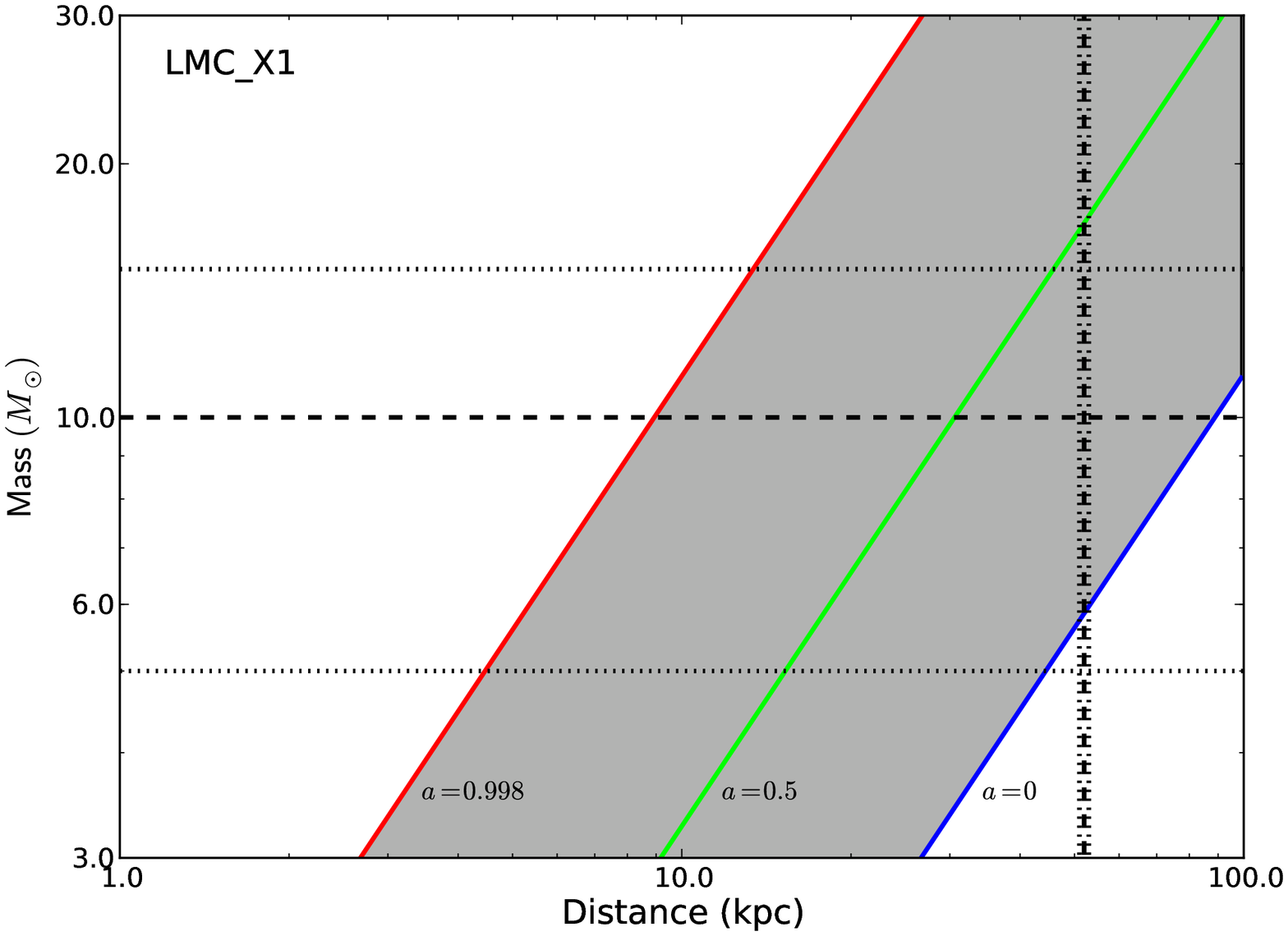}
\includegraphics[width=0.4\textwidth]{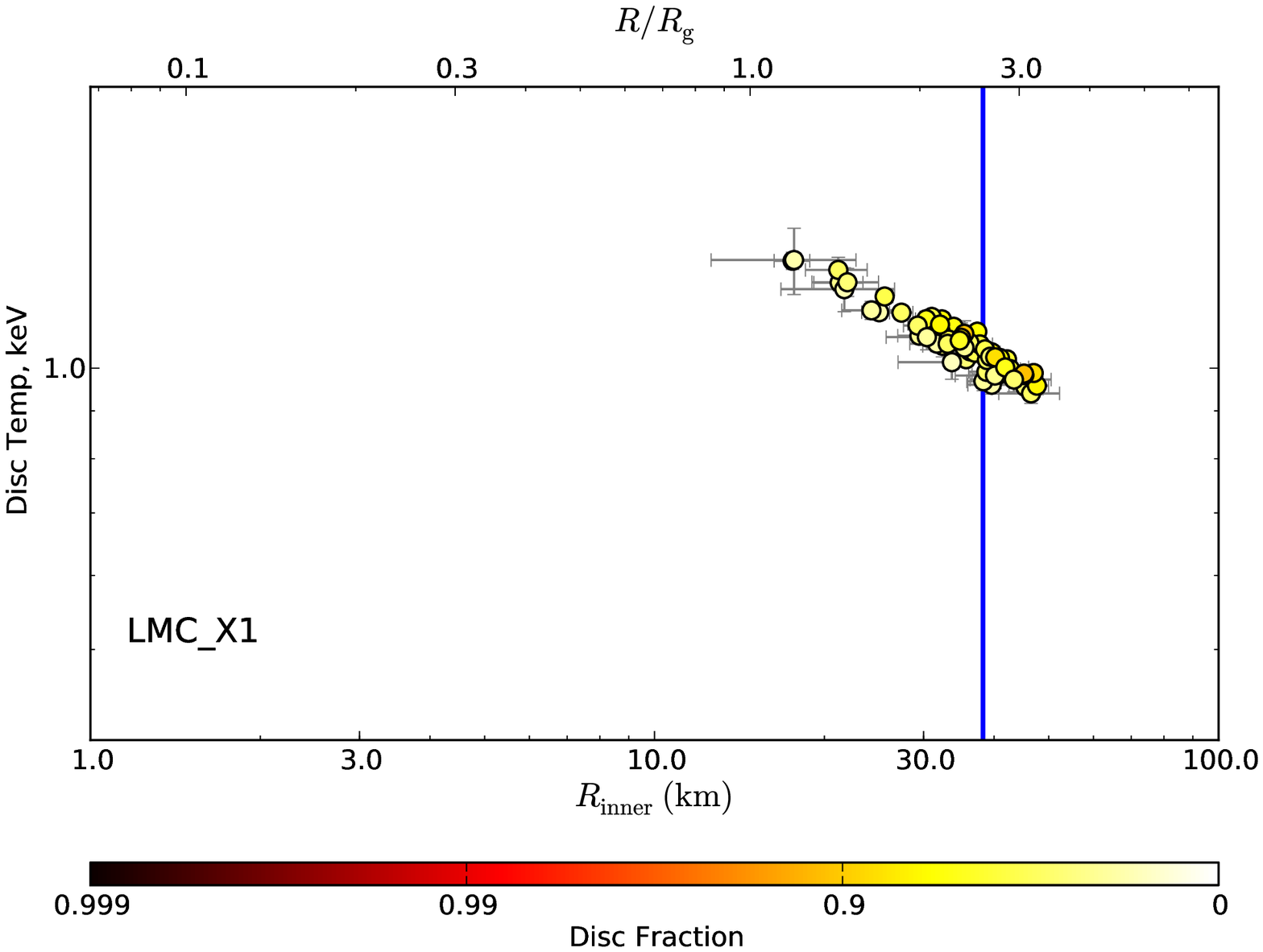}
\includegraphics[width=0.4\textwidth]{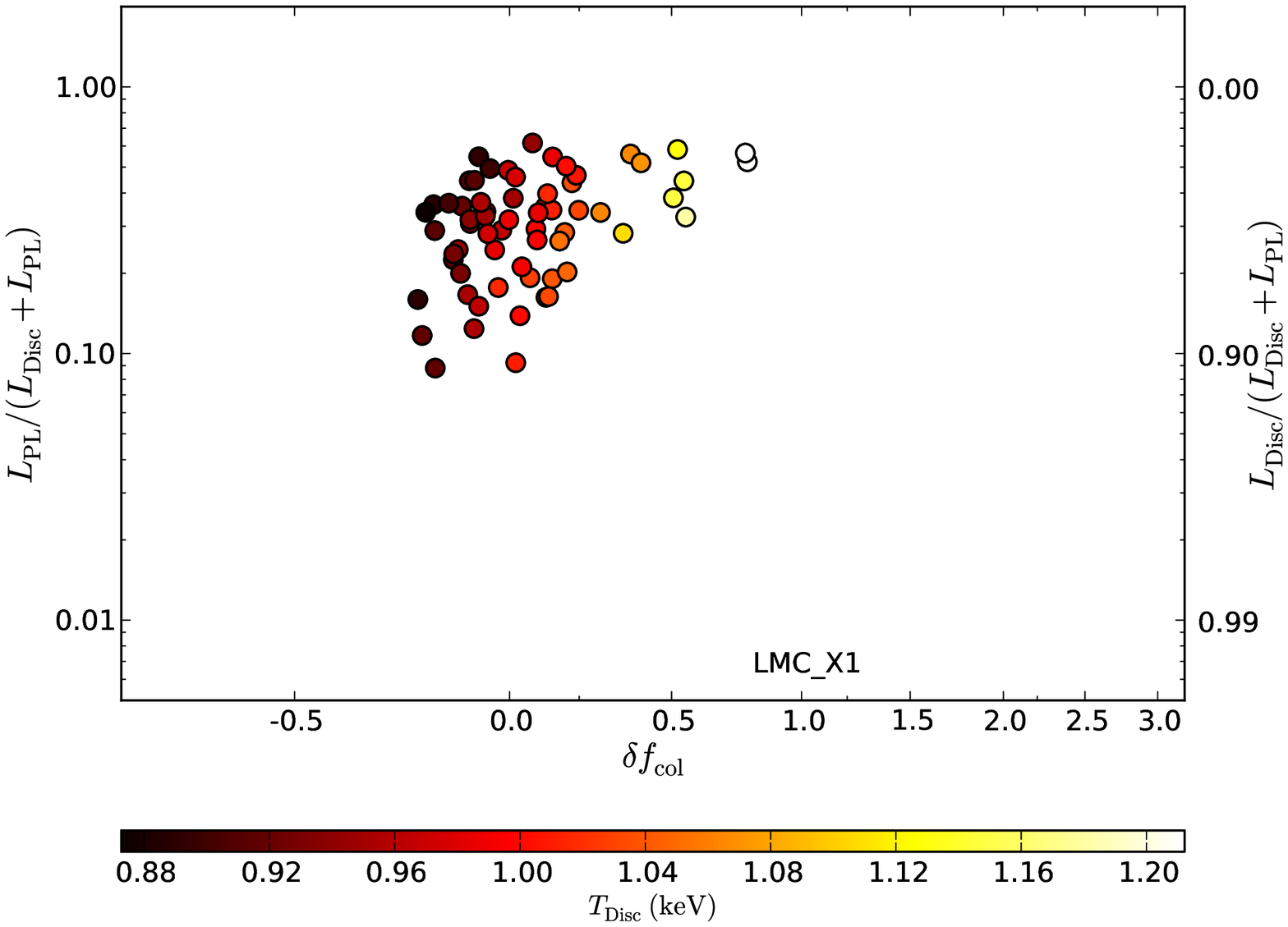}
\caption{(cont) LMC~X-1}
\end{figure*}
\addtocounter{figure}{-1}
\begin{figure*}
\centering
\includegraphics[width=0.4\textwidth]{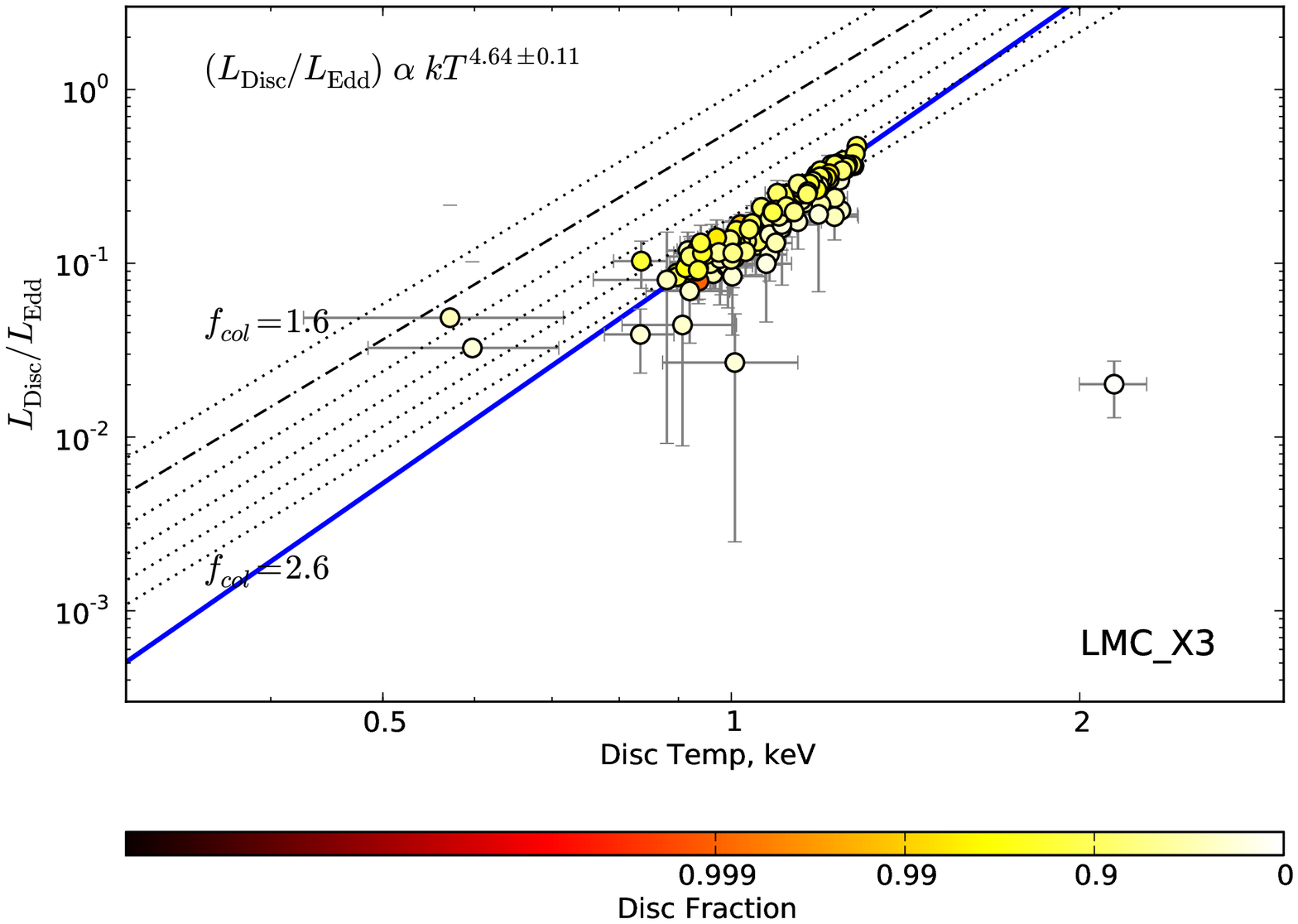}
\includegraphics[width=0.4\textwidth]{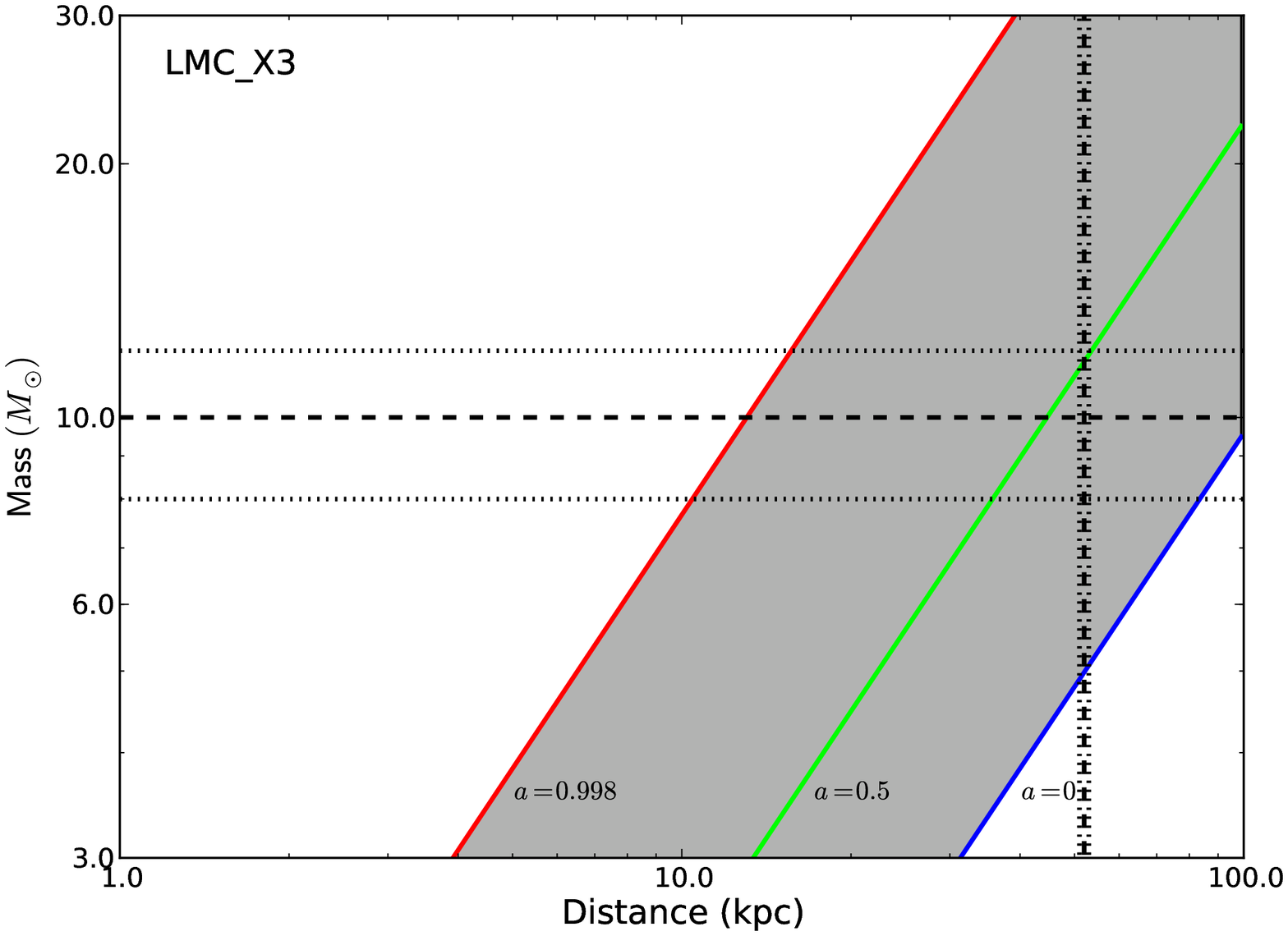}
\includegraphics[width=0.4\textwidth]{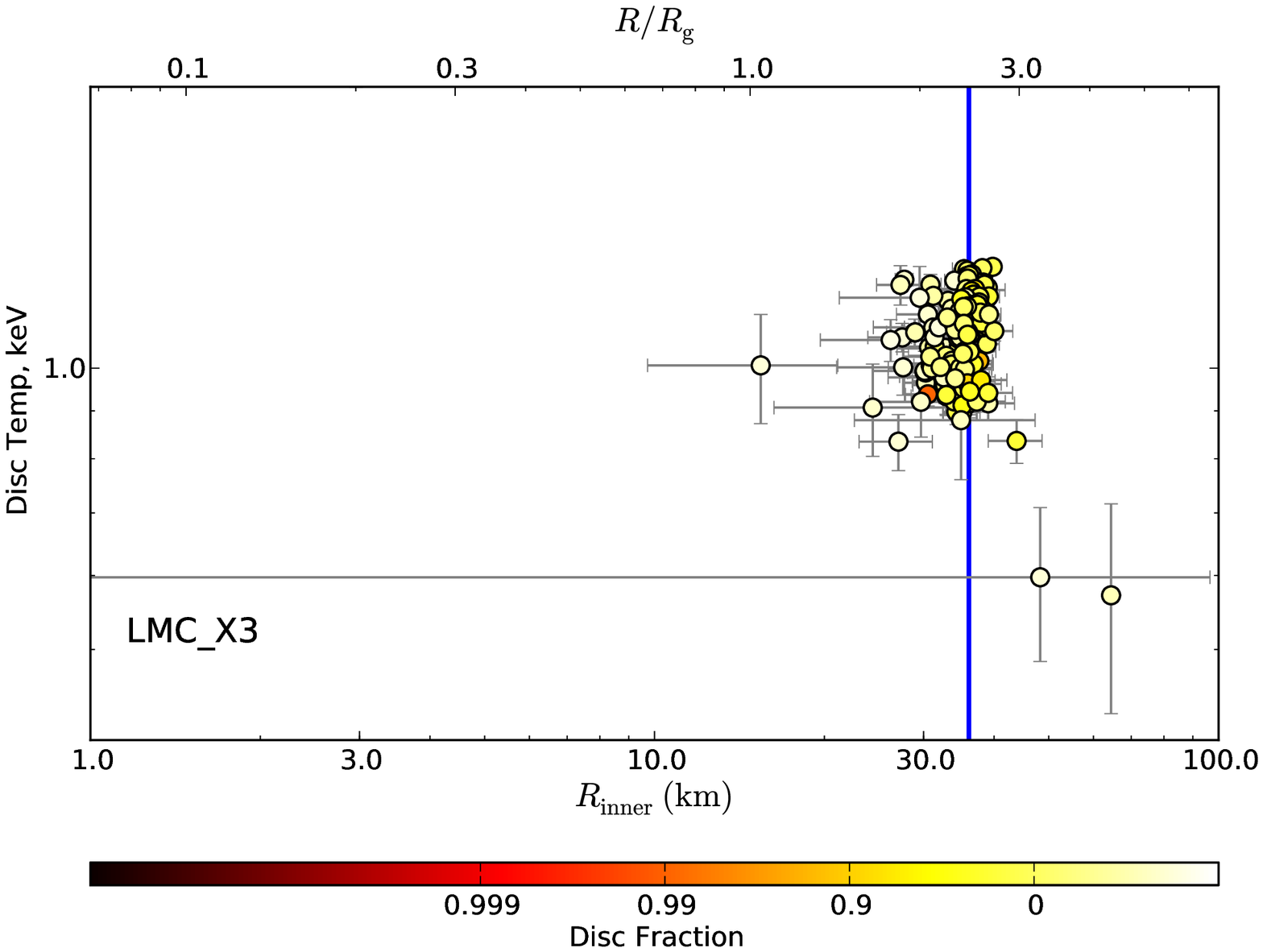}
\includegraphics[width=0.4\textwidth]{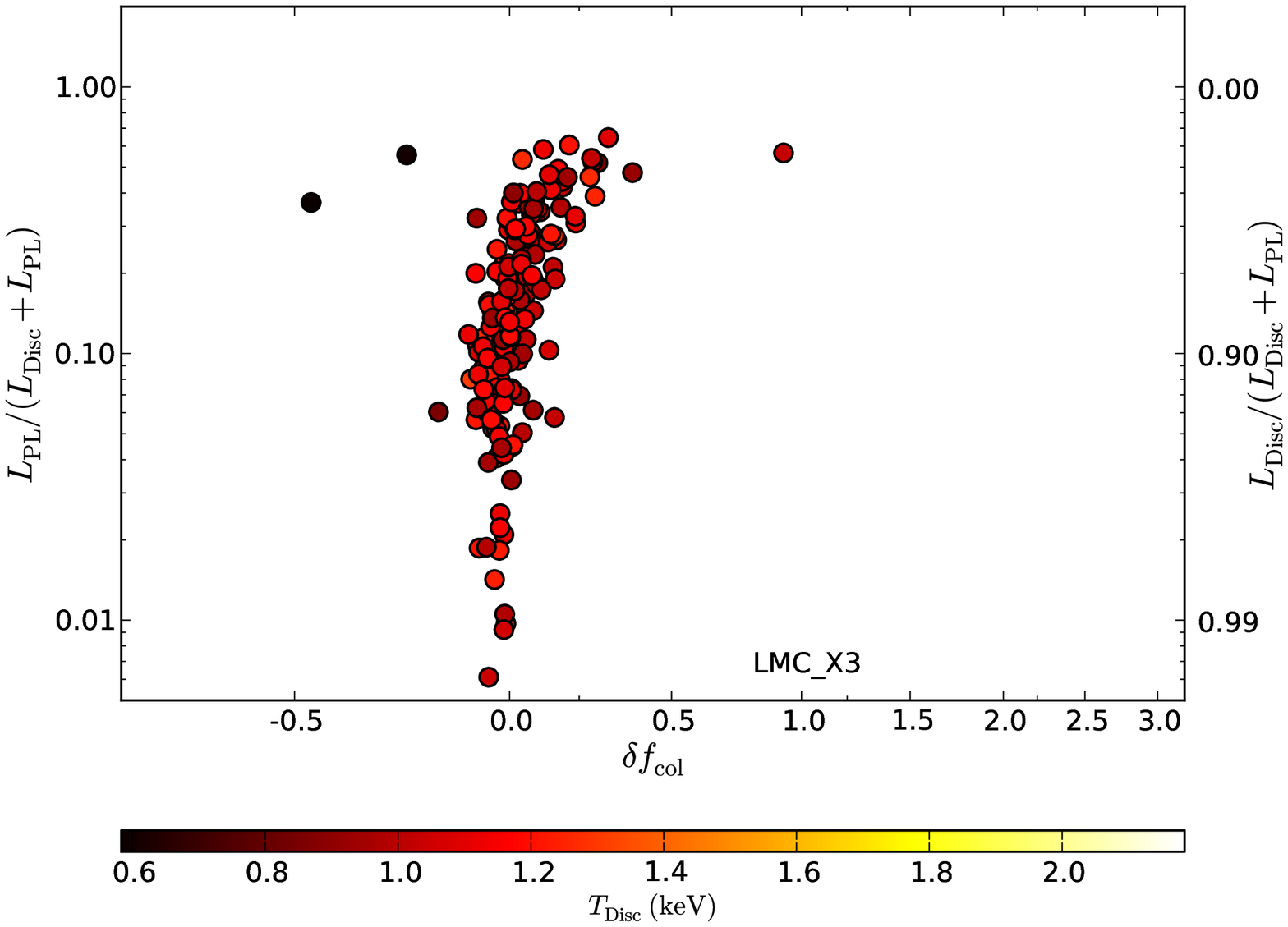}
\caption{(cont) LMC~X-3}
\end{figure*}
\addtocounter{figure}{-1}
\begin{figure*}
\centering
\includegraphics[width=0.4\textwidth]{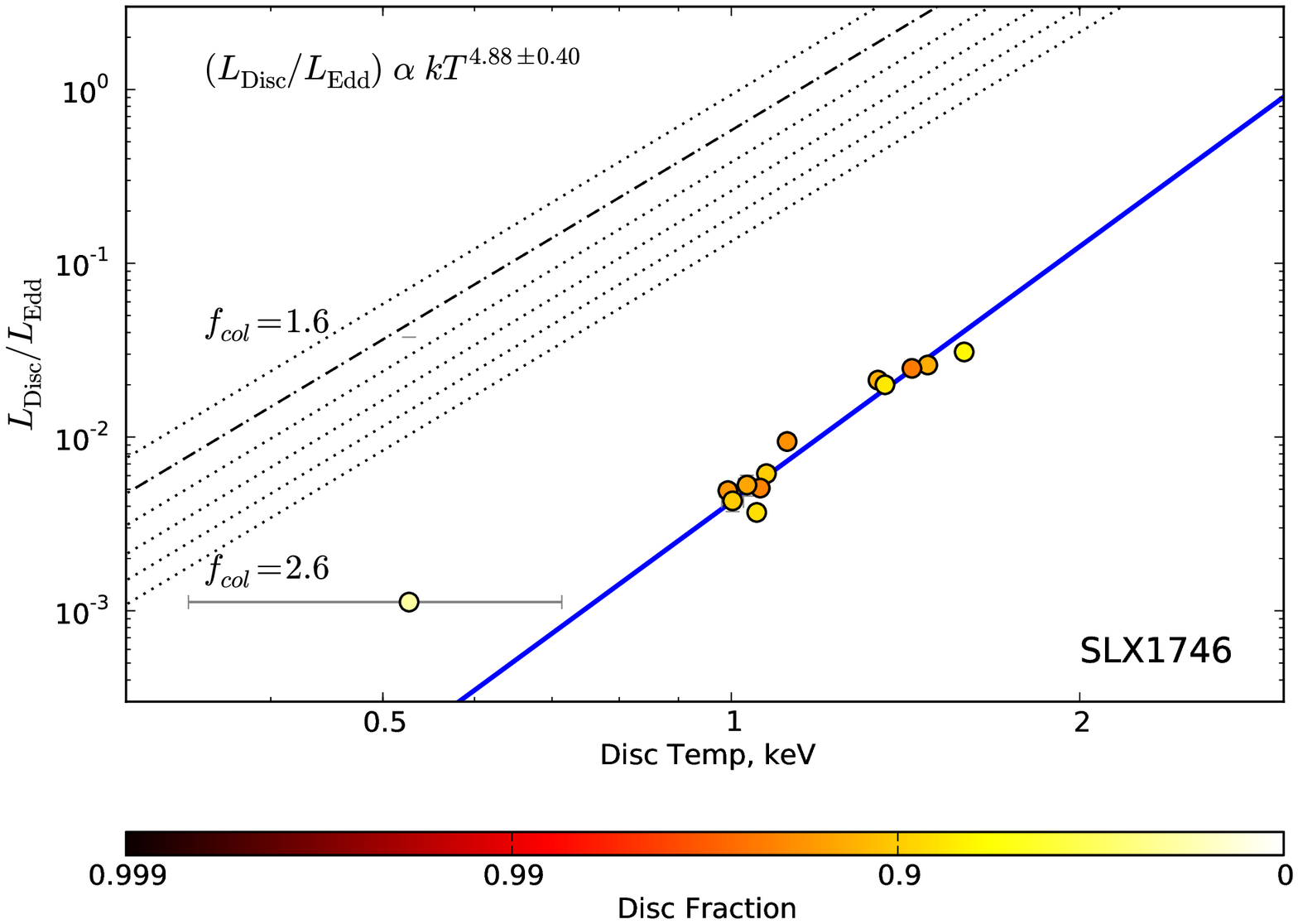}
\includegraphics[width=0.4\textwidth]{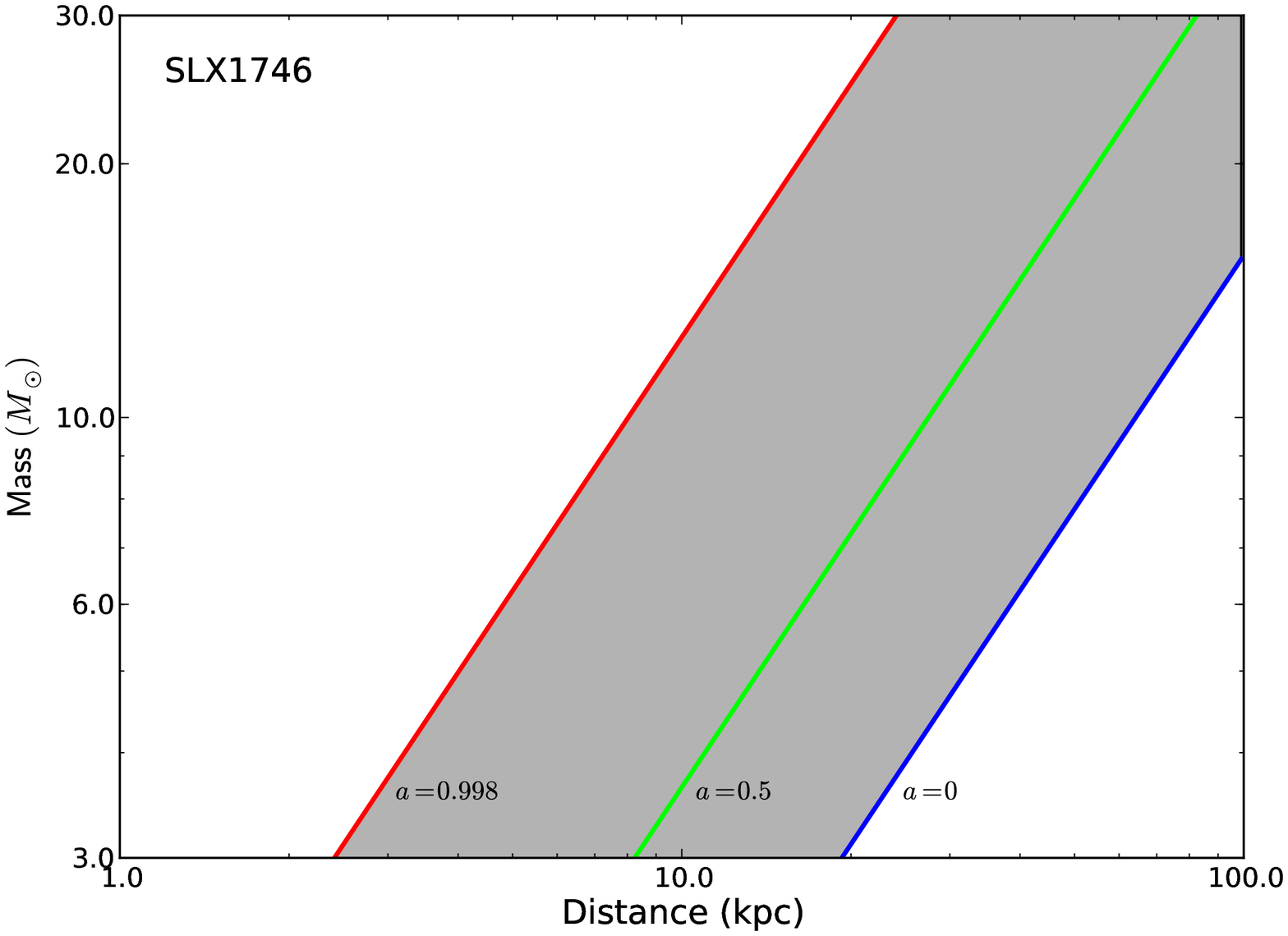}
\includegraphics[width=0.4\textwidth]{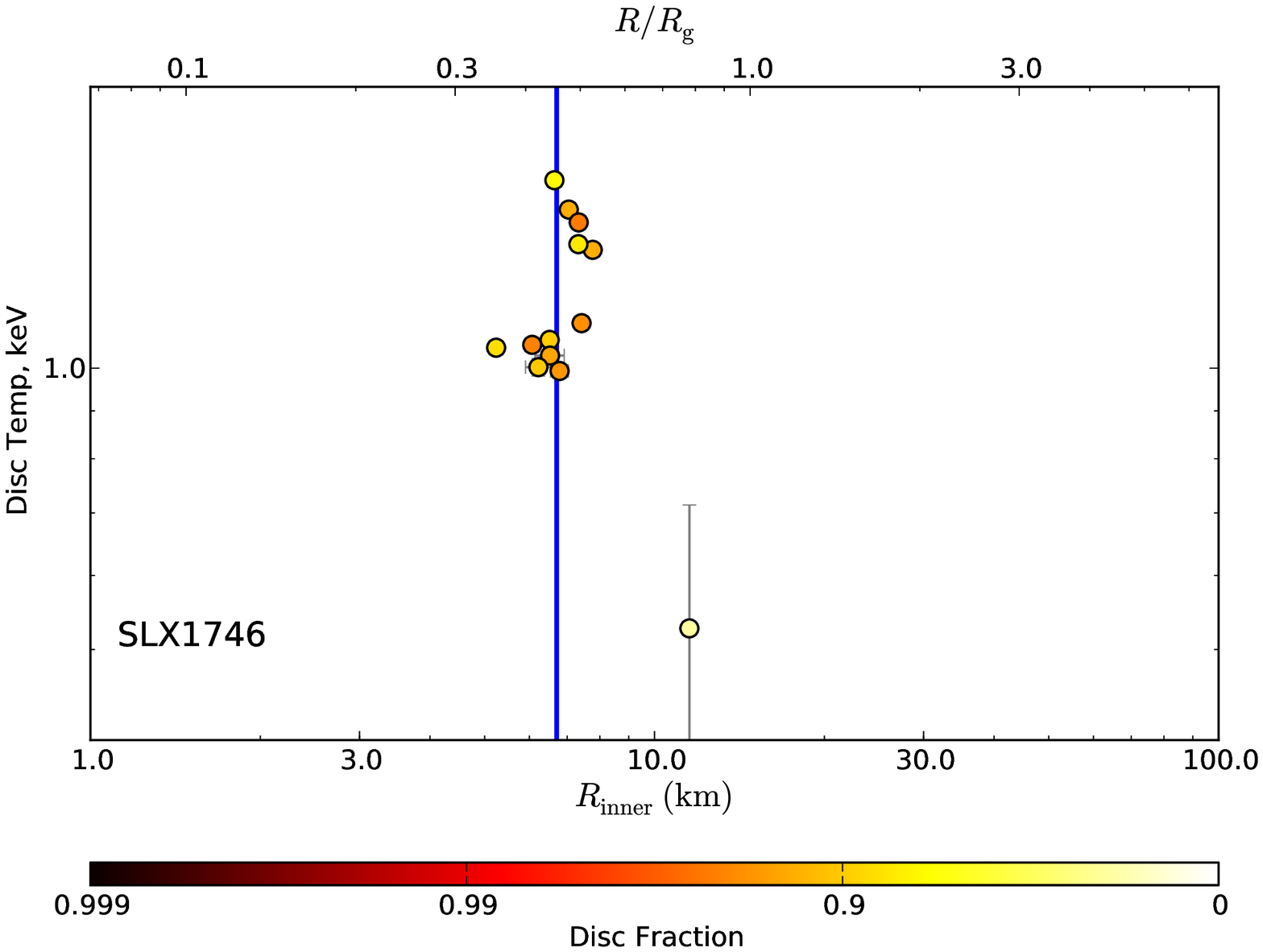}
\includegraphics[width=0.4\textwidth]{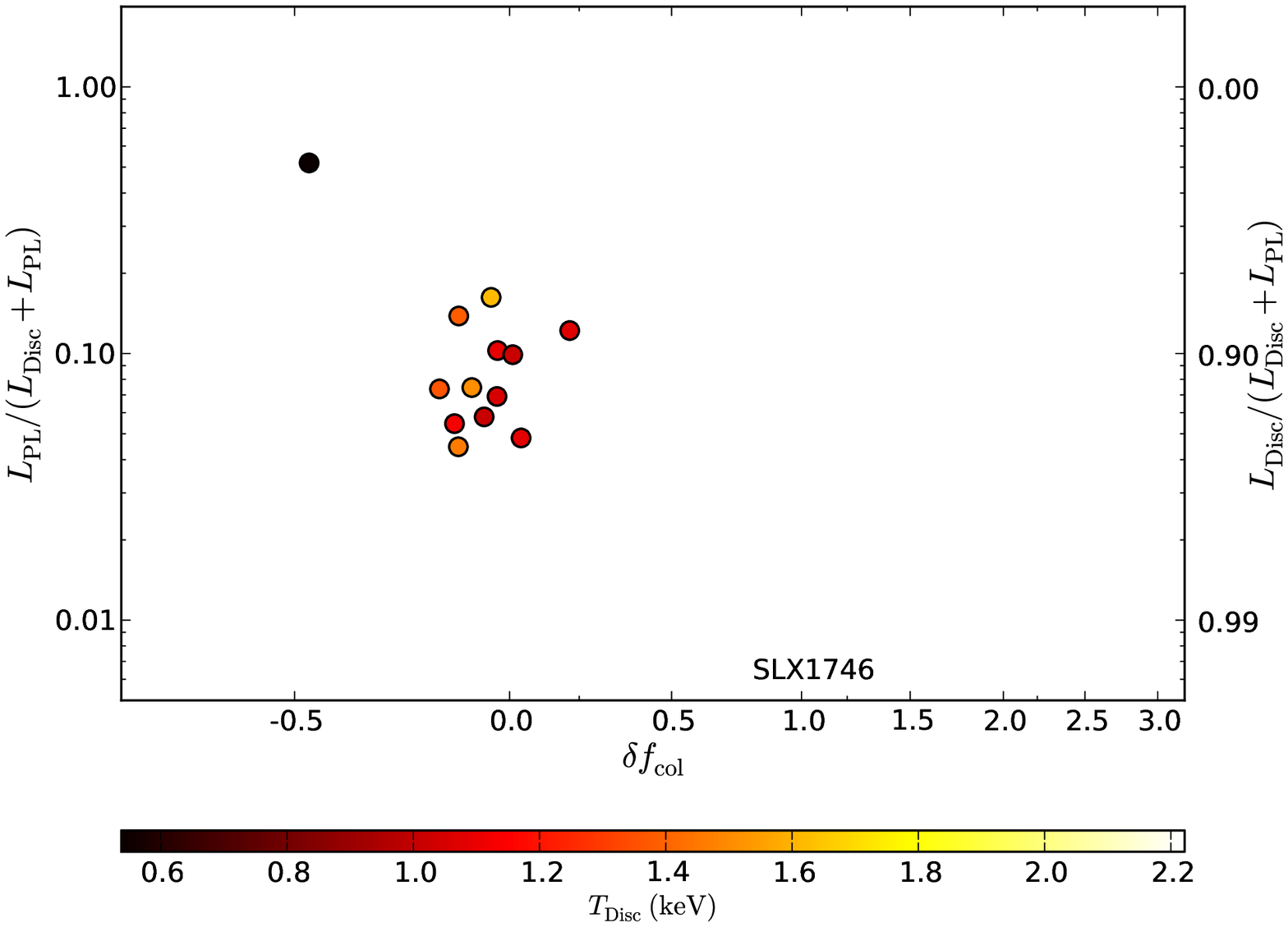}
\caption{(cont) SLX~1746-331}
\end{figure*}

\clearpage

\begin{figure*}
\centering
\includegraphics[width=0.9\textwidth]{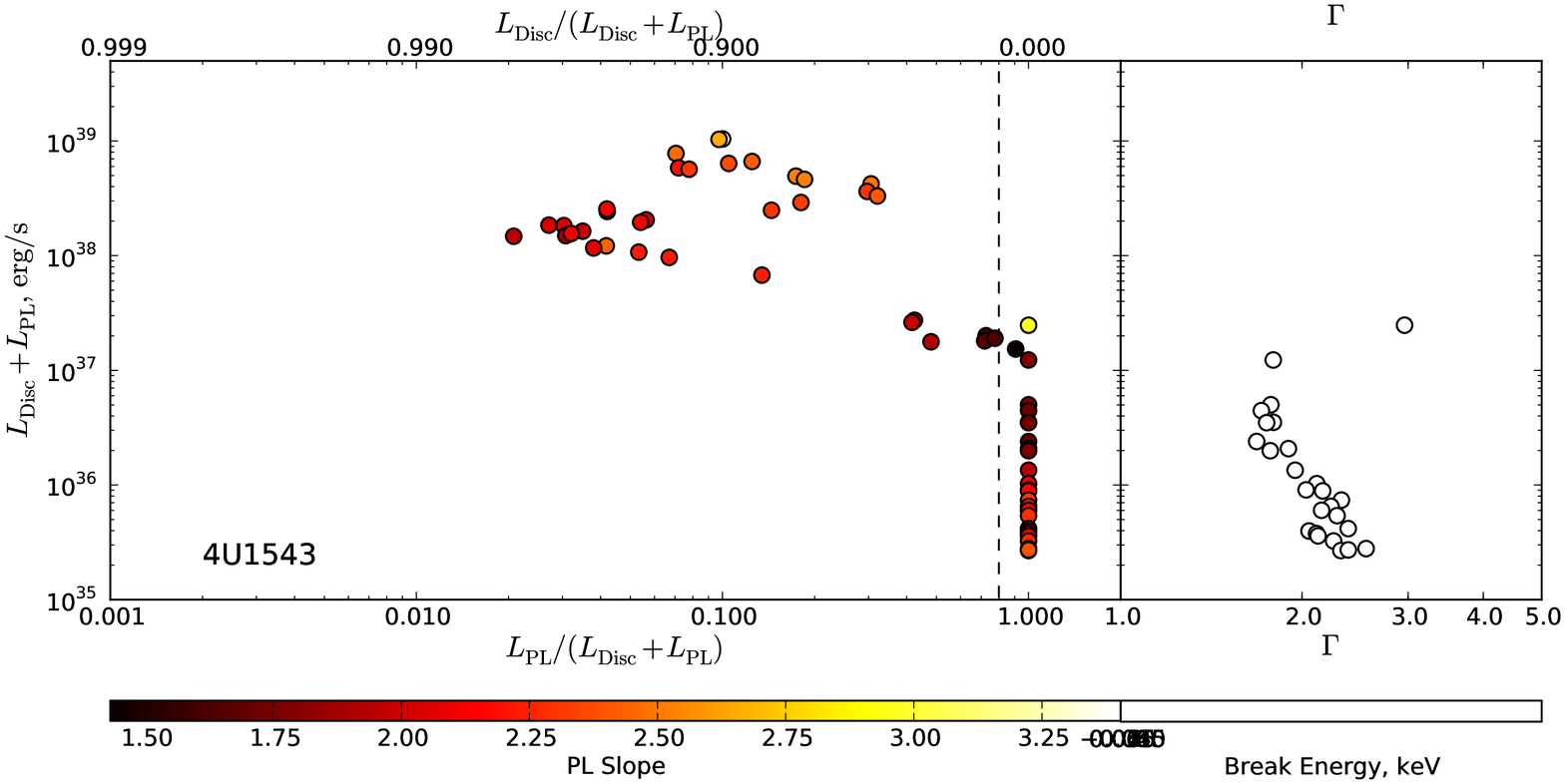}
\includegraphics[width=0.9\textwidth]{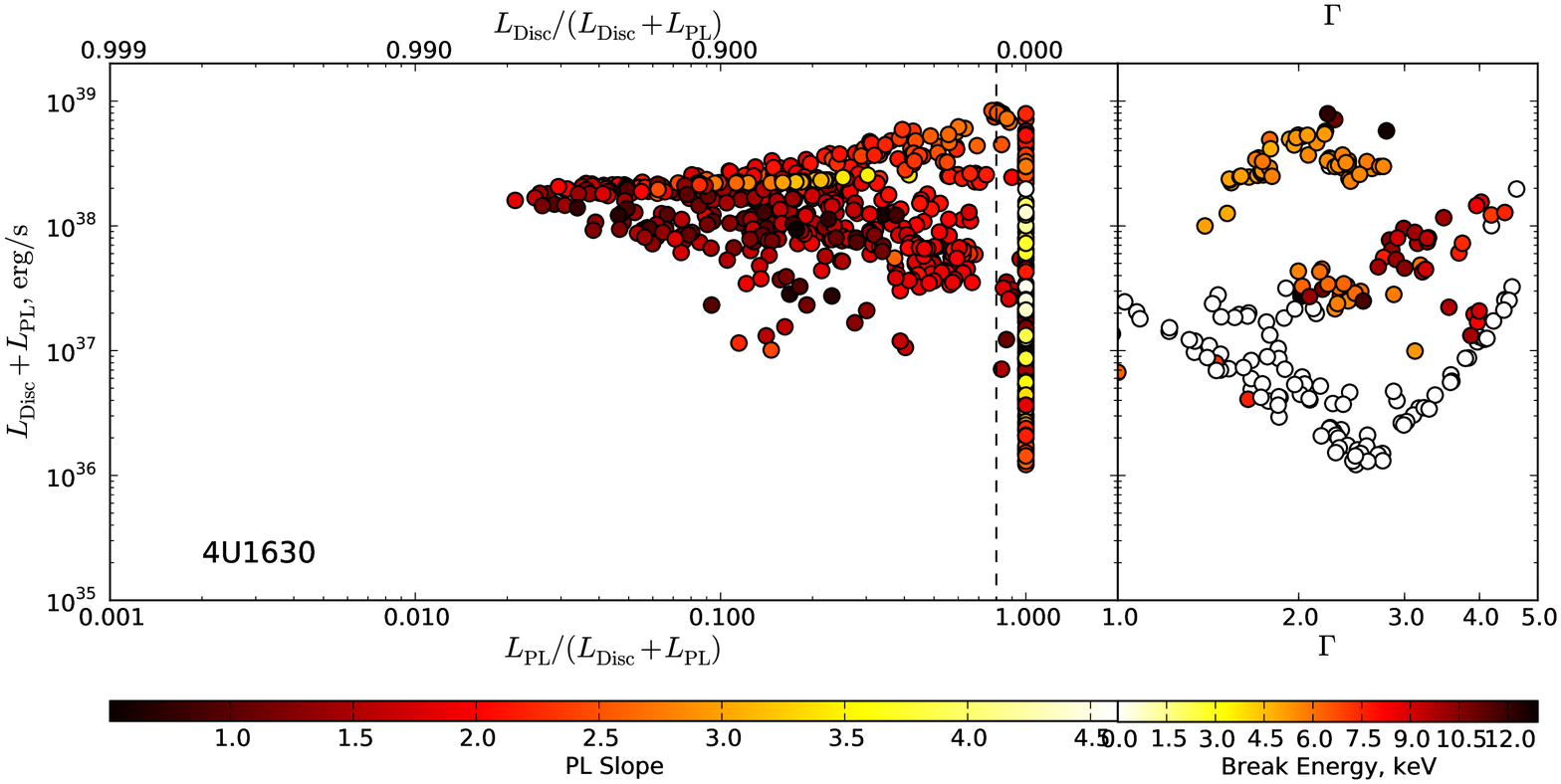}
\includegraphics[width=0.9\textwidth]{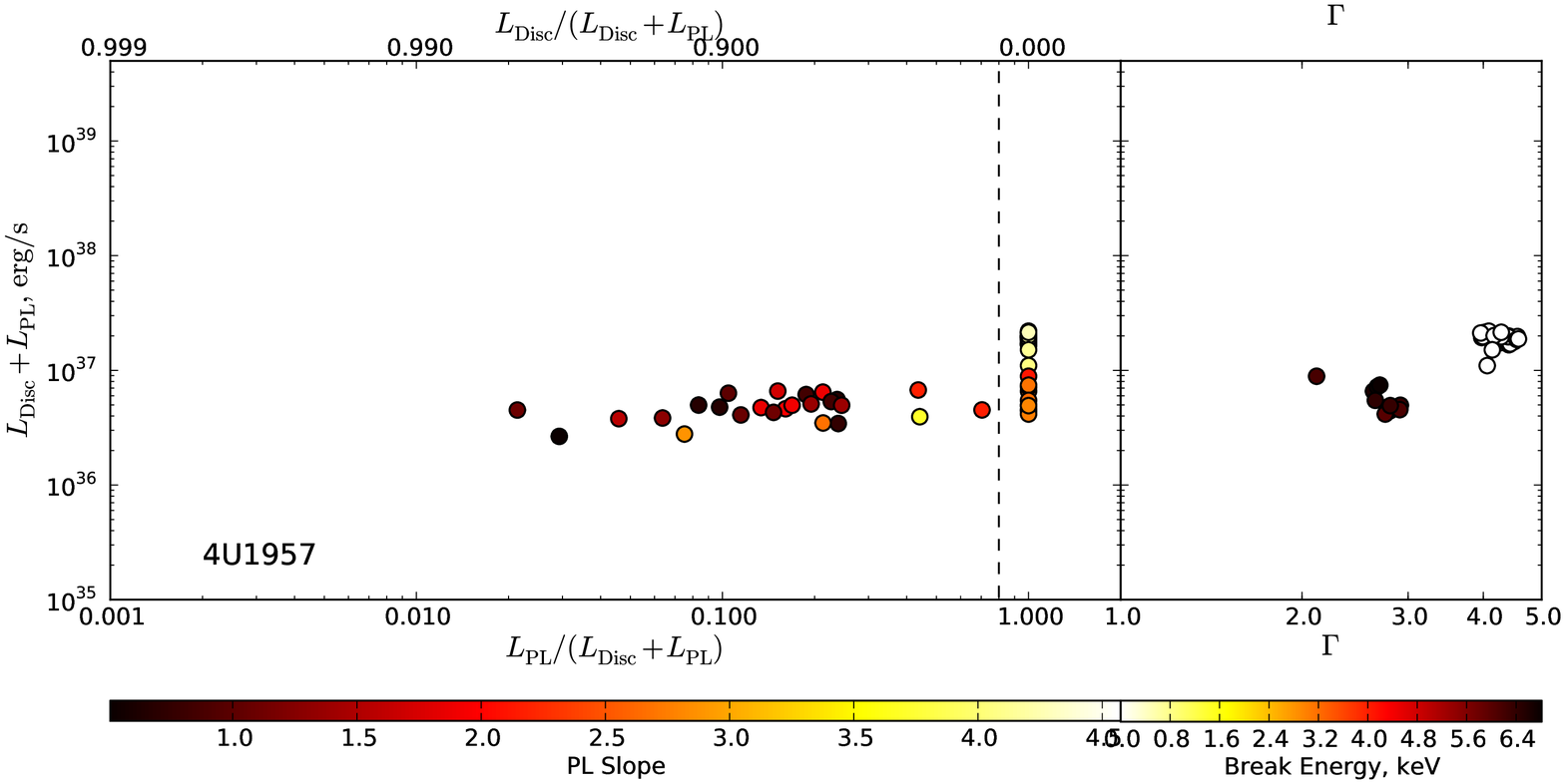}
\caption{\label{fig:DFLD_PLpanel}The DFLDs with the colour scale showing the powerlaw slope.
  Also shown in the side panel is the powerlaw slope (below the break
  where appropriate) for the
  observations with a disc fraction $<0.2$, along with the break
  energy for the broken powerlaw (where it occurs). {\scshape top}:
  4U~1543-47, {\scshape middle}: 4U~1630-47, {\scshape bottom}: 4U~1957+115.  }
\end{figure*}
\addtocounter{figure}{-1}
\begin{figure*}
\centering
\includegraphics[width=0.9\textwidth]{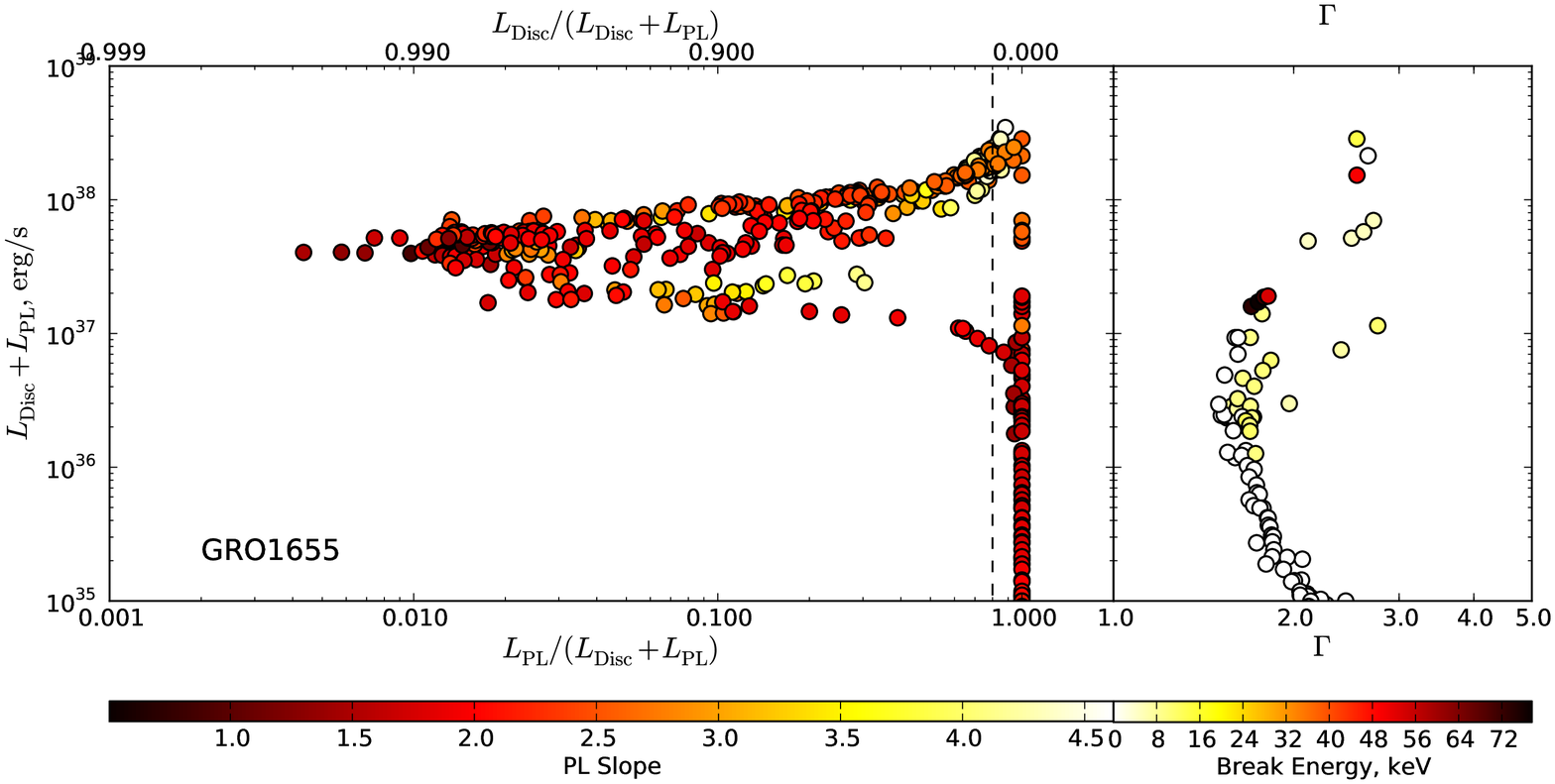}
\includegraphics[width=0.9\textwidth]{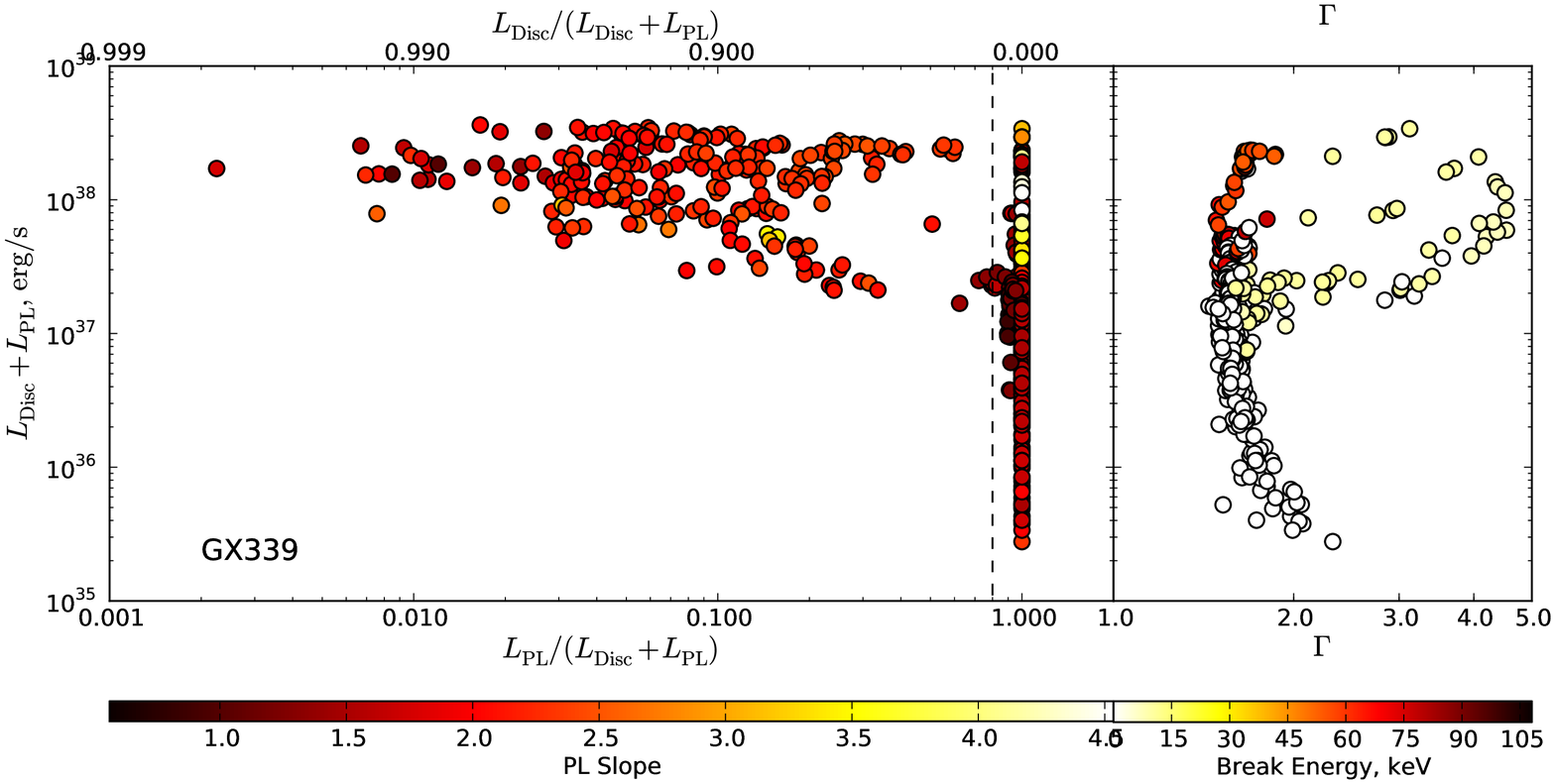}
\includegraphics[width=0.9\textwidth]{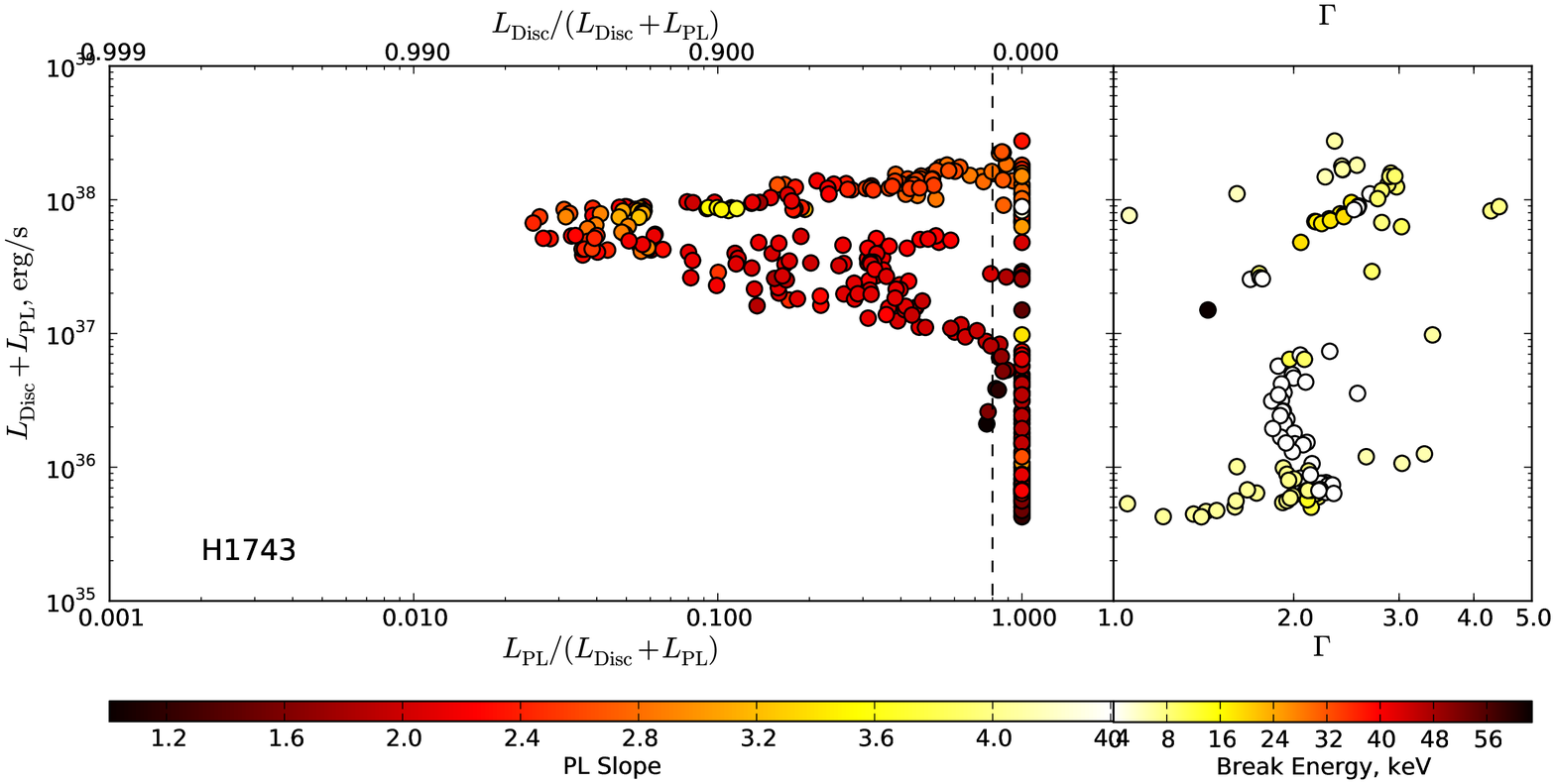}
\caption{(cont) {\scshape top}: GRO~J1655-40, {\scshape middle}: GX~339-4, {\scshape bottom}: H~1743-322.   }
\end{figure*}
\addtocounter{figure}{-1}
\begin{figure*}
\centering
\includegraphics[width=0.9\textwidth]{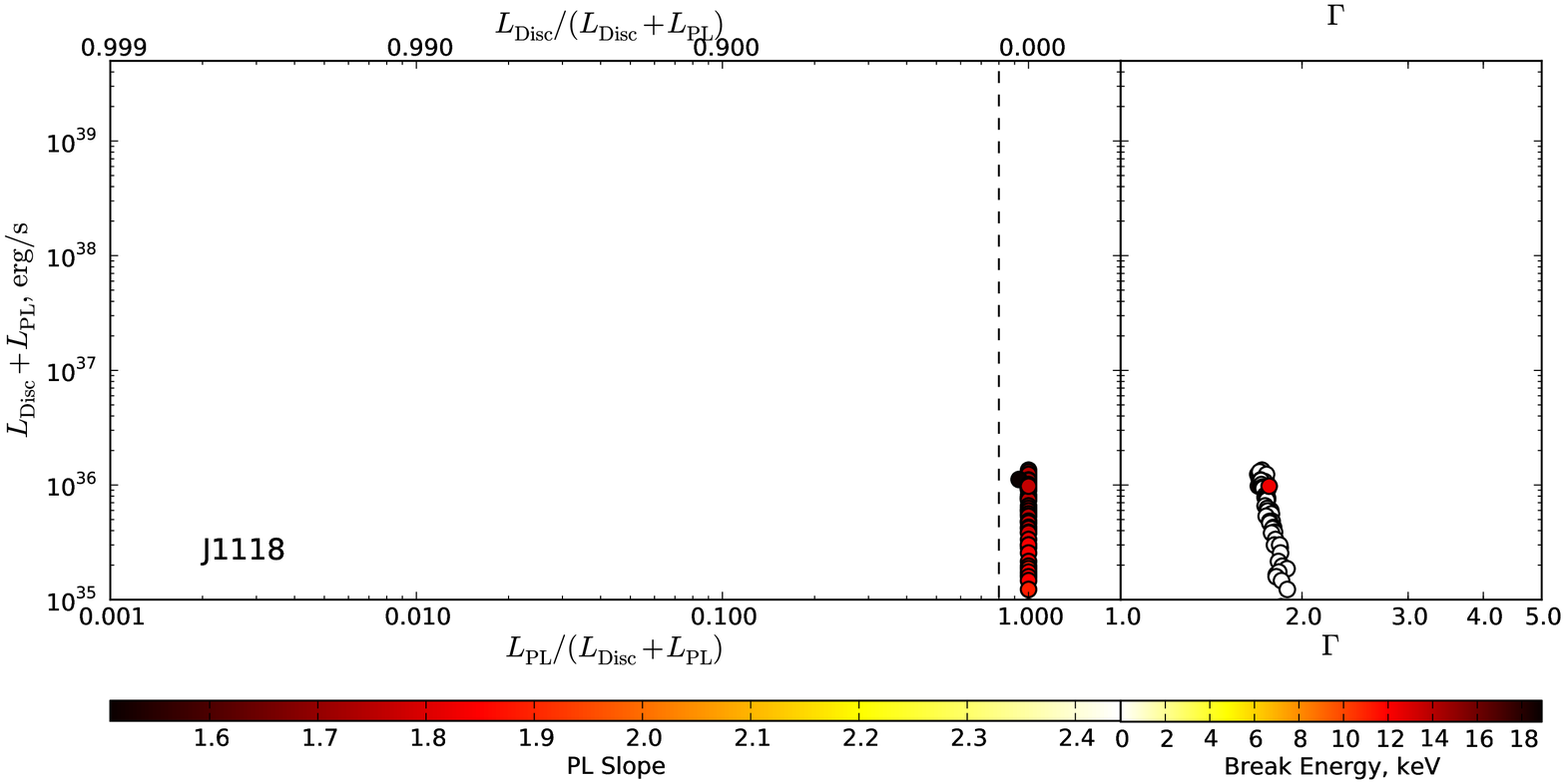}
\includegraphics[width=0.9\textwidth]{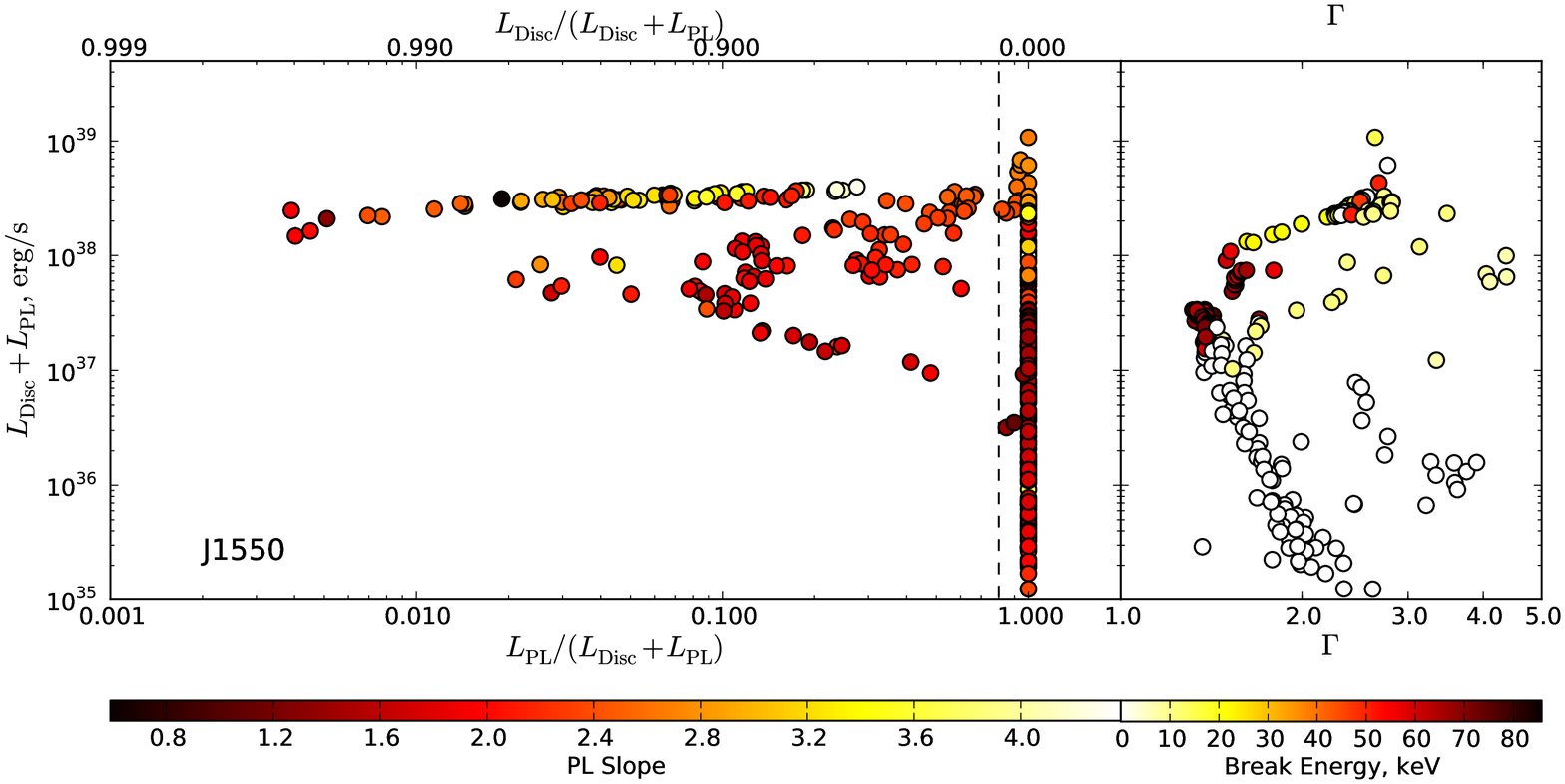}
\includegraphics[width=0.9\textwidth]{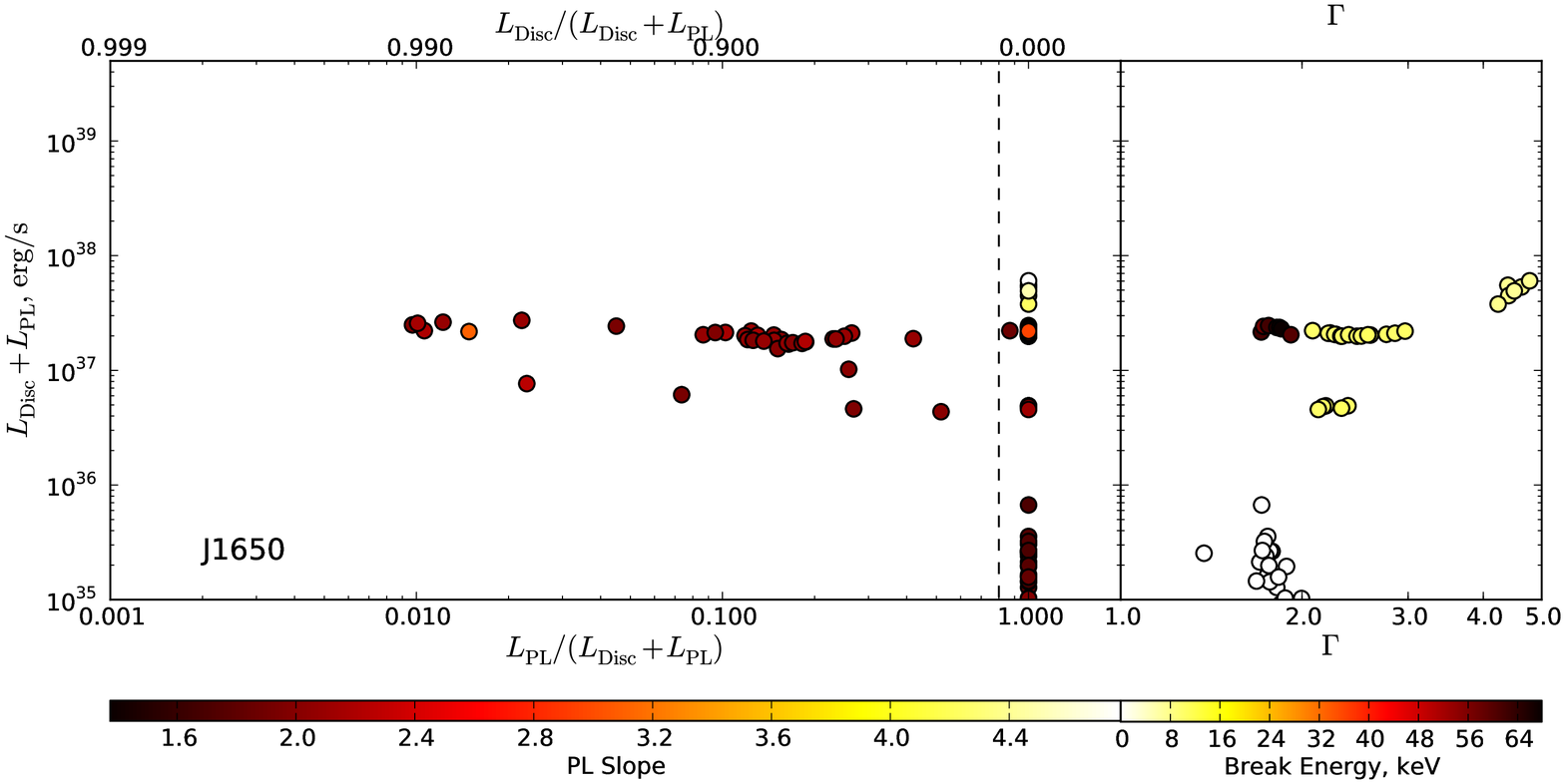}
\caption{(cont)  {\scshape top}: XTE~J1118+480, {\scshape middle}: XTE~J1550-564, {\scshape bottom}: XTE~J1650-500.  }
\end{figure*}
\addtocounter{figure}{-1}
\begin{figure*}
\centering
\includegraphics[width=0.9\textwidth]{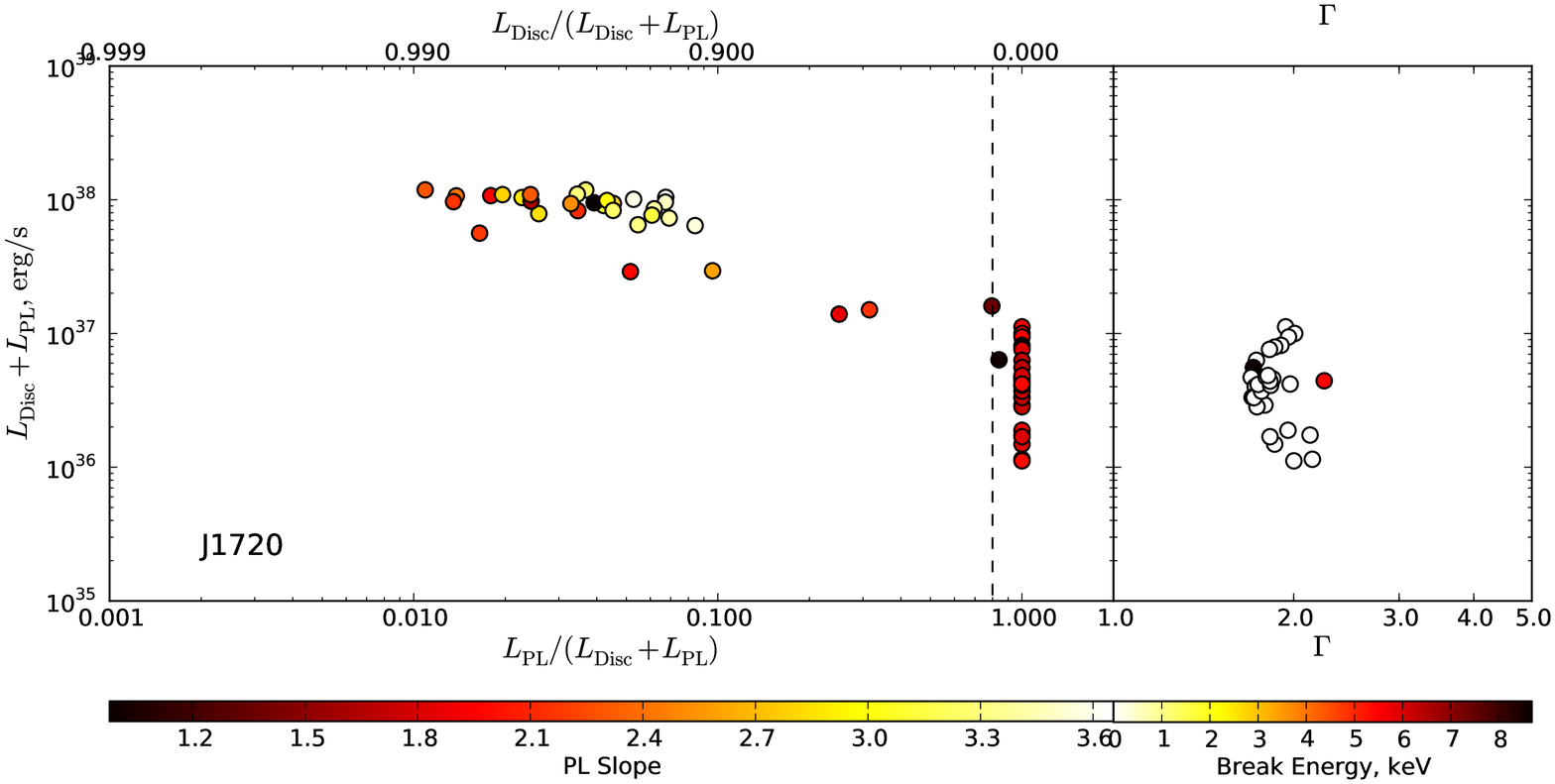}
\includegraphics[width=0.9\textwidth]{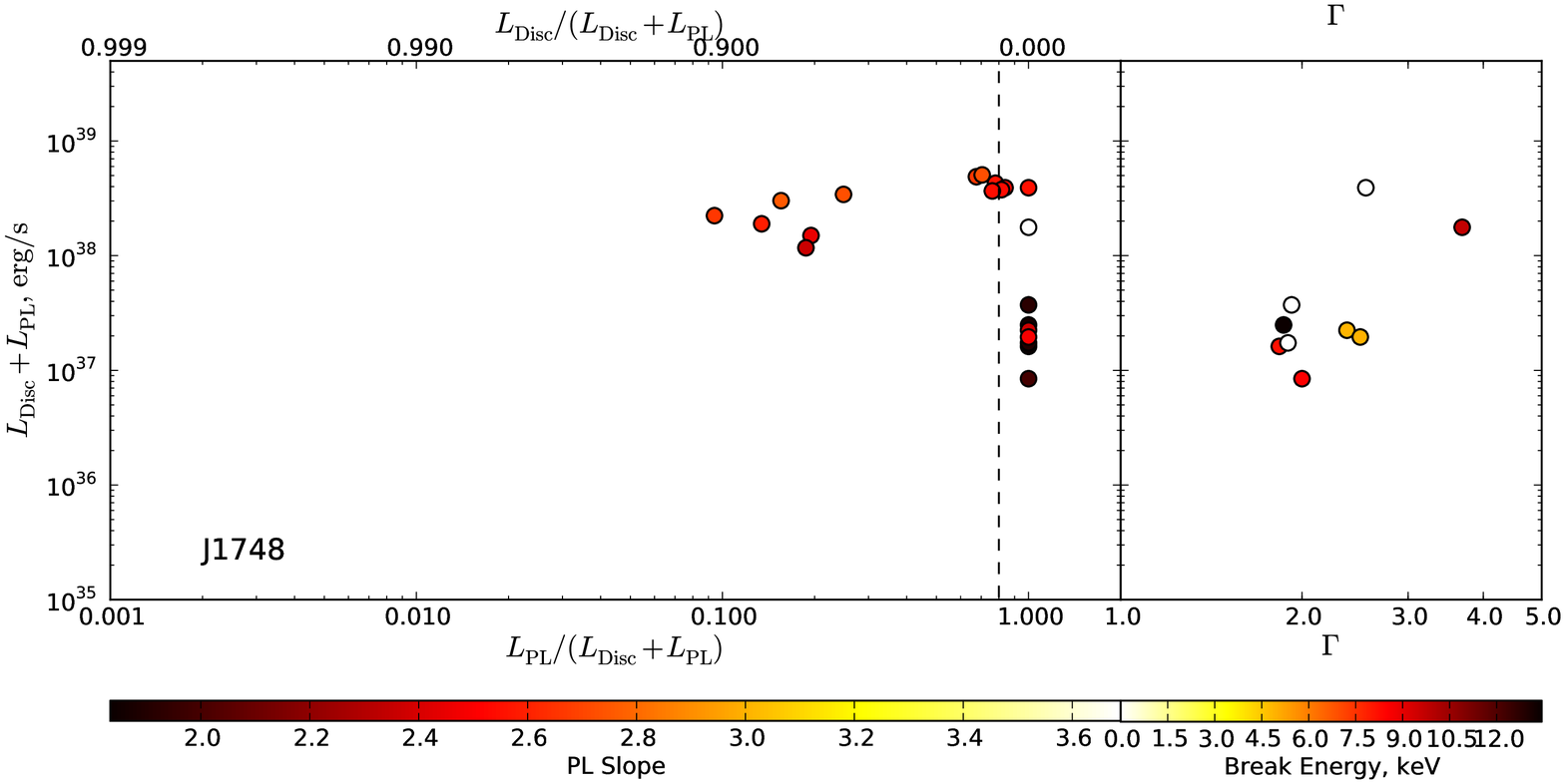}
\includegraphics[width=0.9\textwidth]{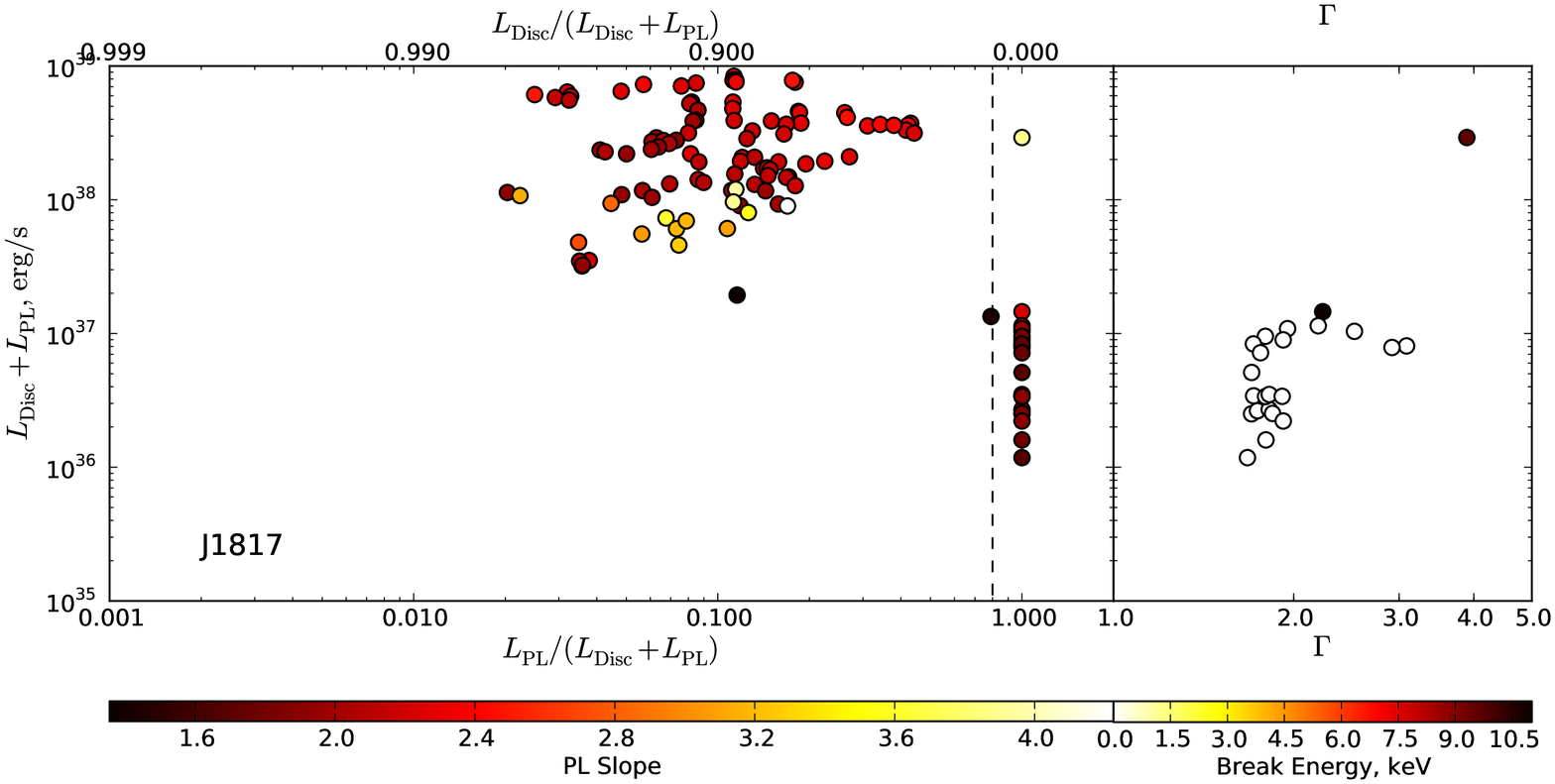}
\caption{(cont)  {\scshape top}: XTE~J1720-318, {\scshape middle}: XTE~J1748-288, {\scshape bottom}: XTE~J1817-330.  }
\end{figure*}
\addtocounter{figure}{-1}
\begin{figure*}
\centering
\includegraphics[width=0.9\textwidth]{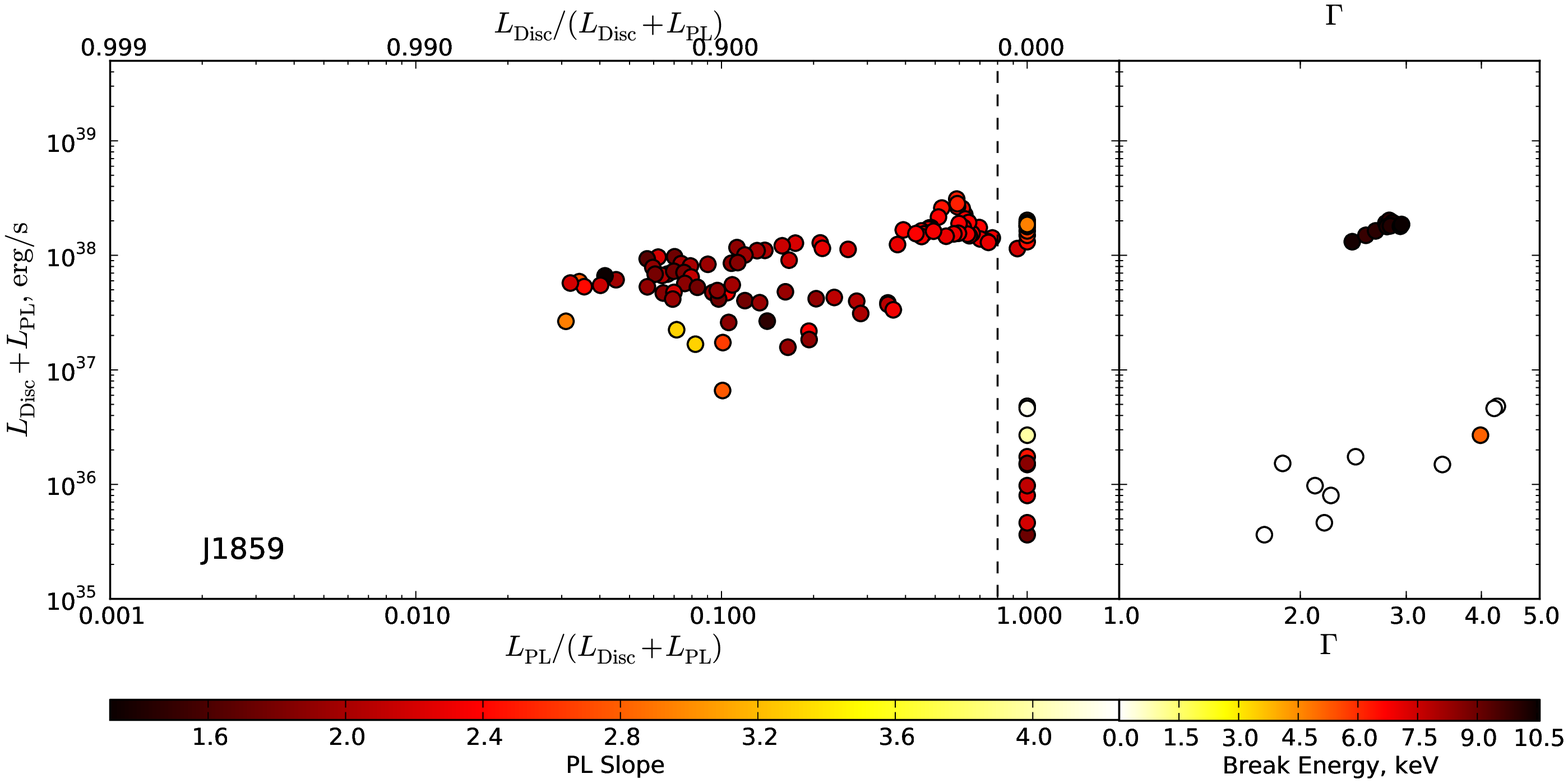}
\includegraphics[width=0.9\textwidth]{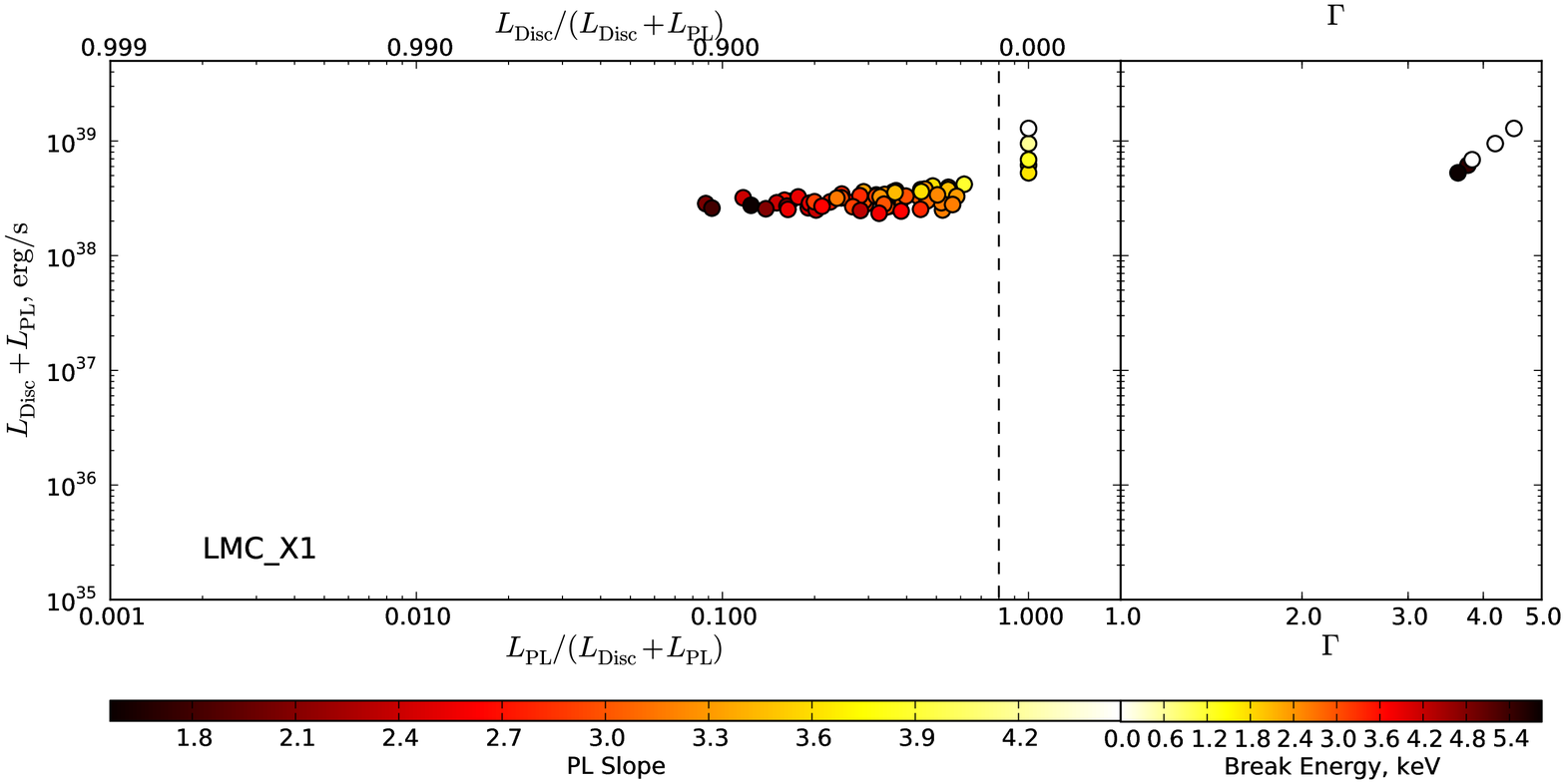}
\includegraphics[width=0.9\textwidth]{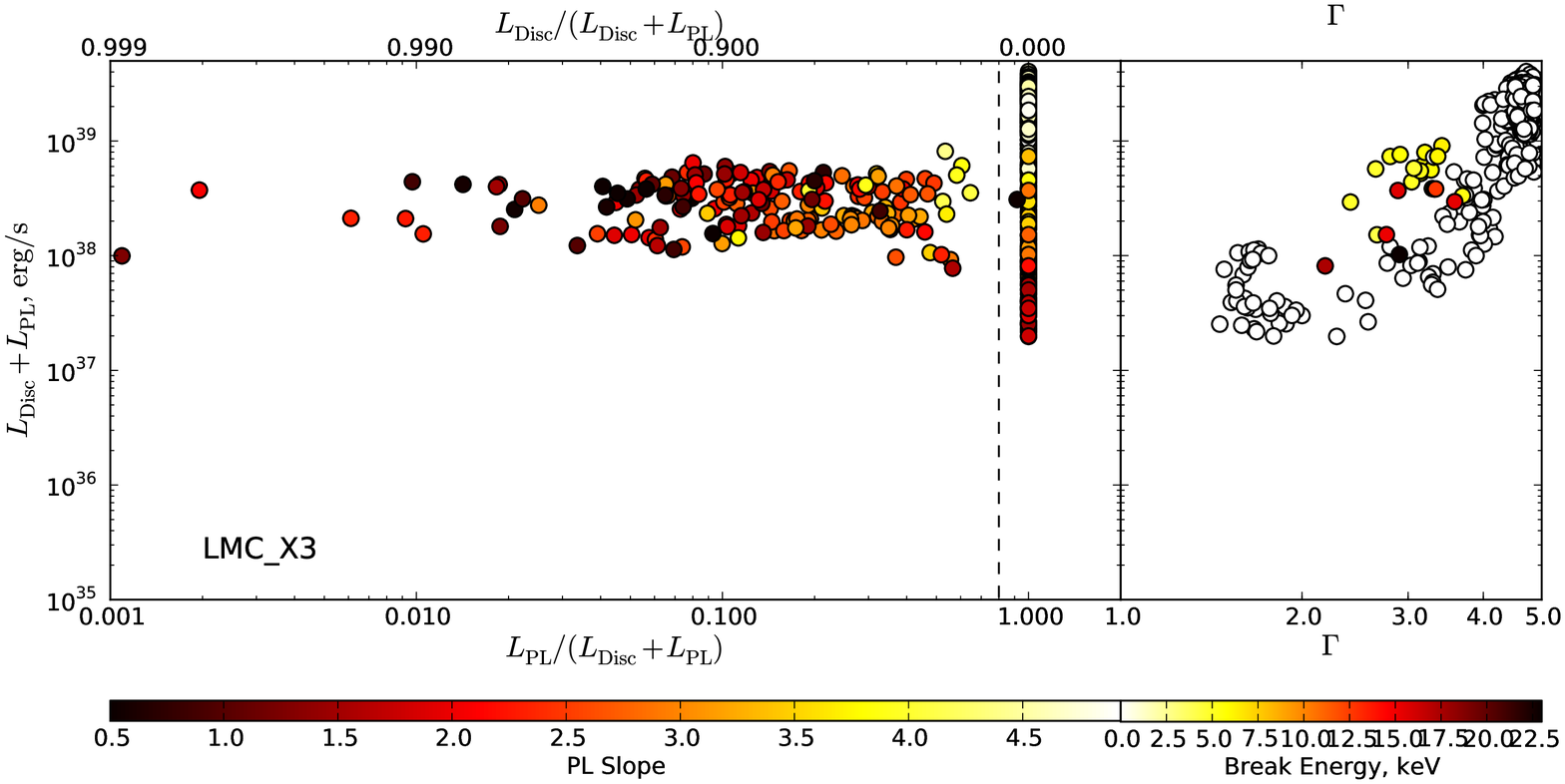}
\caption{(cont)  {\scshape top}: XTE~J1859+226, {\scshape middle}: LMC~X-1, {\scshape bottom}: LMC~X-3.  }
\end{figure*}
\addtocounter{figure}{-1}
\begin{figure*}
\centering
\includegraphics[width=0.9\textwidth]{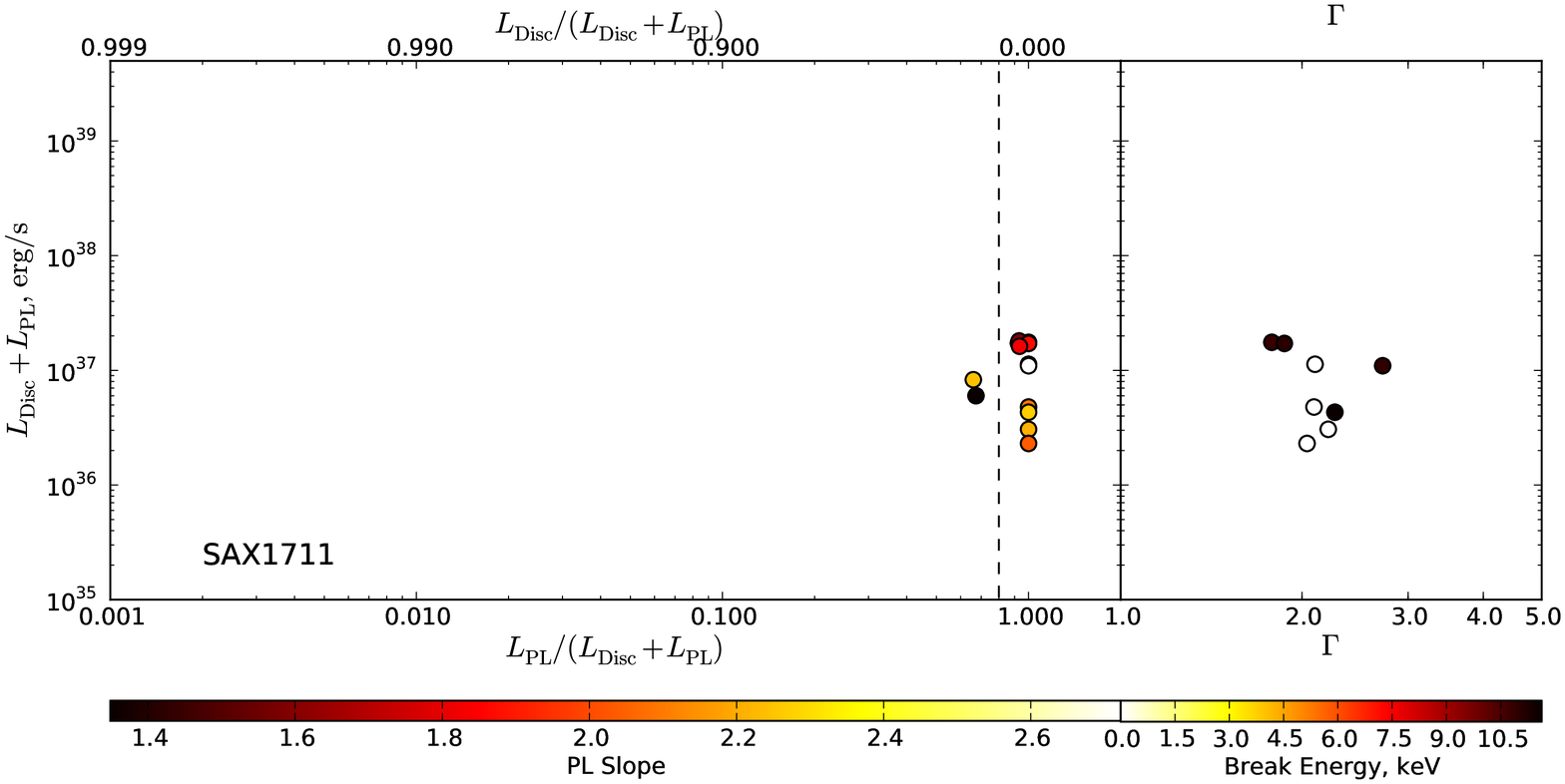}
\includegraphics[width=0.9\textwidth]{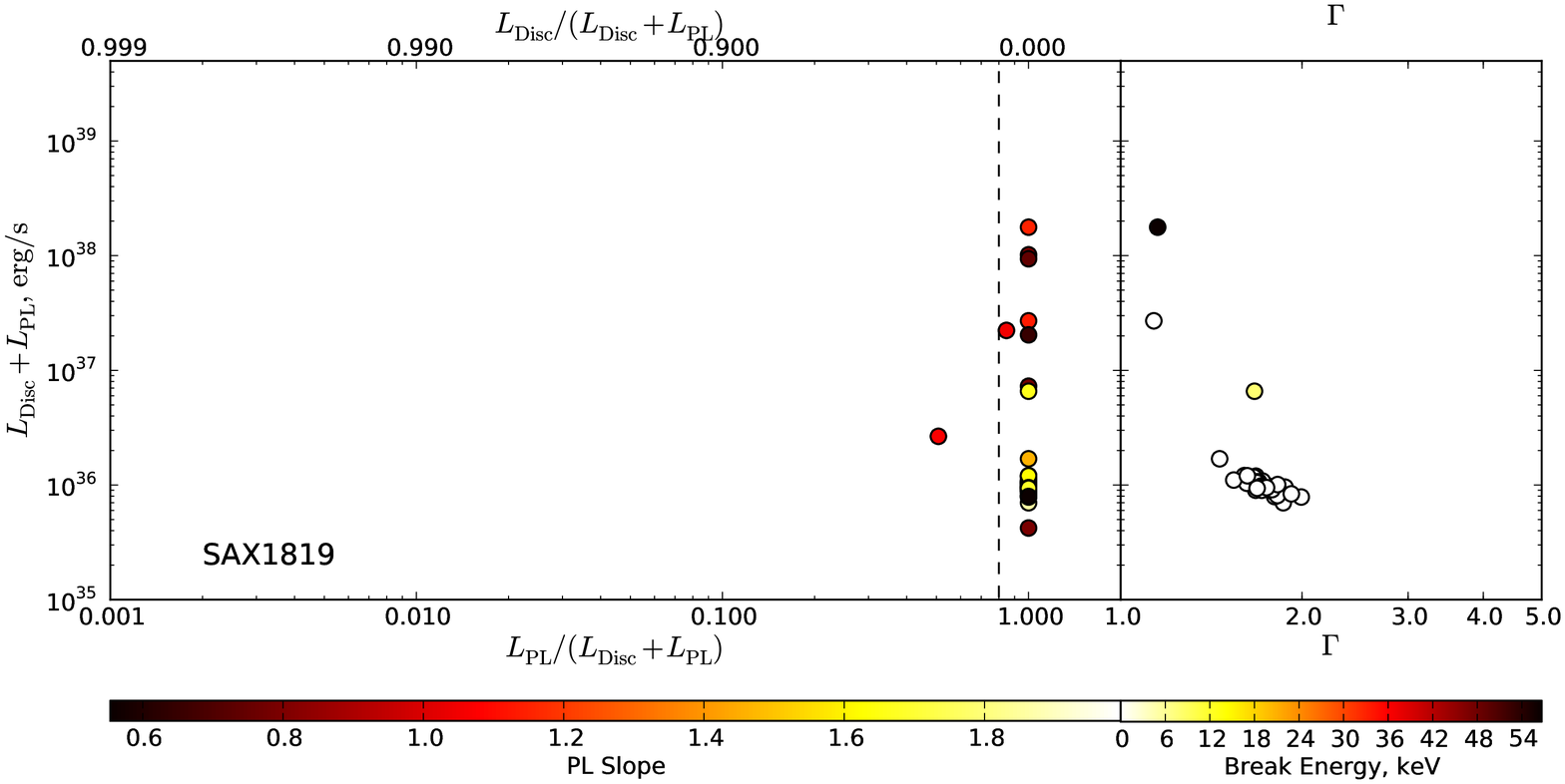}
\includegraphics[width=0.9\textwidth]{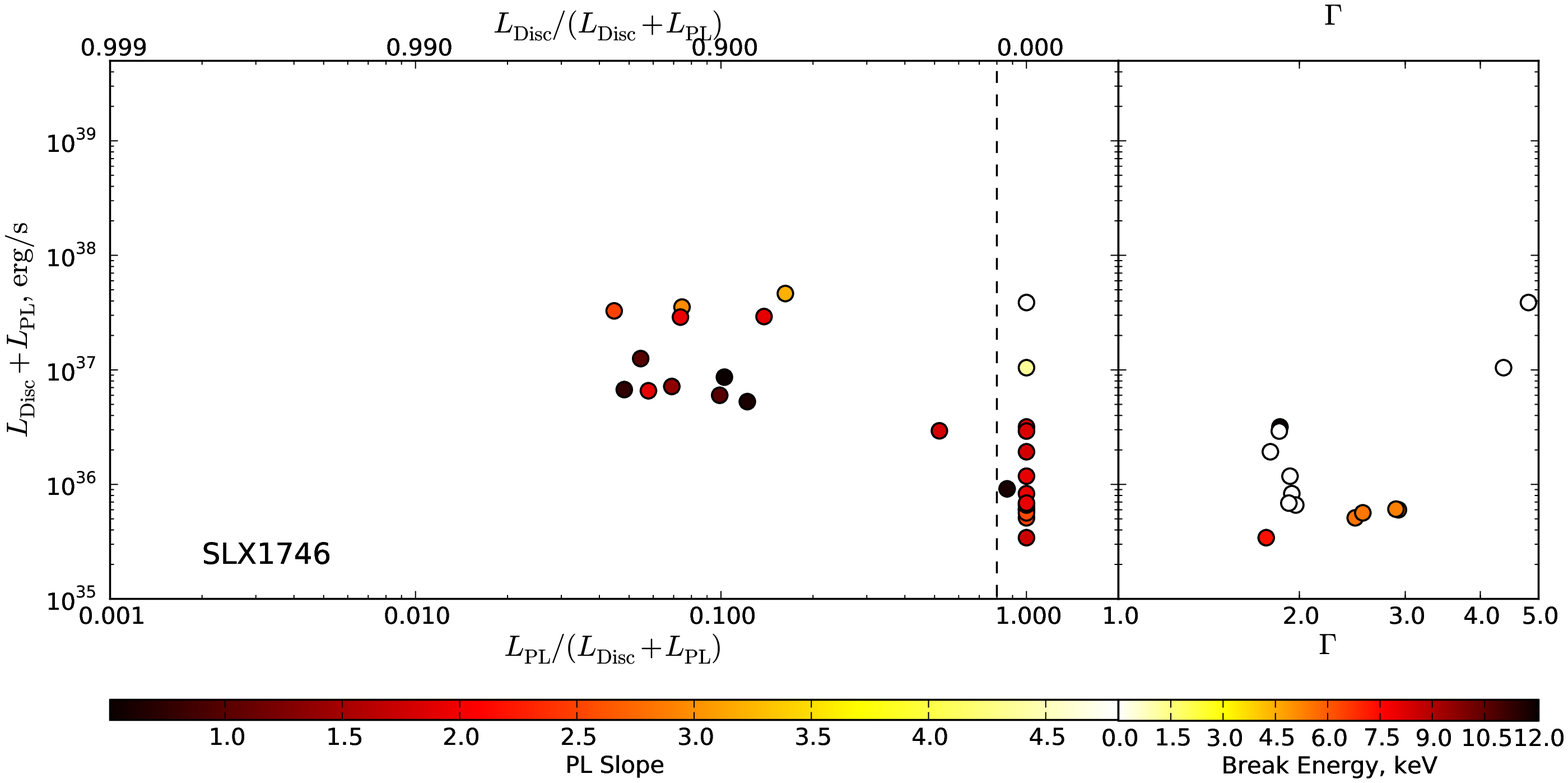}
\caption{(cont)  {\scshape top}: SAX~1711.6-3808, {\scshape middle}: SAX~1918.3-2525, {\scshape bottom}: SLX~1746-331.  }
\end{figure*}

\end{document}